\documentclass[preprint,prd,aps,amsmath,superscriptaddress,nofootinbib,tightenlines]{revtex4}

\usepackage[hypertex]{hyperref}
\usepackage{graphicx}
\usepackage{epstopdf}
\usepackage{bm}
\usepackage{epsfig}
\usepackage{xspace}

  \newcount\hour \newcount\minute
  \hour=\time \divide \hour by 60
  \minute=\time
  \newcommand{\mydate}{\ \today \ - \number\hour :\ifnum \minute<10 0\fi 
\number\minute}

\def\lsim{\mathrel{\raise.3ex\hbox{$<$\kern-.75em\lower1ex\hbox{$\sim$}}}}
\def\gsim{\mathrel{\raise.3ex\hbox{$>$\kern-.75em\lower1ex\hbox{$\sim$}}}}

\def\Dslash{D\!\!\!\!\slash}

\def\nslash{n\!\!\!\slash}
\def\bnslash{\bar n\!\!\!\slash}

\def\OMIT#1{}

\newcommand{\bCH}[2]{\overline\chi_{#1,#2}}

\newcommand{\nn}{\nonumber} 

\newcommand{\bn}{{\bar n}}
\newcommand{\bea}{\begin{eqnarray}}
\newcommand{\eea}{\end{eqnarray}}

\newcommand{\cP}{{\cal P}}

\newcommand{\mcdot}{\!\cdot\!}

\newcommand{\plus}{\ensuremath{\! + \!}}
\newcommand{\minus}{\ensuremath{\! - \!}}

\newcommand{\shat}{\hat{s}}

\begin{document}
\setlength\baselineskip{17pt}


\preprint{
 \vbox{ \hbox{MIT-CTP 3868} \hbox{CALT-68-2625} \hbox{MPP-2007-103}  
 \hbox{arXiv:0711.2079} }}

\title{
  Top Jets in the Peak Region: Factorization\\ Analysis with NLL Resummation }

\vspace*{1cm}

\author{Sean Fleming}
  \affiliation{Department of Physics, University of Arizona, Tucson, AZ 85721 
\footnote{Electronic address: fleming@physics.arizona.edu}}

\author{Andre H. Hoang}
  \affiliation{Max-Planck-Institut f\"ur Physik (Werner-Heisenberg-Institut), 
  F\"ohringer Ring 6, 80805 M\"unchen, Germany
\footnote{Electronic address: ahoang@mppmu.mpg.de}}

\author{Sonny Mantry}
  \affiliation{California Institute of Technology, Pasadena, CA 91125
    \footnote{Electronic address: mantry@theory.caltech.edu}}

\author{Iain W.~Stewart\vspace{0.4cm}}
  \affiliation{Department of Physics, Massachusetts Institute of Technology, 
Boston, MA 02139 \footnote{Electronic address: iains@mit.edu}\vspace*{0.5cm}}



\begin{abstract}
\vspace{0.4cm}

We consider top-quarks produced at large energy in $e^+e^-$ collisions, and
address the question of what top-mass can be measured from reconstruction.  The
production process is characterized by well separated scales: the center-of-mass
energy, $Q$, the top mass, $m$, the top decay width, $\Gamma_t$, and also
$\Lambda_{\rm QCD}$; scales which can be disentangled with effective theory
methods. In particular we show how the mass measurement depends on the way in
which soft radiation is treated, and that this can shift the mass peak by an
amount of order $Q\Lambda_{\rm QCD}/m$.  We sum large logs for $Q\gg m \gg
\Gamma_t> \Lambda_{\rm QCD}$ and demonstrate that the renormalization group ties
together the jet and soft interactions below the scale $m$.  Necessary
conditions for the invariant mass spectrum to be protected from large logs are
formulated.  Results for the cross-section are presented at next-to-leading
order with next-to-leading-log (NLL) resummation, for invariant masses in the
peak region and the tail region.  Using our results we also predict the thrust
distribution for massive quark jets at NLL order for large thrust. We
demonstrate that soft radiation can be precisely controlled using data on
massless jet production, and that in principle, a short distance mass parameter
can be measured using jets with precision better than $\Lambda_{\rm QCD}$.

\end{abstract}

\maketitle

\newpage

\tableofcontents

\newpage  
\section{Introduction}

The top quark is the heaviest known fermion of the Standard Model and couples strongly
to the Higgs sector. The most recent CDF and D\O~measurements obtained a top
mass, $m_t = 170.9\pm 1.8\,{\rm GeV}$~\cite{unknown:2007bxa}, with $\sim 1\%$
uncertainty. For the standard model a precise top mass determination is
important for precision electroweak observables which test the theory at the
quantum level, and which constrain extensions of the theory such as 
supersymmetry.  In Ref.~\cite{Fleming:2007qr} we derived a factorization theorem
for the invariant mass distribution of high energy top jets for $e^+ e^- \to
t\bar{t}$, which allows in principle a determination of $m_t$ with uncertainty
better than $\Lambda_{\rm QCD}$. Such accuracy is possible because the factorization
theorem separates the perturbative and non-perturbative contributions in terms
of field theory Wilson coefficients and matrix elements. A virtue of our
approach is that the non-perturbative matrix elements are universal and in some
cases are straightforward to extract from other processes. In addition the
factorization theorem provides a unique prescription for determining the Wilson
coefficients and perturbative matrix elements at any order in the $\alpha_s$
expansion.  This level of control allows us to make stable predictions for the
invariant mass distribution in terms of a short-distance top quark mass, which
is not limited in precision by $\Lambda_{\rm QCD}$.

Determining the top mass with jet reconstruction methods in general faces issues
such as i) defining an observable that is sensitive to the top mass, ii) soft
gluon interactions and color reconnection, iii) uncertainties from higher order
perturbative corrections, iv) the large top quark width $\Gamma_t^{\rm SM}\simeq
1.4\,{\rm GeV}$, and other finite lifetime effects, v) final state radiation,
vi) initial state radiation, vii) treatment of beam remnants, viii) underlying
events, and ix) parton distributions.  In Ref.~\cite{Fleming:2007qr} we
addressed the definition of a suitable top quark mass $m$ and issues i) through
v) in the framework of electron-positron collisions at high energies $Q\gg m$,
where $Q$ is the center of mass energy and $m$ is the top mass.\footnote{We will
  use $m$ for the top mass when it is not necessary to specify the precise
  scheme which defines this parameter.}  The analysis is suitable for a future
linear collider. Issues vii) through ix) are avoided by treating the $e^+e^-$
initial state, but are important in a hadron collider environment like the
Tevatron or LHC.  Issue vi) is also greatly simplified in $e^+e^-$ annihilation,
since the inclusion of initial state photon radiation mainly shifts $Q$ and thus
has very little impact on our analysis.

Our analysis of top jets uses effective theory techniques to exploit the
hierarchy of scales $Q\gg m \gg \Gamma \gtrsim \Lambda_{\rm QCD}$, and separate
dynamical fluctuations.  This hierarchy provides a systematic power counting in
$m/Q$ and $\Gamma/m$, and gives a clear interpretation to elements in the
factorization theorem.  In Ref~\cite{Fleming:2007qr} we focused on developing
the formalism and describing the main conceptual points in the factorization
theorem for the invariant mass distribution in the peak region.  The same
formalism also yields a factorization theorem for the invariant mass
distribution in the tail region above the peak. Here we use models for the soft
function that are consistent for both the peak and tail regions, and carry out
detailed calculations of perturbative quantities in the factorization theorem.
We verify that the matching conditions which define the Wilson coefficients at
the scales $Q$ and $m$ are infrared safe, compute one-loop perturbative
corrections to the matrix elements, and carry out the next-to-leading-log
renormalization group summation of large logs. For the peak region these are
logs between the scales $Q$, $m$, $\Gamma$, and $\Lambda_{\rm QCD}$, while away
from the peak they are between $Q^2$, $m^2$, and the variables $M^2_{t}-m_t^2$
and $M^2_{\bar t}-m_t^2$ described below.

As an observable sensitive to the top mass, we considered in
Ref.~\cite{Fleming:2007qr} the double differential invariant mass distribution
in the peak region around the top resonance:
\begin{eqnarray}
\label{obs1}
\frac{d^2\sigma}{dM^2_t\>dM^2_{\bar{t}}} \,,
  \qquad\qquad M_{t,\bar{t}}^2 - m^2  \sim m\, \Gamma \ll m^2 \,, 
\end{eqnarray}
where
\begin{align}
 M_t^2 &= \Big( \sum_{i\in X_t} p_i^\mu\Big)^2 \,,
  & M_{\bar t}^2 & = \Big( \sum_{i\in X_{\bar t}} p_i^\mu\Big)^2 \,.
\end{align}
Here $X_t$ and $X_{\bar t}$ represent a prescription to associate final state
hadronic four momenta to top and antitop invariant masses respectively. For
simplicity we call $X_{t,\bar t}$ the top and antitop jets, and $M_{t,\bar t}$
the invariant mass of the top and antitop jets respectively.  The distribution
in Eq.~(\ref{obs1}) has a width $\Gamma\sim \Gamma_t + Q\Lambda_{\rm QCD}/m$
which can be larger than the top quark width $\Gamma_t$. The restriction
$M_{t,\bar{t}}^2 - m^2 \sim m\, \Gamma \ll m^2$ defines the peak region, which
is the region most sensitive to the top quark mass $m$. Here the dynamics is
characterized by energy deposits contained predominantly in two back-to-back
regions of the detector with opening angles of order $m/Q$ associated with the
energetic jets or leptons coming from the top and antitop decays, plus collinear
radiation. The region between the top decay jets is populated by soft particles,
whose momentum is assigned to one of $M_t^2$ or $M_{\bar t}^2$.  The tail region
is defined by invariant masses starting just past the peak where the
cross-section begins to fall off rapidly, namely where $m^2 \gg M_{t,\bar{t}}^2
- m^2$ and either $M_{t,\bar{t}}^2 - m^2\gtrsim m\, \Gamma$ or $M_{t,\bar{t}}^2 -
m^2\gg m\, \Gamma$.  Farther out, when $M_{t,\bar{t}}^2 - m^2\sim m^2$, we have
an ultra-tail region where the cross-section is very small. We do not consider
the region where $M_{t,\bar{t}}^2\sim Qm$. The observable in Eq.~(\ref{obs1})
in the peak and tail regions is the main focus of our analysis.  We also briefly
consider the cross-section in the ultra-tail region.

The result for the double differential cross-section in the peak region to all
orders in $\alpha_s$ is given by~\cite{Fleming:2007qr}
\begin{align} \label{FactThm}
  \frac{d\sigma}{ dM^2_t\, dM^2_{\bar t}} &= 
  \sigma_0 \: H_Q(Q\OMIT{,\mu_Q},\mu_m) 
  H_m\Big(m_J,\frac{Q}{m_J},\mu_m,\mu\Big)\!
  \nn\\
 &\times \int\! d\ell^+ d\ell^- 
   B_+\Big(\hat s_t- \frac{Q\ell^+}{m_J},\Gamma_t,\mu\Big)\:
   B_-\Big(\hat s_{\bar t}-\frac{Q\ell^-}{m_J},\Gamma_t,\mu\Big)  
   S(\ell^+,\ell^-,\mu) 
  \nn\\
  & + {\cal O}\Big(\frac{m\alpha_s(m)}{Q} \Big) 
    + {\cal O}\Big(\frac{m^2}{Q^2} \Big)
     + {\cal O}\Big(\frac{\Gamma_t}{m} \Big)
     + {\cal O}\Big(\frac{s_t,s_{\bar t}}{m^2} \Big) \,,
\end{align}
where, as indicated, power corrections are suppressed by $\alpha_s m/Q$,
$m^2/Q^2$, $\Gamma_t/m$, or $s_{t,\bar t}/m^2$. Here $m_J$ is the short-distance top quark
mass we wish to measure, and for convenience we have defined 
\begin{eqnarray}
  \hat s_t =\frac{s_t}{m_J} = \frac{M_t^2-m_J^2}{m_J} \,,\qquad \qquad 
  \hat s_{\bar t} =\frac{s_{\bar t}}{m_J}  = \frac{M_{\bar t}^2-m_J^2}{m_J}  \,,
\end{eqnarray}
where $\hat s_{t,\bar t}\sim \Gamma$ are of natural size in the peak region.  In
Eq.~(\ref{FactThm}) the normalization factor $\sigma _0$ is the total Born-level
cross-section, the $H_Q$ and $H_m$ are perturbative coefficients describing hard
effects at the scales $Q$ and $m_J$, $B_\pm$ are perturbative jet functions that
describe the evolution and decay of the the top and antitop close to the mass
shell, and $S$ is a nonperturbative soft function describing the soft radiation
between the jets. To sum large logs $B_\pm$ and $S$ will be evolved to distinct
renormalization scales $\mu$, as we discuss in section~\ref{sec:logs} below. For
the tail region Eq.~(\ref{FactThm}) becomes
\begin{align}
    \frac{d\sigma}{ dM^2_t\, dM^2_{\bar t}} &= 
  \sigma_0 \: H_Q\: H_m\: B_+\otimes B_-\otimes S_{\rm part} 
     + {\cal O}\Big(\frac{\Lambda_{\rm QCD} Q}{s_{t,\bar t}} \Big)
   + {\cal O}\Big(\frac{m\alpha_s(m)}{Q},\frac{m^2}{Q^2},\frac{\Gamma_t}{m} \Big) 
   \,,
\end{align}
\begin{figure}
  \centerline{ 
   \includegraphics[width=12cm]{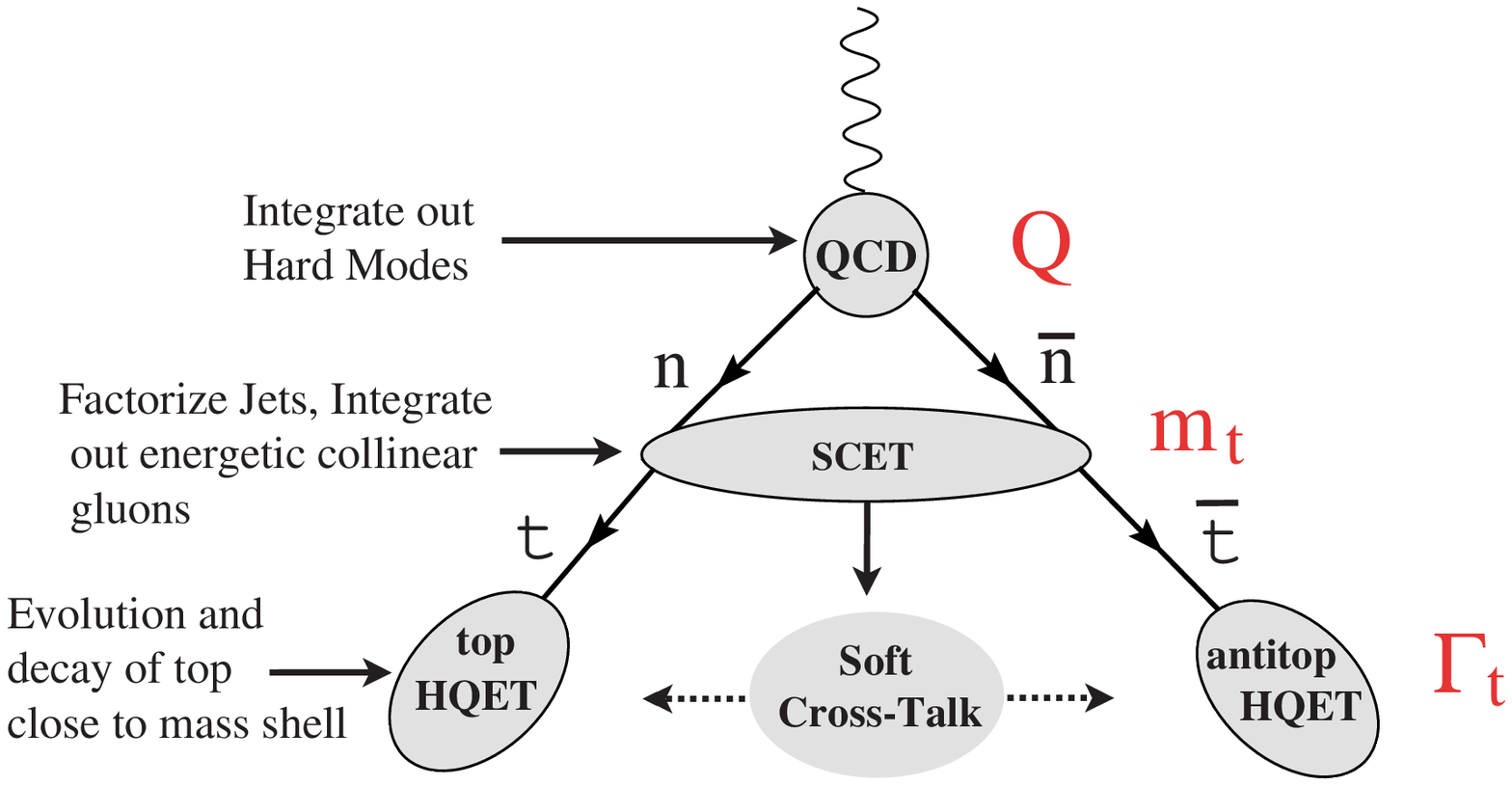}  
  } 
\caption{Sequence of effective field theories used to compute the invariant
mass distribution. }
\label{fig:theoryI}
\end{figure}
so the only changes are that the soft-function $S=S_{\rm
  part}(\ell^+,\ell^-,\mu)$ becomes calculable, and we have an additional ${\cal
  O}(\Lambda_{\rm QCD} Q/s_{t,\bar t})$ nonperturbative correction from the
power expansion of the soft-function which we will include in our analysis.  The
result in Eq.~(\ref{FactThm}) was derived by matching QCD onto the Soft
Collinear Effective
Theory(SCET)~\cite{Bauer:2000ew,Bauer:2000yr,Bauer:2001yt,Bauer:2002nz,Bauer:2001ct}
which in turn was matched onto Heavy Quark Effective
Theory(HQET)~\cite{Manohar:2000dt,Isgur:1989vq,Isgur:1989ed,Grinstein:1990mj,
  Eichten:1989zv,Georgi:1990um} generalized for unstable particles
~\cite{Fadin:1987wz,Beneke:2003xh,Beneke:2004km,Hoang:2004tg} as illustrated in
Fig.~\ref{fig:theoryI}.  The decoupling of perturbative and nonperturbative
effects into the $B_\pm$ jet functions and the $S$ soft function was achieved
through a factorization theorem in SCET and HQET, aspects of which are similar
to factorization for massless event
shapes~\cite{Korchemsky:1994is,Korchemsky:1999kt,Bauer:2002ie,Bauer:2003di}.
The result in Eq.~(\ref{FactThm}) is an event shape distribution for massive
particles, and can be used to determine common event shapes such as thrust or
jet-mass distributions. Note that a subset of our results can also be used to
match results with the event shape cross-sections for massless jets, namely by
using our SCET ultratail cross-section and taking the limit $m\to 0$. 

In general the functions $B_{\pm}$ and $S$ depend on exactly how $M_t$ and
$M_{\bar t}$, or equivalently $X_t$ and $X_{\bar t}$, are defined.  The
factorization theorem in Eq.~(\ref{FactThm}) holds in the form shown when all
the soft radiation is assigned to either $X_t$ or $X_{\bar t}$, and the
probability of radiation being assigned to $X_t$ or $X_{\bar t}$ increases to
unity when we approach the top or antitop direction~\cite{Fleming:2007qr}.
Finally, the definition should be inclusive in the hard jets and leptons from
the top decay.  One possibility for defining $M_{t,\bar t}^2$ in
Eq.~(\ref{FactThm}) is a hemisphere mass definition, where $X_t, X_{\bar{t}}$
contain everything to the left or right of the plane perpendicular to the thrust
axis.  In this case our $S$ is identical to the soft function of
Refs.~\cite{Korchemsky:1999kt,Korchemsky:2000kp,Bauer:2003di} that appears in
the factorization theorem for massless event shapes in the dijet region.  For
studies of soft-functions in massless event shapes see
Refs.~\cite{Korchemsky:1994is,Korchemsky:1998ev,Korchemsky:1999kt,Korchemsky:2000kp,Berger:2003iw,Bauer:2002ie,Bauer:2003di,Lee:2006nr,Lee:2006fn,Berger:2003gr,Berger:2002ig,Belitsky:2001ij}.
The $B_\pm$ are inclusive in the jets from the top decay and collinear radiation
and can be defined by forward matrix elements~\cite{Fleming:2007qr}.  Other
definitions to associate all soft radiation to the top and antitop jets can be
used which modifies the required $S$ function, but for the class of masses
defined above leaves $B_\pm$ unchanged.

The use of a short-distance mass definition in the $B_\pm$ jet function and a
short-distance gap parameter in the soft function $S$~\cite{Hoang:2007vb} are
crucial for obtaining predictions that remain stable when higher order
perturbative corrections are included. In Ref.~\cite{Fleming:2007qr} we showed
that suitable mass schemes for reconstruction measurements can only differ from
the pole mass by an amount $\delta m\sim \Gamma_t \alpha_s$, and we proposed a
jet-mass scheme which satisfies this criteria. We will refine the criteria for
this jet-mass scheme here. In Eq.~(\ref{FactThm}) the jet-mass, $m_J$, only
appears in the calculable Wilson coefficients and jet functions $B_\pm$. The
greatest sensitivity to $m_J$ is in $B_\pm$.  Through these jet functions, $m_J$
influences the spectrum of the mass distribution in the peak region.  The
spectral distribution and location of the peak are also affected by
nonperturbative effects in the soft function $S$. In Ref.~\cite{Hoang:2007vb} a
gap parameter scheme based on moments of the partonic soft function was devised
to avoid perturbative ambiguities in the definition of the partonic endpoint
where the variables $\ell^\pm$ in Eq.~(\ref{FactThm}) approach zero. Methods for
using Eq.~(\ref{FactThm}) to extract $m_J$ are discussed in detail in
Ref.~\cite{Fleming:2007qr}.

In this paper we determine the functions $H_Q$, $H_m$, $B_\pm$ at one-loop order
in $\alpha_s$, and carry out the summation of large logs between the scales
$Q\gg m\gg \Gamma$ in Eq.~(\ref{FactThm}).  The derivation of results for the
top jet-mass scheme are discussed in detail.  We also show that there are
constraints on the allowed soft functions, and implement a consistent method to
include perturbative corrections in $S$. In our numerical analysis we extend the
work in Ref.~\cite{Fleming:2007qr} to one-loop order, including the summation of
the next-to-leading order logarithms using renormalization group (RG) evolution
in effective field theories. Our analysis of the $t\bar t$ jet cross section at
this order includes both invariant masses in the peak region and the tail region
above the peak, and the final results are analytic up to integration over the
soft-function model.

For massless jets there has been a lot of work done on the program of resumming
logs in event shape
variables~\cite{Catani:1991kz,Catani:1992ua,Catani:1991kz,Burby:2001uz,Krauss:2003cr,Gardi:2003iv,Gardi:2001ny,Korchemsky:1993uz,Dokshitzer:1998kz,Catani:1998sf,Berger:2002sv,Berger:2001ns,Banfi:2001pb,Banfi:2001bz,Trott:2006bk}.
In this paper we do not use the traditional approach to resummation, but rather
an approach that sums the same large logs based on the renormalization of
operators in effective field theories, including HQET and
SCET~\cite{Bauer:2000ew,Bauer:2000yr}. The effective theory resummation
technique has the advantage of being free of Landau-pole
singularities~\cite{Manohar:2003vb,Becher:2006mr}, since it only depends on the
evaluation of anomalous dimensions at perturbative scales. This technique can
also be extended in a straightforward manner to arbitrary orders, ${\rm N}^k{\rm
  LL}$ in the resummation~\cite{Neubert:2005nt,Becher:2007ty}. A recent
application of the SCET technique is the resummation for thrust in $e^+e^-$ to
massless jets at NLL order~\cite{Schwartz:2007ib}.

In our log-summation there is an important distinction between large logs which
affect the overall cross section normalization, and large logs that change the
shape of the distribution in $M_{t,\bar t}^2$.  In predicting the normalization
in the dijet region we must sum up a series of double Sudakov logarithms that
occur for $Q\gg m$ and for $m\gg \Gamma$.  However, it turns out that the same
is not true for logs affecting the shape of the invariant mass spectrum.  As we
discuss in detail, the form of the spectrum is protected from large logs below
the scale $Q$ until we reach the fundamental low energy scale governing the
dynamics of either the soft or jet functions.  This conclusion is not affected
by the mass threshold at $m$, and is valid to all orders in perturbation theory
(ie.  for both leading and subleading series of logarithms).  In order for this
cancellation to occur it is important that the invariant mass definition
includes soft radiation at wide angles.  The hemisphere mass definition of $M_t$
and $M_{\bar t}$, as well as other definitions which associate wide angle soft
radiation to both $X_t$ and $X_{\bar t}$, are in this category. In the effective
field theory this protection against the appearance of shape changing large logs
is described by a set of ``consistency conditions''. From our analysis we find
that the only shape changing large logs occur between the low energy scale $\mu
\sim Q\Lambda/m +\Gamma_t$ where logs in the jet functions are minimized, and a
perturbative low energy scale $\mu \gtrsim \Lambda + m\Gamma_t/Q$ where logs in
the soft function are minimized.  Here $\Lambda\sim 0.5\,{\rm GeV}$ is the
hadronic scale where the interactions are non-perturbative. As indicated there
are two scales appearing in each of these functions, and the question of which
dominates depends on the size of these parameters.

The program of this paper is as follows.  In Sec.~\ref{sec:fsummary} we review
the formulation of the factorization theorem for the invariant mass
cross-section from Ref.~\cite{Fleming:2007qr}. In Sec.~\ref{sec:FactGamma} we
show that the finite lifetime effects can be treated as a convolution of $B_\pm$
jet functions for stable top quarks with a Breit-Wigner, and we describe models
for $S$ that are consistent in the presence of perturbative corrections. In
Sec~\ref{sec:logs} we discuss the structure of large logarithms and present the
factorization with log resummation. In section~\ref{sect:RGE} we discuss the
connection between renormalization and the resummation of large logs in SCET and
HQET, derive the consistency conditions, and summarize results for the NLL
renormalization group evolution.  Results for the matching, running, and matrix
elements in SCET including the soft hemisphere function are given in
Sec.~\ref{sect:scet}. In Sec.~\ref{sect:bHQET} we give matching, running, and
matrix element results in HQET. A short-distance jet-mass scheme is discussed in
detail in Sec.~\ref{sect:jetmass}, including its relation to other schemes at
one-loop.  In Sec.~\ref{sec:ModelSoft} we discuss the non-perturbative soft
function model and the scheme for the gap parameter we use in our numerical
analysis. An analysis of the one-loop cross-section with NLL log-summation is
given in Sec.~\ref{sec:analysis} for both the peak and tail regions.
Conclusions are given in section~\ref{sect:conclusion}. Additional computational
details are given in the Appendices~\ref{AppFeyn}--\ref{App:SCETnum}.

\section{Formalism}

\subsection{Invariant Mass Cross-Section} \label{sec:fsummary}

In this section we review the main definitions of effective theory objects
needed for our calculations of terms in the factorization theorem in
Eq.~(\ref{FactThm}). Further details can be found in Ref.~\cite{Fleming:2007qr}.
Starting from QCD, the two-jet cross section $\sigma(e^+e^-\rightarrow
\gamma^*,Z^*\rightarrow j_t\: j_{\bar{t}})$ can be written as
\begin{eqnarray}\label{qcdcrosssection}
\sigma &=& \sum_X^{res.} (2\pi)^4 \, \delta^4(q-p_X) \sum_{i,j=a,v}
L_{\mu\nu}^{ij}\ \langle 0|  {\cal J}^{\dagger\nu}_j(0)  |X\rangle
\langle X |  {\cal J}^\mu_i(0) |0\rangle 
 \, ,
\end{eqnarray}
where $q=p_{e^-}+p_{e^+}$, and $q^2=Q^2$, and $L_{\mu\nu}^{ij}$ is the leptonic
tensor including vector and axial vector contributions from photon and $Z$ boson
exchange. This result is valid to all orders in the QCD coupling but lowest
order in the electroweak interactions. The superscript $res.$ on the summation
symbol denotes a restriction on the sum over final states to the kinematic
situation given in Eq.~(\ref{obs1}).  The QCD top quark currents are 
${\cal J}^\mu_i =\bar\psi(x) \Gamma_i^\mu \psi(x)$, where $\Gamma_v^\mu =
\gamma^\mu$ and $\Gamma_a^\mu = \gamma^\mu \gamma_5 $.  In
Ref.~\cite{Fleming:2007qr} we started with Eq.~(\ref{qcdcrosssection}) and
derived the factorization theorem for the double differential invariant mass
distribution in the peak region in Eq.~(\ref{FactThm}).  There the factor
$\sigma_0$ is the tree-level Born cross-section,
\begin{eqnarray}
\label{sigma0def}
 \sigma_0 &=& N_c \frac{4\pi \alpha^2}{3 Q^2}
   \bigg[\, e_t^2 - 
   \frac{2 Q^2\, v_e v_t e_t}{Q^2-m_Z^2} + 
   \frac{Q^4 (v_e^2+a_e^2)(v_t^2+a_t^2)}{(Q^2-m_Z^2)^2} \bigg] \,,
\end{eqnarray}
where $v_f = (T_3^f-2 Q_f \sin^2\theta_W)/(2\sin\theta_W \cos\theta_W)$ and $a_f
= T_3^f/(2\sin\theta_W \cos\theta_W)$.  Equation~(\ref{FactThm}) can be easily
generalized to include the angular distribution in $\cos(\theta)$ where $\theta$
is the angle between the top jet direction and the $e^-$ momentum:
\begin{align}
  \frac{d^3\sigma}{ dM^2_t\, dM^2_{\bar t}\, d\cos(\theta)} &=
\frac{\sigma_0(\theta)}{\sigma_0}    \frac{d^2\sigma}{ dM^2_t\,
dM^2_{\bar t} } \,,
\end{align}
where 
\begin{align}
  \sigma_0(\theta) &= 
\frac{d\sigma_0}{d\cos(\theta)}
  =  \frac{\pi N_c\alpha^2}{2 Q^2}
   \bigg[ \Big\{\, e_t^2  - 
   \frac{2 Q^2\, v_e v_t e_t}{Q^2-m_Z^2} + 
   \frac{Q^4 (v_e^2+a_e^2)(v_t^2+a_t^2)}{(Q^2-m_Z^2)^2}\Big\}
(1+\cos^2\theta)\nn\\
 & \qquad\qquad\qquad + \Big\{ \frac{4 Q^2 e_t^2 a_e a_t}{Q^2-m_Z^2} - \frac{8Q^4 a_e
v_e a_t v_t }{(Q^2-m_Z^2)^2} \Big\} \cos\theta
   \bigg]  \,.
\end{align}

The remaining functions in Eq.~(\ref{FactThm}) include $H_Q$, a hard-function
that encodes quark-gluon interactions at the production scale $Q$, $H_m$ which
encodes perturbative effects\footnote{The coefficient $H_m$ is also sensitive to
  the ratio $m/Q$ through its anomalous dimension. Here $m/Q$ is the cusp angle
  by which the heavy-quarks are off the light-cone~\cite{Korchemsky:1991zp}. See
  also section~\ref{sect:bHQET} below.} at the scale $m$, $B_\pm$, the jet
functions for the top jet and antitop jet respectively, and the soft function
$S$ which encodes non-perturbative information about soft hadrons radiated
between the hard jets.  The convolution with the soft function causes a
correlation between the two-jet functions and affects the invariant mass
spectrum.  Each of the functions $H$, $H_m$, $B_+$, $B_-$, and $S$ in
Eq.~(\ref{FactThm}) can be defined as matrix elements of operators in an
appropriate EFT, or as matching coefficients between two EFT's.  At the scale
$Q$ the matching of QCD currents onto SCET is given by a convolution
formula~\cite{Bauer:2000yr}
\begin{eqnarray}
\label{currentmatch}
 {\cal J}^\mu_i(0) = \int\!\! d\omega\, d\bar\omega\, C(\omega,\bar\omega,\mu) 
  J^{\mu}_i(\omega,\bar \omega,\mu) \,,
\end{eqnarray}
where $C$ contains short-distance dynamics, while $ J_i^{\mu}$ describes all
scales that are longer distance than $Q$.  After making a field
redefinition~\cite{Bauer:2001yt} the SCET production current at leading order in
$\lambda$ is given by
\begin{eqnarray}
\label{currentscet}
 J^{\mu}_i(\omega,\bar \omega,\mu)
   = [\bar \chi_{n,\omega}  Y_n^\dagger S_n^\dagger \Gamma^\mu_i 
  S_{\bn} Y_\bn \chi_{\bn,\bar\omega}](0) \,,
\end{eqnarray}
where we have collinear fields and Wilson lines defined in the jet-fields
$\chi_{n,\omega}(0) = \delta(\omega- \bn\mcdot \cP) (W_n^\dagger \xi_n)(0)$ and
$\chi_{\bn,\bar\omega}(0) = \delta(\bar\omega- n\mcdot \cP) (W_\bn^\dagger
\xi_\bn)(0)$, as well as soft $Y$-Wilson lines and mass-mode $S$-Wilson lines
to be discussed below. Here the $(0)$ indicates that the fields are at
coordinate $x^\mu=0$; recall that this $x^\mu$ dependence carries information
about residual momenta at scales $\lesssim Q\lambda^2 =m^2/Q$. The dependence on
larger momenta is encoded in the labels of the collinear
fields~\cite{Bauer:2001ct}. For example, $\delta(\omega-\bn\cdot P)$ forces the
total minus-label-momentum of $(W_n^\dagger \xi_n)$ to be $\omega$. In terms of
$C$ defined in Eq.~(\ref{currentmatch}), the hard-function appearing in
Eq.~(\ref{FactThm}) is simply
\begin{align} \label{H}
  H_Q(Q,\mu) = \big| C(Q,-Q,\mu) \big|^2 \,,
\end{align}
and after including RG-evolution we have $H_Q(Q,\mu)=H_Q(Q,\mu_Q)
U_{H_Q}(Q,\mu_Q,\mu)$ where $U_{H_Q}$ is the evolution kernel discussed below in
sections~\ref{sec:logs} and~\ref{sect:RGE}.

We obtain the SCET two-jet cross section by replacing the QCD current in
Eq.~(\ref{qcdcrosssection}) with the SCET current. The resulting expression can
be factorized as discussed in Ref.~\cite{Fleming:2007qr}
\begin{align} \label{SFactThm}
\frac{d^2 \sigma}{dM^2_t dM^2_{\bar{t}}} &= 
\sigma_0 H_Q(Q\OMIT{,\mu_Q},\mu) {\cal M}(m_J,\mu) \\
& \times\!\! \int^\infty_{-\infty}\!\!\! d\ell^+  d\ell^- 
J_n(s_t-Q\ell^+,m_J,\Gamma_t, \mu)
J_{\bar{n}}(s_{\bar{t}}-Q\ell^-,m_J,\Gamma_t,\mu) 
S(\ell^+, \ell^-,\mu,m_J) \,. \nn
\end{align}
This result can be used to compute the cross-section in the ultra-tail region,
where $s_{t,\bar t}\sim m^2$.  Due to the large suppression this region is not
interesting experimentally, however we will still discuss formal aspects of
Eq.~(\ref{SFactThm}) in detail because it is an important step towards deriving
the peak region factorization theorem, and is also important for making the
analogy with massless event shapes.  The soft function
$S(\ell^+,\ell^-,\mu,m_J)$ in Eq.~(\ref{SFactThm}) is the same as the soft
function in Eq.~(\ref{FactThm}), up to perturbative effects due to top-quark
vacuum polarization graphs denoted by the extra argument $m_J$. It can be either
derived by using eikonal Ward identities~\cite{Collins:1988ig} or properties of
the coupling of usoft gluons to collinear particles in SCET~\cite{Bauer:2003di}.
For the case of hemisphere invariant masses it is $S(\ell^+,\ell^-) = S_{\rm
  hemi}(\ell^+,\ell^-)$ where
\begin{align} \label{Slplm}
  S_{\rm hemi}(\ell^+,\ell^-) &\equiv \frac{1}{N_c}\sum _{X_s} 
 \delta(\ell^+ \minus k_s^{+a}) \delta(\ell^- \minus k_s^{-b})
  \langle 0| (\overline {Y}_\bn)^{cd}\,  ({Y}_n)^{ce} (0) |X_s \rangle
\langle X_s| ({Y}^\dagger_n)^{ef}\,  (\overline {Y}_\bn^\dagger)^{df}
(0) |0\rangle 
 \nn\\
 &=  \frac{1}{N_c}
  \big\langle 0 \big| (\overline {Y}_\bn)^{cd}\,  ({Y}_n)^{ce} (0) 
  \delta\big(\ell^+ \minus (\hat P^+_{a})^\dagger \big) \delta\big(\ell^- \minus \hat P^{-}_b\big) 
  ({Y}^\dagger_n)^{ef}\,  (\overline {Y}_\bn^\dagger)^{df}
(0) \big| 0 \big\rangle 
 \,.
\end{align}
The same function $S_{\rm hemi}$ appears in event shapes for massless
two-jet production, and besides the $m_J$ and $\Gamma_t$ dependence,
Eq.~(\ref{SFactThm}) is analogous to the factorization theorem for
massless
dijets~\cite{Bauer:2002ie,Korchemsky:1994is,Korchemsky:2000kp}. 
In Eq.~(\ref{Slplm})
$c$, $d$, $e$, $f$ are color indices, $N_c=3$, and the soft Wilson
lines are
\begin{align} \label{Yn}
 Y_n(x) &=  \overline {\rm P} \:
   \exp\Big(\minus i g\! \int_{0}^\infty \!\!\!\!ds\, n\mcdot A_{s}(ns\!+\! x) \Big) 
    \,,
 & Y_n^\dagger(x) &=   {\rm P} \, 
   \exp\Big(i g\! \int_{0}^\infty \!\!\!\!ds\, n\mcdot A_{s}(ns\!+\!x) \Big) \,,
  \nn\\
 \overline {Y_\bn}^\dagger(x)
  &=   {\rm P} \: \exp\Big( i g\! \int_{0}^{\infty} \!\!\!\!ds\, 
      \bn\mcdot \overline {A}_{s}(\bn s\!+\! x) \Big) 
  \,,  
  &\overline {Y_\bn}(x)  
   &=   \overline {\rm P}\: \exp\Big(\minus i g\! \int_{0}^{\infty} \!\!\!\!ds\, 
      \bn\mcdot \overline {A}_{s}(\bn s\!+\!x) \Big) 
   \,,
\end{align}
with $\overline A_\mu = \overline T^A A_\mu^A$ for the antitriplet
representation, where $\overline T^A= - (T^A)^T$. In Eq.~(\ref{Slplm}) $k_s^a$
is defined as the soft momentum components from the state $X_s$ that are
included in the experimental determination of $M_t$ (and $k_s^b$ for $M_{\bar
  t}$). We also have operators $\hat P_a^{+}=n\cdot \hat P_a$ and $\hat
P_b^{-}=\bn\cdot \hat P_b$ that project out the soft momentum components
$k_s^{+a}$ and $k_s^{-b}$
\begin{align}
  \hat{P}_a\> | X_s\rangle &= k^a_s \>| X_s\rangle,
   & \hat{P}_b \>|X_s\rangle &= k_s^b \>| X_s\rangle \,.
\end{align}
For hemisphere masses the operator $\hat P_a^+$ is defined to project out the
total plus-momentum of soft particles in hemisphere-$a$ (and $\hat P_b^-$ the
minus-momentum in hemisphere-$b$).  In Ref.~\cite{Fleming:2007qr} it was shown
that $S(\ell^+,\ell^-)$ does not depend on the top quark width, and when we pass
below the top quark mass scale is only
modified by a perturbative prefactor,
\begin{align} \label{Smatch}
  S(\ell^+,\ell^-,\mu,m_J) = T_0(m_J,\mu)\, S(\ell^+,\ell^-,\mu) \,.
\end{align}
The matching coefficient $T_0(m_J,\mu)$ is induced by the coupling of $A_s^\mu$
gluons to top-vacuum polarization bubbles at zero-momentum. This result applies
at any order in $\alpha_s$, but at NLL order $T_0=1$.

In Eq.~(\ref{SFactThm}) the mass-mode function ${\cal M}(m_J,\mu)$ contains
virtual perturbative corrections due to gluons $A_m^\mu$ and quarks $\psi_m$
with momenta $p^\mu\sim (m_J,m_J,m_J)$, and is given by
\begin{align}\label{Hmmatch}
  {\cal M}(m_J,\mu) &= \frac{1}{N_c^2} \: \big| \langle 0 | {\overline
    S}_{\bn}^{\,ab} S_n^{ab} | 0\rangle \big|^2 \,.
\end{align}
The definition of these mass-mode $S$-Wilson lines is identical to those in
Eq.~(\ref{Yn}), except that they involve gluon fields $A_m^\mu$ which couple to
massive top-quarks for any momentum, and which have zero-bin subtractions to
avoid double counting the momentum region accounted for by the $A_s^\mu$ gluons.
This implies that ${\cal M}(m_J,\mu)$ only gets contributions from graphs with a
top-vacuum polarization bubble~\cite{Kniehl:1989kz,Burgers:1985qg,Hoang:1995ex}
coupling to the $A_m^\mu$ gluons. At NLL order the function ${\cal
  M}(m_J,\mu)=1$, but is relevant at NNLL order and beyond when considering
virtual top loops. Note that due to the invariant mass constraint $s\ll m Q$,
the $\psi_m$ quarks never appear in the final state.  This is important for the
validity of Eq.~(\ref{SFactThm}).

Matrix elements of top quark collinear fields in SCET give the jet functions
$J_n$ for the top quark jet, and $J_\bn$ for the antitop jet,
\begin{align}
\label{jetfunc2}
J_{n}(Qr_n^+ \minus m_J^2,m_J,\Gamma_t,\mu) 
&= 
\frac{-1}{4\pi N_c Q } \, \textrm{Im}\ \bigg[ i\! \int \!\! d^4 x \, 
  e^{i r_n\cdot x} \,
\langle 0|\text{T}\{ \bCH n Q (0)\slash\!\!\!\bn  \chi_n(x)\}|0 \rangle
  \bigg] \, ,
  \nn\\
 J_{\bn}(Qr_\bn^- \minus m_J^2,m_J,\Gamma_t,\mu) 
&= 
 \frac{1}{4\pi N_c Q} \, \textrm{Im}\ \bigg[ i\! \int \!\! d^4 x \, 
  e^{i r_\bn\cdot x} \,
\langle 0|\text{T}\{ \bCH \bn {-Q} (x) \slash\!\!\! n \chi_\bn(0)\} |0 \rangle
  \bigg] \, .
\end{align}
These jet functions $J_n$ and $J_\bn$ depend on both the mass and width of the
top quarks.  The matrix elements of collinear fields are defined with the
zero-bin subtractions~\cite{Manohar:2006nz}, which avoids double counting the
soft region.

For predictions in the peak region, the $J_n$ and $J_\bn$ functions should be
factorized further by integrating out the top quark mass. This is accomplished
by matching onto jet functions $B_\pm$ in HQET with boosted heavy quarks. The
relevant Feynman rules are given in Appendix~\ref{App:bHQETFeyn}.  The jet
function matching takes the simple form~\cite{Fleming:2007qr}
\begin{align} \label{BFactThm}
 J_n(s_t,m_J,\Gamma_t,\mu_Q) &= T_+(m_J,\mu_Q) B_+(\hat s_t,\Gamma_t,\mu_Q)
  + {\cal O}\Big(\frac{\Gamma}{m}\Big) 
  + {\cal O}\Big(\frac{\hat s_t}{m}\Big) 
 \,,\nn\\[5pt]
 J_\bn(s_{\bar t},m_J,\Gamma_t,\mu_Q) &= T_-(m_J,\mu_Q) B_-(\hat s_{\bar t},\Gamma_t,\mu_Q) 
  + {\cal O}\Big(\frac{\Gamma_t}{m}\Big) 
   + {\cal O}\Big(\frac{\hat s_{\bar t}}{m}\Big)
  \,.
\end{align}
The HQET jet functions $B_+$ and $B_-$ also depend on the residual mass term
$\delta m_J$ that fixes the mass definition in HQET.  They are defined by
\begin{eqnarray}
\label{hqetjet}
B_\pm (\hat s,\Gamma_t,\mu) &=&\,
\textrm{Im} \, \big[ {\cal B}_\pm(\hat s,\Gamma_t,\mu)\big]
\,,
\end{eqnarray}
where the ${\cal B}_\pm$ are vacuum matrix elements of T-products of HQET
operators
\begin{eqnarray}
\label{hqetjet2}
{\cal B}_+(2v_+\mcdot r,\Gamma_t,\mu) & = &
\frac{- i}{4\pi N_c m}  \int\!\! d^4 x \, e^{i r \cdot x} \,
\big\langle 0 \big|T\{\bar{h}_{v_+}(0) W_n(0) W_n^{\dagger}(x) h_{v_+}
(x)\} \big|0 \big\rangle \,,
\nn\\
{\cal B}_-(2v_-\mcdot r,\Gamma_t,\mu) & = &
\frac{i}{4\pi N_c m}  \int\!\! d^4 x \, e^{i r \cdot x} \,
\big\langle 0 \big|T\{\bar{h}_{v_-}(x) W_\bn(x) W_\bn^{\dagger}(0) h_{v_-}
(0)\} \big|0 \big\rangle
\,.
\end{eqnarray}
Here for $B_+(\hat s_t,\Gamma_t,\mu)$ we have $\hat s_t =2 v_+\cdot r$, while
for $B_-(\hat s_{\bar t},\Gamma_t,\mu)$ we have $\hat s_{\bar t}=2 v_-\cdot r$.
The gluons in $W_n$ and $W_\bn^{\dagger}$ and HQET fields $h_{v_\pm}$ are only
sensitive to fluctuations below $m$ and are built of gluons $A_\pm^\mu$
describing low energy fluctuations down to $p^2\sim \Gamma^2$ in the top and
antitop rest frames respectively. In Ref.~\cite{Fleming:2007qr} these gluons
were called ultracollinear. We emphasize that to make the matching consistent,
the collinear gluons in Eq.~(\ref{hqetjet2}) have zero-bin subtractions for the
same region as those in the SCET jet functions. These subtractions ensure that
the $B_\pm$ jet-functions do not double-count the soft region encoded in $S$, and
are critical for ensuring that the functions $B_\pm$ are IR-finite, as we
discuss further in Appendix~\ref{AppFeyn}. In Eq.~(\ref{hqetjet2}) the Wilson
lines are
\begin{align}
\label{bHQETWilsondef}
  W_n^\dagger(x) &= {\rm P} \, 
   \exp\Big(i g\! \int_{0}^\infty \!\!\!\!ds\, \bn\mcdot A_{+}(\bn s\!+\!x)
\Big) \,,
  & W_n(x) &=  \overline {\rm P} \:
   \exp\Big(\minus i g\! \int_{0}^\infty \!\!\!\!ds\, \bn\mcdot
A_{+}(\bn s\!+\! x) \Big) ,
\end{align}
with analogous formulas for $W_\bn$ and $W_\bn^\dagger$ in terms of $n\cdot
A_-$. Note that if the Wilson lines $W_n$ and $W_\bn$ were absent, then ${\cal
  B}_\pm$ would just define the HQET heavy quark/antiquark
propagators~\cite{Manohar:2000dt}.  The Wilson lines, let's say for ${\cal
  B}_+$, encode the color dynamics of gluons that are soft in the top rest frame
and come from the highly boosted antitop quark, and they render this vacuum
matrix element into a gauge-invariant physical object. The analogous situation
with top and antitop switched applies for the vacuum matrix element ${\cal
  B}_-$.  For the SCET jet functions, the Wilson lines appearing in
Eqs.~(\ref{currentscet}) and (\ref{jetfunc2}) have the analogous physical
interpretation where the top quark mass has not yet been integrated out.  

In the final factorization theorem in Eq.~(\ref{FactThm}) we
have the $B_\pm$ functions, as well as the matching condition for the mass
fluctuations, 
\begin{align} \label{Hm}
  H_m(m,\mu)=T_+(m,\mu)T_-(m,\mu) T_0(m,\mu) {\cal M}(m,\mu)
     \,.
\end{align}  
In bHQET all dynamic effects associated with the top-quark mass appear in $H_m$,
and there are no mass-mode quarks or gluons in this theory. Since $T_0$ and
${\cal M}(m,\mu)$ encode finite matching corrections at the scale $\mu\simeq m$
due to top-vacuum polarization, we have $T_0(m,\mu) {\cal M}(m,\mu) =1 + {\cal
  O}(\alpha_s^2)$, and so these factors drop out from our NLL analysis.
Therefore in later sections we simply use $H_m=T_+ T_-$.  This coefficient $H_m$
becomes sensitive to the ratio $Q/m$ through its anomalous dimensions which
depends on a logarithm of $v_+\cdot \bn= v_-\cdot n=Q/m$.  Including the
RG-summation of these logarithms gives the coefficient
$H_m(m,Q/m,\mu_m,\mu)=H_m(m,\mu_m) U_{H_m}(Q/m,\mu_m,\mu)$ appearing in the
factorization theorem, where $U_{H_m}$ is the bHQET current evolution factor
discussed below in sections~\ref{sec:logs} and~\ref{sect:RGE}. Note that in
principle $H_m(m,\mu)$ and the factors in Eq.~(\ref{Hm}) can also have
$Q/m$ dependence at NNLL. For related discussions see
Refs.~\cite{Becher:2007cu,Chiu:2007yn}.

Alternatively, the matching coefficient of SCET and HQET jet functions given by
$H_m$ in Eq.~(\ref{Hm}) can be determined from currents,
\begin{align} \label{Hm2}
  H_m(m,\mu) = \big| C_m(m,\mu) \big|^2 \,,
\end{align}  
 where the boosted HQET current is
\begin{align}
  J_i^\mu(\mu) = C_m(m,\mu) J_{\rm bHQET}^\mu(\mu) \,,
\end{align}
with 
\begin{align}\label{JbHQET}
  J_{\rm bHQET}^\mu = (\bar h_{v_+} W_n) Y_n^\dagger \Gamma_i^\mu Y_\bn
  (W_\bn^{\dagger} h_{v_-}) \,.
\end{align}
The soft Wilson lines $Y$ in this current are the same as those used in the SCET
soft function. The only distinction is that soft gluons in bHQET no longer
couple to massive top-bubbles.

Due to
the large width of the top quarks the $B_\pm$ jet functions can be computed in
perturbation theory. At tree level they are Breit-Wigner distributions
\begin{align}
\label{eq:Btree}
  B^{\rm tree}_\pm(\hat s,\Gamma_t) &
  = {\rm Im} \big[ {\cal B}_\pm^{\rm tree}(\hat s,\Gamma_t) \big]
  = {\rm Im} \bigg[\, \frac{-1}{\pi m}\: \frac{1}{\hat s+i\Gamma_t} \, \bigg]
  =   \frac{1}{\pi m} \: \frac{\Gamma_t}{\hat s^2 + \Gamma_t^2} \:,
\end{align}
where we have adopted a normalization such that 
\begin{align}
  \int_{-\infty}^{+\infty} \!\!\!\!\! ds \  B^{\rm tree}_\pm(\hat s,\Gamma_t) = 1 \,.
\end{align}
The Wilson coefficients in the factorization theorem in Eq.~(\ref{FactThm}) are
also normalized to unity at tree level, $H_Q=1$ and $H_m=1$.

\subsection{Factorization of Lifetime Effects and Soft
  Function Models}
\label{sec:FactGamma}

The leading order bHQET Lagrangian is
\begin{eqnarray} 
\label{bHQETLagrangian}
{\cal L}_{\pm} &=&
\bar{h}_{v_\pm} \big( i v_\pm \cdot D_\pm - \delta m + 
\frac{i}{2} \Gamma_t \big) h_{v_\pm }
\,.
\end{eqnarray}
In light-cone coordinates, $(+,-,\perp)$, we have $v_+^\mu=(m/Q,Q/m,0)$ and
$v_-^\mu=(Q/m,m/Q,0)$ and gluons/residual momenta scaling as $D_+^\mu\sim\Gamma
(m/Q,Q/m,1)$ and $D_-^\mu\sim \Gamma(Q/m,m/Q,1)$. Unlike standard HQET, the
ultracollinear gluon fields in bHQET are defined with zero-bin
subtractions~\cite{Manohar:2006nz} for the soft region.  In
Eq.~(\ref{bHQETLagrangian}) $\Gamma_t$ is a Wilson coefficient obtained by
matching to the full theory and is equal to the top quark total width.  This is
true to leading order in electroweak interactions, to ${\cal O}(m^2/Q^2)$ and
${\cal O}(\Gamma/m)$ in the power counting, and to all orders in
$\alpha_s$.\footnote{Concerning the $m/Q$ expansion this is true because for
  $Q\gg m$ the hemisphere mass definition is inclusive in the top and antitop
  decay products up to ${\cal O}(m^2/Q^2)$ corrections~\cite{Fleming:2007qr}.
  Concerning the $\Gamma/m$ expansion this is related to the fact that finite
  lifetime corrections are related to off-shell corrections that are $\hat
  s/m$-suppressed~\cite{Hoang:2006pd}. Concerning the $\alpha_s$ expansion this
  can be seen by carrying out the matching with free quark states and noting
  that the full theory computation of $t\to bW$ gives the total rate. Now only
  the operator of interest $(i\Gamma_t/2) \bar h_v h_v$ allows for decays in the
  effective theory, but it corresponds to a conserved current and so does not
  get renormalized~\cite{Manohar:2000dt}.}  Finally,
\begin{eqnarray}
\label{deltamdef}
\delta m & = & m_{\rm pole}-m
\end{eqnarray} 
is the residual mass term that fixes the top quark mass definition $m$ that is used
in the HQET computations. It needs to be consistent with the bHQET power
counting~\cite{Fleming:2007qr},
\begin{eqnarray} \label{massscheme}
 \delta m \sim \hat s_t\sim \hat s_{\bar t}\sim\Gamma  \,,
\end{eqnarray}
can be computed perturbatively, and is UV- and IR-finite. Note that the way in
which Eq.~(\ref{bHQETLagrangian}) will be used is to compute a jet-function
where the width smears over a set of states of invariant mass $m\Gamma_t\gg
\Lambda_{\rm QCD}^2$. Thus, for our analysis there are no $\Lambda_{\rm
  QCD}/\Gamma_t$ corrections to Eq.~(\ref{bHQETLagrangian}), just corrections of
${\cal O}(\Lambda_{\rm QCD}/m)$.

In Eq.~(\ref{hqetjet}) the jet functions $B_\pm$ are
expressed in terms of the imaginary part of vacuum matrix elements
${\cal B}_\pm$ in Eq.~(\ref{hqetjet2}). 
From ${\cal L}_\pm$ it is straightforward to see that $B_\pm$ can be obtained
from the imaginary part of the vacuum matrix element ${\cal B}^{\Gamma=0}_\pm$
for (fictitious) stable top quarks by shifting the energy variable $\hat s\to
\hat s+ i\,\Gamma_t$,
\begin{align} \label{Bdisc}
B_\pm(\hat s,\Gamma_t,\mu) &=
\textrm{Im} \, \big[ {\cal B}_\pm(\hat s , \Gamma_t,\mu)\big] 
\nn \\[2mm]
\, &= 
\textrm{Im} \, \big[ {\cal B}^{\Gamma=0}_\pm(\hat s \plus i\,\Gamma_t,\mu)\big]
 \,,
\end{align}
Here we defined results for stable top quarks, namely the jet function
$B_\pm^{\Gamma=0}(\hat s,\mu)\equiv B_\pm(\hat s,0,\mu)$, and a vacuum matrix
element ${\cal B}_\pm^{\Gamma=0}(\hat s,\mu) \equiv {\cal B}_\pm(\hat s,0,\mu)$.
They are related by
\begin{align}
  B_\pm^{\Gamma=0}(\hat s,\mu) 
   = {\rm Im}\big[ {\cal B}_{\pm}^{\Gamma=0}(\hat s,\mu) \big] \,,
\end{align}
and we will refer to $B_\pm^{\Gamma=0}$ as the stable jet function in what
follows.  The result in Eq.~(\ref{Bdisc}) is in complete analogy to the relation
between the production rate of top quark pairs in the nonrelativistic threshold
region, $E_{\rm c.m.}\approx 2m$, where the leading order finite lifetime
effects can be implemented by the shift $E_{\rm c.m.}\to E_{\rm
  c.m.}+i\,\Gamma_t$ prior to taking the imaginary part of the $e^+e^-\to e^+e^-$
forward scattering matrix element~\cite{Fadin:1987wz}.  

To separate the different physical effects in the cross section it is convenient
to derive a factorization theorem for the leading order finite lifetime effects
to all orders in $\alpha_s$.  To do so we define the function
\begin{eqnarray}
g(x) & \equiv & -\frac{i}{2}\,
{\cal B}^{\Gamma=0}_\pm(x,\mu) \, = \,  -\frac{i}{2}\, {\cal B}_\pm(x,0,\mu)
\,.
\end{eqnarray}
It is analytic everywhere in the complex $x$-plane, except along the positive real
axis, $x\ge 0$, where the vacuum matrix elements ${\cal B}^{\Gamma=0}_\pm$,
defined using Eq.~(\ref{hqetjet2}) with $\Gamma_t=0$, has a cut for intermediate
states having invariant masses larger than the top quark mass. Using the residue
theorem for a contour that envelops the cut, it is then straightforward to
derive the dispersion relation
\begin{eqnarray}
g(a) & = & \frac{1}{2\pi i}\,\int_0^\infty\!\! dx \:
\frac{\mbox{Disc}[g(x)]}{x-a}
\,,
\end{eqnarray}
where $a$ is any point in the complex plane not on the positive real axis. With
the choice $a=\hat s+i\,\Gamma_t$ and a change of variable $x=\hat s-\hat s'$,
this dispersion relation can be brought into the form
\begin{eqnarray}
\label{factorizationGamma}
B_\pm(\hat s,\Gamma_t,\mu) & = &
  \int_{-\infty}^{\,\,\mbox{$\hat s$}} \!\!\! d{\hat s}^\prime \
B^{\Gamma=0}_\pm({\hat s}-{\hat s}^\prime,\mu)\:
\frac{\Gamma_t}{\pi\,({\hat s}^{\prime\,2}+\Gamma_t^2)}
\,.
\end{eqnarray}
Note that the upper limit $\hat s$ of the integration can be replaced
by $+\infty$ since the stable jet function only has support for
positive values of its energy variable.
Equation~(\ref{factorizationGamma}) states that the bHQET jet
functions for the physical unstable top quark can be written as a
convolution of the stable jet functions with a Breit-Wigner function
of the width $\Gamma_t$. Thus the leading order finite lifetime
effects can be factorized from the jet function.\footnote{ Note that
subleading finite lifetime effects, which are suppressed by
$\Gamma_t/m$, cannot be factorized as in
Eq.~(\ref{factorizationGamma}) since they are not described by a
simply shift of the energy into the complex
plane~\cite{Hoang:1999zc,Beneke:2004km,Hoang:2004tg}.  }  This means
in particular that the renormalization properties of the jet functions
for stable and unstable top quarks are equivalent - a fact that might
not be obvious since the evolution of the jet functions involves
convolutions with distributions.

Eq.~(\ref{factorizationGamma}) reflects the fact that the top quark
width acts as an infrared cutoff for the jet function through smearing
over a Breit-Wigner function~\cite{Poggio:1975af}. In the
factorization theorem in Eq.~(\ref{FactThm}) additional smearing is
provided by the convolution with the soft function, where the width of
the distribution $S(\ell^+,\ell^-)$ is of order the hadronic scale
$\Lambda$.  Eq.~(\ref{factorizationGamma}) allows us to group both
types of smearing into a common infrared function $R$, with the
following modified version of the factorization theorem,
\begin{align}
\label{bHQETcross-hem3}
 \frac{d^2\sigma }{dM^2_t dM^2_{\bar t}} &= 
    {\sigma_0}
     \> H_Q(Q\OMIT{,\mu_Q},\mu_m) \: H_m\Big(m_J,\frac{Q}{m_J},\mu_m,\mu\Big)\! \\
 &\times
 \int_{-\infty}^{\infty}\!\!\!d\ell^+ d\ell^-  
  \> B^{\Gamma=0}_+\Big(\hat s_t - \frac{Q\ell^{+}}{m_J},\mu\Big) \:
     B^{\Gamma=0}_-\Big(\hat s_{\bar t} - \frac{Q\ell^{-}}{m_J},\mu\Big)\: 
R(\ell^+,\ell^-,\Gamma_t,\mu)
  \,.\nn
\end{align}
The result involves only the stable jet functions and an infrared function
defined as
\begin{align}
\label{modifiedS}
R(\ell^+,\ell^-,\Gamma_t,\mu) & \equiv 
\left(\frac{m\,\Gamma_t}{Q\,\pi}\right)^2\,
\int_0^\infty\!\!\! d\tilde\ell^+ \!\!
\int_0^\infty\!\!\! d\tilde\ell^-\,
\frac{S(\tilde\ell^+,\tilde\ell^-,\mu)}
{\big[(\ell^+ \minus \tilde\ell^+)^2+(\frac{m\,\Gamma_t}{Q})^2 \big]\,
 \big[(\ell^- \minus \tilde\ell^-)^2+(\frac{m\,\Gamma_t}{Q})^2 \big]}
\,.
\end{align}
In the $\Gamma_t \to 0$ limit we have
$R(\ell^+,\ell^-,\Gamma_t=0,\mu)=S(\ell^+,\ell^-,\mu)$, since the Breit-Wigner
factors reduce to delta-functions. If $\hat s_{t,\bar t} \gg \Gamma_t$, as in
the tail region, then $R$ can be simplified with an operator product expansion
whose first term depends on the partonic soft function $S_{\rm part}$ which can
be computed in perturbation theory. In this region $\tilde \ell^\pm \sim
s_{t,\bar t}/Q \gg\Lambda$, and for these momenta $S(\tilde\ell^\pm,\mu)=S_{\rm
  part}(\tilde\ell^\pm,\mu)$ up to power corrections of ${\cal O}(\Lambda
Q/s_{t,\bar t})$.  For the case $\Gamma_t\gg \frac{Q}{m}\Lambda$, $R$ can be
computed using an operator product expansion even in the peak region, taking
Eq.~(\ref{modifiedS}) with $\frac{Q\,\Lambda}{m\,\Gamma_t}\ll 1$, and again the
leading term is determined by $S_{\rm part}$.  This is similar to $B\to
X_s\gamma$ in the multi-scale OPE~\cite{Neubert:2005nt,Bauer:2003pi} where
smearing over the soft function makes it computable in an OPE. On the other
hand, for top-quarks $\Gamma_t\lesssim \frac{Q}{m}\Lambda$, and the infrared
function is significantly effected by nonperturbative effects in $S$. Thus the
cross-section in the peak region cannot be determined entirely from perturbation
theory.

These properties serve as an important guideline for the construction of a
consistent model for the soft function to be used beyond the tree-level
approximation. They require that the perturbative corrections contained in the
partonic soft function {\it must be included} in a viable model in order to
obtain the correct leading order term in the operator product expansion for the
cases $\tilde\ell^\pm\gg\Lambda$ and $\Gamma\gg \frac{Q}{m}\Lambda$ mentioned
above.  As discussed in Ref.~\cite{Hoang:2007vb} one way to give a consistent
implementation of the partonic soft function in $S$ is to use a convolution form
\begin{align}
\label{Smodel1}
S(\ell^+,\ell^-,\mu) & = 
\int_{-\infty}^{+\infty}\!\!\! d\tilde\ell^+
\int_{-\infty}^{+\infty}\!\!\! d\tilde\ell^-\
S_{\rm part}(\ell^+ \minus \tilde\ell^+,\ell^- \minus \tilde\ell^-,\mu,\delta)\,
S_{\rm mod}(\tilde\ell^+,\tilde\ell^- ) 
\,,
\end{align}
where $S^{\rm part}$ is the soft function computed in perturbation theory at
$\mu$, and $S^{\rm mod}$ is a hadronic function satisfying
\begin{eqnarray} \label{Smodelconditions}
  && \int_{-\infty}^{+\infty}\!\!\!\! d\ell^+\!\!
  \int_{-\infty}^{+\infty}\!\!\!\! d\ell^-\,S^{\rm mod}(\ell^+,\ell^-)
   = 1 \,, \nn\\
  && \int_{-\infty}^{+\infty}\!\!\!\! d\ell^+
  \int_{-\infty}^{+\infty}\!\!\!\! d\ell^-\, (\ell^+)^n (\ell^-)^m\: S^{\rm
    mod}(\ell^+,\ell^-)
  \sim (\Lambda_{\rm QCD})^{n+m} \,,
\end{eqnarray}
for $n+m\ge 1$.  An analogous formula to Eq.~(\ref{Smodel1}) was used to
incorporate moment constraints in the study of the soft function in $b\to
s\ell^+\ell^-$ in Ref.~\cite{Lee:2005pw}.  This form ensures that $S$ reduces to
$S_{\rm part}$ for $\ell^\pm\gg\Lambda$, and that for all kinematic regions it
has the proper $\mu$ dependence for the $\overline {\rm MS}$-scheme. It also
gives the proper result for $R$ in taking the limit
$\Gamma\gg\frac{Q\,\Lambda}{m}$ of Eq.~(\ref{modifiedS}). The model in
Eq.~(\ref{Smodel1}) is specified by parameters in $S_{\rm mod}$ which involve
the hadronic scale $\Lambda$, and also by the choice of the scale $\mu$ in
$S_{\rm part}$.  The convolution generates logarithms of the form
$\ln(\ell^\pm/\mu)$ and $\ln(\Lambda/\mu)$ to be discussed in the next section.
There are also complications related to removing a $u=1/2$ renormalon (as
indicated by the subtraction constant $\delta$ in $S_{\rm part}$) and
introducing a renormalon free gap parameter $\bar\Delta$ in the soft
function~\cite{Hoang:2007vb}. We will use the prescription in
Eq.~(\ref{Smodel1}) for our numerical studies of the factorization theorem in
Sec.~\ref{sec:analysis}. Our choice of $S^{\rm mod}$ and a review of how the
renormalon subtractions work are given in Sec.~\ref{sec:ModelSoft}.

\subsection{Summation of Large Logs in SCET and HQET} 
\label{sec:logs}

\begin{figure}
  \centerline{ 
   \includegraphics[width=15.5cm]{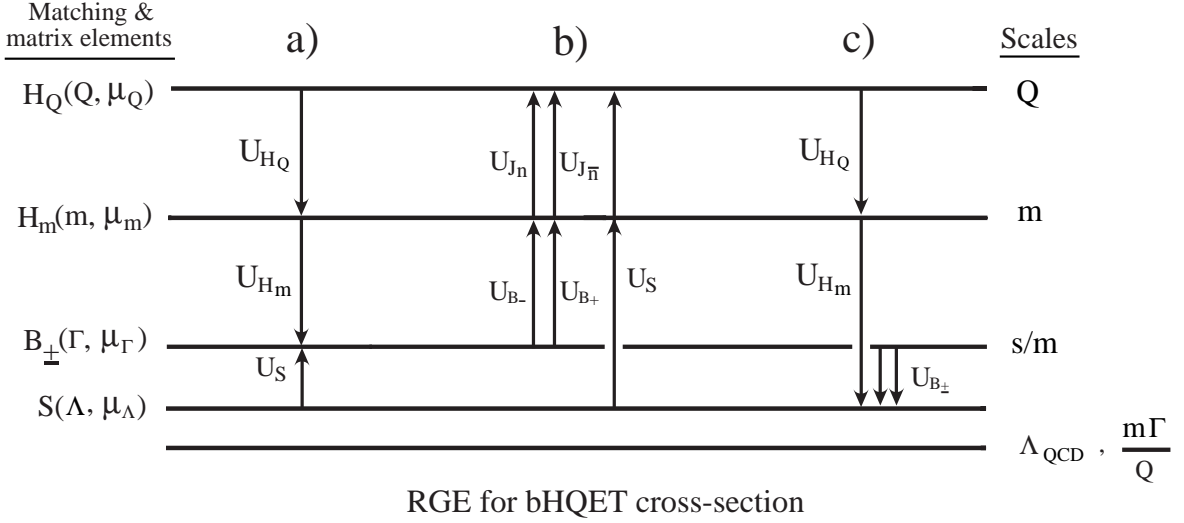}  
  } 
 \vskip-0.1cm
\caption{Matching, running, and matrix elements that determine the functions 
  in the factorization theorem in Eq.(\ref{FactThm}) for the peak region (when
  $s/m\sim\Gamma$) and for the tail region (when $s/m\gtrsim \Gamma$). The running in
  $U_{H_Q}$ and $U_{H_m}$ is local, while that in $U_{J_n}$, $U_{J_\bn}$, $U_S$,
  and $U_{B}$ involves convolutions.  Here the distribution width is $\Gamma\sim
  \Gamma_t+ Q\Lambda/m$. Cases a), b), and c) show three equivalent ways to sum
  large logs with the renormalization group.  The consistency equations
  discussed in the text express the equivalence of running from the top-down in
  case a) and from the bottom-up in case b). Case c) is used for our numerical
  analysis.}
\label{fig:theoryII}
\end{figure}

\begin{figure}
  \centerline{ 
   \includegraphics[width=15cm]{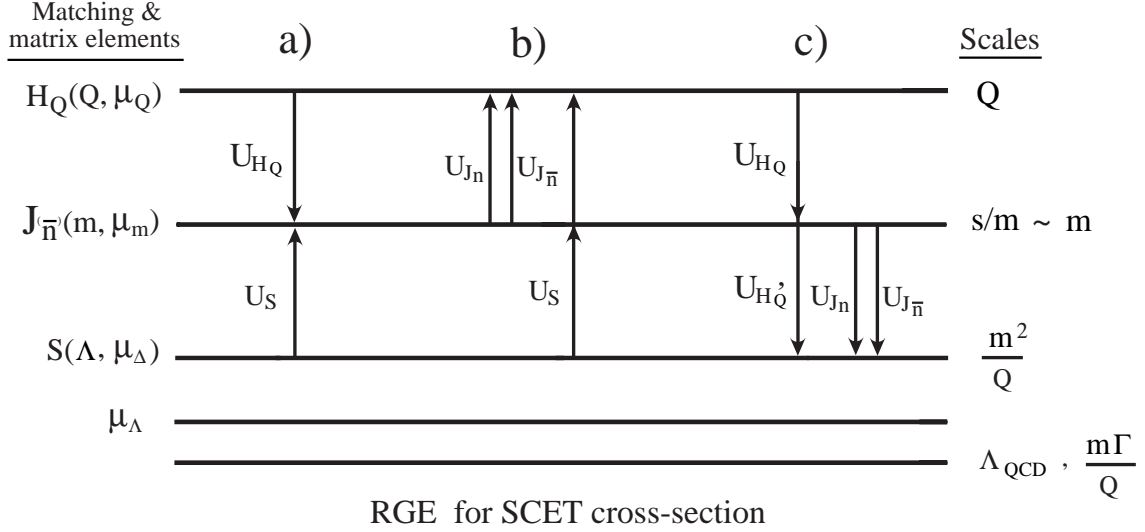}  
  } 
\caption{Matching, running, and matrix elements that determine the functions 
  in the factorization theorem for the ultra-tail region, in
  Eq.(\ref{SFactThm}). The running in $U_{H_Q}$ is local, while that in
  $U_{J_n}$, $U_{J_\bn}$, and $U_S$ involves convolutions.  Cases a), b), and c)
  show three equivalent ways to sum large logs with the renormalization group.
  The consistency equation discussed in the text express the equivalence of
  running from the top-down in case a) and from the bottom-up in case b). }
\label{fig:theoryIII}
\end{figure}

In this section we discuss the summation of large logs between $Q \gg m \gg
\Gamma$ for the peak region, and between $Q\gg m\gg \hat s$ in the tail
region. We also discuss the ultra-tail region where $Q\gg \hat s\sim m \gg m^2/Q$. 
In both SCET and bHQET we can define unitary evolution functions $U_i$,
associated with the renormalization group evolution for hard, jet, and soft
functions in Eqs.~(\ref{FactThm}) and (\ref{SFactThm}).  These $U_i$ factors are
indicated by the arrows in Figs.~\ref{fig:theoryII} and~\ref{fig:theoryIII}.
The figures show that there are two ways of doing the renormalization group
evolution. In the first one, referred to as ``top-down'', we run the SCET and
bHQET production currents, starting with matching at a high scale, and running
toward the low scales. In the second one, referred to as ``bottom-up'', we run
the individual jet and soft functions, starting with initial conditions at the
low scales and running up.

The UV renormalization of the currents in SCET and bHQET generates $U_{H_Q}$ and
$U_{H_m}$ respectively. Since the renormalization of a current does not depend
on the choice of states, these factors do not carry information about the
constraints used to define $M_t$ and $M_{\bar t}$. Instead $U_{H_Q}$ and
$U_{H_m}$ only affect the overall normalization of the invariant mass
distribution. On the other hand the jet and soft functions have evolution
through $U_{J}$, $U_{B}$, and $U_S$ which involve convolutions that change their
respective shape. In general the jet and soft functions also incorporate the
prescriptions to define $M_t$ and $M_{\bar t}$.\footnote{In the class of
  observables we consider this is the case for $S$, while $J_{n,\bn}$ and
  $B_\pm$ are inclusive because they do not depend on the invariant mass
  prescription which only affects radiation at large angles.}  Hence,
although it is expected on general grounds, it is not immediately obvious how
the running of these functions becomes independent of the prescription used to
define these invariant masses.  The equivalence of the top-down and bottom-up
approaches ensures that the changes in shape of the jet and soft function cancel
out in the convolution for any region in $\mu$ where they overlap in
Fig.~\ref{fig:theoryII}, and yield the same result as obtained from $U_{H_Q}$
and $U_{H_m}$.  In the field theory this result is encoded in an SCET
consistency condition between $U_{H_Q}$, $U_{J}$, and $U_s$, and a bHQET
consistency condition between $U_{H_m}$, $U_{B}$, and $U_s$, where as we will
show to all orders in perturbation theory the soft function evolution factorizes
as
\begin{align} \label{US1}
  U_S(\ell^+,\ell^-)=U_s(\ell^+)U_s(\ell^-) \,.
\end{align}
The consistency conditions between the scales $\mu$ and $\mu_0$ are
\begin{align} \label{CE}
 \sqrt{U_{H_Q}(Q,\mu,\mu_0)}\,U_{J}(s,\mu_0,\mu) &=
 \frac{1}{Q}\,U_s\Big(\frac{s}{Q},\mu,\mu_0 \Big)  
 \,,\nn\\
 \sqrt{U_{H_m}\Big(\frac{Q}{m},\mu,\mu_0\Big)}
\,U_{B}(\hat s,\mu_0,\mu) &=
\frac{m}{Q}\,U_s\Big(\frac{m \hat s}{Q},\mu,\mu_0\Big)
\,.
\end{align}
The derivation of Eq.~(\ref{US1}) and (\ref{CE}) will be given in
section~\ref{subsec:consistency} below, while field theory definitions of the
$U_i$ are given in sections~\ref{sssec:scetrenorm} and~\ref{subsec:bHQETrenor}.
Note that this discussion implies that the SCET factorization theorem in
Eq.~(\ref{SFactThm}) can be formulated at any scale $\mu>m$, and the final
factorization theorem in Eq.~(\ref{FactThm}) can be formulated at any scale
$\mu<m$ without affecting the renormalization group evolution. For $\mu>m$ we
have $n_f=6$ flavors, while for $\mu< m$ we have $n_f=5$ flavors for these
evolution factors.

In the peak region factorization theorem in Eq.~(\ref{FactThm}) the
$B_\pm$ and $S$ functions are evaluated at a common scale $\mu$. Since they
involve logarithms of the form $\ln[ (-\hat s-i\Gamma_t)/\mu]$ and
$\ln(\Lambda/\mu)$, where $\Gamma_t$ and the hadronic scale $\Lambda$ differ, it
is natural to consider using different low energy scales $\mu_\Gamma$ and
$\mu_\Lambda$ for the jet and soft functions respectively. In this case we have
a region between $\mu_\Gamma$ and $\mu_\Lambda$ where the consistency conditions
no longer apply. Thus for the general situation shown in Fig.~\ref{fig:theoryII}
the factorization theorem becomes
\begin{align}
\label{bHQETcross-hem2a}
 \frac{d^2\sigma }{dM^2_t dM^2_{\bar t}}
 &= 
  {\sigma_0}
  \> H_Q(Q\OMIT{,\mu_Q},\mu_m) \: H_m\Big(m,\frac{Q}{m},\mu_m,\mu_\Gamma\Big)\! 
   \\
  &\times
  \int_{-\infty}^{\infty}\!\!\!d\ell^+ d\ell^-  
 \int_{-\infty}^{\infty}\!\!\!  d\ell^{\,\prime +}\: d\ell^{\,\prime -} \:
   \: U_{S}(\ell^{\,\prime +}, \ell^{\,\prime -},\mu_\Gamma,\mu_\Lambda)
  \nn\\
 &\times
  \> B_+\Big(\hat s_t - \frac{Q\ell^{+}}{m},\Gamma,\mu_\Gamma\Big) \:
     B_-\Big(\hat s_{\bar t} - \frac{Q\ell^{-}}{m},\Gamma,\mu_\Gamma\Big)\: 
  S(\ell^+\minus \ell^{\,\prime +},\ell^-\minus\ell^{\,\prime -},\mu_\Lambda) 
 \,,
 \nn
\end{align}
or equivalently
\begin{align} \label{bHQETcross-hem2aa}
  \frac{d^2\sigma }{dM^2_t dM^2_{\bar t}}
  &= 
    {\sigma_0}
     \> H_Q(Q\OMIT{,\mu_Q},\mu_m) \: H_m\Big(m,\frac{Q}{m},\mu_m,\mu_\Lambda\Big)\! \\
  &\quad \times
  \int_{-\infty}^{\infty}\!\!\!  d\hat s_t'\: d\hat s_{\bar t}' \:
  \: U_{B_+}(\hat s_t-\hat s_t',\mu_\Lambda,\mu_\Gamma)
  \: U_{B_-}(\hat s_{\bar t}-\hat s_{\bar t}',\mu_\Lambda,\mu_\Gamma)
  \nn\\
 &\quad \times
 \int_{-\infty}^{\infty}\!\!\!d\ell^+ d\ell^-  
  \> B_+\Big(\hat s_t' - \frac{Q\ell^{+}}{m},\Gamma,\mu_\Gamma\Big) \:
     B_-\Big(\hat s_{\bar t}' - \frac{Q\ell^{-}}{m},\Gamma,\mu_\Gamma\Big)\: 
  S(\ell^+,\ell^-,\mu_\Lambda) 
  \,.\nn
\end{align}
We will always take $\mu_\Gamma > \mu_\Lambda$ (although technically these
equations are still valid for the case $\mu_\Lambda > \mu_\Gamma$). The
evolution kernels $U_{B}$ and $U_S$ sum the large logs between $\mu_\Gamma$ and
$\mu_\Lambda$, while the large logs that only affect the overall normalization
are summed into $H_Q$ and $H_m$. In Fig.~\ref{fig:theoryII} we display three
equivalent ways to sum the large logs, labeled cases a), b), and c). In case a)
we run all terms, from the top-down, from $\mu_Q$ down to $\mu_\Gamma$, and we
run the soft function from the bottom-up starting at $\mu_\Lambda$ and ending at
$\mu_\Gamma$. In case b) we run the soft function from $\mu_\Lambda$ all the way
to $\mu_Q$, and the jet functions from $\mu_\Gamma$ to $\mu_m$, and then from
$\mu_m$ to $\mu_Q$.  Applying the consistency equations between $\mu_Q$--$\mu_m$
and $\mu_m$--$\mu_\Gamma$, cases a) and b) are equivalent, and both give the
result shown in Eq.~(\ref{bHQETcross-hem2a}). If we take case a) and apply the
consistency equation between $\mu_\Gamma$ and $\mu_\Lambda$ we obtain another
equivalent result, case c), with the result shown in
Eq.~(\ref{bHQETcross-hem2aa}). We will use case c) for our analysis.

In the previous section we derived a form of the factorization
formula~(\ref{bHQETcross-hem3}), which combines the finite lifetime effects and
the nonperturbative effects into an infrared function $R$. This form gives useful
insights for the proper choice of the scales $\mu_\Gamma$ and $\mu_\Lambda$.  In
terms of stable jet functions and $R$ in Eq.~(\ref{modifiedS}) the resummed
factorization theorem in Eq.~(\ref{bHQETcross-hem2aa}) becomes
\begin{align}
\label{bHQETcross-hem2b}
 \frac{d^2\sigma }{dM^2_t dM^2_{\bar t}} 
  &= 
    {\sigma_0}
     \> H_Q(Q\OMIT{,\mu_Q},\mu_m) \: H_m\Big(m,\frac{Q}{m},\mu_m,\mu_\Lambda\Big)\! 
  \\
  &\quad \times
  \int_{-\infty}^{\infty}\!\!\!  d\hat s_t'\: d\hat s_{\bar t}' \:
  \: U_{B_+}(\hat s_t-\hat s_t',\mu_\Lambda,\mu_\Gamma)
  \: U_{B_-}(\hat s_{\bar t}-\hat s_{\bar t}',\mu_\Lambda,\mu_\Gamma)
  \nn\\
 &\quad \times
 \int_{-\infty}^{\infty}\!\!\!d\ell^+ d\ell^-  
  \> B_+^{\Gamma=0}\Big(\hat s_t' - \frac{Q\ell^{+}}{m_J},\mu_\Gamma\Big) \:
     B_-^{\Gamma=0}\Big(\hat s_{\bar t}' - \frac{Q\ell^{-}}{m_J},\mu_\Gamma\Big)\: 
  R(\ell^+,\ell^-,\Gamma_t,\mu_\Lambda) 
 \,. \nn
\end{align}
From the convolution in this result we see that the smearing with $R$ provides
important information on the infrared cutoff for the fluctuations described by
jet functions, and hence the choice of $\mu_\Gamma$ that minimizes large logs.
Likewise, we see from the definition of $R$ in Eq.~(\ref{modifiedS}) and the form of
the soft function in Eq.~(\ref{Smodel1}) that $\mu_\Lambda$ is affected by a
smearing caused by nonperturbative effects as well as by the scale $m\Gamma/Q$
in the Breit-Wigner functions. Hence in the peak region we should run down to
the scales
\begin{align}
 \mu_\Gamma \simeq {\cal O}\Big(\Gamma_t+\frac{Q\Lambda}{m}+\frac{s_{t,\bar
     t}}{m} \Big) 
   \,,\qquad\qquad
 \mu_\Lambda \simeq {\cal O}\Big( \Lambda+\frac{m\Gamma_t}{Q} 
     + \frac{s_{t,\bar t}}{Q}\Big) \,.
\end{align}
In principle $\mu_\Gamma$ can be substantially larger than $\Gamma_t$ depending
on the $Q/m$ we are interested in. Also with a very large width (which does not
apply for the top quark), $\mu_\Lambda$ could be substantially larger than the
hadronic scale, which would allow for a perturbative prediction of the invariant
mass distribution in the peak region. For the realistic case of $\Gamma_t\sim
1.5$~GeV the scale where the logs would be strictly minimized is in the
nonperturbative regime, and we will specify the soft function at scales
$\mu_\Lambda\sim 1\,{\rm GeV}$ to be close to this regime. In the tail region we
have $\hat s\gtrsim\Gamma$ or $\hat s\gg \Gamma$ and the convolution in
Eq.~(\ref{bHQETcross-hem2b}) sets $\ell^\pm \sim m \hat s_{t,\bar t}/Q\gg
\Lambda$, so to sum the large logs in this region we should instead run down to
the scales
\begin{align}
 \mu_\Gamma \simeq {\cal O}\Big(\frac{s_{t,\bar t}}{m}\Big)  \,,\qquad\qquad
 \mu_\Lambda \simeq {\cal O}\Big( \frac{ s_{t,\bar t}}{Q} \Big) \,.
\end{align}

The results above are designed to study situations where $s\ll m^2$, which is
important for a precision extraction of the top-mass. In our formalism it is
also possible to study the cross section in the ultra-tail region, $|M_{t,\bar
  t}-m_J|\sim m_J$, with renormalization group improvement. This is the closest
analog to the resummation for massless event shapes in regions where the jet
invariant mass $M^2\gg Q\Lambda$.  In this case we use the SCET factorization
theorem in Eq.~(\ref{SFactThm}) which is valid as long as $Q^2\gg m^2,s$. Again
we take different renormalization scales for the jet functions ($\mu_m$) and
soft function ($\mu_\Delta$) as shown in Fig.~\ref{fig:theoryIII}.  The SCET
factorization theorem in the jet mass scheme is [here $m_J=m_J(\mu_m)$]
\begin{align} \label{sigmaMMscet}
  \frac{ d^2\sigma }{dM_t^2\, dM_{\bar t}^2}   
  &  =  \sigma_0\: H_Q(Q\OMIT{,\mu_Q},\mu_m) {\cal M}(m_J,\mu_m)
  \int_{-\infty}^{\infty}\!\!\!d\ell^+ d\ell^-  
 \int_{-\infty}^{\infty}\!\!\!  d\ell^{\,\prime +}\: d\ell^{\,\prime -} \:
   \: U_{S}(\ell^{\,\prime +},\ell^{\,\prime -}\mu_m,\mu_\Delta) \,
 \\[1mm]
 &\, \times 
J_n(s_t-Q\ell^+,m_J, \mu_m)
J_{\bar{n}}(s_{\bar{t}}-Q\ell^-,m_J,\mu_m)
\,S_{\rm part}(\ell^+-\ell^{\prime +}, \ell^--\ell^{\prime -},\mu_\Delta,m_J) 
\nn\\[3mm]
 & \hspace{-1.6cm} 
 = \sigma_0\: H_Q(Q\OMIT{,\mu_Q},\mu_m){\cal M}(m_J,\mu_m)
  U_{H_Q}^{(5)}(Q,\mu_m,\mu_\Delta)\!\!
  \int_{-\infty}^{\infty}\!\!\!\!\!  d s_t'\: d s_{\bar t}' \:
  \: U_{J}(s_t\minus s_t',\mu_\Lambda,\mu_m)
  \: U_{J}(s_{\bar t}\minus s_{\bar t}',\mu_\Lambda,\mu_m) 
 \nn \\[1mm] 
&\hspace{-1.4cm} \, \times 
\int^\infty_{-\infty} d\ell^+  d\ell^-
J_n(s_t^\prime-Q\ell^+,m_J, \mu_m)
J_{\bar{n}}(s_{\bar{t}}^\prime-Q\ell^-,m_J,\mu_m) \,S_{\rm part}(\ell^+, \ell^-,\mu_\Delta,m_J)
  \nn .
\end{align}
With analogy to the discussion above we show three equivalent ways to sum the
large logs in Fig.~\ref{fig:theoryIII}, cases a), b), and c).  Here cases a) and
b) give the first result displayed in Eq.~(\ref{sigmaMMscet}) and are related by
applying the consistency equation between $\mu_Q$ and $\mu_m$. Case c) gives the
second result in Eq.~(\ref{sigmaMMscet}) and is related to case a) by using the
consistency condition between $\mu_m$ and $\mu_\Delta$. In the region below $m$
where top-bubbles have been integrated out we have $n_f=5$ active flavors, so
$n_f=5$ for $U_S$ in case a), and $n_f=5$ for
$U_{H_Q}^{(5)}(\mu_m,\mu_\Delta)$ and $U_J$ in case c). For $H_Q(Q,\mu_m)$
we always have $n_f=6$.  In the ultra-tail region we generically have
$s_{t,\bar t}\sim m^2$, and hence the logs in the jet functions are minimized
with $\mu_m\sim m$.  The convolution with the soft function involves momenta
$\ell^\pm \sim s/Q\sim m^2/Q$, and hence the large logs are summed for
$\mu_\Delta \simeq m^2/Q$.

\section{Renormalization and Anomalous Dimensions} \label{sect:RGE}

In this section we setup our notation and conventions for renormalization of the
quantities defined in Sec.~\ref{sec:fsummary} in the $\overline {\rm MS}$
scheme.  In QCD the vector current is conserved (although in ${\overline {\rm
    MS}}$ one must be careful with the definition~\cite{Collins:2005nj}), but in
the effective theory the currents are renormalized. In the running pictured in
Fig.~\ref{fig:theoryII} we have both local running (anomalous dimensions that
depend only on conserved kinematic variables) and convolution running (anomalous
dimensions that depend on variables that can be changed by dynamics in another
sector). Convolution running involves an integration over anomalous dimensions
that are functions.  The coefficients $H_Q$ and $H_m$ have local running, while
the functions $J_{n,\bn}$, $B_\pm$, and $S$ have convolution running. In this
context an example of local running are the logarithms summed up by the
RG-evolution of gauge couplings, and an example of convolution running are logs
summed by the Altarelli-Parisi evolution equations (which are collinear UV logs
in SCET).  Another important attribute of these functions is whether their
anomalous dimensions involve $\ln(\mu)$ factors and hence sum double Sudakov
logarithms.  These $\ln(\mu)$ factors are induced by cusp angles involving
light-like Wilson
lines~\cite{Polyakov:1980ca,Arefeva:1980zd,Gervais:1978mp,Dotsenko:1979wb,Brandt:1981kf,Aoyama:1981ev}.

Thus, in considering the renormalization group evolution in SCET and bHQET the
physical meaning of the logarithms that are being summed depends on which of
four cases we are in: 1) local single logs, 2) local double logs, 3) convolution
with single logs, 4) convolution with double logs.  In case 1) we have local
running without a $\ln(\mu)$ in the anomalous dimension, and the evolution just
corresponds to the change of a coupling constant $c(\mu)$ from integrating out
virtual effects.  This is the standard case, well known from the running of the
gauge couplings and of the electroweak effective Hamiltonian of four-quark
operators. On the other hand, the UV renormalization in cases 2), 3), 4) are
induced by particular types of phase space restrictions on real radiation that
are built into the effective theory. In case 2) we have local running with a
$\ln(\mu)$ in the anomalous dimension, while in case 3) and 4) we have
convolution running without and with a $\ln(\mu)$ in the anomalous dimension,
respectively. These cases are discussed further in Appendix~\ref{App:genRGE},
and will be mentioned as they arise in the analysis below.

\subsection[1.]{ SCET renormalization}
\label{sssec:scetrenorm}

{\it {Top-down running.} }
In SCET we can renormalize the current $J_i^\mu$ by switching from a bare to
renormalized Wilson coefficient,
\begin{align} \label{CZc}
  C^{\rm bare} = Z_c\: C =  C + (Z_c-1) C \,,
\end{align}
where insertions of $(Z_c-1)C$ are treated as counterterms that render
insertions of the current together with the bare Wilson coefficient
UV-finite. Field, coupling, and mass renormalization are given by
\begin{align} \label{Zscet}
 \xi_n^{\rm bare} &= Z_\psi^{1/2} \xi_n \,,
 & A_n^{\rm bare} &= Z_A^{1/2} A_n \,,
 & m^{\rm bare} &= m + \delta_m\,,
 &g^{\rm bare} & = Z_g \mu^{\epsilon} g \,,
\end{align}
and are all identical to those in
QCD~\cite{Bauer:2000ew,Bauer:2000yr,Chay:2005ck}.\footnote{This is true to all
  orders in $\alpha_s$ because there are no zero-bin
  subtractions~\cite{Manohar:2006nz} for the collinear two-point functions. To
  see this note that all soft loop corrections to these functions vanish in
  Feynman gauge since $n^2=0$. Thus there is no region that is double counted
  and would require a subtraction.}  For later convenience we also write the
mass counterterm as
\begin{align}
 \delta_m = (\delta_m)^{\rm pole} + \delta m \,,
\end{align}
where $(\delta_m)^{\rm pole}$ is the counterterm in the pole-scheme, and for
mass schemes other than the pole scheme the remainder, $\delta m=m^{\rm
  pole}-m$, contains finite perturbative corrections.  Eqs.~(\ref{CZc}) and
(\ref{Zscet}) suffice to cancel all UV divergences involving $J_i^\mu$. The SCET
factorization theorem in Eq.~(\ref{SFactThm}) is generated by a time-ordered
product of two $J_i^\mu$ currents. The factorization theorem shown in
Eq.~(\ref{SFactThm}) only involves renormalized objects.  These depend on the
choice of renormalization scheme in SCET, but this dependence cancels out
between $H_Q$, $J_n$, $J_\bn$, and $S$. The renormalization group equation for
$C$ and $H_Q$ are
\begin{align} \label{gammacgammaH}
  \mu \frac{d}{d\mu} C(Q,\mu) & = \gamma_c(Q,\mu)\, C(Q,\mu) \,,
  & \mu \frac{d}{d\mu} H_Q(Q,\mu) &= \gamma_{H_Q}(Q,\mu)\, H_Q(Q,\mu)\,, 
\end{align}
where from Eq.~(\ref{CZc}) $\gamma_{c} = - Z_{c}^{-1} \mu d/d\mu\, Z_{c}$, and
since $H_Q=|C|^2$ we have $\gamma_{H_Q} =\gamma_c + \gamma_c^*$. Since the
current $J_i^\mu$ involves light-like Wilson lines in the $n$ and $\bn$
direction, the anomalous dimension has a $\ln(\mu/Q)$ cusp anomalous
dimension term. The general form is
\begin{align} \label{gammaHQ}
  \gamma_{H_Q}(Q,\mu) = \Gamma_{H_Q}\big[\alpha_s\big]
  \ln\Big(\frac{\mu^2}{Q^2}\Big) + \gamma_{H_Q}\big[\alpha_s\big] \,.
\end{align}
The running of $H_Q$ is in case 2) and sums local double logs. Here the
current is affected by small invariant mass phase space
restrictions imposed on real radiation, which leads to an incomplete
cancellation of real and virtual contributions from soft and collinear
effects. Once we integrate out virtual effects in the EFT and evolve
the current down to the scale of these restrictions the cancellation
again becomes effective. This is manifest through the elimination of
large logarithms in EFT matrix elements at the low scale. The process
of integrating out virtual effects and performing the RG-evolution
sums double logs between the production scale $Q$ and scale of the
phase space restrictions. For the solution to the RGE equation for
$H_Q$ we write 
\begin{align} \label{UH}
  H_Q(Q\OMIT{,\mu_Q},\mu) = H_Q(Q,\mu_Q)\, U_{H_Q}(Q,\mu_Q,\mu)\,,
\end{align}  
where $H_Q(Q,\mu_Q)$ is the matching condition at the hard scale of
order $Q$ and  $U_{H_Q}(\mu_Q,\mu)$ the evolution factor with 
$\mu_Q>\mu$. The evolution contained in $U_H$ is illustrated in
Fig.~\ref{fig:theoryII}.

{\it Bottom-Up Running.} It is well known that there is an alternative but
equivalent way to renormalize composite operators like $J_i^\mu$, which is often
referred to as operator renormalization (see Ref.~\cite{Buras:1998ra} for a
review). In this approach the UV-divergences in matrix element insertions of the
bare operators are absorbed into the renormalization $Z$-factors multiplying
UV-finite renormalized operators, $(J_i^\mu)^{\rm bare} = Z_J J_i^\mu$. The
equivalence of the two approaches implies that $Z_J=(Z_c^{-1})^T$, where the
transpose is only relevant in case of a multidimensional operator basis.

Here we consider a variant of operator renormalization that introduces
$Z$-factors for the objects $J_n$, $J_\bn$, and $S$ in the SCET factorization
theorem, Eq.~(\ref{SFactThm}). We will refer to this procedure as factorized
operator renormalization. In Sec.~\ref{sec:fsummary} these objects were defined
by matrix elements of time-ordered products of fields, where each involves a
subset of the fields contained in the current $J_i^\mu$. To switch from bare to
renormalized matrix elements we write
\begin{align} \label{ZJJS}
  J_{n,\bar n}^{\rm bare}(s) &= \int\!\! ds'\: Z_{J_{n,\bar n}}(s\minus s',\mu)\:
  J_{n,\bar n}(s',\mu) \,,
\nn\\[3pt]
  S^{\rm bare}(\ell^+,\ell^-) &=\int\!\! d\ell^{\,\prime +} d\ell^{\,\prime -} \:
      Z_S(\ell^+\minus \ell^{\,\prime +},\ell^- \minus \ell^{\,\prime -},\mu )\: 
    S(\ell^{\,\prime +},\ell^{\,\prime -},\mu)\,.
\end{align}
These equations can be inverted using $\int ds\: Z_{J_n}^{-1}(s''-s)
Z_{J_n}(s-s') = \delta(s''-s')$ etc. Note that the $Z$-factors only depend on
differences of momenta because the renormalization is local for position space
fields (as discussed further in Appendix~\ref{App:genRGE}). The RGE's read
\begin{align} \label{rgeJS}
  \mu\frac{d}{d\mu} J_{n,\bn}(s,\mu) &= \int\!\! ds' \: \gamma_{J_{n,\bn}}(s\minus s',\mu)\:
  J_{n,\bn}(s',\mu) ,  \\
   \mu\frac{d}{d\mu} S(\ell^+,\ell^-,\mu) &= \int\!\! d\ell^{\,\prime +}
   d\ell^{\,\prime -} \: 
   \gamma_S(\ell^+\minus \ell^{\,\prime +},\ell^-\minus \ell^{\,\prime -},\mu)
   S(\ell^{\,\prime +},\ell^{\,\prime -},\mu) \,,\nn
\end{align}
with the anomalous dimensions being defined as
\begin{align}
\label{scetanomdims}
  \gamma_{J_{n,\bn}}(s\minus s',\mu) &= - \int\!\! ds''\: Z_{J_{n,\bn}}^{-1}(s\minus s'',\mu)
  \mu\frac{d}{d\mu} Z_{J_{n,\bn}}(s''\minus s',\mu) 
  \,,\\
  \gamma_S(\ell^+\!\minus \ell^{\,\prime +},\ell^-\!\minus \ell^{\,\prime -},\mu)
   &= - \int\!\! d\ell^{''+} d\ell^{''-} 
   Z_S^{-1}(\ell^+\!\minus \ell^{'' +}, \ell^-\! \minus\ell^{'' -},\mu)
    \mu\frac{d}{d\mu} 
   Z_S(\ell^{'' +}\!\minus \ell^{\,\prime+},\ell^{'' -}\!\minus
  \ell^{\,\prime-},\mu) \,.\nn
\end{align}
Renormalizability of the theory requires that they are finite as $\epsilon\to
0$, and the general form for these anomalous dimension is discussed in
Appendix~\ref{App:genRGE}. For the solutions of the RGE's we write
\begin{align} \label{UJS}
 J_{n,\bar n}(s,\mu) &= \int\!\! ds'\ U_{J}(s\minus
 s',\mu,\mu_0)\  J_{n,\bar n}(s',\mu_0)
 \,,\\[3pt]
 S(\ell^+,\ell^-,\mu) &= \int\!\! d\ell^{\,\prime +}  d\ell^{\,\prime -} \: 
   U_S(\ell^+\minus \ell^{\,\prime +},\ell^-\minus \ell^{\,\prime -},\mu,\mu_0)\:
   S(\ell^{\,\prime +},\ell^{\,\prime -},\mu_0) \,. \nn
\end{align}
Note that depending on the set up of scales, as illustrated in
Fig.~\ref{fig:theoryII}, the evolution kernels $U_{J}$ and $U_S$ evolve to
higher scales ($\mu>\mu_0$) or to lower scales ($\mu<\mu_0$).

At any order in $\alpha_s$ the anomalous dimensions in Eq.~(\ref{scetanomdims})
have the general form
\begin{align} \label{gammaF}
  \gamma_F(t\minus t',\mu) & = - \frac{2 \Gamma[\alpha_s]}{j\,\mu^j} \bigg[
  \frac{\mu^j \theta(t\minus t')}{t\minus t'}\bigg]_+ \!\! 
  + \gamma[\alpha_s] \: \delta(t-t')\,,
\end{align}
where $j$ is the dimension of the convolution variable $t'$. Although the soft
function anomalous dimension has two variables, we will show below in
section~\ref{subsec:consistency} that it can be written as
\begin{align} \label{gdgd}
   \gamma_S(\ell^+,\ell^-) = \delta(\ell^+)\, \gamma_s(\ell^-) +
   \delta(\ell^-)\, \gamma_s(\ell^+) \,,
\end{align}
where $\gamma_s$ has the form in Eq.~(\ref{gammaF}). Eq.~(\ref{gammaF}) involves
a plus-function, which we define by the limit in Eq.~(\ref{dim-plus}).  This is
similar to the Altarelli-Parisi kernel in deep-inelastic scattering, except for
the presence of $\mu^j$. This explicit dependence on $\mu$ must appear to make
the plus-function dimensionless, and using Eq.~(\ref{rescale}) can be written as
a a $\ln(\mu)$ factor multiplying a $\delta(t-t^\prime)$. Thus they sum double
logs, making them fall in case 4), which is a combination of case 2) described
above and case 3). For case 3) the real and virtual effects cancel for the soft
contributions, but not for collinear ones, leaving single logarithms to be
summed by the RGE's. The convolution with plus-functions arises because there
are angular restrictions on the radiation such as those that occur when an
energetic proton state absorbs partons in DIS.  Summing logs in this case
involves a convolution since the logarithms generated by the collinear effects
depend on a momentum fraction.

Viewed from the bottom-up each of the jet and soft functions in our
factorization theorem has an evolution equation corresponding to case 4).
However, in SCET when the soft function and collinear-jets are combined in the
factorization theorem their convoluted product no longer has angular
restrictions. So the evolution of the product does not involve a convolution.
The product still restricts the radiation to small invariant mass and so falls
into case 2), of local running with double logs as we mentioned above.

\subsection{bHQET renormalization}
\label{subsec:bHQETrenor}

{\it Top-Down Running.} Next we discuss the renormalization in
bHQET. The renormalization constant for the bHQET current for the
counterterm method is defined as
\begin{align} \label{Zcm}
  C_m^{\rm bare} = Z_{C_m}\: C_m 
    =  C_m + (Z_{C_m}-1) C_m \,.
\end{align}
While gluon field and coupling renormalization in HQET and QCD are equivalent,
the top quark field renormalization differs, with $h_v^{\rm bare} = Z_h^{1/2}
h_v$. The bHQET factorization theorem in Eq.~(\ref{FactThm}) is generated by a
time-ordered product of two $J_{\rm bHQET}^\mu$ currents~\cite{Fleming:2007qr}.
The soft graphs in bHQET are identical to those in SCET up to top-quark vacuum
polarization graphs~\cite{Fleming:2007qr}, and the infrared divergences of the
collinear graphs in SCET exactly match those in bHQET. The mass-mode function
${\cal M}$ is IR finite and just enters in the $H_m$ matching coefficient. Thus,
the same cancellation between collinear and soft graphs that yielded local
running in SCET also occurs in bHQET. So the running of $C_{m}$ is also local.
We will demonstrate this explicitly in the one-loop computations shown below.

Next recall that the $+$ and $-$ bHQET sectors are decoupled, so the anomalous
dimension for $C_{m}$ can only depend on the quantities $\bn\cdot v_- =Q/m$,
$n\cdot v_+=Q/m$, and $n\cdot \bn=2$. With this theory we are interested in
studying small invariant mass fluctuations {\em around} the top quark mass $m$.
Thus the renormalization group evolution is not related to stronger kinematic
restrictions on the magnitude of the overall invariant mass of top plus lighter
degrees of freedom.  Here the evolution falls into case 1) rather than case 2).
However, the anomalous dimension of the bHQET current $J_{\rm bHQET}^\mu$ still
contains a remnant of the $\ln(\mu/Q)$ term in Eq.~(\ref{gammaHQ}) in the form
of a dependence on $\ln(m/Q)$. This $\mu$-independent logarithmic term is
related to a cusp between Wilson lines. This can be made explicit through the
field redefinition $h_{v_\pm}\to W_{v_\pm} h_{v_\pm}^{(0)}$, where $W_{v_\pm}$
are Wilson lines defined in analogy to Eq.~(\ref{bHQETWilsondef}) and
$h_{v_\pm}^{(0)}$ are heavy quark fields that no longer couple to gluon fields
at leading power.  For the operator $\bar h_{v_+}W_n(0)$ that appears for
example in the bHQET current of Eq.~(\ref{JbHQET}) this leads to $\bar
h_{v_+}^{(0)}W_{v_+}^\dagger W_n(0)$.  Insertions of this operator lead to the
anomalous dimension depending on logarithms of the cusp angle $n\cdot
v_+=Q/m$.~\cite{Korchemsky:1991zp,Gatheral:1983cz,Frenkel:1984pz}.  Unlike SCET,
this angle is fixed and independent of $\mu$ because the overall invariant mass
is $\sim m^2$ and does not become parametrically smaller from the RG evolution.

The RG equations for $C_m$ and $H_m=|C_m|^2$ are
\begin{align}\label{hqetrunning}
  \mu \frac{d}{d\mu} C_m\Big(m,\frac{Q}{m},\mu\Big)
   &= \gamma_{C_m}\Big(\frac{Q}{m},\mu\Big)\, C_m\Big(m,\frac{Q}{m},\mu\Big) \,,\nn\\
  \mu \frac{d}{d\mu} H_m\Big(m,\frac{Q}{m},\mu\Big)
   &= \gamma_{H_m}\Big(\frac{Q}{m},\mu\Big)\, H_m\Big(m,\frac{Q}{m},\mu\Big) \,,
\end{align}
where $\gamma_{C_m} = -Z_{C_m}^{-1}\: \mu\, d/d\mu\, Z_{C_m}$ and $\gamma_{H_m}
= \gamma_{C_m}+\gamma_{C_m}^*$, and the general form of the anomalous dimension is
\begin{align} \label{gammaHm}
  \gamma_{H_m}(Q/m,\mu) =  \Gamma_{H_m}\big[\alpha_s\big]
  \ln\Big(\frac{m^2}{Q^2}\Big) + \gamma_{H_m}\big[\alpha_s\big] \,.
\end{align}
We write the solution to Eq.~(\ref{hqetrunning}) as
\begin{align} \label{UHm}
  H_m\Big(m,\frac{Q}{m},\mu_m,\mu\Big) =  H_m(m,\mu_m)\, 
  U_{H_m}\Big(\frac{Q}{m},\mu_m,\mu\Big) \,,
\end{align}  
where $H_m(m,\mu_m)$ is the matching condition of the bHQET current at
the SCET-bHQET matching scale $\mu_m\sim m$ and $U_{H_m}(Q/m,\mu_m,\mu)$
is the evolution factor describing the running to a scale $\mu<\mu_m$.   
The local evolution generated by $U_{H_m}$ is
illustrated in Fig.~\ref{fig:theoryII}. Note that the RHS of Eq.~(\ref{UHm}) is
not $\mu_m$ independent at the order one is working, since part of this
dependence is canceled by the $U_Q(\mu_Q,\mu_m)$ in $H_Q(Q,\mu_m)$. This is
indicated by the $\mu_m$ argument on the LHS of Eq.~(\ref{UHm}).

{\it Bottom-Up Running.} Next consider the equivalent approach of factorized
operator renormalization in bHQET. In this case we introduce $Z$-factors for the
jet functions $B_\pm$ and the soft function $S$ rather than counterterm
contributions for the bHQET current. The resulting evolution equations for the
soft function $S$ agree with those in SCET except for the change from $n_f=6$ to
$n_f=5$, and will not be repeated. To switch from bare to
renormalized HQET jet functions we write
\begin{align} \label{ZB}
  B_\pm^{\rm bare}(\hat s) &= \int\!\! d\hat s'\:
   Z_{B_\pm}(\hat s\minus \hat s',\mu)\: B_\pm(\hat s',\mu) \,,
\end{align}
where $\int d\hat s\: Z_{B_\pm}^{-1}(\hat s''-\hat s,\mu) Z_{B_\pm}(\hat s-\hat s',\mu)
= \delta(\hat s''-\hat s')$.  The RG equations are
\begin{align} \label{rgeB}
  \mu\frac{d}{d\mu} B_{\pm}(\hat s,\mu) &= \int\!\! d\hat s' 
  \: \gamma_{B_\pm}(\hat s\minus \hat s',\mu)\:B_{\pm}(\hat s',\mu) , 
\end{align}
with anomalous dimension
\begin{align} \label{gZB}
  \gamma_{B_\pm }(\hat s\minus \hat s',\mu) &= - \int\!\! d\hat s''\: 
  Z_{B_\pm }^{-1}(\hat s\minus \hat s'',\mu)\:
  \mu\frac{d}{d\mu}\, Z_{B_\pm }(\hat s''\minus \hat s',\mu) \,.
\end{align}
The general form for this anomalous dimension can be found in
Appendix~\ref{App:genRGE}.  For the solutions to the RGE we write
\begin{align} \label{UB}
 B_\pm (\hat s,\mu) &= \int\!\! d\hat s'\ U_{B}(\hat s\minus \hat
 s',\mu,\mu_\Gamma)\ B_\pm(\hat s',\mu_\Gamma)
  \,.
\end{align}
The evolution kernels $U_{B}$ take us from the low-scale $\mu_\Gamma$ to a
scale $\mu$. Depending on the set up of scales, as shown in
Fig.~\ref{fig:theoryII}, we can have $\mu>\mu_\Gamma$ or
$\mu<\mu_\Gamma$.

\subsection{Consistency Conditions in SCET and bHQET}
\label{subsec:consistency}

In this section we derive the factorization of the soft-function evolution factor in
Eq.~(\ref{US1}) and the SCET and bHQET consistency equations quoted above in
Eq.~(\ref{CE}).

Using Eq.~(\ref{ZJJS}) we can obtain a finite result for the SCET factorization
theorem by determining the UV-divergences for the $Z$-factors $Z_{J_{n,\bar n}}$
and $Z_S$ from each individual SCET Feynman diagram contributing to $J_{n,\bar
  n}$ and $S$. If we instead use the counterterm method with the current
renormalization factor $Z_c$ then a consistent form for the counterterm is only
obtained once all collinear and soft vertex graphs that contribute to the
factorization theorem at some order in $\alpha_s$ are added up.  Since the two
methods render UV-finite results and lead to the same predictions, there is a
consistency relation between the renormalization constants for the operator and
the counterterm renormalization method which is very useful for practical
computations.  To derive it we start with Eq.~(\ref{SFactThm}) and switch to
$J_n^{\rm bare}$, $J_\bn^{\rm bare}$, and $S^{\rm bare}$ using first counterterm
renormalization and then factorized operator renormalization. Equating the
results we find that
\begin{align} \label{cons1}
  |Z_c|^2\: \delta(s\minus Q\ell^{\,\prime+})\, \delta(\bar s\minus Q\ell^{\,\prime-})
  &= \!\int\!\! d\ell^+ d\ell^- \:
     Z_{J_n}^{-1}(s\minus Q\ell^+)\: Z_{J_\bn}^{-1}(\bar s\minus Q\ell^-)\:
     Z_S^{-1}(\ell^+\minus \ell^{\,\prime +}, \ell^- \minus \ell^{\,\prime -}) \,.
\end{align}
The consistency condition can also be written in terms of the evolution kernels
that solve the individual RGE's.  To derive this form we consider the
factorization theorem Eq.~(\ref{SFactThm}) at the scale $\mu_0$, and use
$H_Q(Q,\mu_0)= H_Q(Q,\mu)U_{H_Q}(\mu,\mu_0)$. Then we write down the factorization
theorem Eq.~(\ref{SFactThm}) again at the scale $\mu$ and relate the $J_{n,\bar
  n}$ and $S$ at the scale $\mu$ to those evaluated at $\mu_0$ using
Eqs.~(\ref{UJS}). Equating the two results gives the consistency condition
\begin{align} \label{cons2}
  & U_{H_Q}(Q,\mu,\mu_0)  \, \delta(s\minus Q\ell^{\,\prime +}) 
     \, \delta(\bar s\minus Q\ell^{\,\prime -}) \\
  &\qquad =
  \int\!\! d\ell^+ d\ell^-\: U_{J_n}(s\minus Q\ell^+,\mu,\mu_0) 
    U_{J_\bn}(\bar s\minus Q\ell^-,\mu,\mu_0) 
    U_S(\ell^+\minus\ell^{\,\prime +},\ell^-\minus \ell^{\,\prime -},\mu,\mu_0) \,.\nn
\end{align}

Next we multiply Eq.~(\ref{cons2}) by $U_{J}(Q\ell^{\,\prime\pm},\mu_0,\mu)$,
shift $\ell^\pm \to \ell^\pm + \ell^{\prime\pm}$, and integrate over
$\ell^{\,\prime\pm}$ to turn the products of $U_{J}$ factors on the RHS
into delta functions. Carrying out the $\ell^\pm$ integrals then  leaves
\begin{align} \label{Usfactor}
  U_S\Big(\frac{s}{Q},\frac{\bar s}{Q},\mu,\mu_0\Big) &=
   Q^2\, U_{H_Q}(Q,\mu,\mu_0)\, U_{J_n}(s,\mu_0,\mu)\, U_{J_\bn}(\bar
   s,\mu_0,\mu) \,.
\end{align}
This implies a separable structure for $U_S$ to all orders in
perturbation theory, so we write 
\begin{align} \label{UUorg}
  U_S(\ell^+,\ell^-,\mu,\mu_m)
   = U_s(\ell^+,\mu,\mu_m)\, U_s(\ell^-,\mu,\mu_m) .
\end{align}
This result for $U_S$ implies 
\begin{align}
 \mu\frac{d}{d\mu} U_S(\ell^+,\ell^-) &= 
  \Big[ \mu\frac{d}{d\mu} U_s(\ell^+) \Big] U_s(\ell^-)
   +    U_s(\ell^+) \Big[ \mu\frac{d}{d\mu} U_s(\ell^-) \Big] \\
 &= \int\!\! d\ell^{\prime +} d\ell^{\prime -}
  \big[ \gamma_s(\ell^+\!\minus\ell^{\prime +})
   \delta(\ell^-\!\minus\ell^{\prime -})
 + \delta(\ell^+\!\minus\ell^{\prime +})
     \gamma_s(\ell^-\!\minus\ell^{\prime -}) \big]
 U_s(\ell^{\prime +}) U_s(\ell^{\prime -}) \,,\nn
\end{align}  
so the soft function anomalous dimension has the general form shown in
Eq.~(\ref{gdgd}).  Now using the fact that $U_{J_n}=U_{J_\bn}=U_J$,
Eqs.~(\ref{Usfactor}) and (\ref{UUorg}) give the final result for the SCET
consistency equation
\begin{align}
\label{conssimple1}
\sqrt{U_{H_Q}(Q,\mu,\mu_0)}\,U_{J}(s,\mu_0,\mu) &=
\frac{1}{Q}\,U_s\Big(\frac{s}{Q},\mu,\mu_0 \Big)
\,.
\end{align}
This relation expresses the equivalence of running the factorization theorem
between $\mu_Q$ and $\mu_m$ from top-down versus from bottom-up as pictured in
Fig.~\ref{fig:theoryII}.  It also states that when the convolution RGE's for
each of $J_n$, $J_\bn$, and $S$ are combined as shown in the factorization
theorem, the result is local running through $U_{H_Q}$ without a convolution.
This means in particular, that the renormalization group evolution of the soft
function does not depend on the phase space constraints that are imposed
dividing up the soft radiation in $M_t$ and $M_{\bar t}$. This is verified
explicitly at ${\cal O}(\alpha_s)$ in Sec.~\ref{sect:universal} where we show
that the anomalous dimension of the soft function is unchanged if invariant mass
prescriptions are applied that differ from the hemisphere prescription.

Next lets derive the consistency equation in bHQET. Again the use of the
factorized operator renormalization using $Z_{B_\pm}$ and $Z_S$ correspond to
determining the UV-divergences of the individual Feynman diagrams contributing
to $B_+$, $B_-$, and $S$. If we instead use current renormalization via
$Z_{C_m}$ then the consistent form of the counterterm is only obtained once all
vertex graphs contributing to the factorization theorem at a certain order in
$\alpha_s$ are added up. In analogy to SCET this leads to consistency conditions
for the renormalization factors and the solutions of the anomalous dimensions.
The derivation goes along the same lines as in the SCET case, but starting from
Eq.~(\ref{FactThm}). The consistency condition for the renormalization factors is
\begin{align} \label{cons3}
  |Z_{C_m}|^2\: \delta\Big(\hat s\minus \frac{Q\ell^{\,\prime+}}{m}\Big)\, 
  \delta\Big(\hat{\bar s}\minus \frac{Q\ell^{\,\prime-}}{m}\Big)
  &= \!\int\!\! d\ell^+ d\ell^- \:
     Z_{B_+}^{-1}\Big(\hat s\minus \frac{Q\ell^+}{m}\Big)\: 
     Z_{B_-}^{-1}\Big(\hat {\bar s}\minus \frac{Q\ell^-}{m}\Big)\: \nn\\
  &\qquad\times
     Z_S^{-1}(\ell^+\minus \ell^{\,\prime +}, \ell^- \minus \ell^{\,\prime -}) \,,
\end{align}
and for the evolution kernels reads 
\begin{align} \label{cons4}
  & U_{H_m}\Big(\frac{Q}{m},\mu,\mu_0\Big)  \, 
 \delta\Big(\hat s\minus \frac{Q\ell^{\,\prime+}}{m}\Big)\, 
  \delta\Big(\hat{\bar s}\minus \frac{Q\ell^{\,\prime-}}{m}\Big) \\
  &\qquad =
  \int\!\! d\ell^+ d\ell^-\:
    U_{B_+}\Big(\hat s\minus \frac{Q\ell^+}{m},\mu,\mu_0\Big) 
    U_{B_-}\Big(\hat {\bar s}\minus \frac{Q\ell^-}{m},\mu,\mu_0\Big) 
    U_S(\ell^+\minus\ell^{\,\prime +},\ell^-\minus \ell^{\,\prime
      -},\mu,\mu_\Lambda)
   \,.\nn
\end{align}
Removing the $\delta$-functions and integrals in an analogous manner to what we
did for SCET above, we
obtain the final bHQET consistency condition
\begin{align}
\label{conssimple2}
\sqrt{U_{H_m}\Big(\frac{Q}{m},\mu,\mu_0\Big)}
\,U_{B}(\hat s,\mu_0,\mu) &=
\frac{m}{Q}\,U_s\Big(\frac{m \hat s}{Q},\mu,\mu_0\Big)
\,.
\end{align}
This result expresses the equivalence of running the factorization theorem
between $\mu_m$ and $\mu_\Lambda$ using either top-down or bottom-up
approach, as illustrated in Fig.~(\ref{fig:theoryII}). It also states that when
the convolution RGE's for each of $B_+$, $B_-$, and $S$ are combined as shown in
the factorization theorem that the result is local running for $H_{m}$ through
$U_{H_m}$ without a convolution.

These consistency conditions are important phenomenologically because they state
that the RG-evolution from the hard scales down to a common low energy scale for
jet and soft functions does not affect the shape of the invariant mass
distributions. Since we have a consistency condition in both SCET and bHQET the
mass scale $m$ does not affect this protection of the invariant mass shape from
large log modification.  The smooth transition between the SCET and bHQET
consistency conditions is related to a correspondence between geometry and the
dimension of the variables in the factorization theorem, as we discuss in
Appendix~\ref{App:genRGE}. Once the $B_\pm$ jet functions reach the scale
$\mu_\Gamma$ where their logs are minimized, then further evolution of the soft
function to $\mu_\Lambda$ generates logs that affect the shape of the invariant
mass distribution.

\subsection{NLL Resummation} \label{sect:NLL}

To sum large logarithms to NLL we must solve Eq.~(\ref{gammaHQ}) for $U_{H_Q}$,
Eq.~(\ref{gammaF}) for $U_F$, and Eq.~(\ref{gammaHm}) for $U_{H_m}$. As
discussed in Appendix~\ref{App:genRGE} the general solutions  are
\begin{align} \label{UUU}
  U_{H_Q}(Q,\mu_0,\mu) 
   &= e^K \: \Big( \frac{\mu_0^2}{Q^2}\Big)^\omega\,,\qquad\qquad\quad
  U_{H_m}\Big(\frac{Q}{m},\mu_0,\mu\Big) 
   = e^{K_\gamma}  \: \Big( \frac{m^2}{Q^2}\Big)^\omega\,,
  \nn\\
  U_F(t-t',\mu,\mu_0) &= \frac{e^{K}\: \big(
  e^{\gamma_E}\big)^{ \omega}}{\mu_0^j\,\Gamma(- \omega)} \:
    \bigg[ \frac{(\mu_0^j)^{1+\omega}\theta(t\minus t')}{(t\minus
    t')^{1+\omega}}\bigg]_+ \,.
\end{align}
Here $\omega=\omega(\mu,\mu_0)$, $K=K(\mu,\mu_0)$ and
$K_\gamma=K_\gamma(\mu,\mu_0)$ are solutions to the integrals in
Eq.~(\ref{wLfull}). Note that for $\omega$, $K$, and $K_\gamma$ we use the
notation that the first argument $\mu$ is always the final scale of the
evolution while the second argument $\mu_0$ is always the initial scale. This
notation is also chosen for the $U_F$ evolution factors of the soft and jet
functions, but differs from our notation for the evolution factors for the
current Wilson coefficients, $U_{H_Q}$ and $U_{H_m}$ where the opposite ordering
is used.  Thus writing
\begin{align} \label{GammaSeries}
  & \Gamma[\alpha_s] = \frac{\alpha_s(\mu)}{4\pi} \Gamma_0 
     +  \Big[ \frac{\alpha_s(\mu)}{4\pi} \Big]^2\, \Gamma_1 + \ldots \,,
  & \gamma[\alpha_s] &=  \frac{\alpha_s(\mu)}{4\pi} \gamma_0 
     +  \Big[ \frac{\alpha_s(\mu)}{4\pi} \Big]^2\, \gamma_1 + \ldots \,,
\end{align}
we must determine $\Gamma_0$, $\Gamma_1$, and $\gamma_0$. The determination of
$\Gamma_0$ and $\gamma_0$ from one-loop diagrams is discussed in
sections~\ref{sect:scet} and~\ref{sect:bHQET} below, and the results are
summarized in Table~\ref{table:gammas}. To determine $\Gamma_1$ we make use of
the fact that to all orders in perturbation theory $\Gamma[\alpha_s]$ is
proportional to the cusp-anomalous dimension, $\Gamma[\alpha_s]\propto \Gamma^{\rm
  cusp}[\alpha_s]$. The constant of proportionality is also determined by the
one-loop computations and is summarized in Table~\ref{table:gammas}. 
\begin{table}[t!]
\begin{center}
\begin{tabular}{lc|ccccc|cc}
\hline
  & F &  $j$  & $\Gamma_F[\alpha_s]$ & $\Gamma_0$ & $\gamma_0$ & $\Gamma_1$ 
  & $\omega$ & $K$ 
   \\  \hline 
\text{SCET hard function} & $H_Q$ & \ \ $2$\ \
    & \ \ \ $-2\Gamma^{\rm cusp}[\alpha_s]$\ \ \  
    & \  $-8C_F$\ \ \ &  \ $-12C_F$\ \ \ & $-2\Gamma_1^{\rm cusp}$
    & \ \ $\omega_0$\ \   &\ \  $K_0$ \\
 \text{SCET jet function} & $J_{n,\bn}$ & \ \ $2$\ \
    & \ \ \ \ \ $2\Gamma^{\rm cusp}[\alpha_s]$\ \ \  
    & \ \ \  $8C_F$\ \ \ & \ \ $6C_F$ & $2\Gamma_1^{\rm cusp}$
    & \ \ $\omega_1$\ \   &\ \  $K_1$ \\
 \text{Soft hemisphere function} & $S$ & \ \ $1$\ \ 
    &\ \ $-\Gamma^{\rm cusp}[\alpha_s]$
    & $-4C_F$\ \  & $0$ & $-\Gamma_1^{\rm cusp}$
    & \ \ $\omega_2$\ \   &\ \  $K_2$ \\
 \text{bHQET jet function} & $B_\pm$ & \ \ $1$\ \ 
    & \ \ $\phantom{+}\Gamma^{\rm cusp}[\alpha_s]$
    & \ \ $4C_F$\ \  &\ \ $4C_F$ & $\Gamma_1^{\rm cusp}$
    & \ \ $\omega_1$\ \   &\ \  $K_3$ \\
\text{bHQET hard function} & $H_m$ & \ \ $2$\ \
    & \ \ \ $-2\Gamma^{\rm cusp}[\alpha_s]$\ \ \  
    & \ \ $-8 C_F$\ \ \ \ & \ $-8C_F$\ \ \ & $-2\Gamma_1^{\rm cusp}$
    & \ \ $\omega_0$\ \   &\ \  $K_{00}$ \\
\hline
\\[-12pt]
\multicolumn{9}{c}{\hspace{1cm}\vspace{0.1cm} $\Gamma_0^{\rm cusp}=4 C_F$\,,\qquad\qquad 
 $\Gamma_1^{\rm cusp} = 4 C_F \big[ \big(\frac{67}{9}
 -\frac{\pi^2}{3}\big)C_A-\frac{10\, n_f}{9} \big]$} \\[3pt]
\hline
\end{tabular}
\end{center}
\vskip-0.4cm
\caption{Dimension $j$ and anomalous dimensions $\Gamma[\alpha_s]$, 
 $\Gamma_0$, $\gamma_0$, and $\Gamma_1$ for the hard, jet and soft functions in SCET and
 bHQET using the notation in Eqs~(\ref{GammaSeries}) and
 (\ref{GammaCusp}). Values for  the one and two-loop cusp anomalous
 dimensions~\cite{Korchemsky:1987wg} are shown. For
 each case our notation for the resummation functions $\omega$ and
 $K$ is also given.
  \label{table:gammas}}
\end{table}
At two-loop order the cusp anomalous dimension has the form
\begin{align}\label{GammaCusp}
  \Gamma^{\rm cusp}[\alpha_s] &= \frac{\alpha_s(\mu)}{4\pi} \Gamma_0^{\rm cusp} 
     +  \Big[ \frac{\alpha_s(\mu)}{4\pi} \Big]^2\, \Gamma_1^{\rm cusp} + \ldots \,,
\end{align}
with results for $\Gamma_0^{\rm cusp}$ and $\Gamma_1^{\rm cusp}$ also shown in
Table~\ref{table:gammas}. For QCD $C_A=3$, $C_F=4/3$, and we have $n_f$ flavors.
Solving Eq.~(\ref{wLfull}) also requires the two-loop $\beta$-function
\begin{align} \label{beta}
  \mu \frac{d}{d\mu}\alpha_s(\mu)  
   = \beta[\alpha_s] = -2\alpha_s(\mu) \bigg\{ \frac{\alpha_s(\mu)}{4\pi} \beta_0 
   + \Big[ \frac{\alpha_s(\mu)}{4\pi}\Big]^2 \beta_1 + \ldots \bigg\} \,,
\end{align}
where 
\begin{align} \label{b0b1}
  \beta_0=11 C_A/3 - 2 n_f/3\,,\qquad\quad
  \beta_1 = 34 C_A^2/3-10 C_A n_f/3 - 2 C_F
n_f \,.
\end{align}
Defining
\begin{align}
  r = \frac{\alpha_s(\mu)}{\alpha_s(\mu_0)} \,,
\end{align} 
and substituting Eqs.~(\ref{GammaSeries}) and (\ref{beta}) into the integrals in
Eq.~(\ref{wLfull}) gives
\begin{align} \label{wL}
  \omega(\mu,\mu_0) &= 
     -\frac{ \Gamma_0}{j \beta_0} \bigg[ \ln(r)  
     + \bigg(\frac{\Gamma_1}{\Gamma_0}
     -\frac{\beta_1}{\beta_0}\bigg) \frac{\alpha_s(\mu_0)}{4\pi}\, (r-1) \bigg]
     \,, \\
 K_\gamma(\mu,\mu_0) &= -\frac{\gamma_0}{2\beta_0}\ln r \,,
  \nn \\[5pt] 
 K(\mu,\mu_0) &= \frac{-2
         \pi\Gamma_0}{\beta_0^2}\bigg\{ \frac{\big( r\minus 1
       \minus r\ln r \big)}{ \alpha_s(\mu)}  +
     \frac{\gamma_0\,\beta_0}{4\pi\Gamma_0} \ln r
     \plus \bigg(\frac{\Gamma_1}{\Gamma_0}
     -\frac{\beta_1}{\beta_0}\bigg) \frac{(1\minus r\plus \ln r)}{4\pi}
     +\frac{\beta_1}{8\pi\beta_0} \ln^2 r 
     \bigg\}. \nn
\end{align}
Taken together with Eq.~(\ref{UUU}) and the appropriate values for $\Gamma_0$,
$\gamma_0$, and $\Gamma_1$ from Table~\ref{table:gammas}, this determines the
NLL evolution kernels.

\section{SCET Results} \label{sect:scet}

\subsection{Current Matching and Running in SCET} \label{sect:scetcurrent}

\begin{figure}
\begin{center}
\includegraphics[width=3.5in]{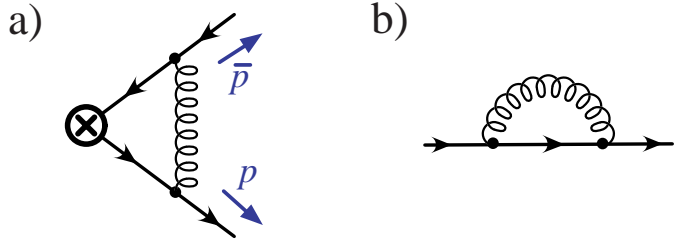}
\caption{One-loop vertex corrections in QCD.}
\label{qcdloops}
\end{center}
\end{figure}

To determine the Wilson coefficient $C$ of Eq.~(\ref{currentmatch}) we match
renormalized QCD and SCET S-matrix elements, which we will simply call
amplitudes in the following.  The QCD vertex graphs are given in
Fig.~\ref{qcdloops} where momenta $p$ and $\bar p$ are defined.  We use
dimensional regularization for UV divergences and small offshell momenta to
regulate the IR divergences, letting $p^2-m^2 = \bar p^2-m^2=\Delta^2\ne 0$.
Since the SCET current should reproduce the infrared physics of the QCD current,
we can perform the matching with arbitrary external states and for any infrared
regulator, as long as the same IR regulators are used in the full and effective
theories.  The values obtained for the Wilson coefficients will be independent
of the choice of IR regulator.

Results for the QCD graphs in Fig.~\ref{qcdloops} are summarized in
Eq.~(\ref{massiveqcdvertex}) of Appendix~\ref{AppFeyn}.  The result for the
amplitude includes the vertex graph, $\overline {\rm MS}$ wavefunction
contributions, and the residue term, $V_{\ref{qcdloops}a} + \Gamma_i^\mu
(Z_\psi-1) + \Gamma_i^\mu (R_\psi-1)$, where the subscript $4a$ on the $V$
indicates that it is the result for Fig.~\ref{qcdloops}a. The form of the
residue term is required to ensure consistency with physical S-matrix elements.
We work in the limit $\Delta^2 \ll m^2 \ll Q^2$. The QCD amplitude is
 \begin{eqnarray}
\label{QCDS-vertex}
\langle p, \bar p | {\cal J}^\mu_i | 0 \rangle_{\rm QCD}
  &=& \Gamma^\mu_i \bigg[ 1+ \frac{\alpha
_sC_F}{4\pi } \bigg\{  2\>\text{ln}^2\Big(\frac{-Q^2}{m^2}\Big) -
4\>\text{ln}\Big(\frac{-Q^2}{m^2}\Big)\>\text{ln}\Big(\frac{Q^2}{\Delta^2}\Big) +
3\>\text{ln}\Big(\frac{-Q^2}{m^2}\Big) \nn \\
&& \qquad\qquad
  + 4\>\text{ln}\Big(\frac{m^2}{-\Delta ^2}\Big) + \frac{2\pi
^2}{3} \bigg \}   \bigg ] \,,
\end{eqnarray}
where for simplicity we use the shorthand notation $\Delta^2$ and $Q^2$ for
$\Delta^2+i0$ and $Q^2+i0$, respectively.
For the SCET computation we have the graphs in Fig.~\ref{scetloops} which are
evaluated in Eqs.~(\ref{vertexgraphsscet},\ref{expvertexgraphsscet}) of
Appendix~\ref{AppFeyn} with non-zero $\Delta^2=p^2-m^2$ and $\bar \Delta^2=\bar
p^2-m^2$.  The sum of collinear and soft vertex graphs, wavefunction
contribution, and residue is $V_{\ref{scetloops}a} +
V_{\ref{scetloops}b}+V_{\ref{scetloops}c}+ \Gamma_i^\mu (Z_\xi-1) +\Gamma_i^\mu
(R_\xi-1)$. For $\bar\Delta =\Delta>0$ and again taking the limit
$\Delta^2 \ll m^2 \ll Q^2$ we obtain
 \begin{eqnarray}
\label{massivescetvertex}
 \langle p, \bar p | J_i^\mu | 0 \rangle_{\rm SCET} &=& \Gamma^\mu _i\bigg[ 1+   
    \frac{\alpha_s C_F}{4 \pi}  \bigg\{
  \frac{2}{\epsilon^2} +\frac{3}{\epsilon}
  + \frac{2}{\epsilon}\ln\Big( \frac{\mu^2}{-Q^2}\Big) 
  +2\ln^2\Big(\frac{\mu^2}{-\Delta^2}\Big) 
  \\
& & \hspace{-15ex}  
  + 2\ln^2\Big(\frac{m^2}{-\Delta^2}\Big)
  - \ln^2\Big(\frac{\mu^2 Q^2}{(-\Delta^2)(\Delta^2)}\Big)
  +4\ln\Big(\frac{m^2}{-\Delta^2}\Big)
 +3\ln\Big(\frac{\mu^2}{m^2}\Big) 
  +8 + \frac{\pi^2}{2} \bigg\} \bigg]\,. \nn
\end{eqnarray}
The remaining divergences in Eq.~(\ref{massivescetvertex}) are
canceled by the current counterterm $Z_C-1$ giving
\begin{equation} \label{Zc}
Z_c = 1 -\frac{\alpha_s C_F}{4 \pi} \bigg[ \frac{2}{\epsilon^2} +\frac{3}{\epsilon}
  +  \frac{2}{\epsilon}\ln\Big(\frac{\mu^2}{-Q^2-i0}\Big) \bigg] \,.
\end{equation}
The running generated by $Z_c$ sums $\ln^2\mu$ terms and falls in case 2) as
defined in section~\ref{sect:RGE}.  The renormalized amplitude in SCET then
reads
\begin{eqnarray}
\label{massivescetvertex1}
 \langle p, \bar p | Z_c{J}_i^\mu | 0 \rangle_{\rm SCET}
   &=& \Gamma^\mu _i\bigg[ 1+    \frac{\alpha_s C_F}{4 \pi}  \bigg\{
 2\ln^2\Big(\frac{\mu^2}{-\Delta^2}\Big) 
  + 2\ln^2\Big(\frac{m^2}{-\Delta^2}\Big)
  - \ln^2\Big(\frac{\mu^2 Q^2}{-\Delta^4}\Big)\nn\\
 && \qquad
  +4\ln\Big(\frac{m^2}{-\Delta^2}\Big)
 +3\ln\Big(\frac{\mu^2}{m^2}\Big) 
  +8 + \frac{\pi^2}{2} \bigg\}
  \bigg]\,.
\end{eqnarray}
Subtracting Eq.~(\ref{massivescetvertex1}) from (\ref{QCDS-vertex}) all
dependence on the IR scales $m$ and $\Delta$ cancels. This is an explicit
demonstration that at one-loop massive SCET has the same IR structure as in QCD.
Evaluating the difference at the scale $\mu=\mu_Q$ gives the matching condition
of the current Wilson coefficient,
\begin{eqnarray} \label{matchcoeff}
C(Q,\mu_Q) 
  = 1 + \frac{\alpha_s C_F}{4 \pi} \left[-\ln^2\Big(\frac{\mu_Q^2}{-Q^2\minus i0}\Big)
    - 3\ln\Big( \frac{\mu_Q^2}{-Q^2\minus i0}\Big) 
      -8 + \frac{\pi^2}{6} \right]\,.
\end{eqnarray}
As expected from the limit $Q\gg m$ the matching condition is mass-independent
and there are no large logarithms for $\mu_Q\simeq Q$.  

\begin{figure}
\begin{center}
\includegraphics[width=6.in]{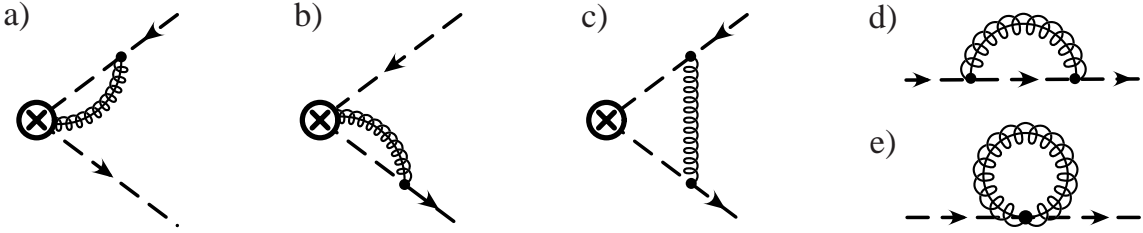}
\caption{One-loop vertex and self-energy corrections in massive SCET. Gluons
  with a line through them are collinear, while those without are soft. The soft
  gluon wavefunction renormalization graph vanishes in Feynman gauge and is not
  shown.}
\label{scetloops}
\end{center}
\end{figure}

The result in Eq.~(\ref{matchcoeff}) is independent of the choice of the IR
regulator and should therefore agree with the matching conditions for the
massless quark production current. In Ref.~\cite{Bauer:2003di} the matching
coefficient was computed using on-shell massless quarks, and
Eq.~(\ref{matchcoeff}) agrees with their result.  With the regulator used in
Ref.~\cite{Bauer:2003di} the SCET diagrams are scaleless and vanish in
dimensional regularization. To see more explicitly how the massless computation
gives the same matching coefficient we repeat the previous computation with an
offshellness $p^2=\bar p^2 \gg m^2$, where $Q^2\gg p^2=\bar p^2$. For this case
the renormalized one loop QCD amplitude is
\begin{align}
\label{Jmatrixqcd}
\langle p,\bar p | {\cal J}^\mu_i | 0 \rangle \Big |_{\rm QCD}&=& \Gamma^\mu_i \bigg[ 1 
   \plus C_F \frac{\alpha_s}{4 \pi} \bigg\{
\minus \ln \Big(\frac{-Q^2}{\mu^2}\Big) \minus 2 \ln^2\Big(\frac{p^2}{Q^2}\Big)
 \minus 4 \ln \Big(\frac{p^2}{Q^2}\Big)  \minus \frac{2\pi^2}{3} \bigg\}  \bigg] \,,
\end{align}
and from Eqs.~(\ref{vertexgraphsscet}) and (\ref{expvertexgraphsscet}) the
renormalized amplitude in SCET has the form
\begin{align}
\label{masslessscetvertex}
 \langle p, \bar p | Z_c {J}_i^\mu | 0 \rangle_{\rm SCET}
   &=& \Gamma^\mu _i\bigg[ 1\plus    \frac{\alpha_s C_F}{4 \pi}  \bigg\{
 2\ln^2\!\Big(\frac{\mu^2}{-p^2}\Big) 
  \minus \ln^2\!\Big(\frac{\mu^2 Q^2}{-p^4}\Big)
  \plus 4\ln\!\Big(\frac{\mu^2}{-p^2}\Big)
  \plus 8 - \frac{5\pi^2}{6} \bigg\}
  \bigg].
\end{align}
To obtain Eq.~(\ref{masslessscetvertex}) the current counterterm in $Z_C$ from
Eq.~(\ref{Zc}) was used. Taking the difference of Eqs.~(\ref{Jmatrixqcd}) and
(\ref{masslessscetvertex}) gives exactly Eq.~(\ref{matchcoeff}), as expected. 

The imaginary parts in $C(Q,\mu_Q)$ and the Z-factor in Eq.(\ref{Zc}) arise from
real QCD intermediate states in the QCD vertex diagram that are not accounted
for in the corresponding SCET diagrams. These SCET graphs account only for
fluctuations associated to sectors for the $n$ and $\bn$ directions, while the
QCD diagrams do not have such a restriction.  Note that the complex $Z$-factor
also means that the anomalous dimension $\gamma_C$ is complex.  However, only
$|C|^2$ appears in the factorization theorem in Eq.~(\ref{FactThm}) and so the
complex phase cancels. This treatment is consistent because in the derivation of
the factorization theorem the part of the phase space integration encoded in the
sum over the $n$ and $\bn$ directions is carried out explicitly prior to the
formulation of the jet and soft functions in SCET.  The matching coefficient
appearing in the factorization theorem therefore reads
\begin{align}
\label{HQmatch}
 H_Q(Q,\mu_Q) = |C(Q,\mu_Q)|^2 = 
1 + \frac{\alpha_s C_F}{4 \pi} \,\bigg[-2\ln^2\Big(\frac{Q^2}{\mu_Q^2}\Big)
    +6\ln\Big( \frac{Q^2}{\mu_Q^2}\Big) 
      -16 + \frac{7\pi^2}{3} \,\bigg] \,.
\end{align}

To evolve the Wilson coefficient to lower scales we need to solve the RG
equation in Eq.~(\ref{gammacgammaH}).  The anomalous dimensions are obtained
from $Z_c$ in Eq.~(\ref{Zc}) and $\mu d/d\mu\: \alpha_s = -2\epsilon\,
\alpha_s + \beta(\alpha_s)$, 
\begin{eqnarray}
\gamma_c(\mu) &=& -Z_c^{-1}(\mu) \mu \frac{d}{d\mu} Z_c(\mu)
  =  -\frac{\alpha_s C_F}{\pi}
     \bigg[\ln\frac{\mu^2}{-Q^2 -i0}+ \frac{3}{2} \bigg]\,,\nn\\
 \gamma_{H_Q}(\mu) &=& \gamma_c(\mu)+\gamma_c^*(\mu) 
  =   -\frac{\alpha_s C_F}{4\pi}
     \bigg[8 \ln\frac{\mu^2}{Q^2}+ 12 \bigg]\,.
\end{eqnarray}
Comparing this result to Eq.~(\ref{gammaHQ}) we find $\Gamma_0^{H_Q}=-8C_F$ and
$\gamma_0^{H_Q}=-12 C_F$ for the coefficients discussed in
section~\ref{sect:NLL}. Also $\Gamma_{H_Q}[\alpha_s]=-2\Gamma^{\rm
  cusp}[\alpha_s]$ and so $\Gamma_1^{H_Q}=-2\Gamma_1^{\rm cusp}$. The solution
for the evolution factor is
\begin{equation}
\label{runcoeff}
 U_{H_Q}(Q,\mu_Q,\mu) =
 e^{K_0} \
 \Big(\frac{\mu_Q^2}{Q^2}\Big)^{\omega_0 }\,,
\end{equation}
where $\omega_0=\omega_0(\mu,\mu_Q)$ and $K_0=K_0(\mu,\mu_Q)$ are determined at
NLL order using Eq.~(\ref{wL}) for ``$\omega$'' and ``$K$''. At LL order the
solutions are
\begin{align}
  \omega_0^{LL}(\mu,\mu_Q)  &= 
     \frac{4C_F}{ \beta_0}\:\ln r
     \,,\quad \qquad
  K^{LL}_0(\mu,\mu_Q) =  
  \frac{16\pi C_F}{\beta_0^2} \, \frac{(r-1-r\ln r)}{\alpha_s(\mu)} \,,
%
\end{align}
with $r=\alpha_s(\mu)/\alpha_s(\mu_Q)$. 
Note that solving the RG-equation directly for $C(Q,\mu)$ leads to an extra phase
factor, 
\begin{equation}
C(Q\OMIT{,\mu_Q},\mu)= \sqrt{ H_Q(Q\OMIT{,\mu_Q},\mu)}  \bigg[
\frac{\alpha_s(\mu)}{\alpha_s(\mu_Q)}\bigg]^{2\pi i \frac{C_F}{\beta_0}} 
 \,,
\end{equation}
which does not, however, appear in the physical cross section.
It's origin is the same as for the phase contained in the current matching
condition $C(Q,\mu_Q)$.

\subsection{SCET Jet Functions and their Running} \label{sect:scetjet}

In this section we compute the SCET jet functions $J_n$ and $J_\bn$, defined in
Eq.~(\ref{jetfunc2}), perturbatively to ${\cal O}(\alpha_s)$.  Due to charge
conjugation symmetry, the results for $J_n$ and $J_\bn$ are identical, so for
simplicity we focus on the former. The purpose of the calculation is two-fold.
First we determine the renormalization factor $Z_{J_n}$, the anomalous dimension
$\gamma_{J_n}$ and evolution kernel $U_{J_n}$ for the jet function. Second, the
renormalized jet function at the scale $\mu_m\simeq m$ is needed to determine
the matching condition of the bHQET jet function, which we work out in
Sec.~\ref{sect:bHQETJet} below.  Since both running and matching do not depend
on infrared effects below $m$ we are free to do the computations for stable top
quarks. Thus in this section we set the electroweak gauge coupling to zero and
neglect finite lifetime effects.

From Eq.~(\ref{jetfunc2}), the tree-level jet function is simply given by the 
imaginary part of the collinear propagator:
\begin{eqnarray}
\label{treejet}
J_{n}(s,m,\Gamma=0,\mu )\, \Big|_{\text{tree}} &=&  \delta (s).
\end{eqnarray}
At one loop, the jet functions are given by the imaginary part of the diagrams
shown in Fig.~\ref{forwardI}, and results for the individual graphs are
summarized in Appendix~\ref{AppFeyn}. We will consider the one-loop jet
function with and without expanding in $s\ll m^2$.

\begin{figure}
\begin{center}
\includegraphics[width=4.5in]{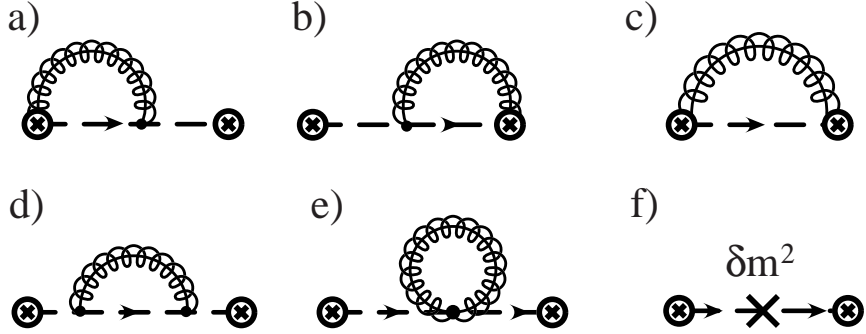}
\caption{SCET graphs for the one-loop top quark jet function. Dashed lines are
  $n$-collinear quarks and springs are $n$-collinear gluons. }
\label{forwardI}
\end{center}
\end{figure}
Prior to taking the imaginary part the tree level graph plus the sum of one-loop
graphs from Eq.~(\ref{Jabc_int}) give
\begin{align} \label{Jabcdesum0}
 & J_{\rm tree} \plus  J_{\ref{forwardI}a} \plus J_{\ref{forwardI}b} 
  \plus J_{\ref{forwardI}c} \plus J_{\ref{forwardI}d} \plus J_{\ref{forwardI}e} 
  \plus J_{\ref{forwardI}f} 
 \\[3pt]
  &= \frac{-1}{\pi s}  - \frac{2 m \delta m}{\pi s^2}
  - \frac{\alpha_s C_F  }{4\pi^2\,s}
  \bigg\{ \frac{4}{\epsilon^2}\plus \frac{4}{\epsilon} \ln\Big(\frac{\mu^2}{-s}\Big)
   \plus \frac{3}{\epsilon} \plus  2\ln^2\Big(\frac{\mu^2}{-s}\Big)
   \minus 2 \ln^2\Big(\frac{m^2}{-s}\Big)
   - 4\, {\rm Li}_2\Big(\frac{-s}{m^2}\Big)  
  \nn\\
 & \qquad
 \plus 3\ln\Big(\frac{\mu^2}{-s}\Big) 
  \plus  4\ln\Big(\frac{m^2}{-s}\Big)\ln\Big(\frac{m^2\plus s}{-s} \Big)
  \plus \frac{m^2(m^2\plus 2 s)}{(s\plus m^2)^2} \ln\Big(\frac{m^2}{-s}\Big)
   - \frac{s}{m^2\plus s} + 8 + \pi^2 \bigg\}
   ,\nn
\end{align}
where $s=s+i0$ and $\delta m = m^{\rm pole} - m$. Hence in the pole mass
scheme $\delta m=0$.  The one-loop massive jet function $J_n$ also appears in
the computation for $B\to X_c\ell\bar\nu$ in the endpoint region studied in
Ref.~\cite{Boos:2005qx}.  Identifying the combination $n_+\cdot p u'$ in
Ref.~\cite{Boos:2005qx} with our variable $s$ we find agreement with their
pole-mass result. Expanding Eq.~(\ref{Jabcdesum0}) in $s \ll m^2$ we find
\begin{align} \label{Jabcdesum}
 &  \big[J_{\rm tree} \plus J_{\ref{forwardI}a} \plus J_{\ref{forwardI}b} 
  \plus J_{\ref{forwardI}c} \plus J_{\ref{forwardI}d} \plus J_{\ref{forwardI}e}
   \plus J_{\ref{forwardI}f}\big]_{s\ll m^2}
  \\[3pt]
  &=  \frac{-1}{\pi s}  - \frac{2 m \delta m}{\pi s^2}
  - \frac{\alpha_s C_F  }{4\pi^2\,s}
  \bigg\{ \frac{4}{\epsilon^2}\plus \frac{4}{\epsilon} \ln\Big(\frac{\mu^2}{-s}\Big)
   \plus \frac{3}{\epsilon} 
   \plus  2\ln^2\Big(\frac{\mu^2}{-s}\Big)
   \plus  2\ln^2\Big(\frac{m^2}{-s}\Big) \plus 3 \ln\Big(\frac{\mu^2}{m^2}\Big)
  \nn\\
 & \qquad 
   \minus  4 \ln\Big( \frac{-s}{m^2}\Big)
   \plus 8\plus \pi^2 \bigg\} 
   \,. \nn
\end{align}
For later convenience we write $s=x \kappa_1^2$ where $x$
is dimensionless and $\kappa_1>0$ is a dummy scale with dimensions of mass.
Taking the imaginary part of Eq.~(\ref{Jabcdesum0}) using the results in
Appendix~\ref{App:Im} we find that the bare SCET jet function is
\begin{align} \label{Jnbare}
J_n^{\text{bare}}(s) &= \delta(s) - 2 m\delta m \,\delta^\prime(s) 
 + \frac{\alpha _s C_F}{4\pi} \bigg[ \delta(s)
  \bigg\{\frac{4}{\epsilon ^2} + \frac{3}{\epsilon} 
  + \frac{4}{\epsilon}\ln \Big(\frac{\mu ^2}{\kappa_1^2}\Big) 
  \plus 2 \ln^2\Big(\frac{\mu ^2}{\kappa_1 ^2} \Big)
  \plus 2 \ln^2\Big(\frac{m ^2}{\kappa_1 ^2} \Big)
  \nn \\
& \plus 3 \ln \Big ( \frac{\mu ^2}{\kappa_1 ^2}\Big ) 
  \plus \ln \Big(\frac{m^2}{\kappa_1 ^2} \Big )
  \plus 8 - \frac{\pi^2}{3} \bigg\}
 \minus \frac{4}{\kappa_1 ^2}\Big[ \frac{\kappa_1 ^2 \theta(s)}{s}\Big ]_+ 
  \:\bigg\{  \frac{1}{\epsilon} + \ln \Big (\frac{\mu ^2}{\kappa_1 ^2}\Big ) 
   \plus \ln \Big (\frac{m^2}{\kappa_1 ^2}\Big ) \plus 1 \bigg\}
  \nn \\
& + \frac{8}{\kappa_1 ^2} \Big[ \frac{\kappa_1^2 \theta(s) \ln(s/\kappa_1^2)}{s}\Big]_+
 + \theta(s) \bigg\{ \frac{s}{(m^2 \plus s)^2} - \frac{4}{s} \ln\Big(1\plus
   \frac{s}{m^2}\Big) \bigg\}
  \bigg] 
  . 
\end{align}
When we expand Eq.~(\ref{Jnbare}) for $s\ll m^2$ all terms have a singular
${\cal O}(s^{-1})$ behavior except for the last two $\theta(s)$ terms which are
${\cal O}(s^0)$ and can be dropped in the peak region.  The result of this
expansion agrees with taking the imaginary part of Eq.~(\ref{Jabcdesum}).

To renormalize $J_n^{\rm bare}$ the required jet function $Z$-factor defined in
Eq.~(\ref{ZJJS}) is
\begin{align} \label{ZJ}
  Z_{J_n}(s\minus s') = \delta(s\minus s') + \frac{\alpha_s C_F}{4\pi}\bigg\{
    \delta(s\minus s') \bigg[\frac{4}{\epsilon^2}\plus 
   \frac{3}{\epsilon} \bigg] 
   \minus \frac{4}{\epsilon\, \mu^2}
    \bigg[ \frac{ \mu^2\theta(s\minus s')}{s\minus s'}\bigg]_+
   \bigg\},
\end{align}
which gives the anomalous dimension
\begin{align} \label{gammaJn}
  \gamma_{J_n}(s-s') & = - \frac{\alpha_s(\mu) C_F}{4\pi}\bigg\{
   \frac{8}{ \mu^2} \bigg[ 
   \frac{ \mu^2\theta(s\minus s')}{s\minus s'} \bigg]_+
   - 6\,  \delta(s\minus s')
   \bigg\} .
\end{align}
Note that $J_n^{\rm bare}(s)$, $Z_{J_n}(s-s')$ as well as $\gamma_{J_n}(s-s')$
are all independent of the choice for $\kappa_1$. The Z-factor and the anomalous
dimension also do not depend on the mass scheme that is being employed
(determined by $\delta m$).  Comparing this result to Eq.~(\ref{gammaF}) we find
$\Gamma_0^{J_{n,\bn}}=8C_F$ and $\gamma_0^{J_{n,\bn}}=6 C_F$ for the
coefficients discussed in section~\ref{sect:NLL}. Also
$\Gamma_{J_{n,\bn}}[\alpha_s]=2\Gamma^{\rm cusp}[\alpha_s]$ and so
$\Gamma_1^{J_{n,\bn}}=2\Gamma_1^{\rm cusp}$. These coefficients give us the NLL
evolution kernel that evolves the jet function from the scale $\mu_0$ to $\mu$:
\begin{align} \label{UJ}
  U_{J_n}(s-s',\mu,\mu_0) &=  \frac{e^{K_1}\: \big(
  e^{\gamma_E}\big)^{\omega_1}}{\Gamma(-\omega_1)\, \mu_0^2} \: 
  \bigg[ \frac{\mu_0^{2+2\omega_1}\,\theta(s\minus s')}{(s\minus
    s')^{1+\omega_1}}\bigg]_+ \,,
\end{align}
where $\omega_1=\omega_1(\mu,\mu_0)$ and $K_1=K_1(\mu,\mu_0)$ are determined at
NLL order from Eq.~(\ref{wL}) with $r=\alpha_s(\mu)/\alpha_s(\mu_0)$. At LL
order they are
\begin{align} \label{wL1}
 \omega_1^{LL}(\mu,\mu_0)  &= 
     -\frac{4C_F}{ \beta_0}\:\ln r
     \,,\qquad \quad
  K_1^{LL}(\mu,\mu_0) = -
  \frac{16\pi C_F}{\beta_0^2} \, \frac{(r-1-r\ln r)}{\alpha_s(\mu)} \,.
%
\end{align}
The resummation induced by $U_{J_n}$
falls in case 4), it sums double logs and involves a convolution. 

Finally, taking into account the counterterm in Eq.~(\ref{ZJ}) the renormalized
jet function for a stable top quark at ${\cal O}(\alpha_s)$ reads
\begin{align} \label{Jren}
 & J_{n}(s,m,\Gamma=0,\mu)   
 =  \delta(s) - 2 m\,\delta m \,\delta^\prime(s) 
  + \frac{\alpha_s(\mu) C_F}{4\pi} \bigg[
   \delta(s) \bigg\{
     2\ln^2\Big(\frac{\mu^2}{\kappa_1^2}\Big)
   \plus  2\ln^2\Big(\frac{m^2}{\kappa_1^2}\Big) 
   \plus \ln\Big( \frac{m^2}{\kappa_1^2}\Big) 
   \nn\\
& \qquad
 \plus 3 \ln\Big(\frac{\mu^2}{\kappa_1^2}\Big)
    \plus 8\minus \frac{\pi^2}{3} \bigg\}
   \plus \frac{8}{\kappa_1^2} \Big[\frac{\theta(x)\ln(x)}{x}\Big]_+  
    - \frac{4}{\kappa_1^2}\Big[\frac{\theta(x)}{x}\Big]_+ \bigg\{ 1 
    \plus \ln\Big(\frac{m^2}{\kappa_1^2}\Big)
     \plus \ln\Big(\frac{\mu^2}{\kappa_1^2}\Big) \bigg\}
   \nn  \\
 & \qquad
   + \theta(s) \bigg\{ \frac{s}{(m^2 \plus s)^2} 
   - \frac{4}{s} \ln\Big(1\plus \frac{s}{m^2}\Big) \bigg\}
   \bigg] .
\end{align}
Here $x=s/\kappa_1^2$ and the result is independent of the choice of $\kappa_1$.
The last term in Eq.~(\ref{Jren}) is regular in the small $x$ limit and can be
dropped in the peak region.  In the ultra-tail region discussed in
Appendix~\ref{App:SCETnum} this term generates the ``non-singular'' contribution
of the SCET function.

From Eq.~(\ref{Jren}) we can see that for the variable range $s\sim m\Gamma$
further matching and RG-evolution is needed for $J_n$: there is no choice of
$\mu$ that minimizes all the logarithmic terms. The particular terms in which
the large logarithms appear are controlled by the choice of $\kappa_1$, but no
choice of $\kappa_1$ removes them completely. For example, with $\kappa_1=m$ and
$\mu=m$ we still have $\ln(x)\sim \ln(\Gamma/m)$; while for $\kappa_1^2=m\Gamma$
and $\mu=\kappa_1$ we have $\ln(m^2/\kappa_1^2) \sim \ln(\Gamma/m)$. This
motivates the matching onto bHQET and RG-evolution between $m$ and $\Gamma$ to
be carried out in section~\ref{sect:bHQET} below.  For later convenience we
quote the leading result for $J_n$ when $s\ll m^2$ using the choice
$\kappa_1=m$,
\begin{align} \label{Jrenm}
 & J_{n}(s,m,\Gamma=0,\mu)   \Big|_{s\ll m^2}
 =\delta(s) - 2 m\,\delta m \,\delta^\prime(s) 
  + \frac{\alpha_s(\mu) C_F}{4\pi} \bigg[
     \delta(s) \bigg\{
     2\ln^2\Big(\frac{\mu^2}{m^2}\Big)
    \plus 3 \ln\Big(\frac{\mu^2}{m^2}\Big)
    \nn \\[3pt]
 & \qquad
    + 8 - \frac{\pi^2}{3} \bigg\}
  \plus \frac{8}{m^2} \Big[\frac{\theta(x)\ln(x)}{x}\Big]_+  
   \minus  \frac{4}{m^2}  \Big[\frac{\theta(x)}{x}\Big]_+ \bigg\{ 1 
   \plus \ln\Big(\frac{\mu^2}{m^2}\Big) \bigg\}
     \bigg] . 
\end{align}

\subsection{Hemisphere Soft Function and its Running} \label{sect:soft}

In this section we determine the ${\cal O}(\alpha_s)$ renormalization group
evolution of the hemisphere soft function, $S_{\rm hemi}(\ell^+,\ell^-,\mu)$ and
its renormalized partonic expression from one-loop perturbation theory, $S_{\rm
  part}(\ell^+,\ell^-,\mu)$, which is needed to construct the soft function
model defined in Eq.~(\ref{Smodel1}). This model builds in the fact that the
full nonperturbative $S$ has the same dependence on $\mu$ as $S_{\rm part}$.
\begin{figure}
\begin{center}
\includegraphics[width=6.4in]{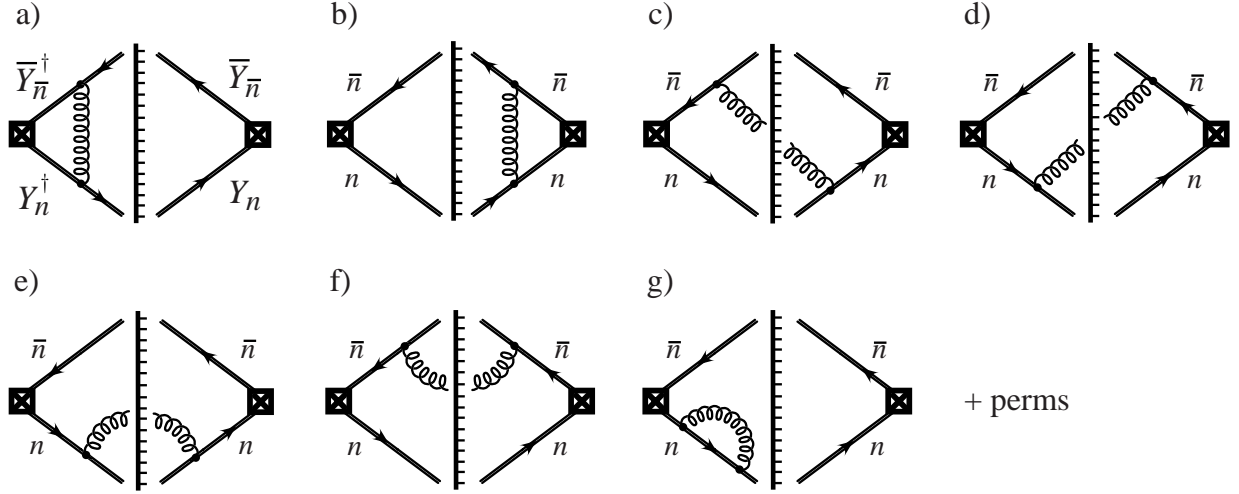}
\caption{Graphs for the hemisphere soft function at one-loop. In this figure the
  solid lines denote $Y$-Wilson lines,and the line with ticks is the final state
  cut.}
\label{softgraphs}
\end{center}
\end{figure}

For the computation we use the squared matrix-element expression in
Eq.~(\ref{Slplm}) for a no-gluon and a single gluon final state. The
corresponding Feynman diagrams are shown in Fig.~\ref{softgraphs}, where the
solid lines denote the four $Y$-Wilson lines. Fig.~\ref{softgraphs}a,b,g are
the virtual graphs with $|X_s\rangle =|0\rangle$, while
Fig.~\ref{softgraphs}c,d,e,f are the real emission graphs with $|X_s\rangle =
|\varepsilon_\mu^A\rangle$. Results for the graphs are summarized in
Eq.~(\ref{softsum1}) of Appendix~\ref{AppFeyn}. Together with the tree-level 
matrix element the bare hemisphere soft function reads
\begin{align} \label{softsum}
  S^{\rm bare}_{\rm part}(\ell^+,\ell^-) 
  &=  \delta(\ell^+)\delta(\ell^-)\\
  & \ + \frac{C_F\,\alpha_s}{\pi}\,
  \frac{e^{\epsilon \gamma_E}}{\epsilon\,\Gamma(1\minus\epsilon)} 
   \frac{\mu^{2\epsilon}}{\kappa_2^{2\epsilon}}
  \bigg[ \frac{\delta(\ell^-)\theta(\ell^+)}{\kappa_2} 
    \Big(\frac{\kappa_2}{\ell^+}\Big)^{1\plus 2\epsilon}
    \!+ \frac{\delta(\ell^+)\theta(\ell^-)}{\kappa_2}
     \Big(\frac{\kappa_2}{\ell^-}\Big)^{1\plus 2\epsilon} \bigg] . \nn
\end{align}
Note that $S^{\rm bare}_{\rm part}(\ell^+,\ell^-)$ is independent of the dummy
mass scale $\kappa_2>0$ introduced here. However $\kappa_2$ facilitates the
application of the standard distribution relation for dimensionless variables
\begin{align}
  \frac{\theta(x)}{x^{1+2\epsilon}} =  -\frac{\delta(x)}{2\epsilon} 
     + \Big[\frac{\theta(x)}{x}\Big]_+   - 2\epsilon \Big[\frac{\theta(x)\ln
       x}{x}\Big]_+ + {\cal O}(\epsilon^2) \,.
\label{standarddistribution}
\end{align}
The relation leads to the following expression for the bare hemisphere soft
function:
\begin{align} \label{softsum2}
   S^{\rm bare}_{\rm part}(\ell^+,\ell^-) 
  &= \delta(\ell^+)\delta(\ell^-)
  + \frac{C_F\,\alpha_s}{\pi}\,
    \bigg\{ \minus \frac{\delta(\ell^+)\delta(\ell^-)}{\epsilon^2} 
  \plus \frac{\delta(\ell^-)}{\epsilon\:\kappa_2}
   \Big[\frac{\kappa_2 \theta(\ell^+)}{\ell^+}\Big]_+
  \plus \frac{\delta(\ell^+)}{\epsilon\: \kappa_2}
   \Big[\frac{\kappa_2\theta(\ell^-)}{\ell^-}\Big]_+
  \nn \\[3pt]
  & \qquad\qquad\qquad\qquad\qquad\quad
 \minus
 \frac{\delta(\ell^+)\delta(\ell^-)}{\epsilon}\ln\Big(\frac{\mu^2}{\kappa_2^2}\Big)
 \plus G_S(\ell^+,\ell^-) \bigg\},
\end{align}
where $G_S(\ell^+,\ell^-)$ contains the finite terms 
\begin{align}
\label{Gfunc}
 & G_S(\ell^+,\ell^-,\mu) =  \frac12 \delta(\ell^+)\delta(\ell^-) \Big[\frac{\pi^2}{6} \minus
 \ln^2\Big(\frac{\mu^2}{\kappa_2^2}\Big) \Big] 
  \plus  \frac{\delta(\ell^-)}{\kappa_2} \ln\Big(\frac{\mu^2}{\kappa_2^2}\Big) 
     \Big[\frac{\kappa_2\theta(\ell^+)}{\ell^+}\Big]_+ 
    \\[3pt]
 &\ \  \plus  \frac{\delta(\ell^+)}{\kappa_2} \ln\Big(\frac{\mu^2}{\kappa_2^2}\Big) 
     \Big[\frac{\kappa_2\theta(\ell^-)}{\ell^-}\Big]_+
   \minus \frac{2 \delta(\ell^-)}{\kappa_2}
  \Big[\frac{\theta(\ell^+)\ln(\ell^+/\kappa_2)}{\ell^+/\kappa_2}\Big]_+
   \minus \frac{2 \delta(\ell^+)}{\kappa_2}
   \Big[\frac{\theta(\ell^-)\ln(\ell^-/\kappa_2)}{\ell^-/\kappa_2}\Big]_+
\,. \nn
\end{align}
The renormalization factor for the hemisphere soft function then reads
\begin{align} \label{Zs}
  & Z_s(\ell^{\,\prime +}\minus \ell^+,\ell^{\,\prime -}\minus \ell^-) =
    \delta( \ell^+ \!\minus\ell^{\,\prime +})\delta(\ell^- \!\minus\ell^{\,\prime -})
    - \frac{C_F\,\alpha_s}{\pi}\,
    \bigg\{ \frac{\delta(\ell^+\!\minus \ell^{\,\prime +})
       \delta(\ell^- \!\minus\ell^{\,\prime -})}{\epsilon^2} 
   \\[3pt]
  &\quad - \frac{\delta(\ell^- \!\minus\ell^{\,\prime -})}{\epsilon\:\mu}
   \bigg[\frac{\mu\,\theta(\ell^+\!\minus \ell^{\,\prime +})}
    {\ell^+\!\minus \ell^{\,\prime +}}\bigg]_+ \!\!
   - \frac{\delta(\ell^+\minus \ell^{\,\prime +})}{\epsilon\: \mu}
   \bigg[\frac{\mu\,\theta(\ell^- \!\minus\ell^{\,\prime -})}
     {\ell^- \!\minus\ell^{\,\prime -}}\bigg]_+
\bigg\},\nn
\end{align}
while the renormalized soft function is
\begin{align} 
\label{Sren}
    S_{\rm part}(\ell^+,\ell^-,\mu) 
  &= \delta(\ell^+)\delta(\ell^-)
  + \frac{C_F\,\alpha_s(\mu)}{\pi}\,
      G_S(\ell^+,\ell^-,\mu) \,.
\end{align}
We caution once more that the soft function is in general dominated by 
non-perturbative effects, so the partonic and perturbative result in
Eq.~(\ref{Sren}) can only be used in the framework of the soft function model
of Eq.~(\ref{Smodel1}) or in situations where an operator product expansion
can be carried out. We also note that to ${\cal O}(\alpha_s)$ the renormalized
partonic soft function can be factored in the form 
\begin{align}
\label{factored}
S_{\rm part}(\ell^+,\ell^-,\mu) &=
S_{{\rm part}}(\ell^+,\mu)\,S_{{\rm part}}(\ell^-,\mu)
\,,
\end{align}
where the partonic soft function with one kinematic variable is
\begin{align}
\label{Sdefsing}
S_{{\rm part}}(\ell^\pm,\mu) &=
\delta(\ell^\pm)
  + \frac{C_F\,\alpha_s(\mu)}{\pi}\,
    \bigg\{ 
 \delta(\ell^\pm) \, \Big[\frac{\pi^2}{24} \minus
 \frac{1}{4}\ln^2\Big(\frac{\mu^2}{\kappa_2^2}\Big) \Big]
 \plus  \frac{1}{\kappa_2} \ln\Big(\frac{\mu^2}{\kappa_2^2}\Big) 
     \Big[\frac{\kappa_2\theta(\ell^\pm)}{\ell^\pm}\Big]_+ 
\nn\\[2mm] & \hspace{3cm}
   \minus \frac{2}{\kappa_2}
  \Big[\frac{\theta(\ell^\pm)\ln(\ell^\pm/\kappa_2)}{\ell^\pm/\kappa_2}\Big]_+
\, \bigg\}
\,.
\end{align} 
The factored form in Eq.~(\ref{factored}) was expected at ${\cal O}(\alpha_s)$
because the single gluon in the real graphs of Fig.~\ref{softgraphs} can make
either $\ell^+$ or $\ell^-$ nonzero but not both. Because there is in general
more than one real parton in the real graphs at ${\cal O}(\alpha_s^2)$ and
beyond, the factored form of $S_{\rm part}$ is not expected to hold in
general, nor for the full non-perturbative soft function $S$. 

From $Z_s$ in Eq.~(\ref{Zs}) one obtains the anomalous dimension
\begin{align} \label{gammaS}
  \gamma_S( \ell^+,\ell^-) 
   & = \delta(\ell^-) \gamma_s(\ell^+) + \delta(\ell^+)\gamma_s(\ell^-)\,,
   \nn\\[5pt]
   \gamma_s(\ell^\pm) &= \frac{2C_F\,\alpha_s}{\pi}\,
   \frac{1}{\mu}
   \Big[\frac{\mu\, \theta(\ell^\pm)}{\ell^\pm}\Big]_+
  \,.
\end{align}
As anticipated from the form of the consistency condition discussed in
Sec.~\ref{sssec:scetrenorm}, it has a separable structure in the light cone
variables $\ell^+$ and $\ell^-$. This separation for $\gamma_S(\ell^+,\ell^-)$
holds to all orders in $\alpha_s$ as discussed in
section~\ref{subsec:consistency}. Comparing this result to Eq.~(\ref{gammaF}) we
find $\Gamma_0^{s}=-4C_F$, $\gamma_0^{s}=0$, and infer
$\Gamma_1^{s}=-\Gamma_1^{\rm cusp}$ for the coefficients discussed in
section~\ref{sect:NLL}. The anomalous dimension of various soft-functions in
SCET were studied in Ref.\cite{Chay:2004zn}, and in particular the one-loop
anomalous dimension for the jet-energy soft-function in $e^+e^-\to {\rm dijets}$
was derived.  This anomalous dimension has the same form but opposite sign of
the anomalous dimension in Eq.~(\ref{gammaS}). However this is inconsequential,
since there is no simple relation between the jet energy soft function and the
hemisphere soft function studied here. The hemisphere soft function's $\gamma_s$
is the same as the anomalous dimension for the soft function for Drell-Yan in
the endpoint region at one-loop~\cite{Korchemsky:1993uz}.

Before solving the soft-function anomalous dimension we pause to consider the
number of active flavors in the RGE. For the SCET computation we can consider
the five lightest flavors to be massless. In addition because we are above the
scale of the top-mass, we have a top-quark loop contribution to the soft-gluon
induced $\beta$-function. The top-quark bubble couples to the soft-gluon with a
multipole expansion since the soft-gluon has $p^2\ll m^2$ for the purpose of
power counting. Thus the bubble enters as an insertion of the vacuum
polarization function at zero-momentum on the gluon line, $\Pi(0)$. The
renormalization of this $\Pi(0)$ means that for $\mu>m$ there are a total of
$n_f=6$ flavors for the SCET running of the soft-function.  For $\mu< m$ the
contributions from these top-bubbles is integrated out, and the soft-function
runs with $n_f=5$ flavors. Thus, in particular we always have $n_f=5$ for the
soft-function in bHQET.

To solve the RG-equation in Eq.~(\ref{rgeJS}) we can use that the same equation
holds for the evolution kernel $U_S(\ell^+,\ell^-,\mu,\mu_0)$ defined in
Eq.~(\ref{UJS}) that describes the running of the soft function from $\mu_0$ to
the scale $\mu$. Using the separable form $ U_S(\ell^+,\ell^-,\mu,\mu_0) =
U_s(\ell^+,\mu,\mu_0) U_s(\ell^-,\mu,\mu_0)$ of Eq.~(\ref{UUorg}) one obtains
from Eq.~(\ref{gammaS}) the relations
\begin{align}
 \mu \frac{d}{d\mu} U_s(\ell^\pm,\mu,\mu_0) = \pm K U_s(\ell^\pm,\mu,\mu_0)
    + \int\!\! d\ell^{\,\prime \pm}\: \gamma_s(\ell^\pm\!\minus
    \ell^{\,\prime\pm})\: U_s(\ell^{\,\prime\pm},\mu,\mu_0) \,.
\end{align} 
Here $K$ is a separation constant that can be set to zero.\footnote{Note that
  keeping the $\pm K$ term simply adds a multiplicative factor of
  $(\mu/\mu_\Lambda)^{\pm K}$ to the solutions, and cancels in the product
  $U_S(\ell^+,\ell^-)=U_s(\ell^+)U_s(\ell^-)$.} The solution for
$U_s(\ell^\pm,\mu,\mu_\Lambda)$ is given by Eq.~(\ref{UUU}),
\begin{align} \label{US}
  U_s(\ell^\pm,\mu,\mu_0) &= \frac{e^{K_2}\: (
  e^{\gamma_E})^{\omega_2} }{\mu_0\: \Gamma(-\omega_2)} \:
  \bigg[ \frac{\mu_0^{1+\omega_2}\theta(\ell^\pm)}  {(\ell^\pm)^{1+\omega_2}}\bigg]_+ 
\,,\nn\\[2mm]
  U_S(\ell^+,\ell^-,\mu,\mu_0) &= \frac{e^{2K_2}\: (
  e^{\gamma_E})^{2\omega_2} }{\mu_0^2\: \Gamma(-\omega_2)^2} \:
  \bigg[ \frac{\mu_0^{1+\omega_2}\theta(\ell^+)}  {(\ell^+)^{1+\omega_2}}\bigg]_+\:
   \bigg[ \frac{\mu_0^{1+\omega_2}\theta(\ell^-)}  {(\ell^-)^{1+\omega_2}}\bigg]_+  \,,
\end{align}
where at NLL order we use Eq.~(\ref{wL}) for $\omega_2=\omega_2(\mu,\mu_0)$ and
$K_2=K_2(\mu,\mu_0)$ with the values of $\Gamma_{0,1}$ and $\gamma_1$ determined
above. At LL order these are
\begin{align} \label{wLS}
  \omega_2^{LL}(\mu,\mu_0) &= 
     \frac{4 C_F}{\beta_0}\:\ln\Big[\frac{\alpha_s(\mu)}{\alpha_s(\mu_0)}\Big]
     \,,\qquad\quad
   K_2^{LL}(\mu,\mu_0) = 
  \frac{8\pi C_F}{\beta_0^2} \, \frac{(r-1-r\ln r)}{\alpha_s(\mu)} \,,
%
\end{align}
where $r=\alpha_s(\mu)/\alpha_s(\mu_0)$.  The running generated by
Eq.~(\ref{US}) falls in case 4). Note that for the jet function
$\omega_1(\mu,\mu_0)>0$ for $\mu>\mu_0$, while for the soft function
$\omega_2(\mu,\mu_0) < 0$. Although this affects the behavior of the jet and the
soft functions for large values of their arguments, the convolution of the jet
functions and soft function always remains finite.

The factorization of the soft function evolution,
$U_S(\ell^+,\ell^-)=U_s(\ell^+)U_s(\ell^-)$, is necessary for the consistency
equation in Eq.~(\ref{conssimple1}) to hold since it allows the cancellation to
independently occur for the two jets, which are each convoluted with one of the
variables of the soft function. Using Eqs.~(\ref{runcoeff}), (\ref{UJ}), and
(\ref{US}) and the relations
\begin{align}
  \omega_0(\mu_0,\mu) = - \omega_1(\mu_0,\mu) = - \omega_2(\mu,\mu_0) \,,\qquad
  e^{\frac12 K_0(\mu_0,\mu)} e^{K_1(\mu_0,\mu)} 
   = e^{K_2(\mu,\mu_0)} \Big[\frac{\mu}{\mu_0}\Big]^{-\omega_2(\mu,\mu_0)},
\end{align}
we find
\begin{align} \label{CEver1}
 \big[U_{H_Q}(Q,\mu,\mu_0) \big]^{1/2} U_J(s,\mu_0,\mu) &=  
    \Big[\frac{\mu}{Q}\Big]^{-\omega_2}\,
     \Big[\frac{\mu}{\mu_0}\Big]^{-\omega_2}\,
    \frac{e^{K_2} (e^{\gamma_E})^{\omega_2}}{\mu^2 \Gamma(-\omega_2)}
  \bigg[ \Big(\frac{\mu^2}{\mu_0 Q}\Big)^{1\plus\omega_2}\,
  \frac{\theta(s)\mu_0^{1\plus \omega_2}} {(s/Q)^{1\plus\omega_2}}\bigg]_+
   \nn\\
  &= \frac{1}{Q} \, U_s\Big(\frac{s}{Q},\mu,\mu_0\Big) \,.
\end{align}
This verifies that the SCET consistency condition is satisfied for the NLL
evolution factors. The consistency equations can also be verified at the level
of the distributions in Eq.~(\ref{cons2}) using the results in
Appendix~\ref{App:genRGE}. Between $\mu_Q=5 m$ and $\mu_m=m=172\,{\rm GeV}$ we
find $0 \le \omega_0(\mu_Q,\mu)\lesssim 0.14$ at LL and NLL order.

\subsection{Universal Running for a Class of $M_{t,\bar t}$ Definitions} 
\label{sect:universal}

In the last section we showed that the soft function for hemisphere invariant
masses satisfies the SCET consistency condition. Since the renormalization is
not sensitive to low energy properties of the soft function, like the mass
definitions, one should expect that there is a broader class of soft functions
that are consistent with the RGE in our factorization theorem.  In this section
we demonstrate this explicitly by working with a broader set of mass definitions
and calculating the corresponding soft functions to ${\cal O}(\alpha_s)$.

In Ref.~\cite{Fleming:2007qr} it was shown that the form of the factorization
theorem in Eqs.~(\ref{FactThm}) and (\ref{bHQETcross-hem3}) is retained for any
$M_{t,\bar t}$ prescription that assigns the hard top and antitop decay jets
unambiguously to $M_t$ and $M_{\bar t}$, and the momentum of every soft particle
to {\em either} $M_t$ or $M_{\bar t}$.  The former condition ensures that the
jet functions $B_\pm$, which are fully inclusive for the top decay products and
collinear radiation, remain unchanged. The latter condition ensures that
concentrating on $M_{t,\bar t}$ in the peak region automatically selects events
in the dijet region for which the SCET-bHQET setup can be applied.  The first
condition is satisfied by reconstruction methods since for $Q\gg m$ the hard
jets and collinear radiation are collimated to two back-to-back regions of the
detector and the decay products only have a power-suppressed probability of
${\cal O}(m^2/Q^2)$ to show up in the opposite hemisphere~\cite{Fleming:2007qr}.
The second condition restricts us to invariant mass definitions that incorporate
all soft radiation.  Examples include the hemisphere definition used in the last
section, and particle recombination methods such as those based on $k_T$ jet
algorithms~\cite{Catani:1991hj} with a $y_{\rm cut}$ parameter chosen so that
all soft radiation is assigned to the hard jets from the top/antitop decay.
Based on the equivalence of the top-down and the bottom-up approach to the
renormalization of quantities in the factorization theorem, and the fact that
the renormalization of the top-antitop production currents can only depend on
virtual corrections, it was concluded in Ref.~\cite{Fleming:2007qr} that the
renormalization properties of this class of soft functions does not depend on
the prescription how the soft gluon momenta are assigned to $M_t$ and $M_{\bar
  t}$.

Lets now extend the soft-function analysis beyond hemisphere masses, by setting
up a more general definition for $M_t$ and $M_{\bar t}$ and hence for the matrix
element defining $S(\ell^+,\ell^-,\mu)$.  Since the contributions from the
virtual graphs in Fig.~\ref{softgraphs}a,b,g are unaffected by phase space
constraints it is sufficient to consider the graphs \ref{softgraphs}c,d,e,f
describing real soft gluon final states. It is useful to write the gluon phase
space integral given in Eq.~(\ref{Sabcdintegrals}) in terms of the perp-momentum
$q_\perp$ and the angular variable
\begin{equation}
x\, \equiv\, \tan\frac{\theta}{2} \, = \, e^{-\eta}
\,,
\end{equation}
where $\theta$ is the gluon angle and $\eta$ the rapidity with respect to the
top momentum direction. This gives
\begin{align}
\tilde \mu^{2\epsilon} \int \frac{d^{d-1}q}{(2\pi)^{d-1}}\frac{1}{q^+q^- (q^+ + q^-)}
&= \tilde \mu^{2\epsilon} \,\frac{(4\pi)^{-2+\epsilon}}{\Gamma(1-\epsilon)} 
   \int_0^\infty d q^+ \int_0^\infty d q^- (q^+ q^-)^{-1-\epsilon}
\nn\\[2mm]
&= 2\,\tilde \mu^{2\epsilon}\, \frac{(4\pi)^{-2+\epsilon}}{\Gamma(1-\epsilon)}
  \int_0^\infty \frac{d x}{x} \, \int_0^\infty \frac{d q_\perp}{q_\perp^{1+2\epsilon}}
\,,
\end{align}
where $\tilde\mu$ is given in terms of $\mu$ in Eq.~(\ref{mutilde}).
For the hemisphere invariant mass prescription gluons in hemisphere-a ($0\le
x\le 1$) are assigned to the top, and gluons in hemisphere-b ($1\le x\le
\infty$) are assigned to the antitop. One can interpret this hemisphere
prescription as a crude jet algorithm. A prescription such as the $k_T$ jet
algorithm~\cite{Catani:1991hj}, that is tuned such that the total number of final
jets equals the number of hard jets from the top and antitop quark decays, leads
to a more complicated pattern since it depends on the particular momentum
configuration of the hard jets.

However, the situation is simplified since at leading order in the power
counting the hard jets are assigned unambiguously to the top and antitop
invariant masses. Thus upon averaging over all hard jet configurations the jet
algorithm assigns a soft gluon to either $M_t$ or $M_{\bar t}$ according to a
probability function, $f(x)$, that depends only on the angle $\theta$.
Expressing the phase space integration of the ${\cal O}(\alpha_s)$ soft function
for this general jet algorithm we have
\begin{align}
\lefteqn{S_{6c}+S_{6d} \nn}\\[2mm]
&=
2\,\frac{C_F\alpha_s}{\pi}\,\frac{(4\pi)^\epsilon\,\tilde \mu^{2\epsilon}}
  {\Gamma(1\minus\epsilon)}
 \!\int_0^\infty \!\frac{d x}{x} \! \int_0^\infty\!\! \frac{d q_\perp}{q_\perp^{1+2\epsilon}}
 \Big\{
f(x)\,\delta(\ell^-)\,\delta(\ell^+ \minus q_\perp x) 
+ \left[1\minus f(x)\right]\,\delta(\ell^- \minus q_\perp/x)\,\delta(\ell^+)
 \Big\}
\nn\\[2mm]
&= 
2\,\frac{C_F\alpha_s}{\pi}\,\frac{(4\pi)^\epsilon\tilde\mu^{2\epsilon}}{\Gamma(1\minus\epsilon)} 
\,\bigg\{\,
\frac{\delta(\ell^-)}{(\ell^+)^{1+2\epsilon}}\! \int_0^\infty\! 
 \frac{dx}{x^{1-2\epsilon}}\,f(x)
\, +
\frac{\delta(\ell^+)}{(\ell^-)^{1+2\epsilon}}\! \int_0^\infty\!
  \frac{dx}{x^{1+2\epsilon}}\,
\left[1\minus f(x)\right]
\,\bigg\}
\,,
\end{align}
where $f(x)$ gives the probability that a soft gluon with $x$ is assigned to
$M_t$. For the hemisphere masses we have $f(x)=\Theta(1-x)$.  Consistency at
leading order in the $m/Q$ power counting requires that $f(0)=1$ and
$f(\infty)=0$, i.e.\,\,the soft gluon is assigned with unit probability to
$M_t$ ($M_{\bar t}$) if it is radiated in exactly the top (antitop) momentum
direction.

Using the identity of Eq.~(\ref{standarddistribution}) and the scaling variable
$\kappa_2$ from the previous subsection it is then straightforward to determine
the bare soft function:
\begin{align} \label{softbaregeneral}
   S^{\rm bare}(\ell^+,\ell^-) 
  &= \delta(\ell^+)\,\delta(\ell^-)
  + \frac{C_F\,\alpha_s}{\pi}\,
    \bigg\{ \minus \frac{\delta(\ell^+)\delta(\ell^-)}{\epsilon^2} 
  + \frac{\delta(\ell^-)}{\epsilon\:\mu}
   \Big[\frac{\mu\, \theta(\ell^+)}{\ell^+}\Big]_+
  + \frac{\delta(\ell^+)}{\epsilon\: \mu}
   \Big[\frac{\mu\, \theta(\ell^-)}{\ell^-}\Big]_+
  \nn \\[3pt]
  & \qquad\qquad\qquad\qquad\qquad\quad
 + \tilde G_S(\ell^+,\ell^-) \bigg\},
\end{align}
where it is the finite terms that depend on the arbitrary probability function
$f(x)$
\begin{align}
 \tilde G_S(\ell^+,\ell^-) &=  
\frac12\, \delta(\ell^+)\,\delta(\ell^-)\, 
  \Big[\,\frac{\pi^2}{6} -8\,f_1 
\, \Big]
  +
2\,f_0\, 
   \frac{\delta(\ell^-)}{\mu}
   \,\Big[\frac{\mu\, \theta(\ell^+)}{\ell^+}\Big]_+ 
  -2\,f_0\,
   \frac{\delta(\ell^+)}{\mu}
   \, \Big[\frac{\mu\, \theta(\ell^-)}{\ell^-}\Big]_+
\nn\\[2mm] &
   - \frac{2 \delta(\ell^-)}{\mu}
  \Big[\frac{\theta(\ell^+)\ln(\ell^+/\mu)}{\ell^+/\mu}\Big]_+
   - \frac{2 \delta(\ell^+)}{\mu}
   \Big[\frac{\theta(\ell^-)\ln(\ell^-/\mu)}{\ell^-/\mu}\Big]_+
\,, 
\end{align}
and
\begin{align}
  f_n & \equiv \int_0^\infty\!\! dx\: \left(\frac{\ln^n x}{x}\right)_+\,f(x)
  \,.
\end{align}
For the hemisphere masses $f_n=0$ for any $n$, and $\tilde G_S(\ell^+,\ell^-)$
reduces to $G_S(\ell^+,\ell^-)$ in Eq.~(\ref{Gfunc}). For a more general
prescription for the soft gluon assignments that is symmetric under exchange of
top and antitop, i.e.\,\,has $f(x)=1-f(1/x)$, then one still has $f_0=0$, while
$f_1$ is in general non-vanishing. 

The result in Eq.~(\ref{softbaregeneral}) demonstrates that the UV-divergences
and the RG-evolution of the soft function are not affected by the phase space
constraints imposed on the soft gluons, whereas the UV-finite contributions
depend on them. This demonstrates to ${\cal O}(\alpha_s)$ that the form of the
factorization theorem in Eqs.~(\ref{FactThm},\ref{bHQETcross-hem3}) is retained for
the class of invariant mass definitions described above, and that different mass
prescriptions only affect the form of the soft function, but not its
renormalization scale-dependence.

\section{HQET Results}  \label{sect:bHQET}

To describe scales below the top mass we need to integrate out $m$ by switching
from SCET to bHQET. Here we describe the bHQET analogs of the matching, running,
and matrix element results given in the previous section on SCET.

\subsection{bHQET Current Matching and Running}\label{sect:bHQETcurrent}

In this section we determine the matching and the running of the $t\bar{t}$
current in bHQET at ${\cal O}(\alpha_s)$. For this we need to consider the
one-loop graphs in Fig.~\ref{bHQETloops}. For convenience, the relevant bHQET
Feynman rules have been collected in Appendix~\ref{App:bHQETFeyn}, as are the
results for the individual graphs.  We use dimensional regularization for UV
divergences and offshell momenta to regulate the IR divergences. For the top and
antitop quark momenta we take $p^\mu = m v_+^\mu + r_+^\mu$ and $\bar p^\mu = m
v_-^\mu + r_-^\mu$, respectively, and then let $p^2-m^2 = 2 m v_+\cdot r_+ =
\Delta^2$, and $\bar p^2-m^2 = 2 m v_-\cdot r_- = \Delta^2$, with $\Delta\ne 0$.
\begin{figure}
\begin{center}
\includegraphics[width=5.8in]{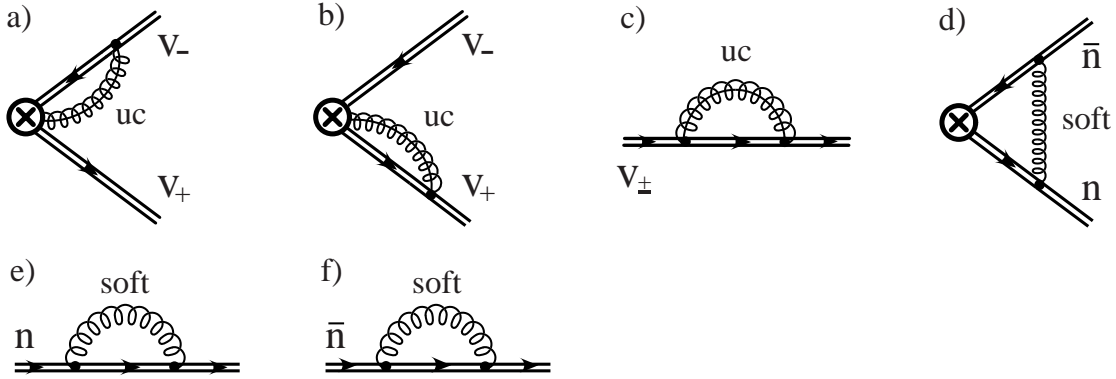}
\caption{Nonzero one-loop vertex and wavefunction corrections in boosted HQET. Graphs a), 
b), c) involve heavy quark fields $h_{v_\pm}$, while graphs d), e), and f) only involves the Wilson 
lines $Y_n^\dagger$ and $Y_\bn$.}
\label{bHQETloops}
\end{center}
\end{figure}

The sum of the three vertex contributions, the wavefunction contribution and the
residue is $V_{7a} + V_{7b} +V_{7c} + \Gamma^\mu_i(Z_h-1) + \Gamma^\mu_i(R_h-1)
$ and gives the bare amplitude
\begin{eqnarray}
\label{hqetvertex}
 \langle  p, \bar p | {\cal J}_i^\mu | 0 \rangle_{\rm bHQET} 
&=& \bar{h}_{v_+} \gamma^\mu_{\perp } h_{v_-}  \bigg[ 1+    \frac{\alpha_s C_F}{4 \pi} \bigg(
\frac{2}{\epsilon} \ln\frac{m^2}{-Q^2} + \frac{2}{\epsilon} 
\nn \\
& & \qquad - \ln^2\frac{\mu^2 Q^2}{-\Delta^4} + 4 \ln^2 \frac{m \mu}{-\Delta^2}
+ 4 \ln\frac{m\mu}{-\Delta^2}
+4 + \frac{\pi^2}{3} \bigg) \bigg]\,,
\end{eqnarray}
where $Q^2=Q^2+i0$ and $\Delta^2=\Delta^2+i0$.
The UV divergences in the bHQET current are subtracted by the counterterm for
the Wilson coefficient 
\begin{equation} \label{ZCm1}
Z_{C_m} = 1 -\frac{\alpha_s C_F}{4 \pi} \bigg[ \frac{2}{\epsilon}
\ln\frac{m^2}{-Q^2} + \frac{2}{\epsilon}  \bigg]  \,,
\end{equation}
giving the renormalized bHQET amplitude
\begin{align}
\label{hqetvertexren}
 \langle  p, \bar p | Z_{C_m}{\cal J}_i^\mu | 0 \rangle_{\rm bHQET} 
&= \bar{h}_{v_+} \gamma^\mu_{\perp } h_{v_-} 
\\
& \times \bigg[ \,1 + \frac{\alpha_s C_F}{4 \pi} \bigg(
 - \ln^2\frac{\mu^2 Q^2}{-\Delta^4} + 4 \ln^2 \frac{m \mu}{-\Delta^2}
+ 4 \ln\frac{m\mu}{-\Delta^2}
+4 + \frac{\pi^2}{3} \bigg) \bigg]\,.\nn
\end{align}
As discussed already in Sec.~\ref{subsec:bHQETrenor} the current
renormalization constant contains a term $\ln(m/Q)/\epsilon$ with a coefficient that
agrees with the coefficient of the $\ln(\mu/Q)/\epsilon$ term in the
renormalization constant of the SCET current in Eq.~(\ref{Zc}). Also, the
anomalous dimension of the bHQET current only exhibits single $1/\epsilon$
poles and thus sums only single $\ln(\mu)$ contributions. Although this
running sums double logarithmic terms of the form  $[\alpha_s(\mu) \ln(m/Q)\ln(\mu/m)]^k$,
it formally belongs to class 1).    

From the difference of the renormalized bHQET amplitude, $\langle p,\bar p|Z_{C_m}
{\cal J}_i^\mu | 0 \rangle$, and the renormalized SCET amplitude in
Eq.~(\ref{massivescetvertex1}), we obtain the bHQET current matching
conditions at the scale $\mu_m$: 
\begin{eqnarray}
C_{m}(m,\mu_m) &=& 1 + \frac{\alpha_s C_F}{4 \pi} \left(
\ln^2\frac{\mu_m^2}{m^2} + \ln\frac{\mu_m^2}{m^2} +4+ \frac{ \pi^2}{6} \right)\,
  \,.
\end{eqnarray}
The matching coefficient $H_m(m,\mu_m) = |C_m(m,\mu_m)|^2$ that appears in the
factorization theorem reads
\begin{eqnarray} \label{bhqetmatchcoeff}
 H_{m}(m,\mu_m) &=& 1 + \frac{\alpha_s C_F}{2 \pi} \left(
\ln^2\frac{\mu_m^2}{m^2} + \ln\frac{\mu_m^2}{m^2} +4+ \frac{ \pi^2}{6} \right)\,
  .
\end{eqnarray}
This matching result only depends on the parameter $m$, and at the scale
$\mu_m\sim m$ there are no large logarithms in $H_m(m,\mu_m)$.

The anomalous dimension is obtained from $Z_{C_m}$ using Eq.~(\ref{gZB}) and gives
\begin{eqnarray}
\gamma_{C_m}(\mu) &=& -Z_{C_m}^{-1}(\mu)\: \mu \frac{d}{d\mu} Z_{C_m}(\mu)
= \frac{\alpha_s
  C_F}{\pi}\bigg[\ln\frac{-Q^2-i 0}{m^2}-1\bigg] \,,
\nn \\
\gamma_{H_m}(\mu)&=&   \gamma_{C_m}(\mu) + \gamma_{C_m}(\mu)^* =  \frac{\alpha_s
  C_F}{4\pi}\bigg[8\ln\frac{Q^2}{m^2}-8\bigg]\, .
\end{eqnarray}
Comparing this result to Eq.~(\ref{gammaHm}) we find $\Gamma_0^{H_m}=-8C_F$,
$\gamma_0^{H_m}=-8 C_F$, and infer $\Gamma_1^{H_m}=-2\Gamma_1^{\rm cusp}$ for the
coefficients discussed in section~\ref{sect:NLL}.  The solution for the
evolution factor for the mass scale coefficient $H_m$ in Eq.~(\ref{UHm}) reads
\begin{align}
\label{hqetrun}
U_{H_m}\Big(\frac{Q}{m},\mu_m, \mu\Big)
&=  e^{K_{00}} \ \Big( \frac{m^2}{Q^2} \Big)^{\omega_0}
  \,,
\end{align}
where at NLL order we use the expressions in Eq.~(\ref{wL}) for
$\omega_0=\omega_0(\mu,\mu_m)$ and $K_{00}=K_{00}(\mu,\mu_m)$.  At LL order we
have
\begin{align}
  K_{00}^{LL} (\mu,\mu_m) &= 
  \frac{4 C_F}{\beta_0} \ln\bigg[ \frac{\alpha_s(\mu)}{\alpha_s(\mu_m)}
    \bigg] \,,
\end{align}
and just as in the running with $U_{H_Q}$, $\omega_0^{LL}(\mu,\mu_m) = (4
C_F/\beta_0)\, \ln[ \alpha_s(\mu)/\alpha_s(\mu_m) ]$.
Note that as in the case of the SCET current the RGE solution for the current
Wilson coefficient $C_m(m,\mu_Q,\mu)$ contains an extra phase factor,
\begin{equation}\label{cmrun}
C_m(m,\mu) = \sqrt{H_m(m,\mu) } 
\bigg[ \frac{\alpha_s(\mu)}{\alpha_s(\mu_m)}\bigg]^{2 \pi i \frac{C_F}{ \beta_0}} \,,
\end{equation}
that does not, however, appear in the cross section.~\footnote{It is interesting
  to note that the result in Eq.~(\ref{cmrun}) can be obtained from running the
  heavy-to-heavy current in HQET~\cite{Manohar:2000dt}, analytically continuing
  to the production region~\cite{Falk:1990cz}, and expanding in $m/Q$.}  The
origin of this phase, and the reason it drops out of the final predictions, is
the same as for the SCET current Wilson coefficient discussed in
Sec.~\ref{sect:scetcurrent}.
%

\subsection{ bHQET Jet functions Matching and Running} 
\label{sect:bHQETJet}

In this subsection we determine the bHQET jet functions $B_\pm$ defined in
Eq.~(\ref{hqetjet}) at ${\cal O}(\alpha_s)$, obtaining one-loop corrections to
the Breit-Wigner distributions in Eq.~(\ref{eq:Btree}). The results for $B_+$
and $B_-$ are identical by charge conjugation. We also determine the bHQET jet
function renormalization factor $Z_B$, the jet anomalous dimension $\gamma_B$,
the NLL evolution kernel $U_{B}$, and finally $B_\pm$ at NLL order.  By
comparing the jet functions in bHQET and SCET we confirm that their IR
divergences agree.  Finally, we demonstrate that the matching condition for
$H_m(m,\mu_Q)$, already obtained for the top-antitop currents in
Eq.~(\ref{bhqetmatchcoeff}), is reproduced by jet function matching. This is a
reflection of the statement that we have the same soft function in the SCET and
bHQET theories to ${\cal O}(\alpha_s)$.  Thus the soft function computations in
sections~\ref{sect:soft} and \ref{sect:universal} apply equally well for bHQET.

For the computation it is convenient to use the formulae from
Sec.~\ref{sec:FactGamma} which determine the jet function for the unstable top
quark from the results for a stable bHQET theory. To do this one can either use
the relation Eq.~(\ref{Bdisc}) which shifts the invariant mass variable into the
complex plane, or use the convolution relation in
Eq.~(\ref{factorizationGamma}).

\begin{figure}
\begin{center}
\includegraphics[width=5in]{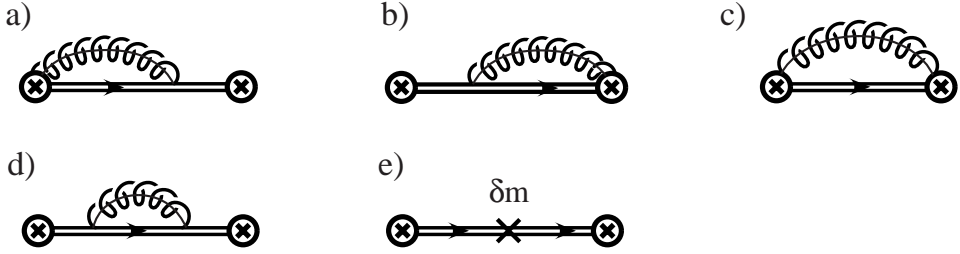}
\caption{bHQET graphs for the top quark jet function.}
\label{forwardII}
\end{center}
\end{figure}

The bHQET jet functions are given by the imaginary part of the vacuum matrix
elements ${\cal B}_\pm$ defined in Eq.~(\ref{hqetjet2}). At tree level
they are just given by the the HQET propagator,
\begin{equation} \label{cBtree}
{\cal B}_\pm(\shat,\Gamma_t=0) = 
-\frac{1}{\pi m}\:  \frac{1}{\hat s+ i 0}
\,.
\end{equation}
At one loop the diagrams contributing to the vacuum
matrix elements ${\cal B}_\pm$ are shown in Fig.~(\ref{forwardII}). 
Results for individual graphs are given in the appendix. The sum of the one-loop
graphs in an arbitrary mass scheme gives the bare expression
\begin{eqnarray}\label{bhqetpredisc}
{\cal B}_\pm^{\text{bare}} (\hat{s},\Gamma_t=0,\mu,\delta m) &=&  
 -\frac{1}{ \pi m}\:  \frac{1}{\hat s+ i  0} \bigg \{ 1 + 
\frac{\alpha _s C_F}{4\pi}  \bigg [ 
\frac{2}{\epsilon ^2}  +
  \frac{4}{\epsilon }\ln\Big( \frac{ \mu}{-\hat s-i  0}\Big)+ \frac{2}{\epsilon} 
\nn \\
& &\hspace{-3cm}
 + 4 \ln^2\Big( \frac{ \mu}{-\hat s - i  0} \Big)
  + 4\ln\Big( \frac{\mu}{-\hat s-i  0} \Big) + 4  + \frac{5\pi^2}{6} \bigg] \bigg\} 
 - \frac{1}{\pi m}\,\frac{2 \delta m}{(\hat s+i 0)^2} \,.
\end{eqnarray}
In general the residual mass term $\delta m=m_{\rm pole}-m$ is nonzero and
uniquely fixes the mass scheme $m$ that is being employed in HQET. In an
arbitrary mass scheme we have ${\cal B}_\pm(\hat s,\Gamma_t,\mu,\delta m)={\cal
  B}_\pm(\hat s-2\delta m,\Gamma_t,\mu)$. Here $\delta m$ is computed as a
perturbative series in $\alpha_s$. We have $\delta m \sim \alpha_s$ at lowest
order, and the $\delta m$ in the HQET Lagrangian, Eq.~(\ref{bHQETLagrangian}),
should be included as a perturbative insertion. This yields the term shown in
Eq.~(\ref{bhqetpredisc}).

The result in Eq.~(\ref{bhqetpredisc}) can be compared to the computation of
initial state radiation from a heavy color scalar resonance produced by the
collision of massless colored and neutral scalars in Ref.~\cite{Beneke:2004km}.
At leading order in the $1/m$ expansion the HQET gluon interactions are spin
independent and so only affect the normalization.  Furthermore to ${\cal
  O}(\alpha_s)$ there is no difference between initial state and final state
radiation, the signs of the $i0$ terms in the eikonal propagators do not modify
the result. In our calculation the analog of the initial state colored scalar in
Ref.~\cite{Beneke:2004km} is given by the final state Wilson lines in our
jet-function. Thus, we expect that the linear combination of terms in
Eq.~(\ref{bhqetpredisc}) in the pole mass scheme where $\delta m=0$ should be
the same as obtained in the scalar computation~\cite{Beneke:2004km}, and we have
checked that this is indeed the case. The scalar analysis of
Ref.~\cite{Beneke:2004km} was based on diagrams rather than deriving a
factorization theorem, so an operator analogous to the one we give for our
jet-functions was not given.

 The renormalization of the vacuum matrix element and the jet functions for the
stable or the unstable bHQET theory is equivalent, so one can obtain the
renormalization factor for the jet functions from
Eq.~(\ref{bhqetpredisc}). The result reads
\begin{equation}
\label{ZBp}
Z_{B_\pm} (\hat s-\hat s') = 
\delta(\hat s-\hat s') + \frac{\alpha _s C_F}{4\pi } 
 \bigg \{ \delta(\hat s-\hat s')
\bigg [  \frac{2}{\epsilon^2}
  + \frac{2}{\epsilon}  \bigg ]  
- \frac{4}{\mu\,\epsilon}\bigg [ \frac{\mu\, \theta (\hat s- \hat
  s')}{\hat s -\hat s'}\bigg]_+ \bigg \} \,.
\end{equation}
Note that the bHQET jet functions Z-factor and their anomalous dimension do not
depend on the mass scheme that is being used.  The renormalized vacuum matrix
element has the form
\begin{align}\label{bhqetprediscren}
{\cal B}_\pm (\hat{s},0,\mu,\delta m) &=
 -\frac{1}{\pi m}\:  \frac{1}{\hat s\plus i  0} \bigg \{ 1 + 
\frac{\alpha _s C_F}{4\pi}  \bigg [
4  \ln^2\bigg( \frac{ \mu}{-\hat s \minus i  0} \bigg) \plus 4\ln\bigg( \frac{
  \mu}{-\hat s\minus i  0} \bigg) \plus 4  \plus \frac{5\pi^2}{6} \bigg] \bigg\}
\nn\\
 &  - \frac{1}{\pi m}\,\frac{2\delta m}{(\hat s+i 0)^2} \,.
\end{align}
The renormalized jet function accounting for the large top quark width is then
either given by taking the imaginary part of Eq.~(\ref{bhqetprediscren})
upon the shift $\hat{s}\to\hat{s}+i\Gamma_t$ (see Eq.~(\ref{Bdisc})) or by applying
the convolutions formula of Eq.~(\ref{factorizationGamma}). The result is
\begin{align}\label{bhqetjetwidth}
B_\pm(\hat{s},\Gamma_t,\mu,\delta m) &=
%
 \frac{1}{\pi m}\: \frac{\Gamma}{\hat s^2 \plus \Gamma^2} \bigg\{
1+  \frac{\alpha _s C_F}{4\pi} \bigg [\,
    \text{ln}^2 \Big( \frac{  \mu^2}{\hat s^2 \plus \Gamma^2} \Big ) 
      \plus 2 \> \text{ln} \Big ( \frac{  \mu^2}{\hat s^2 \plus \Gamma ^2} \Big ) 
      \minus 4\, \text{arctan}^2\Big ( \frac{\Gamma}{\hat s} \Big ) 
\nn \\[2mm]       
&\hspace{-1.5cm}
 \mbox{}\quad 
+\, 4\, \frac{\hat s}{ \Gamma} \,\text{arctan}\Big ( \frac{\Gamma}{\hat s} \Big ) 
   \Big\{  \text{ln} \Big ( \frac{  \mu^2}{\hat s^2 \plus \Gamma^2} \Big) + 1
   \Big\}        
         + 4 +  \frac{5\pi^2}{6}\, \bigg ] \bigg \}  
 +  \frac{1}{\pi m} \, \frac{(4\,\hat s\,\Gamma_t)\, \delta m}{(\hat s^2 +
   \Gamma^2)^2} 
  \,,
\end{align}
where the $\arctan$ is evaluated in the third quadrant. In practice, in the
presence of the width it is more convenient to not bother evaluating the
imaginary part explicitly, and simply use $B_\pm(\hat{s},\Gamma_t,\mu,\delta m)
= \textrm{Im}\big[ {\cal B}_\pm(\hat s \plus i\,\Gamma_t,0,\mu,\delta m)\big] $.
We also present the stable bHQET jet function,
\begin{align}\label{bhqetjetstable}
B_\pm^{\Gamma=0}(\hat{s},\mu,\delta m) &=
\textrm{Im} \, \big[ {\cal B}_\pm(\hat s,0,\mu,\delta m)\big] 
\nn \\[3mm]
 & =  \delta(s) 
  + \frac{\alpha_s(\mu) C_F}{\pi} \bigg\{
     \frac{2}{m\kappa_3} \bigg[\frac{\theta(z)\ln(z)}{z}\bigg]_+  
    - \frac{1}{m\kappa_3} \bigg[ 1 + 
    2\,\ln\Big(\frac{\mu}{\kappa_3}\Big) \bigg]
     \bigg[\frac{\theta(z)}{z}\bigg]_+
   \nn  \\[2mm]
 &\mbox{}\hspace{1cm}   + \delta(s) \bigg[
     \ln^2\Big(\frac{\mu}{\kappa_3}\Big)
   \plus  \ln\Big(\frac{\mu}{\kappa_3}\Big) 
   + 1 - \frac{\pi^2}{8} \bigg] \bigg\} 
  -\,\frac{2\,\delta m}{m} \,\delta^\prime(\hat s) ,
\end{align}
where we have allowed for an arbitrary rescaling of $\hat s = \kappa_3\, z$.  A
convenient choice for the parameter $\kappa_3$ is $\kappa_3=\mu$, where $z=\hat
s/\mu$ and
\begin{align}\label{bhqetjetstable2}
B_\pm^{\Gamma=0}(\hat{s},\mu,\delta m) &
=  \delta(s) 
  + \frac{\alpha_s(\mu) C_F}{\pi} \bigg\{
     \frac{2}{m\mu} \bigg[\frac{\theta(z)\ln(z)}{z}\bigg]_+  
    - \frac{1}{m\mu} 
     \bigg[\frac{\theta(z)}{z}\bigg]_+
    + \delta(s) \bigg[
     1 - \frac{\pi^2}{8} \bigg] \bigg\}  \nn  \\[2mm]
 &\mbox{}\hspace{1cm} 
  -\,\frac{2\,\delta m}{m} \,\delta^\prime(\hat s) \,.
\end{align}

To determine $H_m$ we can match the bHQET and SCET jet function results for
$\hat{s}=s/m\ll m$. For the matching the top width only takes the role of an IR
parameter, and the computation is most conveniently carried out for stable top
quarks. If the same mass scheme is used in SCET and bHQET, then the $\delta m$
terms are the same, and cancel in the matching. The jet-function matching
coefficient can be obtained from matching either the jet functions or the vacuum
matrix elements.  For this computation it is convenient to pick $\kappa_3=m$ in
Eq.~(\ref{bhqetjetstable}) and subtract it from Eq.~(\ref{Jrenm}) to obtain
\begin{eqnarray}
\label{matchbhqetoneloop}
T_{\pm}(m,\mu_m) =
1 +  \frac{\alpha _sC_F}{4\pi }\bigg ( \text{ln}^2\frac{m^2}{\mu_m^2} 
 -\>\text{ln}\frac{m^2}{\mu_m^2}
 + 4 + \frac{\pi^2}{6} \bigg ) \,.
\end{eqnarray}
Using $H_m = T_+ T_-$ this agrees with Eq.~(\ref{bhqetmatchcoeff}) at ${\cal
  O}(\alpha_s)$.

The anomalous dimension for the jet function is determined from
Eq.~(\ref{ZBp}) and reads 
\begin{equation}
\gamma _{B} (\hat s-\hat s', \mu ) = 
-\frac{\alpha _s C_F}{4\pi }  \bigg \{
\frac{8}{\mu}  \bigg[ \frac{\mu\,
 \theta (\hat s- \hat s')}{\hat s - \hat s'}\bigg]_+
-  4 \delta (\hat s-\hat s')  
\bigg \} \,. 
\end{equation}
Comparing this result to Eq.~(\ref{gammaF}) we find $\Gamma_0^{B}=4C_F$,
$\gamma_0^{B}= 4 C_F$, and infer $\Gamma_1^{B}= \Gamma_1^{\rm cusp}$ for the
coefficients discussed in section~\ref{sect:NLL}.  The solution for the
evolution equation~(\ref{UB}) is
\begin{equation}
\label{UB2}
U_{B} (\hat s-\hat s', \mu , \mu_0) = \frac{e^{K_3}(
  e^{\gamma_E})^{\omega_1}}{\mu_0\, \Gamma(- \omega_1)}
\bigg [ \frac{\mu_0^{1+\omega_1} \theta (\hat s - \hat s')}{(\hat s-\hat s')^{1+\omega_1}} \bigg ]_+ \,,
\end{equation}
where $\omega_1=\omega_1(\mu,\mu_0)$ and $K_3=K_3(\mu,\mu_0)$ are determined at
NLL order using Eq.~(\ref{wL}) with $r=\alpha_s(\mu)/\alpha_s(\mu_0)$. At LL
order
\begin{align} \label{wL12}
 \omega_1^{LL}(\mu,\mu_0)  &= 
     -\frac{4C_F}{ \beta_0}\:\ln\Big[\frac{\alpha_s(\mu)}{\alpha_s(\mu_0)}\Big]
     \,,\quad\qquad
 K_3^{LL}(\mu,\mu_0) = -
  \frac{8\pi C_F}{\beta_0^2} \, \frac{(r-1-r\ln r)}{\alpha_s(\mu)} 
  \,.
\end{align}

In comparing the bHQET jet function to that in SCET, the most striking
difference is that the dimension-1 variable $\hat s$ is natural for bHQET,
whereas in SCET we had a natural dimension-2 variable $s$. This causes a
difference in the convolution of jet and soft functions in the two theories.
However, comparing the bHQET jet function evolution function $U_{B}$ to the SCET
evolution function $U_{J}$ in Eq.~(\ref{UJ}), one notices that we have the same
function $\omega_1$ of $\alpha_s$.  This is crucial to the fact that the
spectrum remains protected against large logs as we evolve below $\mu=m$, with
the consistency condition in SCET carrying over to a consistency condition in
bHQET, as given in Eq.~(\ref{CE}).  In particular, the relation between the soft
and jet evolution factors, $\omega_1(\mu_0,\mu)=\omega_2(\mu,\mu_0)$, remains
valid. This ensures that the plus functions in $U_J$ and $U_s$ match, which was
the key to verifying the SCET consistency condition in Eq.~(\ref{CEver1}) above.
The fact that $\omega_1$ is unchanged can be viewed as a compensation between a
change in the {\em geometry} of the physical color flow governing the QCD
dynamics in the two theories, and a change in the {\em dimension} $j$ of the
natural jet function variable used in Appendix~\ref{App:genRGE}.  Here the
geometric properties are encoded in the cusp anomalous dimensions $\Gamma_0$,
which are determined by the cusp angle between Wilson lines in the jet
functions. In particular the equality of the $\omega_1$'s follows from the
equality of the ratio
\begin{align}
  \frac{(\Gamma_0)^{J_{n,\bn}}}{(j)^{J_{n,\bn}}}
    =  \frac{ (\Gamma_0)^{B_\pm}}{(j)^{B_\pm}} \,,
\end{align}
as can be seen from Eq.~(\ref{wL}) in Sec.~\ref{sect:NLL}.  To verify the
bHQET consistency condition  to NLL order we use
$\omega_0(\mu_0,\mu)=-\omega_1(\mu_0,\mu)=-\omega_2(\mu,\mu_0)$ and note that
\begin{align}
  e^{\frac12 K_{00}(\mu_0,\mu)} e^{K_3(\mu_0,\mu)} 
   = e^{\frac12 K_0(\mu_0,\mu)} e^{K_1(\mu_0,\mu)} 
  = e^{K_2(\mu,\mu_0)} \Big[ \frac{\mu}{\mu_0} \Big]^{-\omega_2(\mu,\mu_0)}
   \,,
\end{align}
which allows us to obtain the desired result in Eq.~(\ref{CE}):
\begin{align} \label{CEver2}
 \Big[U_{H_m}\Big(\frac{Q}{m},\mu,\mu_0\Big) \Big]^{1/2} U_B(\hat s,\mu_0,\mu) &=  
    \Big[\frac{m}{Q}\Big]^{-\omega_2}
     \Big[\frac{\mu}{\mu_0}\Big]^{-\omega_2}\,
    \frac{e^{K_2} (e^{\gamma_E})^{\omega_2}}{\mu \Gamma(-\omega_2)}
  \bigg[ \Big(\frac{\mu\, m}{\mu_0\, Q}\Big)^{1\plus\omega_2}
  \frac{\theta(\hat s)\mu_0^{1\plus \omega_2}} 
   {(m\,\hat s/Q)^{1\plus\omega_2}}\bigg]_+
   \nn\\
  &= \frac{m}{Q} \, U_s\Big(\frac{m\hat s}{Q},\mu,\mu_0\Big) \,.
\end{align}
Again the equivalent bHQET consistency equation with distributions given in
Eq.~(\ref{cons4}) can be verified using results from Appendix~\ref{App:genRGE}.
Between $\mu_m=172\,{\rm GeV}$ and $\mu = 1\,{\rm GeV}$ we find $0 \le
\omega_0(\mu_Q,\mu)\le 0.93$ at LL and NLL order. Interestingly, from the
factorization theorem in Eqs.~(\ref{bHQETcross-hem2a}) and
(\ref{bHQETcross-hem2aa}) the $U_{B}$ factor is run down from $\mu_\Gamma$ to
$\mu_\Lambda$ giving $\omega_1<0$, or the $U_S$ factor is run up from
$\mu_\Lambda$ to $\mu_\Gamma$ giving $\omega_2<0$. Thus we never exceed the bound
$\omega<1$ on the range of validity of the convolution resummation formulas
given in Sec.~\ref{sect:NLL}.

Finally, we quote analytic results for the resummed bHQET jet functions.  Using
the tree level result for bHQET propagator ${\cal B}_\pm$ in Eq.~(\ref{cBtree})
as the initial condition at the scale $\mu_0$ it is straightforward to determine
the LL result at the scale $\mu$ by carrying out the integral in Eq.~(\ref{UB}).
For the vacuum matrix element this gives
\begin{equation}
 {\cal B}_\pm^{\textrm{LL}}(\hat s,0,\mu\OMIT{,\mu_0}) = \frac{1}{m\pi} \: 
 e^{K_3^{\rm LL}(\mu , \mu_0)}\big(\mu_0 e^{\gamma_E})^{\omega_1^{\rm LL}}
  \frac{\Gamma(1+\omega_1^{\rm LL})}{(-\hat s - i  0)^{1+\omega_1^{\rm LL}}} \,,
\end{equation}
where $\omega_1^{\rm LL}$ and $K_3^{\rm LL}$ are given in Eq.~(\ref{wL12}).  Now
using Eq.~(\ref{factorizationGamma}) gives the LL bHQET jet function at the
scale $\mu$ with the tree level jet function $B_{\pm}^{\rm tree}=\Gamma/[\pi
m(\hat s^2+\Gamma^2)]$ as the input at the scale $\mu_0$:
\begin{equation}
\label{BLL2}
 B_\pm^{\textrm{LL}}(\hat s,\Gamma_t,\mu\OMIT{,\mu_0}) = \frac{1}{m\pi} \: 
 e^{K_3^{\rm LL}(\mu , \mu_0)}\, \big(\mu_0 e^{\gamma_E})^{\omega_1^{\rm LL}}\,
 \Gamma(1+\omega_1^{\rm LL})\, \textrm{Im}\,\bigg[\ \frac{1}{(-\hat s - i
 \Gamma)^{1+\omega_1^{\rm LL}}} 
\,\bigg]\,.
\end{equation}
Since the boundary condition is specified at tree level the result does not
involve $\delta m$. When the ${\cal O}(\alpha_s)$ jet function is taken as the
initial condition at the scale $\mu_0$ with NLL evolution we obtain what we will
call the NLL result for the jet function. This result depends on $\delta m$ and
is given by the analytic result
\begin{align} \label{BNLL}
 B_\pm^{\textrm{NLL}}(\hat s,\Gamma_t,\mu\OMIT{,\mu_0},\delta m) &= \frac{1}{m\pi} \: 
 e^{K_3(\mu , \mu_0)}\, \big(\mu_0 e^{\gamma_E})^{\omega_1}\,
 \Gamma(1+\omega_1)\, \textrm{Im}\,\Bigg[ \frac{1}{(-\hat s - i
   \Gamma_t)^{1+\omega_1}}  \Bigg\{ 1  +   \nn\\ 
 & \hspace{-2cm}
 \frac{C_F \alpha_s(\mu_0)}{\pi} \bigg[ 1
 \plus\frac{\pi^2}{24} \minus \ln\Big(\frac{-\hat s\minus i\Gamma_t}{\mu_0}\Big)\plus 
 H({\omega_1}) \plus \Big\{  \ln\Big(\frac{-\hat s\minus i\Gamma_t}{\mu_0}\Big)
 \minus H({\omega_1})\Big\}^2 \nn\\
 & \hspace{-2cm}
  \plus  \Psi'(1\plus \omega_1) \bigg] 
  - \frac{2(1\plus \omega_1) \delta m}{(-\hat s -i\Gamma_t) }
 \Bigg\}\Bigg],
\end{align}
where $H(\omega)$ is the harmonic number function and $\Psi^\prime(z)= d/dz
[\Gamma^\prime(z)/\Gamma(z)]$ is the derivative of the polygamma function. This
result was derived using Eq.~(\ref{Gn}) from Appendix~\ref{App:Gfunc}.


\section{Short-Distance Top Jet Mass and the Top Quark Pole} 
\label{sect:jetmass}


One of the main goals of the factorization based analysis of $d^2\sigma/dM_t^2
dM_{\bar t}^2$ is to facilitate a high precision determination of the top-quark
mass. To do so it is important to explore the correspondence between the
top-mass and the peak in the predicted $M_{t,\bar t}$ invariant mass
distribution.  Eq.~(\ref{FactThm}) shows that the peak position of the
top/antitop invariant mass distributions is affected by perturbative corrections
in the jet functions $B_\pm(\hat s,\Gamma_t,\mu)$ and by nonperturbative effects
through the convolution with the soft function $S(\ell^+,\ell^-,\mu)$.  In this
section we analyze effects of the jet-functions on the peak, and define a
consistent short distance jet-mass scheme. The jet functions were computed to
${\cal O}(\alpha_s)$ in section~\ref{sect:bHQET}. In these computations the top
quark width $\Gamma_t$ provides an IR-cutoff that makes the perturbative
determination of the jet function line shape valid for any value of $\hat s$,
and in particular for the peak region.

As can be seen from Fig.~\ref{forwardII}, the Feynman diagrams for $B_\pm$
contain the HQET heavy quark self-energy. The other diagrams are generated from
the Wilson lines $W_{n,\bn}$ and render the jet functions gauge-invariant.  In
the pole mass scheme the HQET self-energy develops a linear sensitivity to IR
momenta~\cite{Beneke:1994sw}. This is caused by an ambiguity of ${\cal
  O}(\Lambda_{\rm QCD})$ in the pole mass, and hence in the pole-scheme
invariant mass variable $\hat s$, i.e.  in $\hat s=(M^2_{t,\bar t}-m_{\rm
  pole}^2)/m_{\rm pole}$ where $\delta m=0$.  In perturbation theory this
ambiguity is associated to an asymptotic behavior $\propto \mu
\alpha_s(\mu)^{n+1} \beta_0^n\, n!$, where $\mu$ is the scale employed for
$\alpha_s(\mu)$ in the jet function.  The situation is similar to the total
cross section for top-antitop pair production in the threshold region $Q\sim
2m$~\cite{Hoang:2000yr,Hoang:2001mm}, where using the pole mass the peak
position can not be rendered stable in perturbation theory. Thus, in the pole
mass scheme $B_\pm$ is expected to have a poorly behaved perturbation series,
indicating that it is not the pole mass that can be accurately determined from
the measured invariant mass distribution.  This is because the pole mass is
defined order by order to be a zero of the inverse heavy quark two-point
function that is not observable physically.  This feature is demonstrated for
the one-loop jet function below and will be analyzed at higher orders in
Refs.~\cite{FHMSmass:inprep,JSS:inprep}.

It is therefore advantageous and even mandatory to switch to a short-distance
mass scheme that can stabilize the location of the jet function peak location in
perturbation theory. In a general mass scheme the location of the peak is
determined by
\begin{align}
\label{speak}
\frac{d {B}_\pm (\hat{s},\Gamma_t,\mu,\delta m)}{d \hat{s}}\Big
|_{\hat{s}=\hat{s}_{\rm peak}} =
\frac{d {B}_\pm (\hat{s}-2\delta m,\Gamma_t,\mu)}{d \hat{s}}\Big
|_{\hat{s}=\hat{s}_{\rm peak}}
= 0 \,.
\end{align}
At tree level $\delta m=0$ and the jet functions are equal to the Breit-Wigner
functions in Eq.~(\ref{eq:Btree}), so $\hat s_{\rm peak}=0$. At ${\cal
  O}(\alpha_s)$ the jet functions ${B}_\pm$ are given by the expressions in
Eqs.~(\ref{bhqetjetwidth}) and solving the condition in Eq.~(\ref{speak})
perturbatively gives
\begin{align} \label{dmjetpeak}
 {\hat s}_{\rm peak}^{\text{NLO}} =
  2 \delta m - \frac{\alpha_s(\mu)\, C_F}{2} \,\Gamma_t\,
 \Big[ \ln\Big(\frac{\mu}{\Gamma_t}\Big) + \frac{3}{2}  \,\Big]
  \,.
\end{align}
As explained in Ref.~\cite{Fleming:2007qr} a viable short-distance mass scheme
must have $\delta m\sim \alpha_s\Gamma_t$ in order not to violate the power
counting.  This condition rules out the ${\overline{\rm MS}}$-mass as a
candidate since $\delta m_{\overline {\rm MS}}\sim \alpha_s m_t$, and hence
violates the factorization theorem for the cross-section. A viable
short-distance mass scheme can be defined using Eq.~(\ref{speak}) by demanding
that $\hat s_{\rm peak}=0$ order-by-order in perturbation theory.
Eq.~(\ref{dmjetpeak}) then determines $\delta m$ at NLO. This mass satisfies the
power counting criteria, but unfortunately has a complicated dependence on the
renormalization scale.  This can be seen from the peak position at LL order,
derived using Eq.~(\ref{BLL2}) for $B_{\pm}$ with the initial condition of a
Breit-Wigner at the scale $\mu_0$.  This LL result is independent of the mass
scheme and we find
\begin{align} \label{speakLL}
  \hat s_{\rm peak}^{\rm LL}(\mu,\mu_0) &= - \Gamma_t \,
    \cot\Big[\frac{\pi}{2+\omega_1^{LL}(\mu,\mu_0)}\Big]  
  = \, \Gamma_t\, \frac{C_F\pi}{\beta_0} 
    \ln\Big[\frac{\alpha_s(\mu)}{\alpha_s(\mu_0)}\Big] + \ldots \,.
\end{align}
Here $\omega_1^{LL}(\mu,\mu_0)$ is given in Eq.~(\ref{wL12}), and in the second
equality we show the leading term for small $\omega_1$. 
%
%
Due to the non-linear
nature of the cotangent in Eq.~(\ref{speakLL}
) a mass scheme determined by $\hat s_{\rm peak}=0$ is not transitive, in the
sense that $\hat s_{\rm peak}^{\rm LL}(\mu,\mu_1) + \hat s_{\rm peak}^{\rm
  LL}(\mu_1,\mu_2) \ne \hat s_{\rm peak}^{\rm LL}(\mu,\mu_2)$.  This makes a
mass definition based on the peak position awkward to use.  The problem occurs
because the peak position is a local feature of $B_\pm(\hat s,\mu)$, while
$B_\pm$ requires a convolution for its RG-evolution as seen in Eq.~(\ref{UB}).
If Eq.~(\ref{speak}) is evolved to a different renormalization scale then it
involves an integral over $B_\pm$.  In perturbation theory this nonlocal feature
is reflected by terms $[C_F\, \alpha_s \ln]^k$ in the expansion of $\hat s_{\rm
  peak}^{\rm LL}$.

To define a transitive short-distance jet-mass $m_J(\mu)$, which is still
closely related to the peak-position, we will use the first moment of
$B_+^{\Gamma=0}$. To LL order it suffices to simply use an upper cutoff $L_m$ on
this moment, and define $\delta m_J$ so that this moment vanishes
\begin{align} \label{mJmomentdef}
  0 = \int_{-\infty}^{L_m} \!\! d\hat s \ \hat s\  B_+^{\Gamma=0}(\hat s,\mu,\delta m_J)  
    =\int_{-\infty}^{L_m} \!\! d\hat s \ \hat s\  B_+^{\Gamma=0}(\hat s -2\,\delta m_J,\mu)
  \,.
\end{align}
Different choices of $L_m$ define different schemes for the mass.  As
indicated, it also suffices to define the mass scheme using the zero-width jet
function.  As shown in Eq.~(\ref{factorizationGamma}) the jet function for a
non-zero width is related to the stable one by
\begin{eqnarray}
\label{factorizationGamma2}
B_\pm(\hat s,\delta m_J,\Gamma_t,\mu) & = &
  \int_{-\infty}^{\,\,\mbox{$\hat s$}} \!\!\! d{\hat s}^\prime \
B^{\Gamma=0}_\pm({\hat s}-{\hat s}^\prime,\delta m_J,\mu)\:
\frac{\Gamma_t}{\pi\,({\hat s}^{\prime\,2}+\Gamma_t^2)}
\,,
\end{eqnarray}
and so the stability of $B_\pm^{\Gamma=0}$ is directly transferred to $B_\pm$.
We can solve Eq.~(\ref{mJmomentdef}) keeping only the linear term in $\delta
m_J$, thus
\begin{align}
  0
  &=  \int_{-\infty}^{L_m} \!\! d\hat s \ \hat s\  B_+^{\Gamma=0}(\hat s,0)
   -2 \delta m_J \int_{-\infty}^{L_m} \!\! d\hat s \ \hat s\  \frac{d}{d\hat s}
     B_+^{\Gamma=0}(\hat s,0)  + {\cal O}\big[(\delta m_J)^2 \alpha_s\big]
  \,.
\end{align}
Integrating by parts gives the solution
\begin{align}
  \delta m_J = -\:
   \frac{  \int_{-\infty}^{L_m}  d\hat s \ \hat s\  B_+^{\Gamma=0}(\hat s,0)  }
  { \big[ 2\int_{-\infty}^{L_m}  d\hat s \:  B_+^{\Gamma=0}(\hat s,0) \big]
  -2 L_m B_+^{\Gamma=0}(L_m,0) } 
 \,.
\end{align}
Expanding in $\alpha_s$ to one-loop order we find
\begin{align} \label{dmjet}
  \delta m_J(\mu) = L_m\: \frac{\alpha_s(\mu) C_F}{\pi}\,
   \Big[ \ln\Big(\frac{\mu}{L_m}\Big) + \frac{3}{2} \Big]\,.
\end{align}
To obtain a consistent mass for top-quark jets we must choose the scheme
parameter $L_m\sim \Gamma_t$.  We will adopt $L_m=1\,{\rm GeV}$ for our analysis
since then $\delta m_J$ gives $|\hat s_{\rm peak}^{\rm NLO}| \le 32\,{\rm MeV}$
for $\mu=2$--$10\,{\rm GeV}$ and hence a very stable peak position.  The use of
a first moment to define a mass scheme as in Eq.~(\ref{mJmomentdef}) has been
applied in a similar way earlier to inclusive B-decays~\cite{Bosch:2004th}, to give what is known
as the shape-function scheme. However the shape-function scheme does result in a
different mass definition from the jet-scheme defined here.

To achieve cancellation of the $O(\Lambda_{\rm QCD})$ renormalon ambiguity in
the jet mass scheme the scale $\mu$ in $\delta m_J(\mu)$ needs to agree with the
renormalization scale used for the strong coupling in the corrections of the jet
function. Using Eq.~(\ref{deltamdef}) the one-loop relation between the pole and
jet mass $m_J$ is:
\begin{align}
\label{mjetpole}
m_J^{\text{NLO}}(\mu) = m_{\rm pole} - L_m \frac{\alpha_s(\mu)\,
  C_F}{\pi} \Big[
\ln\Big(\frac{\mu}{L_m}\Big) + \frac{3}{2} 
\,\Big]\,.
\end{align}
We can also derive a LL result for the running jet-mass.  From the NLO $\delta
m_J$ in Eq.~(\ref{dmjet}) we can compute a renormalization group equation for
$m_J(\mu)$, whose LL solution is
\begin{align} \label{mjetLL}
  m_J^{\rm LL}(\mu) = m_J(\mu_0) + L_m\: \frac{2 C_F }{\beta_0} \ln\bigg[
  \frac{\alpha_s(\mu)}{\alpha_s(\mu_0)}\bigg] \,.
\end{align}
To verify that this result contains all the leading-logs, we use
Eq.~(\ref{BLL2}) to determine $(B_+^{\Gamma=0})^{\rm LL}$ with evolution from
$\mu_1$ up to $\mu$. Using the LL jet function in Eq.~(\ref{BLL2}) and solving
Eq.(\ref{mJmomentdef}) without expanding in $\delta m$, we find a solution
$\delta m(\mu,\mu_1)$ that contains all leading-logs between $\mu_1$ and
$\mu$. Then by taking $m_J^{\rm LL}(\mu) = m_J(\mu_0) + \delta m(\mu_0,\mu_1) -
\delta m(\mu,\mu_1)$ we obtain a $\mu_1$-independent result that reproduces
exactly Eq.~(\ref{mjetLL}).  The jet-mass $m_J(\mu)$ has a standard series of
$[\beta_0 \alpha_s\ln]^k$ terms, and as far as its RG-evolution is concerned
behaves very similar to an $\overline {\rm MS}$ mass. In particular, this
jet-mass is transitive at LL order.

\begin{figure}
  \centerline{ 
   \includegraphics[width=8.5cm]{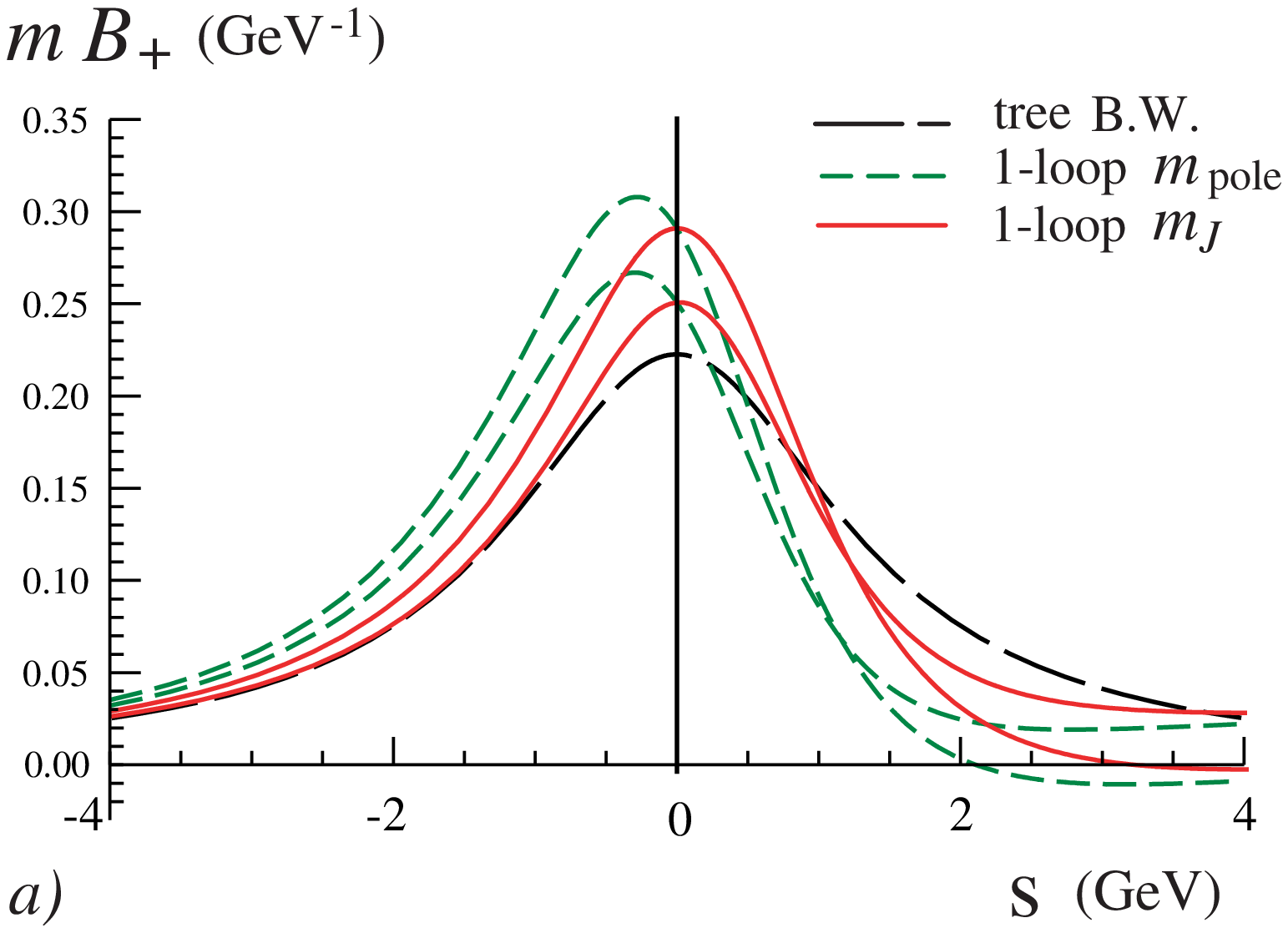} 
  \hspace{0.3cm}
   \includegraphics[width=7cm]{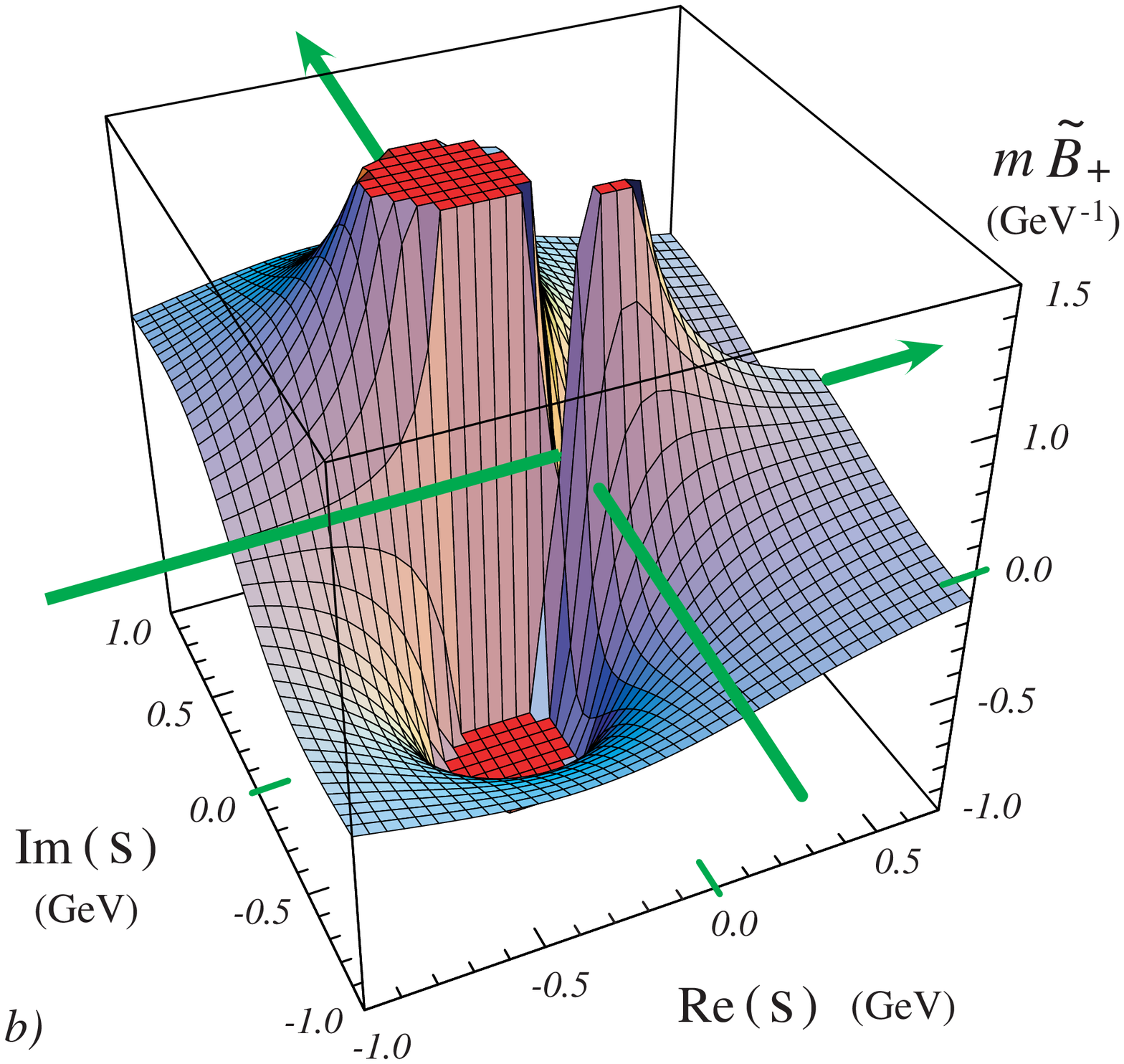} 
  } 
\caption{ (a) The bHQET jet function $m B_\pm(\hat
  s,\Gamma_t,\mu)$ as a function of $\hat s$ at tree level (black long dashed
  line) and ${\cal O}(\alpha_s)$ in the pole mass scheme (green short dashed
  lines) and the jet mass scheme (red solid lines) for $\mu=2$ (lower lines) and
  $5$~GeV (upper lines). (b) Imaginary part of $m {\cal B}_\pm(\hat s,0,\mu)$ in
  the pole mass scheme for $\mu=2$~GeV plotted in the complex $\hat s$-plane.
  Solid green lines indicate ${\rm Re}(\hat s)=0$ or ${\rm Im}(\hat s)=0$ in the
  plane where ${\rm Im}(m{\cal B}_\pm)=0$.  For the strong coupling we used
  $\alpha_s=0.262,0.203$ for $\mu=2,5$~GeV.  }
\label{fig:shortmass}
\end{figure}

Results for the jet functions are shown in Fig.~\ref{fig:shortmass}a, where
we have plotted $mB_\pm(\hat s,\Gamma_t,\mu)$ at tree-level (long dashed black
line), and at NLO in the pole mass scheme (green short dashed lines) and in the
jet-mass scheme (solid red lines). We take $\Gamma_t=1.43$~GeV.  At ${\cal
  O}(\alpha_s)$ we use Eq.~(\ref{bhqetjetwidth}), taking $\delta m=0$ in the
pole mass scheme, and taking $\delta m_J$ with $L_m=1\,{\rm GeV}$ from
Eq.~(\ref{dmjet}) in the jet-mass scheme.  For each ${\cal O}(\alpha_s)$
prediction we show two curves, one for $\mu=2\,{\rm GeV}$ (lower lines) and one
for $\mu=5\,{\rm GeV}$ (upper lines).  While the resonance peak is located at
$\hat s=0$ at tree-level, in the pole scheme at one-loop it is shifted by
$250$~MeV towards smaller masses.  In the jet-mass scheme the peak is located at
$\hat s\simeq 0$.

One may wonder how the shift of the jet function in the pole scheme arises,
given that ${\cal B}_\pm$ in Eq.~(\ref{bhqetjetwidth}) obviously has a pole at
$\hat s+i\Gamma_t=0$. The reason this pole is not visible in
Fig.~\ref{fig:shortmass}a is that jet-function is modified by powers of
$\ln[(-\hat s-i\Gamma_t)/\mu]$. To illustrate this, consider the inverse of the
stable vacuum matrix element in the pole mass scheme
\begin{align}\label{bhqetprediscrenresum}
  \big[ \tilde {\cal B}_\pm (\hat{s},0,\mu) \big]^{-1} &\equiv -\pi m \, (\hat s
  \plus i0) \,\bigg \{ 1 - \frac{\alpha _s C_F}{4\pi} \bigg [ 4 \ln^2\Big(
  \frac{ \mu}{-\hat s \minus i 0} \Big) \plus 4\ln\Big( \frac{ \mu}{-\hat
    s\minus i 0} \Big) \plus 4 \plus \frac{5\pi^2}{6} \bigg] \bigg\} \,.
\end{align}
In Fig.~\ref{fig:shortmass}b the imaginary part of $m\tilde {\cal B}_\pm$ is
plotted in the complex $\hat s$-plane for $\mu=2$~GeV. The small positive peak
visible at $\hat s=0$ is related to the zero of $(\tilde {\cal B}_\pm)^{-1}$ at
$\hat s=0$ and thus connected to the pole mass. However, it is inaccessible
physically when the finite top quark width is accounted for, i.e.  when $\tilde
{\cal B}_\pm$ is evaluated in the upper complex half-plane at $\hat
s+i\Gamma_t$.  Moreover, it has the wrong causality structure since it leads to
a small negative dip when approached from the upper complex half-plane. The
vacuum matrix element is instead dominated by the pole on the negative real
axis, that is visible as a large peak to the left of the smaller peak at $\hat
s=0$.  Conceptually this means that the physical pole of the jet function is not
located at the pole mass and that the pole mass {\it per se} is not tied to a
physical object. This conclusion is fully compatible with previous work on the
consequences of the pole mass renormalon
problem~\cite{Beneke:1998ui,Hoang:1998nz,Beneke:1998rk,Uraltsev:1998bk,Hoang:1998ng,Hoang:1998hm},
from which it is already known that the pole mass is an unphysical parameter.
The analysis of the calculable ${\cal B}_\pm$ provides a surprisingly direct
view on the mechanism of how this is achieved within perturbation theory.

From Eq.~(\ref{mjetpole}) we see that the jet mass, $m_J(\mu)$, depends on the
renormalization scale. This dependence arises because the jet functions have an
anomalous dimension.  In the jet-mass scheme we induce additional
$\mu$-dependence in the cross-section that appears through $m_J(\mu)$ in the
variable $\hat s$ and through the $\delta m_J(\mu)$ term in $B_\pm$. This
$\mu$-dependence cancels out order by order in perturbation theory.  To
implement the jet mass in the factorization theorem we proceed as follows. We
take the value of the jet mass at a certain reference scale, $m_J(\mu_0)$, as
the parameter one would like to determine from fitting the cross-section to
data. In terms of this parameter one determines the jet mass $m_J(\mu_\Gamma)$
via Eq.~(\ref{mjetLL}) where $\mu=\mu_\Gamma$ is the scale for which the jet
function is to be determined perturbatively.  In the jet functions
$B_\pm(\hat{s},\Gamma_t,\mu_\Gamma,\delta m)$ we then use $\delta m=\delta
m_J(\mu_\Gamma)$ and the invariant mass variables $\hat s_{t,\bar t}$ are
determined via
\begin{eqnarray}
\hat s_{t,\bar t} & = & \frac{M^2_{t,\bar t}-m_J^2(\mu_\Gamma)}{m_J(\mu_\Gamma)}\,.
\end{eqnarray}
We emphasize again that it is crucial that the renormalization scale in $\delta
m$ and in the explicit logs in the jet functions $B_\pm$ agree, in order to
ensure the cancellation of the ${\cal O}(\Lambda_{\rm QCD})$ renormalon
ambiguity. This is because asymptotically (for large order $n$) the terms
causing the poorly behaved perturbative behavior in the pole mass scheme at
${\cal O}(\alpha_s^{n+1})$ are proportional to
$[\mu_\Gamma\,\alpha_s^{n+1}(\mu_\Gamma)\beta_0^n\, n!]$, and only cancel if the
same renormalization scale is used.

{\it Relation to other mass schemes.}  It is useful to relate the jet-mass to
the top $\overline {\rm MS}$ mass, $\overline{m}_t(\mu)$. This facilitates using
a top mass measurement from jets in other computations, such as electroweak
precision tests. Typically one is interested in $\overline{m}_t(\overline{m}_t)$, since the
renormalization group evolution in $\overline {\rm MS}$ makes sense only above
the mass of the particle. To relate the two mass schemes we take the measured
$m_J(\mu_0)$ and use the solution to the jet-mass RGE equation in
Eq.~(\ref{mjetLL}) to run it up to, lets say, $\mu=\overline{m}_t$, obtaining $m_J(\overline{m}_t)$.  Now we use
the relations to the scale independent pole mass at $\mu=\overline{m}_t$:
\begin{align} \label{cc}
  \overline {c}(\overline{m}_t)\: \overline {m}_t(\overline{m}_t) =
  m_t^{\rm pole}= m_J(\overline{m}_t) + c_J(\overline{m}_t,L_m)\, L_m ,
\end{align}
where $\overline{c} (\mu)=1 + C_F \alpha_s(\mu)/\pi(1+\frac{3}{2}\ln(\mu/\overline{m}_t))
+\ldots$ and $c_J$ is given to one-loop by Eq.~(\ref{mjetpole}). Recall that the
choice of $L_m$ determines a scheme, so $m_J(\mu)$ also depends on this
parameter.  Expanding the relation in Eq.~(\ref{cc}) to one-loop order we obtain
a translation of the jet-mass to the $\overline {\rm MS}$ scheme that is free of
the ${\cal O}(\Lambda_{\rm QCD})$ renormalon contained in the pole mass:
\begin{align} \label{Mbartojet}
    \overline {m}_t(\overline {m}_t) =  m_J(\overline{m}_t) 
  - \frac{\alpha_s(\overline{m}_t)C_F}{\pi} m_J(\overline{m}_t)
 + \frac{\alpha_s(\overline{m}_t)C_F}{\pi} L_m \Big[
 \ln\Big(\frac{\overline{m}_t}{L_m}\Big)+\frac{3}{2} \Big]  \,.
\end{align}
Note again that it is essential to strictly expand the series on the RHS of
Eq.~(\ref{Mbartojet}) and to use the strong coupling constant at the 
{\it same} scale everywhere to ensure the proper cancellation of the 
${\cal O}(\Lambda_{\rm QCD})$ renormalon ambiguity~\cite{Hoang:2005zw}.
We also note that $\overline {c}(m_t)$ is known to three-loop
order~\cite{Gray:1990yh,Melnikov:2000qh,Chetyrkin:1999qi}, while
$c_J(m_t,\Gamma_t)$ is only known to one-loop at this time.
Due to the small size of $\Gamma_t$ the one-loop contribution of
$c_J(m_t,\Gamma_t)$ causes only a shift of about $\simeq 250\,{\rm
  MeV}$ in the determination of the $\overline {\rm MS}$ mass
$\overline {m}_t(m_t)$. 
Thus this correction may in many cases not be of critical concern when
converting a top-mass determination from jets into an $\overline{\rm
  MS}$ mass at one-loop order. However, we emphasize that mistaking a
jet mass measurement as a pole mass value beyond the one-loop order
can lead to a significant error in precision quantities that have a
strong dependence on the top quark $\overline{\rm MS}$ mass and which
have been computed to high order in QCD. This is due to the  ${\cal
  O}(\Lambda_{\rm QCD})$ renormalon inherent to the pole mass
definition. It is therefore an important task to determine the higher
order contributions in the jet mass definition of
Eq.~(\ref{mjetpole}). 

Another important class of top quark masses are the so-called threshold
masses~\cite{Hoang:2000yr} which can be determined to very high precision from a
threshold-scan of the total top pair production cross section at a future
$e^+e^-$ linear collider. Based on theoretical predictions at the
next-to-next-to leading level in
QCD~\cite{Hoang:2000yr,Hoang:2001mm,Pineda:2006ri} and through dedicated
experimental studies it is expected that a threshold top mass such as the
1S-mass~\cite{Hoang:1998ng,Hoang:1998hm,Hoang:1999zc} can be determined with
theoretical and experimental uncertainties at the level of
$100$~MeV~\cite{Martinez:2002st,Juste:2006sv}.  It is therefore useful to relate
the jet mass to the top 1S mass.  To establish this relation one should note
that the 1S mass is defined in the framework of nonrelativistic QCD and
incorporates effects which are associated to soft $\sim m_t\alpha_s$ and
ultrasoft $\sim m_t\alpha_s^2$ scales. Since ultrasoft effects are not
responsible for the nonrelativistic binding effects that define the 1S mass
definition and since the ${\cal O}(\Lambda_{\rm QCD})$ renormalon contribution
in the 1S-pole mass relation are associated to the soft
scale~\cite{Hoang:1998nz}, the relation between the jet and 1S mass has to be
determined for the soft scale $\mu_S\sim m_t\alpha_s$. To obtain the relation
one can use an approach similar to the one described above and first evolve the
jet mass to $\mu_S$ using Eq.~(\ref{mjetLL}). It is then straightforward to
relate the jet mass to the 1S mass using the known results for the 1S-pole mass
relation, see e.g.  Refs.~\cite{Hoang:2001rr,Hoang:2001mm,Pineda:2001ra} for
three-loop results (see also Refs.~\cite{Pineda:2002bv,Hoang:2002yy}) accounting
also for summation of large logarithmic terms and
Refs.~\cite{Kniehl:2002br,Beneke:2005hg} for four-loop fixed order expressions.
At one-loop order the relation reads
\begin{align} \label{M1Stojet}
    m_t ^{\rm 1S}=  m_J(\mu_S) - \frac{\alpha_s(\mu_S)C_F}{8}\,
\Big[\alpha_s(\mu_S)\,C_F\, m_J(\mu_S)\Big]
 + \frac{\alpha_s(\mu_S)C_F}{\pi} L_m \Big[
 \ln\Big(\frac{\mu_S}{L_m}\Big)+\frac{3}{2} \Big]  \,.
\end{align}
Note that the same principles for treating the perturbative series discussed
above for the $\overline {\rm MS}$-jet mass relation have to be applied here to
ensure the proper cancellation of the ${\cal O}(\Lambda_{\rm QCD})$ renormalon
contributions. In addition it is necessary to treat the terms in the
perturbative series in the 1S-pole mass relation in the so-called Upsilon
expansion, where terms that are of order $\alpha_s^{n+1}$ are formally treated
of order $\alpha_s^n$~\cite{Hoang:1998ng,Hoang:1998hm}. This is because the
physical scale that governs this series is the inverse Bohr radius $C_F
m_t\alpha_s$ (which is the analog of $L_m$ in Eq.~(\ref{mjetpole})). Note that
the one-loop corrections from the jet-pole mass relation have a larger numerical
impact in the one-loop relation of Eq.~(\ref{M1Stojet}) than in
Eq.~(\ref{Mbartojet}) because the 1S-pole mass corrections are an order of
magnitude smaller than the corrections in the $\overline {\rm MS}$-pole mass
relation. We will give a more detailed discussion on the higher order structure
of Eqs.~(\ref{Mbartojet},\ref{M1Stojet}) in Ref.~\cite{FHMSmass:inprep}.

\section{Soft Function Models with Perturbative Corrections} \label{sec:ModelSoft}

The soft function at a scale $\mu\sim \mu_\Lambda$ is written as
\begin{align}
\label{Smodel1v2}
S(\ell^+,\ell^-,\mu) & = 
\int_{-\infty}^{+\infty}\!\!\! d\tilde\ell^+
\int_{-\infty}^{+\infty}\!\!\! d\tilde\ell^-\
S_{\rm part}(\ell^+ \minus \tilde\ell^+,\ell^- \minus \tilde\ell^-,\mu,\delta)\,
S_{\rm mod}(\tilde\ell^+,\tilde\ell^- ) 
\,.
\end{align}
This combines the partonic perturbative result for the soft function $S^{\rm
  part}$ (given in Eqs.~(\ref{Sren}) for the hemisphere prescription), with a
model hadronic function $S^{\rm mod}$ satisfying the moment constraints in
Eq.~(\ref{Smodelconditions}). As explained in Ref.~\cite{Hoang:2007vb}, this
form encodes the features we require for an appropriate soft-function $S$ for
our analysis. In particular it works equally well for the peak region where the
soft-function in non-perturbative, and for the tail region where the
soft-function is perturbatively calculable at leading power. $S$ in
Eq.~(\ref{Smodel1v2}) has $\mu$ dependence consistent with its anomalous
dimension and the $\overline {\rm MS}$ scheme. And finally it should be totally
free from the ${\cal O}(\Lambda_{\rm QCD})$ soft-function renormalon ambiguity
identified in Ref.~\cite{Hoang:2007vb}, which is also known to appear in event
shapes for massless jets~\cite{Gardi:2000lr}.

For the analyses in this work we will use the exponential model $f_{\rm exp}$ of
Ref.~\cite{Korchemsky:2000kp}, with the addition of a gap parameter $\Delta$, so
that
\begin{align} \label{SM1}
  S_{\rm mod}(\ell^+,\ell^-,\Delta) &= f_{\rm
    exp}\big(\ell^+-\Delta,\ell^--\Delta\big) \,, 
  \\
  f_{\rm exp}(\ell^+,\ell^-) &= \theta(\ell^+)\theta(\ell^-) \frac{ {\cal
      N}(a,b) }{\Lambda^2} \Big( \frac{\ell^+\ell^-}{\Lambda^2}\Big)^{a-1}
  \exp\Big( \frac{-(\ell^+)^2-(\ell^-)^2-2 b \ell^+\ell^-}{\Lambda^2} \Big) \,.
   \nn
\end{align}
Here the normalization constant ${\cal N}(a,b)$ is defined so that $\int d\ell^+
d\ell^- S(\ell^+,\ell^-) = 1$. The parameter $\Lambda \sim \Lambda_{\rm QCD}$
sets the width of the hadronic function and hence the scale for $\ell^\pm$ and
the soft radiation. The dimensionless parameter $a$ controls how fast the
function vanishes at the origin, and the dimensionless parameter $b>-1$ controls
the correlation of energy flow into the two hemispheres. Any $b\ne 0$ implies
cross-talk between the two hemispheres.\footnote{In
  Ref.~\cite{Korchemsky:2000kp} the values $a=2$ and $b=-0.4$ were obtained from
  a fit to LEP data. The analysis used a different scheme for including
  perturbative corrections in the soft-function than the one advocated here.}
The gap parameter $\Delta$ enforces $\ell^\pm\ge\Delta$ and encodes the minimal
hadronic energy deposit due to soft radiation.

As explained in Ref.~\cite{Hoang:2007vb}, there is a renormalon in $S_{\rm
  part}(\ell^\pm-\tilde\ell^\pm)$ that corresponds to an ${\cal O}(\Lambda_{\rm
  QCD})$ ambiguity in the partonic threshold where $\ell^\pm-\tilde\ell^\pm=0$,
and a corresponding ambiguity in the non-perturbative gap-parameter $\Delta$.
It can be removed by shifting to a renormalon free gap parameter $\bar\Delta$,
using $= \Delta = \bar \Delta(\mu) + \delta(\mu)$,
\begin{align}
\label{S2}
S(\ell^+,\ell^-,\mu) & = 
\int_{-\infty}^{+\infty}\!\!\! d\tilde\ell^+
\int_{-\infty}^{+\infty}\!\!\! d\tilde\ell^-\
S_{\rm part}(\ell^+ \minus \tilde\ell^+ ,
  \ell^- \minus \tilde\ell^-,\mu)\,
f_{\rm exp}(\tilde\ell^+ \minus\Delta,\tilde\ell^- \minus\Delta)
\\
 & = 
\int_{-\infty}^{+\infty}\!\!\! d\tilde\ell^+
\int_{-\infty}^{+\infty}\!\!\! d\tilde\ell^-\
S_{\rm part}(\ell^+ \minus \tilde\ell^+ \minus \delta,
  \ell^- \minus \tilde\ell^-\minus \delta,\mu)\,
f_{\rm exp}(\tilde\ell^+ \minus \bar\Delta,\tilde\ell^- \minus \bar\Delta) \,.
 \nn
\end{align}
Here $\delta=\sum_{i=1}^\infty \delta_i$ is a perturbative series with
$\delta_i\sim {\cal O}(\alpha_s^i)$ that defines the scheme for $\bar\Delta$.
Expanding $S_{\rm part}(\ell^\pm-\tilde\ell^\pm-\delta)$ in perturbation theory
the $\delta_i$'s remove the renormalon ambiguity from $S_{\rm part}$ order by
order.  Up to ${\cal O}(\alpha_s)$ this gives
\begin{align} \label{Spartd}
  S_{\rm part}(\ell^+,\ell^-,\mu,\delta_i) = S^0_{\rm part}(\ell^+,\ell^-) + 
  \bigg[ S_{\rm part}^1(\ell^+,\ell^-,\mu) - \delta_1
  \Big(\frac{d}{d\ell^+}\plus \frac{d}{d\ell^-}\Big)  S_{\rm part}^0(\ell^+,\ell^-)
  \bigg],
\end{align}
where defining ${\cal L}^1(\ell)= 1/\mu\,
\big[\theta(\ell)\ln(\ell/\mu)/(\ell/\mu)\big]_+$ we have
\begin{align}
  S_{\rm part}^0(\ell^+,\ell^-) &=\delta(\ell^+)\delta(\ell^-)\,, 
  \qquad
  S_{\rm part}^1(\ell^+,\ell^-,\mu) =\delta(\ell^+) S_{\rm part}^1(\ell^-,\mu) 
   + \delta(\ell^-) S_{\rm part}^1(\ell^+,\mu) \,,
  \nn \\[5pt]
  S_{\rm part}^1(\ell,\mu) &= \frac{C_F\alpha_s(\mu)}{\pi} \Big[
   \frac{\pi^2}{24}\, \delta(\ell) - 2
    {\cal L}^1(\ell) \Big] \,. 
\end{align}
A renormalon free scheme for the gap $\bar\Delta$ can be
defined~\cite{Hoang:2007vb} using a first moment of the soft function with upper
cutoff $L_\Delta$, similar to the jet-mass in Eq.~(\ref{mJmomentdef}). This
definition can be written
\begin{align}
 0 = \int_{-\infty}^{L_{\Delta}} \!\!\! d\ell^+ \!\!
    \int_{-\infty}^{L_{\Delta}} \!\!\! d\ell^- \ \ell^+\ S_{\rm
      part}(\ell^+-\delta,\ell^--\delta,\mu) \,,
\end{align}
and at ${\cal O}(\alpha_s)$ gives~\cite{Hoang:2007vb}
\begin{align} \label{delta1}
  \delta_1 = - 2 L_{\Delta} \frac{C_F\alpha_s(\mu)}{\pi} 
  \bigg[\ln\Big(\frac{\mu}{L_\Delta}\Big) + 1 \bigg] \,.
\end{align}
Because $\Delta = \bar\Delta(\mu) + \delta(\mu)$ is RG-invariant, this gives an
anomalous dimension equation
\begin{align}
 \mu\frac{d}{d\mu} \bar\Delta(\mu) = 2 L_{\Delta} \frac{C_F\alpha_s(\mu)}{\pi}\,,
\end{align}
with a LL solution
\begin{align}\label{DeltaLL}
  \bar\Delta(\mu) = \bar\Delta(\mu_0)  - L_\Delta \frac{4 C_F}{\beta_0}
  \ln\bigg[ \frac{\alpha_s(\mu)}{\alpha_s(\mu_0)}\bigg] \,. 
\end{align}

\begin{figure}
  \centerline{ 
   \includegraphics[width=17cm]{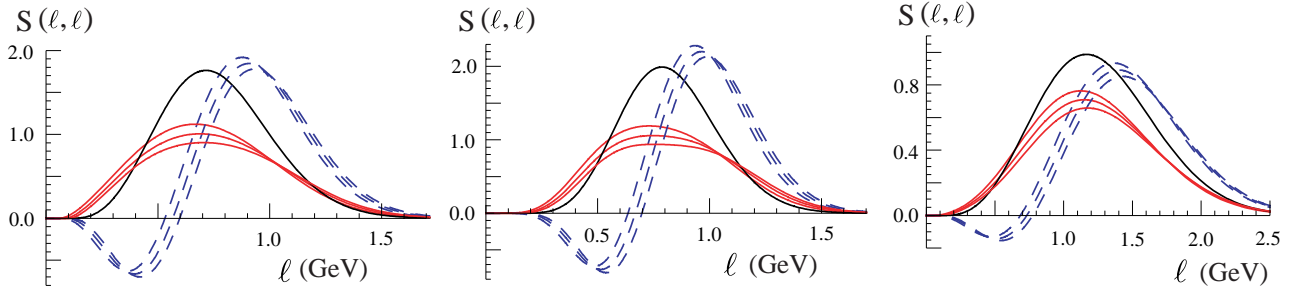} 
  } 
\caption{ 
  Soft function models based on Eq.~(\ref{Sfinal}) with the hadronic model
  function $S^{\rm mod}$ given in Eq.~(\ref{SM1}) for $\Lambda=0.55\,{\rm GeV}$,
  $\bar\Delta(\mu=1\,{\rm GeV})=0.1\,{\rm GeV}$ and $(a,b)=(2.5,-0.4)$ (left
  panel), $(3.5,-0.2)$ (middle panel), and $(a,b)=(2.5,-0.8)$ (right panel). The
  curves are tree-level (black solid line), ${\cal O}(\alpha_s)$ with
  $\delta_1=0$ (blue dashed lines) and ${\cal O}(\alpha_s)$ with a renormalon
  free gap (red light solid lines). The blue and red curves are shown for
  $\mu=0.8,0.9,1.0\,{\rm GeV}$, with the higher curves correspond to lower
  values of the renormalization scale.  }
\label{fig:softmodels}
\end{figure}
Using Eq.~(\ref{Spartd}) in (\ref{SM1}) and integrating by parts we obtain a
suitable soft-function for our NLL analysis
\begin{align} \label{Sfinal}
  S(\ell^+,\ell^-,\mu) &= S_{\rm mod}(\ell^+,\ell^-,\bar\Delta) 
   - \delta_1 \Big(\frac{d}{d\ell^+}\plus \frac{d}{d\ell^-}\Big)
     S_{\rm mod}(\ell^+,\ell^-,\bar\Delta) \\
  &+ \int_{-\infty}^{+\infty}\!\!\! d\tilde\ell^+
\int_{-\infty}^{+\infty}\!\!\! d\tilde\ell^-\
S_{\rm part}^1(\ell^+ \minus \tilde\ell^+,\ell^- \minus \tilde\ell^-,\mu)\,
S_{\rm mod}(\tilde\ell^+,\tilde\ell^-,\bar\Delta) \nn\\
 & \equiv \int_{-\infty}^{+\infty}\!\!\! d\tilde\ell^+
\int_{-\infty}^{+\infty}\!\!\! d\tilde\ell^-\ \tilde S_{\rm
  part}(\ell^+\minus\tilde\ell^+,\mu,\delta_1)\,\tilde S_{\rm
  part}(\ell^-\minus\tilde\ell^-,\mu,\delta_1)\,
  S_{\rm mod}(\tilde\ell^+,\tilde\ell^-,\bar\Delta) \,, \nn
\end{align}
where the modified one-dimensional partonic soft-function is
\begin{align} \label{Stilde}
 \tilde S_{\rm part}(\ell,\mu,\delta_1) & = \delta(\ell) 
   -\delta_1(\mu)\: \delta^\prime(\ell) 
   + S_{\rm part}^1(\ell,\mu)  \,.
\end{align}
With Eq.~(\ref{SM1}) the result in Eq.~(\ref{Sfinal}) involves logarithmic terms
$\ln(\ell^\pm/\mu)$ and $\ln(\Lambda/\mu)$, that arise from the convolution of
the partonic plus functions with the smooth hadronic functions. In the peak
region there is a possible tension between the convergence of the perturbative
series in $\alpha_s(\mu)$ for $S_{\rm part}$ and the size of the
$\ln(\Lambda/\mu)$ terms.  In Ref.~\cite{Hoang:2007vb} the log-series in the
soft-function was analyzed and it was concluded that for scales $\mu\simeq
1\,{\rm GeV}$ in the peak region this tension is not an issue.  In the tail
region the cross-section is dominated by $\ell^\pm \sim \hat s m/Q$. Thus
$\ell^\pm$ grows, and it becomes necessary to increase $\mu$ so that $\mu\sim
\hat s m /Q$ to avoid large logs from the $\ln(\ell^\pm/\mu)$ terms.

To demonstrate the importance of the renormalon subtraction we show
$S(\ell^+,\ell^-,\mu)$ in Fig.~\ref{fig:softmodels}, plotted with
$\ell=\ell^+=\ell^-$ for $\Lambda=0.55\,{\rm GeV}$ and three sets of the
remaining parameters $(a,b,L_\Delta/\Lambda)=(2.5,-0.4,0.8)$, $(3.5,-0.2,0.7)$,
and $(2.5,-0.8,0.8)$ respectively.  The tree-level soft function is $S= S_{\rm
  mod}$ in Eq.~(\ref{SM1}) where we take $\bar\Delta=100\,{\rm MeV}$, and is
shown by the black solid lines. The three dashed blue lines denote the ${\cal
  O}(\alpha_s)$ soft function obtained from Eqs.~(\ref{Sfinal}) without the
renormalon subtraction ($\delta_1=0$), and for $\mu=0.8,0.9,1.0\,{\rm GeV}$. The
three light solid red lines denote the ${\cal O}(\alpha_s)$ soft function
obtained from Eqs.~(\ref{Sfinal}) with a renormalon free gap using $\delta_1$
from Eq.~(\ref{delta1}), and $\mu=0.8, 0.9, 1.0\,{\rm GeV}$. For both the blue
and red curves we use $\bar\Delta(\mu)=60,82,100\,{\rm MeV}$ for these three
$\mu$'s respectively, which corresponds to implementing the LL running from
Eq.~(\ref{DeltaLL}).  Compared to the tree level result, the blue dashed ${\cal
  O}(\alpha_s)$ curves show a significantly shifted location of their maximum
and become negative for small values of $\ell$. The red light solid curves show
that the renormalon subtraction stabilizes the peak location and removes the
negative dip. These features are generic for any choice of a hadronic model
parameters $(a,b)$.

\section{Numerical Analysis up to Next-to-Leading Log Order} 
\label{sec:analysis}

In Ref.~\cite{Fleming:2007qr} we carried out a numerical analysis of the
top-invariant mass distribution concentrating on nonperturbative effects caused
by the soft function and on the dependence of the invariant mass distributions
on the parameters used for the soft function model. The analysis was based on
the tree-level results for the jet functions in Eq.~(\ref{eq:Btree}), without
summation of logarithms, and on the hemisphere soft function as given by the
model of Ref.~\cite{Korchemsky:2000kp} which had been obtained from fits to
event shapes in $e^+e^-$ annihilation. The soft-function caused a positive shift
in the peak position of the invariant mass distribution, $M_{t,\bar t}> m_J$,
where the shift is parametrically $\sim \Lambda_{\rm QCD} Q/m$, and it was
demonstrated that the peak shift and peak width grow linearly with $Q/m$.

In this section we will extend the analysis to include radiative corrections and
examine the perturbative convergence of predictions for the invariant mass
distribution. This amounts to a full NLL analysis (i.e.\,\,one-loop matrix
elements plus NLL summation of logarithms). Recall from Fig.~\ref{fig:theoryI}
that there are four relevant scales for the log-summation, $\mu_Q\simeq Q$,
$\mu_m\simeq m$, $\mu_\Gamma\simeq \hat s +\Gamma_t+ Q\Lambda/m_t$ and
$\mu_\Lambda \gtrsim \Lambda_{\rm QCD}+m\Gamma_t/Q +\hat s m/Q$. For the peak
region we use $\mu_\Lambda \simeq 1\,{\rm GeV}$. Our analysis is performed in
several steps. After setting up the cross-section formula, we proceed in
section~\ref{sec:runJS} to consider the summation of large logs for the
perturbative corrections, and analyze the scale and scheme dependence. We show
that the summation of logs between $\mu_\Gamma\simeq \Gamma_t+ Q\Lambda/m_t$ and
$\mu_\Lambda\simeq 1\,{\rm GeV}$ have a significant impact on stabilizing the
cross-section.  Then in section~\ref{sec:sigpeak} we convolute the perturbative
corrections with the soft-function model and analyze the cross-section in the
peak region.  In section~\ref{sect:away} we analyze the cross-section in the
tail region, and plot combined peak and tail results. Finally in
section~\ref{sec:eventshapes} we use our results to determine the thrust
distribution at NLL order.

For the numerical analysis it is convenient to write the invariant mass
cross-section in the top jet-mass scheme in terms of dimension one invariant
mass variables
\begin{align} \label{sigmaMM}
  \frac{ d^2\sigma }{dM_t\, dM_{\bar t}} 
    & = \, \frac{ \sigma_0 }{ \Gamma_t^{\,2}}\
    F\Big(M_t,M_{\bar t},m_J,\frac{Q}{m_J}\Big) \,,
\end{align}
where the prefactor $\sigma_0$ is given in Eq.~(\ref{sigma0def}). Here
$m_J$ is the jet-mass and the dimensionless function $F$ is
\begin{align} \label{Fdef}
  F\Big(M_t, M_{\bar t}, m_J, \frac{Q}{m_J}\OMIT{;\mu_Q,\mu_m,\mu_\Gamma,\mu_\Lambda}\Big) 
  &= \! \int_{-\infty}^\infty\!\!\!\! d\ell^+\, d\ell^- \:
  {\rm P}\Big(\hat s_t - \frac{Q\ell^+}{m_J},\hat s_{\bar t} -
  \frac{Q\ell^-}{m_J},  \mu_\Lambda \Big)
 \,S^{\rm mod}\big(\ell^+,\ell^-,\bar\Delta(\mu_\Lambda)\big) 
 \\[4pt]
 & \hspace{-2.5cm}
 = \! \int_{-\infty}^\infty\!\!\!\! d\ell^+\, d\ell^- \:
  {\rm P}\Big(\hat s_t - \frac{Q\ell^+}{m_J}-\frac{Q\bar\Delta(\mu_\Lambda)}{m_J},
  \hat s_{\bar t} -\frac{Q\ell^-}{m_J}-\frac{Q\bar\Delta(\mu_\Lambda)}{m_J}, 
  \mu_\Lambda \Big)
 \,S^{\rm mod}\big(\ell^+,\ell^-,0\big) 
  \,, \nn
\end{align}
with $S^{\rm mod}$ the hadronic model function given in Eq.~(\ref{SM1}). In the
second line we shifted the integration variables to put all $\mu$-dependent
factors into ${\rm P}$. In terms of $M_t$ and $M_{\bar t}$ the invariant mass
variables $\hat s_{t,\bar t}$ in Eq.~(\ref{Fdef}) are
\begin{align} \label{ssM}
  \hat s_t = \frac{M_t^2 -m_J^2}{m_J} \,,\qquad\quad 
  \hat s_{\bar t} = \frac{M_{\bar t}^2 -m_J^2}{m_J}
  \,.
\end{align}
All the perturbatively computable contributions in Eq.~(\ref{Fdef}) are grouped
into the dimensionless function
\begin{align} \label{Pdef}
{\rm P}\big(\hat s_t,\hat s_{\bar t},\mu_\Lambda\big) 
 &=  4 M_t M_{\bar t}\,\Gamma_t^{\,2}\,
  H_{Q}(Q,\mu_Q)\, U_{H_Q}(Q,\mu_Q,\mu_m) \,
  H_m(m_J,\mu_m)\, U_{H_m}\Big(\frac{Q}{m_J},\mu_m,\mu_\Lambda\Big)
 \nn\\
  &\quad \times 
  G_+\Big(\hat s_t, \frac{Q}{m_J}, \Gamma_t,\mu_\Lambda \Big)
  G_-\Big(\hat s_{\bar t} , \frac{Q}{m_J}, \Gamma_t,\mu_\Lambda \Big)
  \,. 
\end{align}
${\rm P}$ also depends on $Q$, $m_J$, and $\Gamma_t$, but for simplicity we have
not shown this dependence in its arguments. For the hard coefficients we used
Eqs.~(\ref{UH}) and (\ref{UHm}) to write them in terms of the one-loop matching
coefficients $H_Q(Q,\mu_Q)$ and $H_m(m,\mu_m)$ in
Eqs.~(\ref{HQmatch},\ref{bhqetmatchcoeff}) and the NLL evolution factors
$U_{H_Q}$ and $U_{H_m}$ given by Eq.~(\ref{UUU}).  The functions $G_\pm$ in
Eq.~(\ref{Pdef}) contain perturbative corrections that modify the shape of the
cross-section.  Using Eqs.~(\ref{bHQETcross-hem2b},\ref{modifiedS},\ref{Sfinal})
and a few trivial changes of integration variables, these functions are
\begin{align} \label{Gpm}
& G_\pm\Big(\hat s, \frac{Q}{m_J},\Gamma_t\OMIT{,\mu_\Gamma},\mu_\Lambda\Big) \, \equiv \,
\int_{-\infty}^{+\infty}\!\! d\hat s^\prime\, d\hat s^{\prime\prime}\, d\ell^\prime
 \ \ U_{B}(\hat s-\hat s^\prime,\mu_\Lambda,\mu_\Gamma) 
\nn\\[3mm]
 &\hspace{1cm} \times
B_\pm^{\Gamma=0}\Big(\hat s^\prime-\hat
  s^{\prime\prime}-\frac{Q}{m_J}\,\ell^\prime,\mu_\Gamma,\delta m\Big)
\,\tilde S_{{\rm part}}(\ell^\prime,\mu_\Lambda,\delta_1)\,
\frac{\Gamma_t}{\pi(\hat s^{\prime\prime\, 2}+\Gamma_t^2)}
\,.
\end{align}
This result depends on $B_\pm^{\Gamma=0}$, the jet function for stable quarks in
Eq.~(\ref{bhqetjetstable}), and $\tilde S^{\rm part}$ the modified partonic soft
function of Eq.~(\ref{Stilde}). The form in Eq.~(\ref{Fdef}) is derived from the
factorization theorem given in Eq.~(\ref{bHQETcross-hem2b}), where the
renormalization scales $\mu_\Gamma$ and $\mu_\Lambda$ were distinguished. This
leads to the presence of the evolution factor $U_B$ in Eq.~(\ref{Gpm}), which is
given at NLL in Eq.~(\ref{UUU}).  The functions $G_\pm$, and hence all the
ingredients in ${\rm P}$, can be computed in perturbation theory, and analytic
results for $G_\pm$ are given in Appendix~\ref{App:Gfunc}.
 
When quoting results at LL order we take $U_B$, $U_{H_Q}$, and $U_{H_m}$ at LL
order, and use tree-level results for $B_\pm^{\Gamma=0}$ and $\tilde S_{\rm
  part}$, including $\delta m=\delta_1=0$. The results quoted at NLL order use
NLL-evolution for $U_B$, $U_{H_Q}$, and $U_{H_m}$. They also include the ${\cal
  O}(\alpha_s)$ results for matching coefficients and matrix elements, including
$B_\pm^{\Gamma=0}$, $\tilde S_{\rm part}$, $\delta m$, $\delta_1$,
$H_Q(Q,\mu_Q)$ and $H_m(m,\mu_m)$. These ${\cal O}(\alpha_s)$ terms have
no-large logs, and in our numerical analysis we strictly drop all terms of
${\cal O}(\alpha_s^2)$ or higher in the product of these matching and matrix
element terms that appear in ${\rm P}$.  We also make use of the two-loop
solution for the running coupling
\begin{align}
  \frac{1}{\alpha_s(\mu)} = \frac{1}{\alpha_s(\mu_0)}
  +\frac{\beta_0}{2\pi}\ln\Big(\frac{\mu}{\mu_0}\Big)
  +\frac{\beta_1}{4\pi\beta_0}  \ln\Big[
  1+\frac{\beta_0}{2\pi}\alpha_s(\mu_0)\ln\Big(\frac{\mu}{\mu_0}\Big)\Big] \,,
\end{align}
with $\alpha_s(\mu_0=m_Z)=0.118$ as our reference value, and with $\beta_0$ and
$\beta_1$ from Eq.~(\ref{b0b1}). For the running above $\mu_m$ we take $n_f=6$,
while for running below $\mu_m$ we take $n_f=5$ (hence neglecting the $b$-quark
threshold). 

Since there are many features of the cross-section formulae in
Eqs.~(\ref{sigmaMM}-\ref{Gpm}) that we wish to explore, it is useful to have a
default set of parameters to use at both LL and NLL order.  When not otherwise
specified, we use the following values for our analysis below. Our default
$Q/m_J=5$, and the default renormalization scales are $\mu_Q=5*172\,{\rm GeV}$,
$\mu_m=172\,{\rm GeV}$, $\mu_\Gamma=5\,{\rm GeV}$, and $\mu_\Lambda=1\,{\rm
  GeV}$.  For results that involve the running jet mass we take as a reference
value $m_J(\mu_0=2\,{\rm GeV})=172\,{\rm GeV}$ in the $L_m=1\,{\rm GeV}$ scheme,
and evolve to other scales using Eq.~(\ref{mjetLL}) for $m_J(\mu)$. For results
in the pole mass scheme we also use $m_t^{\rm pole}=172\,{\rm GeV}$.  When the
running gap is included we take as our reference value $\bar\Delta(\mu_0
\!=\!1\,{\rm GeV})=100\,{\rm MeV}$ with $L_\Delta=0.44\,{\rm GeV}$, and evolve to
other scales using Eq.~(\ref{DeltaLL}) for $\bar\Delta(\mu)$. We refer to this
as the $\bar\Delta$-scheme.  For results quoted without a renormalon free gap
parameter we use $\Delta=100\,{\rm MeV}$, and refer to this as the
$\Delta$-scheme.  Finally when studying perturbative aspects of the
cross-section our default model for the soft function $S_{\rm
  mod}(\ell^+,\ell^-,\bar\Delta)$ in Eq.~(\ref{SM1}) is
$(\Lambda,a,b)=(0.55\,{\rm GeV},2.5,-0.4)$.  Note that the dependence of $S_{\rm
  mod}$ on the model parameters $\Lambda$, $a$ and $b$ is not shown explicitly
in its arguments. We will explore deviations from these default parameters on a
case by case basis.

\subsection{Analysis of Perturbative Corrections and Log Summation} 
\label{sec:runJS}

We begin our analysis by studying the perturbative corrections contained in the
function
\begin{align} \label{P}
    \tilde {\rm P}(M_t,M_{\bar t})\, \equiv\,
   {\rm P}\Big(\frac{M_t^2 \!- m_J^2-\! Q\bar\Delta(\mu_\Lambda)}{m_J},\,
  \frac{M_{\bar t}^2 \!- m_J^2-\! Q\bar\Delta(\mu_\Lambda)}{m_J},
  \mu_\Lambda \Big)\,.
\end{align}
This function $\tilde {\rm P}$ is convoluted with the soft-function model to give
the cross-section, as shown in the second line of Eq.~(\ref{Fdef}). All of the
dependence on the scales $\mu_Q$, $\mu_m$, $\mu_\Gamma$, and $\mu_\Lambda$
cancels out in this function to the order we are working, and we can analyze the
residual scale dependence of $\tilde {\rm P}$ to obtain an estimate of the
remaining uncertainties from higher order corrections.

\begin{figure}
  \centerline{ 
   \includegraphics[width=17cm]{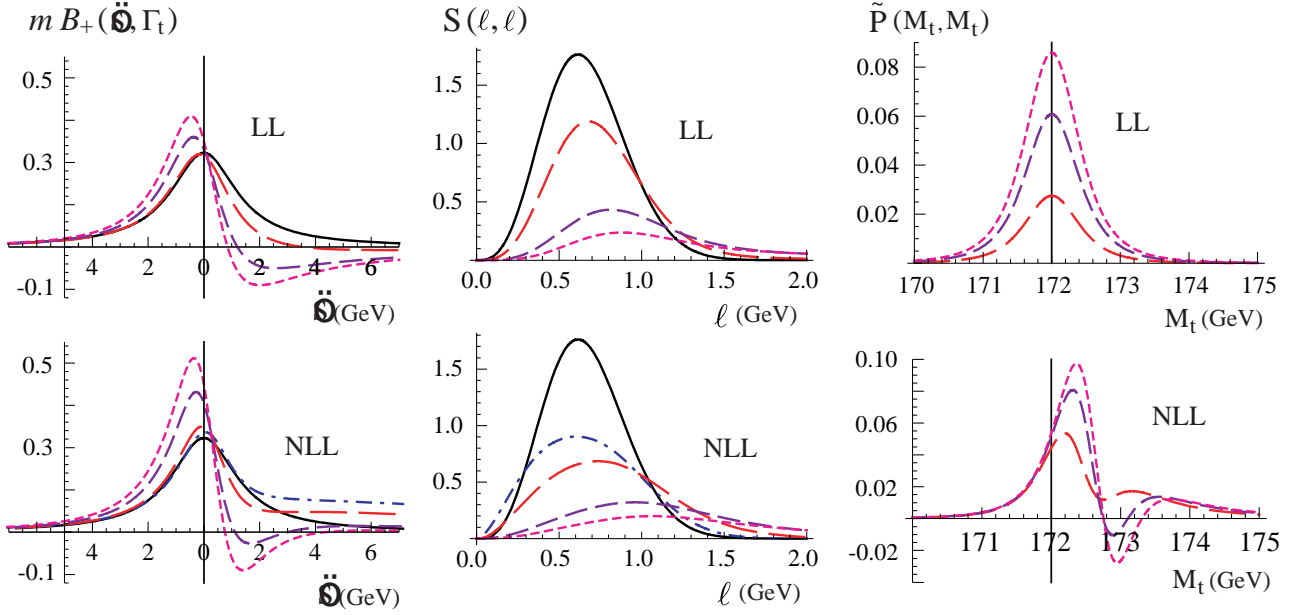} 
  } 
\vskip-0.4cm
\caption{ 
  Renormalization group evolution of the jet function, $m_J B_+(\hat
  s,\Gamma_t,\mu)$ (left panels), the diagonal soft function $S(\ell,\ell,\mu)$
  (center panels), and the function $\tilde P(M_t,M_t)$ in Eq.~(\ref{P}) (right
  panels) with $\mu=\mu_\Gamma=\mu_\Lambda$. The top three panels show LL
  results, while the bottom three are NLL results.  For the left and center
  panels the curves are tree-level (black solid), NLO at $\mu=1\,{\rm GeV}$
  (blue dot-dashed lines), and curves which evolve at LL or NLL order from
  $\mu_0=1$ to $\mu=1.5,4.0,7.0\,{\rm GeV}$ (red, purple, magenta, with
  decreasing dash sizes respectively). The right panels show only these last
  three curves.  }
\label{fig:BSrun}
\end{figure}
In Fig.~\ref{fig:BSrun} we demonstrate the effect of summing logs for the
functions $m_J B_+(\hat s,\Gamma_t,\mu)$ (left panels, see Eq.~(\ref{UB})),
$S(\ell,\ell,\mu)$ (center panels, see Eq.~(\ref{UJS})), and $\tilde {\rm
  P}(M_t,M_t)$ (right panels, see Eq.~(\ref{P})). For this plot we take
$\mu=\mu_\Gamma=\mu_\Lambda$. The three top panels show LL results using the
pole-mass and $\Delta$-schemes. They demonstrate that increasing $\mu$ changes
the shape of both $B_+$ and $S$.  However, in the convolution that gives ${\rm
  P}$ these changes just reduce to a shift in the overall normalization, due to
the consistency condition in Eq.~(\ref{conssimple2}). The three bottom panels
show NLL results using the jet-mass and $\bar\Delta$-schemes. At LL and NLL
order the peak of $B_+$ moves to the left as we increase $\mu$ (top-left panel
and bottom-left panel). At both LL and NLL order the peak of $S$ moves to the
right for increasing $\mu$ (central panels).  The right panels show that at LL
and NLL we still have a strong residual scale dependence in ${\rm P}$, and that
there is still a peak shift at NLL. This occurs because we have taken
$\mu_\Gamma=\mu_\Lambda$ and not yet summed the large logs between $\mu_\Gamma$
and $\mu_\Lambda$, where $\mu_\Gamma/\mu_\Lambda\simeq Q/m$. These remaining
large logs can be seen explicitly in the formula in Eq.~(\ref{Fpmfin}).

\begin{figure}[t!]
  \centerline{ \includegraphics[width=17cm]{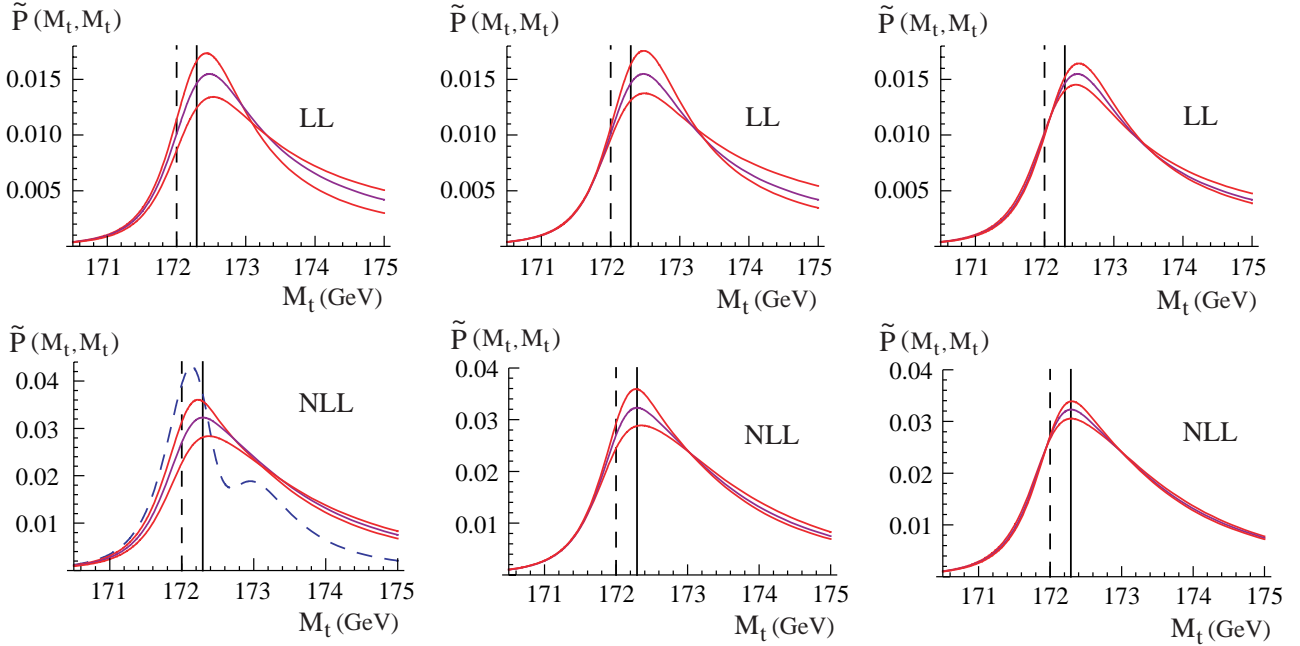} } 
\vskip-0.3cm
\caption{ The
  $\mu_\Gamma$ and $\mu_\Lambda$ scale dependence of the perturbative
  contributions, $\tilde{\rm P}(M_t,M_t)$.  The top-panels show LL results,
  while bottom panels show NLL results.  Central values are $\mu_\Gamma=5\,{\rm
    GeV}$ and $\mu_\Lambda=1\,{\rm GeV}$. In the left panels the solid curves
  are for $\mu_\Gamma =3.3,5,7.5\,{\rm GeV}$ (from top to bottom at the peak),
  while the blue-dashed line shows the result when
  $\mu_\Gamma=\mu_\Lambda=1\,{\rm GeV}$.  In the central panels the solid curves
  are for $\mu_\Lambda =0.8,1.0,1.2\,{\rm GeV}$ (from bottom to top at the peak).
  The right panels are the same as the central panels, except that we also
  change $\mu_\Gamma$ so that $\mu_\Gamma/\mu_\Lambda=Q/m=5$ remains fixed.  }
\label{figFmu}
\end{figure}
\begin{figure}[t!]
  \centerline{ 
   \includegraphics[width=17cm]{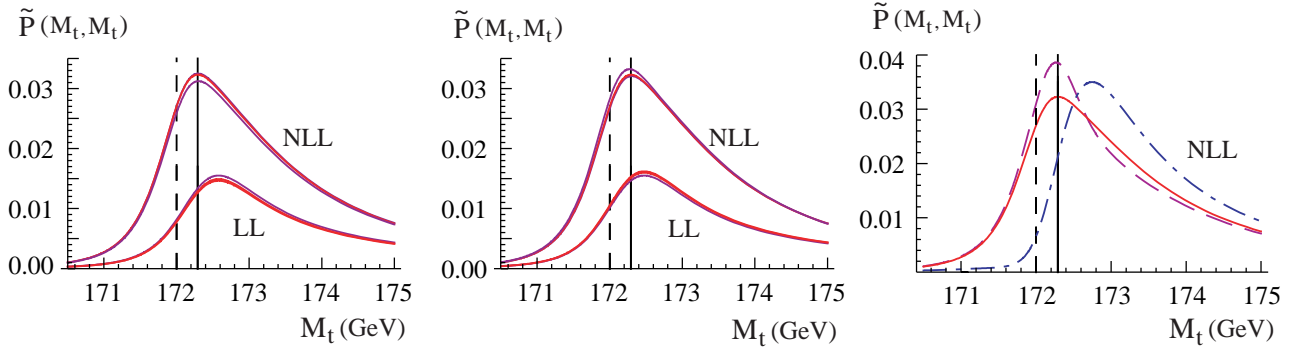} 
  } 
\vskip-0.4cm
\caption{ $\tilde{\rm P}(M_t,M_t)$ versus $M_t$.
  Left panel shows the $\mu_m$ dependence at LL (bottom curves) and NLL (top
  curves) taking $\mu_m=86,172,344\,{\rm GeV}$. Central panels shows the $\mu_Q$
  dependence at LL (bottom curves) and NLL (top curves) taking
  $\mu_Q=430,860,1720\,{\rm GeV}$. The right panel shows the effect of using
  renormalon free gap and mass parameters, where the red solid curve includes both.
  The purple dashed curve turns off the renormalon subtractions for the mass (thus
  using the pole mass scheme), and the blue dot-dashed curve turns off the renormalon
  subtraction for the gap.  }
\label{figFmu2}
\end{figure}

This situation is rectified in Fig.~\ref{figFmu}, where we show results for
$\tilde{\rm P}(M_t,M_t)$ with separated scales $\mu_\Gamma$ and $\mu_\Lambda$.
Again the top three panels are LL, and bottom three are NLL. In the left two
panels we vary $\mu_\Gamma$ about $\mu_\Gamma=5\,{\rm GeV}$ holding
$\mu_\Lambda=1\,{\rm GeV}$ fixed. We use the range $\mu_\Gamma=3.3$--$7.5\,{\rm
  GeV}$ to estimate the scale uncertainty because of the importance of not
upsetting the $\mu_\Gamma/\mu_\Lambda\simeq Q/m$ relation too severely. (For
contrast the blue-dashed curve in the lower-left panel shows the result for
$\mu_\Gamma=\mu_\Lambda=1\,{\rm GeV}$.) In the center panels we hold
$\mu_\Gamma=5\,{\rm GeV}$ and instead vary $\mu_\Lambda = 0.8$--$1.2\,{\rm GeV}$
(where variation below $0.8\,{\rm GeV}$ is not advisable since $\alpha_s$ has
grown to $0.45$ at this scale).  The dashed vertical line shows the input value
of the short-distance mass, $m_J(\mu_0=2\,{\rm GeV})=172\,{\rm GeV}$, while the
solid vertical line shows the shift due to the gap, $M_{\bar t}= [m_J^2(\mu_0)+
Q\bar\Delta]^{1/2}\simeq 172.25\,{\rm GeV}$, with $\bar\Delta(1\,{\rm
  GeV})=0.1\,{\rm GeV}$. Finally in the two right panels we show how the
$\mu$-variation of the center panels is reduced if we vary $\mu_\Lambda$ in the
same range, but simultaneously change $\mu_\Gamma$ so that
$\mu_\Gamma/\mu_\Lambda=Q/m$ is fixed. Because of this sizeable correlation the
overall uncertainty is smaller than from naively summing the uncertainties from
the left and center panels in quadrature. We believe that an uncertainty in the
shape that is of order the $\mu_\Gamma$ variation shown in the left panels gives
a reasonable error estimate. From comparing the percent change at LL and the
percent change at NLL we see that there is a reduction to the $\mu$-variation in
all cases, particularly in the cross-section above the peak.

In Fig.~\ref{figFmu2} we show results for the $\mu_m$ and $\mu_Q$ scale
dependence of $\tilde{\rm P}(M_t,M_t)$ (left and central panels respectively).
We increase and decrease $\mu_m$ and $\mu_Q$ by a factor of two, and both
variations exhibit rather small scale uncertainty. Here we show LL and NLL as
the bottom and top three curves inside each panel. As is often the case in
jet-physics, we note that there is a sizeable change to the normalization of the
cross-section in going from LL to NLL order. The vertical lines are
the same as in Fig.~\ref{figFmu}.  In the right most panel of Fig.~\ref{figFmu2}
we show the effect of using the renormalon free jet mass and renormalon free gap
parameter $\bar \Delta$ (solid red curve) in contrast to turning off the
renormalon subtraction for the mass, i.e. when using the top quark pole scheme
(purple dashed curve), and turning off the renormalon subtraction for the gap
(blue dot-dashed curve). Even at ${\cal O}(\alpha_s)$ the importance of having a
renormalon free soft function with $\bar \Delta$ is clearly visible.

\subsection{Cross Section in the Peak Region} \label{sec:sigpeak}

\begin{figure}
  \centerline{ 
   \includegraphics[width=10cm]{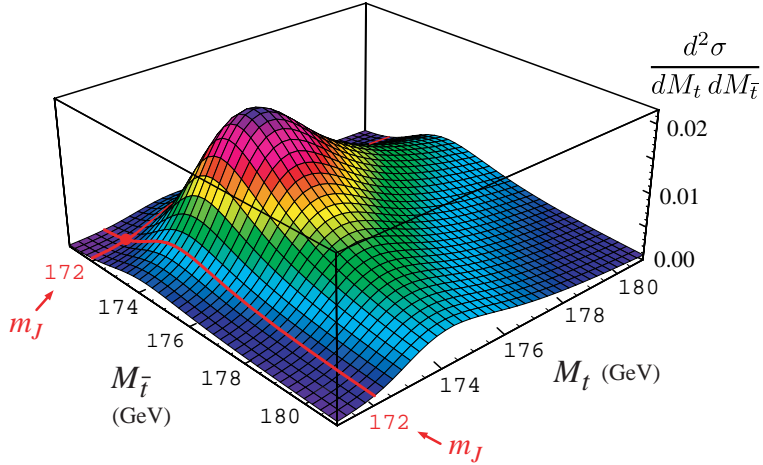} 
  } 
\vskip-0.3cm
\caption{${\rm F}(M_t,M_{\bar t})$, the differential cross-section in units of
  $\sigma_0/\Gamma_t^2$, versus $M_t$ and $M_{\bar t}$. The result is shown at
  NLL order.  }
\label{fig3D}
\end{figure}

Having examined the scale dependence of the perturbative corrections we now turn
to the convolution with the soft-function that gives the normalized
cross-section $F(M_t,M_{\bar t},m_J,Q/m_J)$ in Eq.~(\ref{Fdef}).  For most of
our plots we keep the soft function model fixed, having in mind that it can be
extracted from LEP data. In
Fig.~\ref{fig3D} we show $F$ at NLL for our default parameter set as a function
of the two invariant mass variables $M_t$ and $M_{\bar t}$. The underlying
short-distance quark mass is $m_J(\mu=2\,{\rm GeV})=172\,{\rm GeV}$, and the
peak of the cross-section occurs for $M_t$ and $M_{\bar t}$ values which are
$\simeq 2.4\,{\rm GeV}$ larger. This peak shift occurs due to the presence of
the low energy radiation described by the soft function as discussed in
Ref.~\cite{Fleming:2007qr}. At LO the shift is in the positive direction to
$M_t^{\rm peak}\simeq m_J + Q S_{\rm mod}^{[1,0]}/(2 m_J)$, where here
$S^{[1,0]}=\int d\ell^+d\ell^-\:\ell^+ S_{\rm mod}(\ell^+,\ell^-)\sim
\Lambda_{\rm QCD}$ is the first moment of the underlying soft-function
model~\cite{Fleming:2007qr}. As described below, this linear behavior with $Q/m$
persists at NLL order, although the slope is no longer simply $S_{\rm
  mod}^{[1,0]}$. Above the peak one sees in Fig.~\ref{fig3D} the perturbative
tails from gluon radiation, and that the tails are largest if we fix one of
$M_t$ or $M_{\bar t}$ at the peak.
\begin{figure}
  \centerline{ 
   \includegraphics[width=17cm]{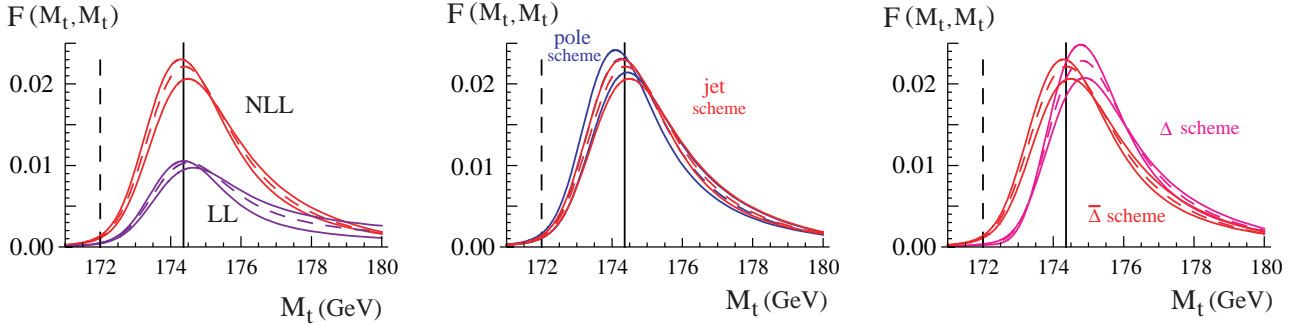} 
  } 
\caption{Normalized peak cross-section, ${\rm F}(M_t,M_t)$ versus $M_t$. 
  The dashed curves have $\mu_\Gamma=5\,{\rm GeV}$, and the solid curves have
  $\mu_\Gamma=3.3,7.5\,{\rm GeV}$.  The left panel shows results at LL (lower
  purple curves) and NLL (upper red curves) with the jet and $\bar\Delta$
  schemes. The center panel shows results in the jet-mass scheme (red) versus
  the pole-mass scheme (blue), where in both cases we use the $\bar\Delta$
  scheme. The right panel shows results in the $\bar\Delta(\mu)$ scheme for the
  gap parameter (red) versus the $\Delta$ scheme (magenta), where in both cases
  we use the jet-mass scheme.}
\label{figFmu2x}
\end{figure}

In order to analyze the parameter dependence of the cross-section we will now
consider the diagonal ${\rm F}(M_t,M_t,m_J,Q/m_J)$, which we simply referred to as ${\rm
  F}(M_t,M_t)$ in the analysis that follows.  In Fig.~\ref{figFmu2x}, in the
left panel, we show LL curves (bottom three lines) and NLL curves (top three
lines) using $\mu_\Gamma=3.3, 5.0, 7.5\,{\rm GeV}$ in the jet-mass and
$\bar\Delta$-scheme. We find that the peak of the cross-section is very stable
to the variation of $\mu_\Gamma$, and changes very little from LL to NLL order.
As explained above, by far the dominant contribution of the shift of the peak
away from the input short-distance jet-mass is due to the underlying
soft-function, shown here by the difference between the dashed and solid lines.
In the central panel we show again the NLL order cross sections in the jet-mass
and $\bar\Delta$-scheme (red curves) and compare it to the NLL predictions in
the pole-mass scheme for the same three $\mu_\Gamma$ values (blue curves). The
results show that in the pole-mass scheme there is more variation of the peak
position than in the jet-mass scheme. Finally in the right panel we show
variations of the cross-section in comparing the renormalon free
$\bar\Delta$-scheme (red curves) and the gap with a renormalon ambiguity in the
$\Delta$-scheme (magenta curves).  This figure demonstrates that the effect of
the switching to a renormalon free gap-scheme is larger than the residual
$\mu_\Gamma$ dependence at NLL order.

\begin{figure}
  \centerline{ 
\epsfxsize=9cm
\includegraphics[width=16cm]{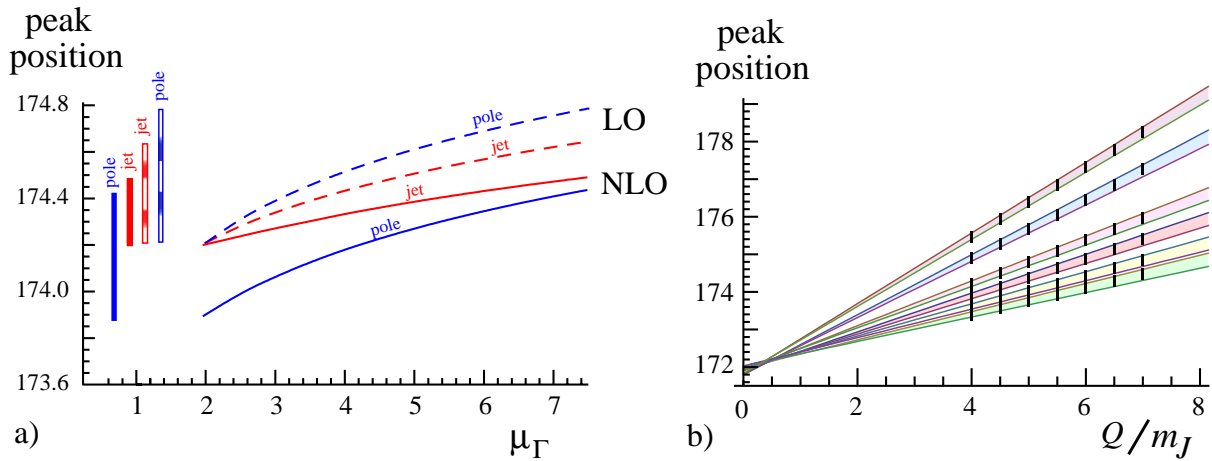}  
  } 
\vskip-0.3cm
\caption{a) Peak position of the single differential distribution
  $d\sigma/d M_t d M_{\bar t}(M_t,M_t)$ as a function of $\mu_\Gamma$. Red
  curves use the jet-mass and $\bar\Delta$ scheme and blue curves use the
  pole-mass and $\Delta$ scheme. Dashed curves are LL order, and solid curves
  are NLL order. The bars to the left show the size of the scale variation and
  left-to-right correspond to the curves from bottom-to-top.  b) The solid
  curves show the peak position versus $Q/m$ for six different models, which from
  top to bottom are
  $(a,b)=(3.5,-0.8)$ (purple), $(a,b)=(2.5,-0.8)$ (blue), $(a,b)=(3.5,-0.4)$,
  (magenta), $(a,b)=(2.5,-0.4)$ (red), $(a,b)=(3.5,0.4)$ (yellow),
  $(a,b)=(2.5,0.4)$ (green). The solid curves show a linear fit using the
  values at $Q/m=4$ and $5$.  Extrapolated to $Q/m=0$ any line converges
  on the underlying short-distance mass, independent of the soft-radiation
  model, yielding $m_t(\mu=5\,{\rm GeV})=171.9\pm 0.1\,{\rm GeV}$.  }
\label{fig:peak_pos_1}
\end{figure}

The $\mu_\Gamma$ dependence of the peak position is shown more explicitly in the
left panel of Fig.~\ref{fig:peak_pos_1}. In the pole-scheme (blue curves) we see
that there is very little change to the $\mu_\Gamma$ dependence in going from LL
(dashed blue curve) to NLL (solid blue curve).  In contrast in the jet-mass
scheme (red curves) the $\mu_\Gamma$ dependence is already smaller at LL order
(dashed red curve), and is significantly reduced by the NLL results (solid red
curve). In the right panel of Fig.~\ref{fig:peak_pos_1} we plot as black ticks
results for the peak position versus $Q/m$ using six different models for the
soft function with $\mu_\Gamma=5\,{\rm GeV}$. It is clearly visible that the
peak is shifted in a linear fashion with $Q/m_J$, with a slope that is model
dependent. For each model the solid lines show a fit to the the peak position
for $\mu_\Gamma=3.3\,{\rm GeV}$ and $\mu_\Gamma=7.5\,{\rm GeV}$ with a solid
band to show the uncertainty from this $\mu$-dependence.  The fits are done
using the points at $Q/m=4$ and $5$.
Extrapolating back to $Q/m=0$ removes the dependence on the soft-radiation, and
we see that for any soft function model the intercept determines the
short-distance mass parameter $m_J(\mu=5\,{\rm GeV})=171.9\,{\rm GeV}$. From the
spread of the curves we have $\simeq 0.13\,{\rm GeV}$ theoretical uncertainty in
this determination of the short-distance mass. This provides a method for
determining the short-distance mass even if the soft-function is unknown. In
order to maintain the perturbative stability of the relation of this intercept
with the top-mass it is important to use the jet-mass scheme.



\subsection{Cross Section in the Tail Region} \label{sect:away}

\begin{figure}
  \centerline{ 
   \includegraphics[width=17cm]{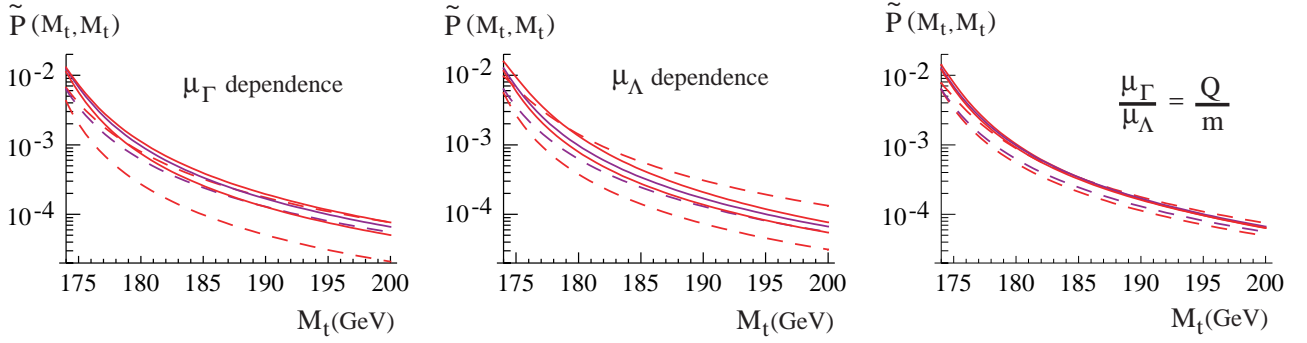} 
  } 
\vskip-0.4cm
\caption{ 
  Perturbative contributions, $\tilde{\rm P}(M_t,M_t)$ in the tail region at LL
  (dashed curves) and NLL (solid curves).  In the left panel we take
  $\mu_\Lambda=\mu_\Lambda^0$ and plot three curves with $\mu_\Gamma
  =\{0.5,1.0,1.5\}\mu_\Gamma^0$.  In the center panel we take
$\mu_\Gamma=\mu_\Gamma^0$ and show three curves with $\mu_\Lambda
=\{0.5,1.0,1.5\}\mu_\Lambda^0$. In the right panel we show $\mu_\Gamma
=\{0.5,1.0,1.5\}\mu_\Gamma^0$ with $\mu_\Lambda=\mu_\Gamma\: m_J(2\,{\rm GeV})/Q
$. Here $\mu_\Gamma^0$ and $\mu_\Lambda^0$ are given in Eq.~(\ref{mu0Gmu0L}).
}
\label{figPmutail}
\end{figure}
The tail region of the cross-section is characterized by invariant masses where
$\hat s \gg \Gamma$. In this region we are varying $M_t$ and hence $\hat s$ over a
large range, and it becomes necessary to scale $\mu_\Gamma\sim \hat s$ and
$\mu_\Lambda\sim \hat s m_J/Q$ to avoid having large logarithms that spoil the
perturbative expansion. We therefore define reference scales for the tail region,
\begin{align} \label{mu0Gmu0L}
  \mu^0_{\Gamma} =\sqrt{ \left[\frac{M^2\minus m_J^2(2\,{\rm GeV})}{m_J(2\,{\rm
      GeV})}\right]^2 \plus  \big(5\,{\rm GeV}\big)^2} \,,\qquad
  \mu_\Lambda^0 =\sqrt{ \left[\frac{M^2\minus m_J^2(2\,{\rm GeV})}{Q}
    \right]^2 \plus  \big(1\,{\rm GeV}\big)^2}
  \,,\qquad\qquad
\end{align}
where $m_J(2\,{\rm GeV})=172\,{\rm GeV}$, and we study the scale dependence by
varying $\mu_\Gamma$ and $\mu_\Lambda$ about these results.
In Fig.~\ref{figPmutail} we show the perturbative function $\tilde{\rm P}$ from
Eq.~(\ref{P}) at LL order (dashed curves) and NLL order (solid curves). The left
panel varies $\mu_\Gamma$ by 50\% about $\mu_\Gamma^0$ holding $\mu_\Lambda=\mu_\Lambda^0$
fixed, while the central panel varies $\mu_\Lambda$ holding
$\mu_\Gamma=\mu_\Gamma^0$ fixed. In contrast to the peak region we now plot the
cross-section over a log-scale.  Note that the LL results exhibit larger
uncertainty in this tail region, which is again substantially improved by the
NLL results. In the right most panel we vary $\mu_\Lambda$ as in the central
panel, but now take $\mu_\Gamma/\mu_\Lambda=Q/m$ as fixed. Just as in the peak
region this choice substantially reduces the scale uncertainty, indicating once
again that simply adding the individual variations of $\mu_\Gamma$ and
$\mu_\Lambda$ very likely overestimate the size of higher order perturbative
corrections. Finally since $\hat s$ increases in the tail region, the
uncertainty from the power expansion also increases as we require $v\cdot k/m =
\hat s/(2m) \simeq (M_t-m_J)/m_J\ll 1$. At $M_t=185\,{\rm GeV}$ this is an
expansion in $1/12$ and by $M_t=200\,{\rm GeV}$ its an expansion in $1/6$, both
of which are larger than the ratio $\simeq 1/100$ that we have in the peak
region.

\begin{figure}[t!]
  \centerline{ 
   \includegraphics[width=17cm]{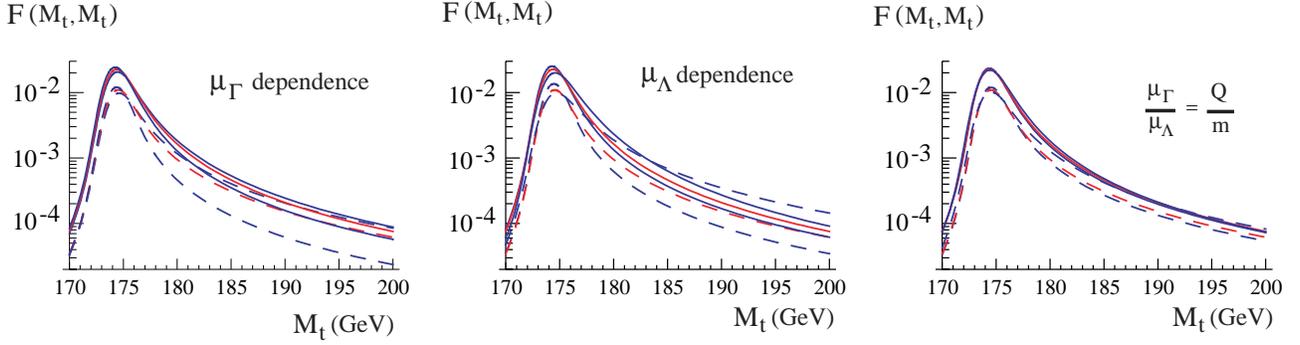} 
  } 
\vskip-0.4cm
\caption{
 Cross-section plotted over both the peak and tail regions at LL order (dashed)
 and NLL order (solid). In the left panel we take
  $\mu_\Lambda=\mu_\Lambda^0$ and plot three curves with $\mu_\Gamma
  =\{0.5,1.0,1.5\}\mu_\Gamma^0$.  In the center panel we take
$\mu_\Gamma=\mu_\Gamma^0$ and show three curves with $\mu_\Lambda
=\{0.5,1.0,1.5\}\mu_\Lambda^0$. In the right panel we show $\mu_\Gamma
=\{0.5,1.0,1.5\}\mu_\Gamma^0$ with $\mu_\Lambda=\mu_\Gamma\: m_J(2\,{\rm GeV})/Q$.
}
\label{figPmutail2}
\end{figure}
\begin{figure}[t!]
  \centerline{ 
   \includegraphics[width=12cm]{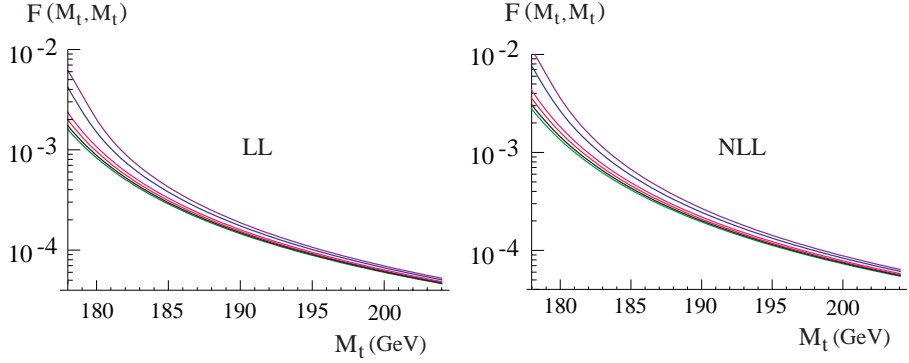} 
  } 
\vskip-0.4cm
\caption{ 
  Variation of the tail cross-section using different soft function models at LL
  (left panel) and NLL (right panel). The curves are $(a,b)=(3.5,-0.8)$
  (purple), $(a,b)=(2.5,-0.8)$ (blue), $(a,b)=(3.5,-0.4)$, (magenta),
  $(a,b)=(2.5,-0.4)$ (red), $(a,b)=(3.5,0.4)$ (black), $(a,b)=(2.5,0.4)$
  (green).
}
\label{figPmutail3}
\end{figure}
In Fig.~\ref{figPmutail2} we convolute ${\rm P}$ with the soft-function model as
in Eq.~(\ref{Fdef}), and plot the normalized cross-section $F$ over both the
peak and tail regions. The three panels show the same $\mu$-variations as
Fig.~\ref{figPmutail}. These plots show one of the attractive features of our
treatment of the soft-function. In the tail region there are perturbative
corrections in the soft-function that are important for determining the
cross-section, and there is also an important power correction due to the first
moment of the model function $S_{\rm mod}$. Both of these are included in our
analysis by using the full $S(\ell^+,\ell^-,\mu)$ function from
Eq.~(\ref{Smodel1v2}) as explained in Ref.~\cite{Hoang:2007vb}. In the peak
region these terms naturally interpolate into a full model soft-function in a
consistent manner.  Once again we see from the third panel of
Fig.~\ref{figPmutail2} that taking $\mu_\Gamma/\mu_\Lambda=Q/m_J$ leads to quite
small $\mu$-dependence for the entire cross-section.  Finally in
Fig.~\ref{figPmutail3} we show the effect that variation of the soft-function
model has on the cross-section in the tail region. The effect of the soft
function becomes larger as we get closer to the peak region, as expected. Since
the cross-section has already dropped by two-orders of magnitude by
$M_t=200\,{\rm GeV}$ we have not bothered to analyze it in the ultra-tail
region, $M_t-m_J\sim m_J$, where it is further suppressed by several more
decades. However, in appendix~\ref{App:SCETnum} we do give formulas for the
cross-section in the ultra-tail region, which are analogous to the ones used for
our analysis of the peak and tail cross-sections.  These formulae could be
useful as a means of estimating top-quark backgrounds from $t\bar t$ events for
other processes in the ultra-tail region.


\subsection{Thrust} \label{sec:eventshapes}

Starting from the two-dimensional distribution, $d^2\sigma/dM_t^2dM_{\bar t}^2$
in Eq.~(\ref{sigmaMM}) it is straightforward to derive results for other event
shape variables for massive particles. For example, for the thrust $T$, we have
$1-T\equiv\tau=(M_t^2+M_{\bar t}^2)/Q^2$, so
\begin{align} \label{Tfactorization}
  \frac{1}{\sigma_0} \frac{d\sigma}{dT}
 &= \int_0^\infty\!\!\! dM_t^2\! \int_0^\infty\!\!\! dM_{\bar t}^2\ \:
  \delta\Big( \tau - \frac{M_t^2+M_{\bar t}^2}{Q^2} \Big) \:
  \frac{1}{\sigma_0}
  \frac{d^2 \sigma}{dM_t^2 dM_{\bar t}^2}  \\[5pt]
 &= \int_0^\infty\!\!\! d\ell\ \: {\rm P_T}\bigg(\frac{Q^2\tau-2m_J^2-Q\ell-2
   Q\bar\Delta(\mu_\Lambda)}{m_J},\mu_\Lambda\bigg)
   S_{\rm mod}^{\rm symm}(\ell,0)
  \,.\nn
\end{align}
The perturbative contributions are grouped into the dimensionless function
${\rm P_T}$ which is a projection of our function ${\rm P}$, 
\begin{align} \label{PT}
  {\rm P_T}(\hat s,\mu_\Lambda)
   &\equiv \int_{-\infty}^{+\infty}\!\!\!\!\! d\hat s_d \
    \frac{m_J Q^2}{8 M_t M_{\bar t}\, \Gamma_t^2}\: \:
    {\rm P}\Big(\frac{\hat s+\hat s_d}{2},
    \frac{\hat s-\hat s_d}{2}, \mu_\Lambda\Big)\,.
\end{align}
Here under the $\hat s_d$ integral $M_t^2= m_J^2 + m_J(\hat s+\hat s_d)/2$
and $M_{\bar t}^2= m_J^2 + m_J(\hat s-\hat s_d)/2$. An analytic formula for
${\rm P_T}$ is derived in Appendix~\ref{app:PT}. The appropriate soft-function
for thrust, $S_{\rm mod}^{\rm symm}(\ell)$ in Eq.~(\ref{Tfactorization}), is
also simply a projection of the model for the hemisphere soft function, $S_{\rm
  mod}(\ell^+,\ell^-,\Delta)$, where
\begin{align} \label{Sthrust}
  S_{\rm mod}^{\rm symm}(\ell,0) & = \int_0^\infty\!\!\! d\ell^+\: d\ell^-
    \delta\big(\ell- \ell^+ - \ell^-\big)
   S_{\rm mod}(\ell^+,\ell^-,0)
  \,.
\end{align}
For the exponential model in Eq.~(\ref{SM1}) this projection gives
\begin{align} \label{SM2}
  S_{\rm mod}^{\rm symm}(\ell,0) &= \frac{{\cal N}(a,b)}{\Lambda}
\frac{\sqrt{\pi}\Gamma(a)}{\Gamma(a\plus \frac12 )}
\Big(\frac{\ell}{2\Lambda}\Big)^{2a-1} \:
{}_1F_1\Big(\frac12,\frac12+a,\frac{(b-1)\ell^2}{2\Lambda^2}\Big) \:
 e^{-(1\plus b)\ell^2/(2\Lambda^2)} \,,
\end{align}
where $\{a,b,\Lambda\}$ are the model parameters and ${\cal N}(a,b)$ is the same
normalization constant as in Eq.~(\ref{SM1}).

\begin{figure}
  \centerline{ 
   \includegraphics[width=17cm]{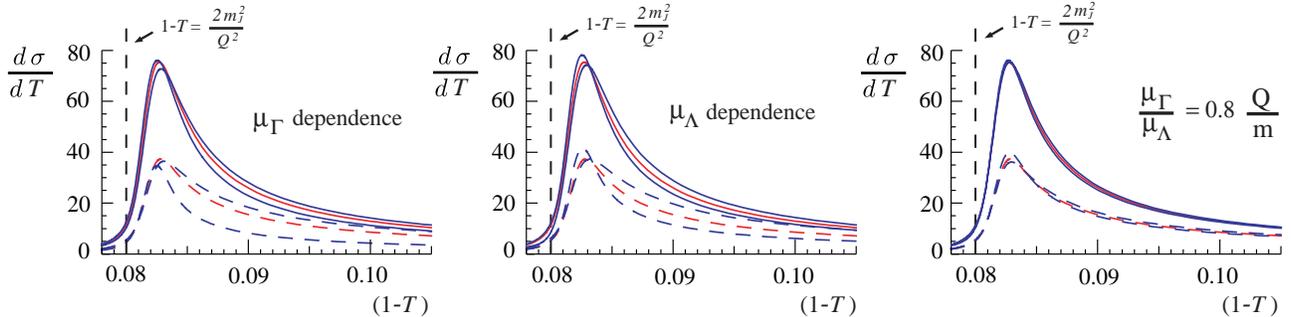} 
  } 
\caption{ 
  Thrust distribution, $d\sigma/dT$ in units of $\sigma_0$, plotted versus $1-T$
  at LL (dashed curves) and NLL (solid curves).  In the left panel we take
  $\mu_\Lambda=\mu_\Lambda^0$ and plot three curves with $\mu_\Gamma
  =\{0.5,1.0,1.5\}\mu_\Gamma^0$.  In the center panel we take
  $\mu_\Gamma=\mu_\Gamma^0$ and show three curves with $\mu_\Lambda
  =\{0.7,1.0,1.5\}\mu_\Lambda^0$. In the right panel we show $\mu_\Lambda
  =\{0.7,1.0,1.5\}\mu_\Lambda^0$ with $\mu_\Gamma= \mu_\Lambda\:[0.8\,Q/
  m_J(2\,{\rm GeV})]$.}
\label{figThrust}
\end{figure}
In Fig.~\ref{figThrust} we plot the thrust distribution at LL order (dashed
curves) and NLL order (solid curves) for events which were initiated by the
massive unstable top-quarks in $e^+e^-$ collisions.  Since the plot includes
values in the tail region we use the reference scales
\begin{align}
 \mu_\Gamma^0 = \sqrt{
   \frac{Q^4}{4m_J^2}\Big(\tau-\frac{2m_J^2}{Q^2}\Big)^2  \plus  \big(5\,{\rm GeV}\big)^2} \,,
\qquad
 \mu_\Lambda^0 = \frac{1}{0.8}\: \frac{\mu_\Gamma^0\, m_J}{Q} \,,
\end{align}
where $\tau=1-T$. Taking $\mu_\Gamma\simeq\mu_\Gamma^0$ and
$\mu_\Lambda\simeq\mu_\Lambda^0$ ensures that the logs involving these
parameters do not grow substantially over the region plotted. Our choice for
$\mu_\Lambda$ here is slightly larger than the ones used earlier. This is
because of the effective doubling of the anomalous dimensions for the thrust
cross-section (see Appendix~\ref{app:PT}), which necessitates using slightly
larger values for $\mu_\Lambda$ to avoid the region where large values for
$\alpha_s$ cause a break down in perturbation theory.

The threshold for thrust for two-massive particles is given by $1-T=2m_J^2/Q^2$
and is shown by the vertical dashed lines in Fig.~\ref{figThrust}.  Just as for
the invariant mass-distribution, there is a peak in the thrust cross-section and
it is shifted above the massive particle threshold due to soft-radiation effects
by an amount $\simeq 2\Lambda_{\rm QCD}/Q$. The analog of this for massless jets
is a peak in the thrust distribution at values $1-T\simeq 2\Lambda_{\rm
  QCD}/Q$ (see for example~\cite{Korchemsky:2000kp}), which is a shift above the
massless dijet threshold at $1-T=0$. The three panels in Fig.~\ref{figThrust}
show the $\mu$-dependence of our NLL results, varying $\mu_\Gamma$ in the left
panel, $\mu_\Lambda$ in the center panel, and $\mu_\Lambda$ with
$\mu_\Gamma/\mu_\Lambda = 4$ fixed in the right panel (since here $Q/m_J=5$).
Again we see that there is very small $\mu$-dependence when $\mu_\Gamma$ and
$\mu_\Lambda$ are varied in a correlated fashion. We believe the left panel
gives a reasonable estimate of the perturbative uncertainties in the shape of
the thrust distribution. An analysis of the thrust-distribution peak for
different values of $Q$ could also be used to extract the short-distance
top-mass parameter.



\section{Conclusion} \label{sect:conclusion}

Precise measurements of the top quark mass $m_t$ belong to the most important
standard measurements carried out at the Tevatron and the LHC. The most
sensitive method relies on the reconstruction of the top quark invariant mass
distribution through measurements of the energies and momenta of jets from the
top decay. While considerable work has and is being invested to control
experimental systematic effects, very little theoretical work exists which
studies both perturbative and nonperturbative QCD aspects of the resulting
invariant mass distribution. Also, to our knowledge, there has been no
theoretical work on how the shape and the resonance mass of this distribution
are related to a short-distance top mass parameter in the QCD Lagrangian.

In Ref.~\cite{Fleming:2007qr} we derived the factorization theorem in
Eq.~(\ref{FactThm}) which describes the simpler environment of $e^+e^-$
collisions. It predicts the double differential invariant mass distribution
$d^2\sigma/dM_t^2dM_{\bar t}^2$ in the resonance region for the large
c.m.\,energies $Q\gg m_t$, where $M_{t,\bar t}$ are the total invariant masses
of all particles in the two hemispheres determined with respect to the event
thrust axis. The factorization represents the leading order result in a power
expansion in $m/Q$ and $\Gamma/m$, and these corrections are indicated in
Eq.~(\ref{FactThm}). Here $\Gamma$ is the width of the invariant mass
distribution, which is larger than the underlying total width of the top-quarks
$\Gamma_t$. The derivation was based on the hierarchy $Q\gg
m_t\gg\Gamma,\Lambda_{\rm QCD}$, where $\Lambda_{\rm QCD}$ is the hadronization
scale and uses the effective theories Soft-Collinear-Effective-Theory (SCET) and
Heavy-Quark-Effective-Theory to achieve a separation of different physical
effects associated to $Q$, $m_t$, $\Gamma_t$ and $M_{t,\bar t}$ and
$\Lambda_{\rm QCD}$.  For the systematic inclusion of mass and width effects the
use of both effective theories was crucial.  The factorization theorem separates
perturbative from non-perturbative effects and represents the leading order term
in the power expansion, but is valid to all orders in the expansion in
$\alpha_s$.

In this paper we extended the presentation given in Ref.~\cite{Fleming:2007qr}
and presented detailed computations of the different pieces entering the
factorization theorem in the peak region at NLL order.  We also presented NLL
predictions for the tail of the invariant mass distribution, where $M_{t,\bar
  t}$ are above the resonance peak. The double invariant hemisphere mass
distribution is itself an event-shape distribution that can be related to other
event-shape variables such as thrust or jet masses in a straightforward way. The
factorization formula consists of several functions that can be computed
perturbatively order-by-order in $\alpha_s$, including hard coefficients for the
scales $Q$ and $m$, and two jet functions for the top and antitop quarks which
depend strongly on the top quark Lagrangian mass. It also involves a
nonperturbative soft function that describes the momentum distribution of soft
final state radiation. Using alternative invariant mass prescriptions, for which
the soft particles are assigned differently to $M_t$ and $M_{\bar t}$, the same
factorization formula applies, but with a different soft function. In the tail
region the soft function also contains perturbatively calculable corrections.

Our analysis uses effective theory techniques. In particular we calculated
Wilson coefficients that arise from matching QCD onto SCET, and from matching
SCET onto bHQET to NLO. In addition we calculated the NLL running: first between
the scales $Q$ and $m$ using anomalous dimensions of operators in SCET, and then
between the scales $m$ and $\Gamma$ and between $\Gamma$ and $\Lambda$ using
anomalous dimensions of operators in bHQET.  The perturbative corrections,
including resummation, are given by simple analytic functions, and our strategy
for computing these functions naturally generalizes for future analytic
computations at NNLL order. One important result of our analysis is that in the
peak region the running between the scales $Q$, $m$, and $\Gamma$ is local, it
only changes the overall normalization and not the shape of the invariant mass
distribution. Thus only large logs between the scales $\Gamma$ and $\Lambda$ can
shift the shape of the distribution. This is encoded in consistency equations
for the renormalization-group evolution functions entering the factorization
theorem.

The observable invariant mass distribution is obtained from a convolution of the
jet functions with the soft function. Through this convolution the energy of the
peak and the width of the observed distribution are dependent on the center of
mass energy $Q$. In particular, non-perturbative effects described by the soft
function shift the resonance peak position towards larger masses, and broaden
the distribution.  In Ref.~\cite{Fleming:2007qr} we demonstrated that the soft
function for the hemisphere mass prescription can be determined from event shape
distributions for massless dijet events, and that the dependence on the mass
scheme is controlled through the perturbative expansion of the jet function.
This allows in principle for a top quark mass determination that is free of
hadronization uncertainties. Even if the soft-function is not known from
measurements of massless jets, one can still extract the short-distance mass
parameter using an analysis like the one described in Fig.~\ref{fig:peak_pos_1}.
We have demonstrated that these statements remain true in the presence of
perturbative corrections and with the summation of large logs.  We also
introduced the so-called jet-mass scheme to define the Lagrangian top mass
parameter.  This is a top quark short-distance mass scheme that is particularly
suited to mass determinations related to the resonance peak position, since it
makes the peak position stable to the inclusion of perturbative corrections.
One-loop relations of this jet-mass to the ${\overline {\rm MS}}$-mass and
1S-mass schemes were given in section~\ref{sect:jetmass}, as well as the LL
evolution formula for the jet-mass.  For the construction of the soft function
at NLL order we used results from Ref.~\cite{Hoang:2007vb} where a soft model
function is convoluted with the soft function contributions determined from
fixed order perturbation theory. This soft function works equally well in the
peak and tail regions. To avoid large higher order corrections it was necessary
to introduce a gap parameter in the soft-function that accounts for the fact
that there is a minimal hadronic energy for the soft-radiation between the jets.
This parameter allows us to make the perturbative corrections in the soft
function free from an ${\cal O}(\Lambda_{\rm QCD})$ renormalon ambiguity.

In our numerical analysis we analyzed the hemisphere mass distribution on the
peak and away from the peak, and the thrust distribution. We demonstrated that
NLL order corrections are important and need to be accounted for to make viable
predictions. We also showed that it is important to sum the large logs between
$\mu_\Gamma\simeq \Gamma$ and $\mu_\Lambda\gtrsim \Lambda$ to avoid sizeable
scale uncertainties.  We also studied the impact of the jet mass scheme, and
showed that it improves predictions for the resonance peak position in
comparison to the pole mass scheme. For mass measurements this result implies
that the jet mass can be determined from mass reconstruction more accurately
than the pole mass which is known to suffer from renormalon ambiguities.  In our
NLL analysis we demonstrated that the perturbative corrections associated to the
gap parameter improve significantly the perturbative behavior of predictions in
the peak and in the tail region. This result implies that soft functions that
account for a gap as proposed in Ref.~\cite{Hoang:2007vb} are crucial for
precise measurements of the top quark mass and the model parameters from
experimental data.  Finally we also presented a NLL prediction for the thrust
distribution for top pair production in the region of large thrust.  To our
knowledge this NLL result presents for the first time a full resummed
event-shape distribution for massive quarks. The thrust distribution has a
strong dependence on the mass and can serve as an alternative way of measuring
heavy quark masses.  Our numerical analysis can be extended to make predictions
for bottom quark production by taking the limit $\Gamma_t\to 0$. It would also
be interesting to study the $\alpha_s m_t/Q$ and $m_t^2/Q^2$ power corrections,
which can be accomplished in our effective field theory setup.

Through our detailed calculations of the jet invariant mass distributions and
their relation to the top quark mass, we have demonstrated the viability of
extracting the top mass with high precision at a future linear collider such as
the ILC. In principle, a precision of better than $\Lambda _{\textrm{QCD}}$ can
be achieved since there is a clear relation between the top mass Lagrangian
parameter and the physically observed jet invariant mass distribution. In the
future we intend to extend the work presented here and in
Ref.~\cite{Fleming:2007qr} to the study of top-mass reconstruction at the LHC.

\acknowledgments{We would like to thank B.~Holdom, A.~Juste, V.~Khoze, S.~Kluth,
  J.~H.~K\"uhn, S.~Menke, M.~Peskin, G.~Salam for helpful discussions and
  communication.  We thank the Max-Planck Institute for Physics for hospitality
  while parts of this work were completed. This work was supported in part by
  the Department of Energy Office of Nuclear Science under the grants
  DE-FG02-94ER40818 and DE-FG03-92ER40701, and in part by the EU network
  contract MRTN-CT-2006-035482 (FLAVIAnet).}

\appendix

\section{Summary of Feynman diagrams in SCET} \label{AppFeyn}


In this appendix we list results for the individual Feynman diagrams used in the
body of the paper in section~\ref{sect:scet}. 

{\it QCD graphs.} For the QCD current at one-loop we have vertex and
wavefunction graphs. We use free quark external states with an offshellness
IR regulator, $\Delta^2=p^2-m^2 = \bar p^2-m^2$ where $\Delta^2\ll m^2$. We use
dimensional regularization with $d=4-2\epsilon$ for the UV divergences. The
graphs in Fig.~\ref{qcdloops} are 
\begin{align} \label{massiveqcdvertex}
  V_{\ref{qcdloops}a} &= \Gamma^\mu_i \frac{\alpha_sC_F}{4\pi }
   \bigg(\frac{1}{\epsilon} + 2\>\text{ln}^2\frac{-Q^2}{m^2} -
    4\>\text{ln}\frac{-Q^2}{m^2}\>\text{ln}\frac{Q^2}{\Delta ^2} +
    3\>\text{ln}\frac{-Q^2}{m^2} 
    + \text{ln}\frac{\mu^2}{m^2} + \frac{2\pi ^2}{3} \bigg )
    \,,\nn\\
  V_{\ref{qcdloops}b} &=  
   \frac{-i\alpha_s C_F}{4 \pi} \bigg[
  m \bigg\{ \frac{4}{\epsilon} \plus 4 \ln\Big(\frac{\mu^2}{m^2}\Big) \plus  6 + 
   \frac{4\Delta^2}{m^2} \ln\Big(\frac{m^2}{-\Delta^2}\Big) \bigg\} \nn\\ 
   & \qquad\qquad\quad +
    \slash\!\!\! p   \bigg\{ \frac{-1}{\epsilon} \minus \ln\Big(\frac{\mu^2}{m^2}\Big) 
    \minus 2 \plus \frac{\Delta^2}{m^2} +\frac{2\Delta^2}{m^2}
     \ln\Big(\frac{-\Delta^2}{m^2}\Big) \bigg\} \bigg] \,. 
\end{align}


{\it SCET vertex graphs.} For the SCET current we have collinear vertex and wavefunction
graphs and a soft vertex graph.  We use dimensional regularization for the UV
divergences and the offshellness
IR regulators $\Delta^2=p^2-m^2$ and $\bar \Delta^2=\bar p^2-m^2$. To compare IR
divergences to the full theory results we should set $\Delta^2=\bar \Delta^2$
and expand in $\Delta^2\ll m^2$. For later convenience we will first quote
results prior to making this expansion.  The massive collinear quark Lagrangian
is given by~\cite{Leibovich:2003jd,Rothstein:2003wh}
\begin{eqnarray} \label{Lscet}
{\cal L}^{(0)}_{qn} = \bar{\xi}_n\Big [ i n\cdot D_s + gn \cdot
A_n + (i \Dslash _c^\perp \!-\! m) W_n\frac{1}{\bn\mcdot {\cal P}}W_n^\dagger
(i\Dslash _c^\perp \!+\! m) \Big ] \frac{\bnslash}{2} \xi_n  \,.
\end{eqnarray}
Using the corresponding massive SCET Feynman rules the graphs in
Fig.~\ref{scetloops} are given by
$\text{Fig}.\ref{scetloops}\{a,b,c\}=(\bar{\xi}_n \Gamma ^\mu _i  \xi_\bn
)V_{\ref{scetloops}\{a,b,c\}}$  and  
$\text{Fig}.\ref{scetloops}\{d,e\}=(\frac{-i\bnslash
}{2})V_{\ref{scetloops}\{d,e\}}$  where the intgerals
$V_{\ref{scetloops}a,b,c,d,e}$ read
\begin{align}
  V_{\ref{scetloops}a} &=  -2i g^2 C_F \tilde\mu ^{2\epsilon}
   \int\!\! \frac{d^dk}{(2\pi)^d} \bigg\{ \frac{ \bn\mcdot(k\plus p)}
    {[\bn\mcdot k][k^2][(k\plus p)^2\minus m^2] }-  \frac{ \bn\mcdot p}
    {[\bn\mcdot k][k^2][\bn\mcdot p\, n\mcdot(k\plus p)\minus m^2]  } \bigg\},
   \nn\\
  V_{\ref{scetloops}b} &=   V_{\ref{scetloops}a} \text{ with } n\leftrightarrow \bn \,,
    p \rightarrow \bar p  \,,\nn\\  
  V_{\ref{scetloops}c} &=  -2i g^2 C_F  \tilde\mu ^{2\epsilon}
   \int\!\! \frac{d^dk}{(2\pi)^d} \frac{ (\bn\mcdot p)(-n\mcdot\bar p)}
    {[k^2][\bn\mcdot p\, n\mcdot k + \Delta^2 +i0][-n\mcdot \bar p\, \bn\mcdot k + \bar\Delta^2 ] } \,,
  \nn\\
 V_{\ref{scetloops}d} &= -2 g^2 C_F    \tilde\mu ^{2\epsilon}
   \int\!\! \frac{d^dk}{(2\pi)^d} \frac{ \bn\mcdot(k\plus p)}
    {[k^2 ][(k\plus p)^2\minus m^2] }  \bigg\{
   \frac{ 4 m^2(1\minus\frac{\epsilon}{2})}{[\bn\mcdot p ] [\bn\mcdot (k\plus p)]} 
   \plus \frac{(k_\perp^2\minus m^2)(1\minus\epsilon)}{[\bn\mcdot(k\plus p)]^2}
   \minus \frac{ m^2(1\minus\epsilon)}{[\bn\mcdot p]^2}
   \bigg\},
   \nn\\
 V_{\ref{scetloops}e} &= 2 g^2 C_F  \tilde\mu ^{2\epsilon} 
   \int\!\! \frac{d^dk}{(2\pi)^d} \frac{1}{[k^2]} \frac{(1\minus\epsilon)}
    {[\bn\mcdot(k\plus p)]} 
    \,,
\end{align}
where 
\begin{eqnarray} \label{mutilde}
  \tilde\mu^2\equiv\frac{\mu^2\,e^{\gamma_E}}{4\pi} \,,
\end{eqnarray}
and where all terms in the denominator with square brackets are defined with the
$+i0$ prescription. To evaluate 
Fig.~\ref{scetloops}a,b we included the zero-bin minimal
subtraction~\cite{Manohar:2006nz} to avoid double counting the region encoded in
Fig.~\ref{scetloops}c.  For $V_{\ref{scetloops}d}$ and $V_{\ref{scetloops}e}$
we have a singularity for $\bn\cdot (k+p)\to 0$ with fixed $k_\perp^2$, but it
cancels in the sum of the two diagrams:
\begin{align}
 V_{\ref{scetloops}d} \plus V_{\ref{scetloops}e} 
  &= -2 g^2 C_F  \tilde\mu ^{2\epsilon} \!
   \int\!\! \frac{d^dk}{(2\pi)^d} \frac{(1\minus\epsilon)}
    {[k^2 ][(k\plus p)^2\minus m^2] }  \bigg\{
   \frac{ 4 m^2(1\minus \frac{\epsilon}{2})}{(1\minus\epsilon)[\bn\mcdot p ] } 
   \minus n\mcdot (k\plus p)\:
   \minus \frac{ m^2\,\bn\mcdot(k\plus p)}{[\bn\mcdot p]^2}
   \bigg\} . \nn
\end{align}
Setting $\bn\cdot p=n\cdot\bar p=Q$ and computing the integrals we find
\begin{align}\label{vertexgraphsscet}
  V_{\ref{scetloops}a} &= \: \frac{\alpha_s C_F}{4 \pi} \bigg[
  \frac{2}{\epsilon^2} + \frac{2}{\epsilon} \ln\Big(\frac{\mu^2}{-\Delta^2}\Big)
  + \frac{2}{\epsilon} +\ln^2\Big( \frac{\mu^2}{-\Delta^2}\Big) -
  \ln^2\Big(\frac{m^2}{-\Delta^2}\Big) -2 {\rm
    Li_2}\Big(\frac{-\Delta^2}{m^2}\Big)
  \nn\\
  &\ \ +
  2\ln\Big(\frac{m^2}{-\Delta^2}\Big)\ln\Big(\frac{m^2\plus \Delta^2}{-\Delta^2}
  \Big) + 2\ln\Big(\frac{\mu^2}{-\Delta^2}\Big) - \frac{2m^2}{m^2\plus \Delta^2}
  \ln\Big(\frac{m^2}{-\Delta^2}\Big)
  +4 +  \frac{\pi^2}{2} \bigg] ,
  \nn\\
  V_{\ref{scetloops}b} &= V_{\ref{scetloops}a}[\Delta^2\to \bar \Delta^2] ,
  \nn\\
  V_{\ref{scetloops}c} &=\: \frac{\alpha_s C_F}{4 \pi} \bigg[
  -\frac{2}{\epsilon^2} - \frac{2}{\epsilon} \ln \Big(\frac{Q^2
    \mu^2}{(-\Delta^2)\bar\Delta^2} \Big) - \ln^2 \Big(\frac{Q^2
    \mu^2}{(-\Delta^2) \bar \Delta^2}\Big)
  - \frac{\pi^2}{2} \bigg] \,,
  \nn\\
  V_{\ref{scetloops}d} \plus V_{\ref{scetloops}e} &=
 \frac{1}{Q} \frac{\alpha_s C_F}{4 \pi} \bigg[
  \frac{6 m^2\minus \Delta^2}{\epsilon} + (6m^2\minus
  \Delta^2)\ln\Big(\frac{\mu^2}{-\Delta^2}\Big) - \frac{m^4(6 m^2\plus 7\Delta^2)}
   {(m^2\plus \Delta^2)^2} \ln\Big(\frac{m^2}{-\Delta^2}\Big) \nn\\ 
  &\ \   + 8 m^2 -\frac{\Delta^4}{m^2\plus\Delta^2} \bigg] \,.
\end{align} 
Note that the soft graph, $V_{\ref{scetloops}c}$, is independent of $m$.  Expanding
$V_{\ref{scetloops}a}$ and $V_{\ref{scetloops}d} + V_{\ref{scetloops}e}$ for
$\Delta^2\ll m^2$ gives
\begin{align}\label{expvertexgraphsscet}
V_{\ref{scetloops}a} 
 & =  \frac{\alpha_s C_F}{4 \pi}   \bigg[ \frac{2}{\epsilon^2} \plus
 \frac{2}{\epsilon} \ln\Big(\frac{\mu^2}{-\Delta^2}\Big) \plus \frac{2}{\epsilon}
  \plus \ln^2\Big(\frac{\mu^2}{-\Delta^2}\Big)
  \plus \ln^2\Big(\frac{m^2}{-\Delta^2}\Big) \plus 2 \ln\Big(\frac{\mu^2}{m^2}\Big) 
 \plus 4 \plus  \frac{\pi^2}{2}\bigg] ,
  \nn\\
 V_{\ref{scetloops}d} \plus V_{\ref{scetloops}e} &=
\frac{1}{Q} \frac{\alpha_s C_F}{4 \pi} \bigg[
  m^2 \bigg\{ \frac{6}{\epsilon} \plus 6 \ln\Big(\frac{\mu^2}{m^2}\Big) \plus  8\bigg\} - 
   \Delta^2 \bigg\{ \frac{1}{\epsilon} \plus \ln\Big(\frac{\mu^2}{m^2}\Big) 
   \plus 4 \ln\Big(\frac{-\Delta^2}{m^2}\Big) \bigg\} \bigg] .
\end{align} 
{}From Eq.~(\ref{Lscet}) with Eq.~(\ref{Zscet}) we have a mass counterterm, $-i
\bnslash/(2\bn\mcdot p) 2 m \delta_m$. In the pole mass scheme 
\begin{eqnarray}
 (\delta_m)^{\rm pole} = -m \frac{\alpha_s C_F}{4\pi}
  \bigg[\,\frac{3}{\epsilon} + 3 \ln\frac{\mu^2}{m^2} +4 \bigg] \,,
\end{eqnarray}
which exactly cancels the entire $m^2$ term in the self energy graphs
$V_{\ref{scetloops}d} \plus V_{\ref{scetloops}e}$. In a
general mass scheme we have $\delta_m=(\delta_m)^{\rm pole}+\delta m$, and 
\begin{align} \label{SCETse_nom}
 V_{\ref{scetloops}d} \plus V_{\ref{scetloops}e}
    \plus \frac{2 m}{\bn\mcdot p} \delta_m 
  =
   \frac{2 m\delta m}{Q} -  
  \frac{\Delta^2}{Q}\, \frac{\alpha_s C_F}{4 \pi} 
  \bigg\{ \frac{1}{\epsilon} \plus \ln\Big(\frac{\mu^2}{m^2}\Big) 
   \plus 4 \ln\Big(\frac{-\Delta^2}{m^2}\Big) \bigg\} \,.
\end{align}
Using the $\overline{\rm MS}$ subtraction, the wavefunction renormalization
$Z_\xi$ removes the 
remaining $1/\epsilon$ divergence in Eq.~(\ref{SCETse_nom}), and leaves a finite
correction to the residue of the collinear quark propagator, $iR_\xi\, (\nslash/2)\,
\bn\mcdot p/(p^2 \minus m^2)$, with
\begin{align}\label{wfgraphscet} 
  Z_\xi = 1 - \frac{\alpha_s C_F}{4\pi\epsilon}\,,\qquad
  R_\xi = 1+ \frac{\alpha_s
    C_F}{4 \pi} \bigg[ - \ln\Big( \frac{\mu^2}{m^2}\Big) -4 \ln\Big(
  \frac{-\Delta^2}{m^2}\Big) \bigg] \,.
\end{align}
This is identical to results from the QCD self energy, $V_{(\ref{qcdloops}b)}$
in Eq.~(\ref{massiveqcdvertex}), namely $R_\psi=R_\xi$ and $Z_\psi=Z_\xi$.


{\it SCET jet function.} The computation of the stable massive SCET jet function
is given by the imaginary part of the graphs in Fig.~\ref{forwardI}.  The
integrals are identical to those for the vertex graphs above, except that now
there is true external momentum, $r_n^\mu$, that is routed through the diagram,
and the invariant mass, $s$, takes the place of the offshellness, therefore
these diagrams do not require an IR regulator. Here $s = r_n^2 - m^2 = Q
r_n^+-m^2$ is defined for convenience. The diagrams in Fig.~\ref{forwardI}
contribute to the SCET jet functions. Taking the spin and color trace, but not
yet the imaginary part, the graphs give
\begin{align} \label{Jabc_int}
  J_{\ref{forwardI}a} = J_{\ref{forwardI}b} 
 &= \Big[\frac{i}{\pi}\,\frac{i }{s\plus i0}\Big]\: 
   V_{\ref{scetloops}a}(\Delta^2\to s) 
   \\
 &= \frac{-\alpha_s C_F  }{4\pi^2\,s}
  \bigg\{ \frac{2}{\epsilon^2}\plus \frac{2}{\epsilon} \ln\Big(\frac{\mu^2}{-s}\Big)
   \plus \frac{2}{\epsilon} \plus  \ln^2\Big(\frac{\mu^2}{-s}\Big)
   \minus  \ln^2\Big(\frac{m^2}{-s}\Big)
   \minus 2 {\rm Li}_2\Big(\frac{-s}{m^2}\Big) 
    \nn\\
 & \ \  \plus  2\ln\Big(\frac{m^2}{-s}\Big)\ln\Big(\frac{m^2\plus s}{-s} \Big)
  \plus 2\ln\Big(\frac{\mu^2}{-s}\Big) \minus \frac{2m^2}{m^2\plus s}
  \ln\Big(\frac{m^2}{-s}\Big)
    \plus 4 \plus \frac{\pi^2}{2}  \bigg\}
  \,,\nn \\
   J_{\ref{forwardI}c} &= 0
  \,,\nn\\
  J_{\ref{forwardI}d} \plus J_{\ref{forwardI}e} \plus J_{\ref{forwardI}f}
     &=  \Big[\frac{i}{\pi}\,\frac{(i Q)}{(s\plus i0)^2}\Big]\: \Big[
    V_{\ref{scetloops}d}(\Delta^2\to s) \plus V_{\ref{scetloops}e}(\Delta^2\to
    s) +\: \frac{2 m}{Q} \delta m \Big] \nn\\
  &= \frac{-2 m\delta m}{\pi s^2} + \frac{\alpha_s C_F}{4 \pi^2 s} 
  \bigg\{ \frac{1}{\epsilon} + \ln\Big(\frac{\mu^2}{-s}\Big) 
   \minus \frac{m^2 (5m^2 \plus 6s )}{(m^2\plus s)^2} \ln\Big(\frac{m^2}{-s}\Big) 
   + \frac{s}{ m^2 \plus s} \bigg\}
  \,, \nn
\end{align}
where $1/\epsilon$ terms are UV divergences and all $s$ factors are $s+i0$. An
internal $Z_\psi$ counterterm is not needed since these factors cancel between
the propagator and vertices.  The sum of the terms in Eq.~(\ref{Jabc_int}) is
quoted as Eq.~(\ref{Jabcdesum0}) in the text. Expanding Eq.~(\ref{Jabc_int}) to
leading order in $s/m^2\ll 1$ gives
\begin{align} \label{Jabcde}
  J_{\ref{forwardI}a} =  J_{\ref{forwardI}b} &=  \frac{-\alpha_s C_F  }{4\pi^2\,s}
  \bigg\{ \frac{2}{\epsilon^2}\plus \frac{2}{\epsilon} \ln\Big(\frac{\mu^2}{-s}\Big)
   \plus \frac{2}{\epsilon} \plus  \ln^2\Big(\frac{\mu^2}{-s}\Big)
   \plus  \ln^2\Big(\frac{m^2}{-s}\Big) \plus 2 \ln\Big(\frac{\mu^2}{m^2}\Big)
   \plus 4 \plus \frac{\pi^2}{2}  \bigg\}
  , \nn\\
 J_{\ref{forwardI}d} \plus J_{\ref{forwardI}e} \plus J_{\ref{forwardI}f}
 &= \frac{-2 m\delta m}{\pi s^2} +\frac{\alpha_s C_F  }{4\pi^2\,s} \bigg\{
     \frac{1}{\epsilon} + \ln\Big(\frac{\mu^2}{m^2}\Big) 
   + 4 \ln\Big( \frac{-s}{m^2}\Big) 
   \bigg\}\,,
\end{align}
and the sum of these five terms gives Eq.~(\ref{Jabcdesum}).


{\it Soft function graphs.} Next we summarize the computation of the hemisphere
soft function at one-loop given by the graphs in Fig.~\ref{softgraphs}. We use
dimensional regularization for both UV and IR divergences. For
Fig.~\ref{softgraphs}a,b 
we have a loop integral, and for Fig.~\ref{softgraphs}c,d a phase space
integral:
\begin{align}
\label{Sabcdintegrals}
  S_{\ref{softgraphs}a} &= \frac{-2 ig^2 C_F \tilde\mu^{2\epsilon}}{(2\pi)^d} \int\!\!d^d q\:
  \frac{\delta(\ell^+)\delta(\ell^-)}{(q^+ \plus i0)(q^-\minus i0)(q^2\plus i0)}
   \,,\\
  S_{\ref{softgraphs}b} &=\frac{ 2 ig^2 C_F\tilde\mu^{2\epsilon} }{(2\pi)^d} \int\!\!d^d q\:
  \frac{\delta(\ell^+)\delta(\ell^-)}{(q^+ \minus i0)(q^-\plus i0)(q^2\minus i0)}
   \,,\nn\\
  S_{\ref{softgraphs}c} &= \frac{- 2 g^2 C_F \tilde\mu^{2\epsilon}}{(2\pi)^{d-1}}
  \int\!\!\frac{d^{d\minus 1} q}{(q^+ \plus q^-)} \:
  \frac{[\theta(q^-\minus q^+)\delta(\ell^+\minus q^+)\delta(\ell^-)\plus 
  \theta(q^+\minus q^-)\delta(\ell^+)\delta(\ell^-\minus q^-)]}
  {(q^- \plus i0)(-q^+\plus i0)}
   \,,\nn\\
   S_{\ref{softgraphs}d} &=\frac{- 2 g^2 C_F \tilde\mu^{2\epsilon}}{(2\pi)^{d-1}}
  \int\!\! \frac{d^{d\minus 1} q}{(q^+ \plus q^-)} \:
  \frac{[\theta(q^-\minus q^+)\delta(\ell^+\minus q^+)\delta(\ell^-)\plus 
  \theta(q^+\minus q^-)\delta(\ell^+)\delta(\ell^-\minus q^-)] }
  {(q^+ \plus i0)(-q^-\plus i0)} 
  \nn \,,\\
  S_{\ref{softgraphs}e} &= S_{\ref{softgraphs}f} = S_{\ref{softgraphs}g} =0
  \,.\nn
\end{align}
Here $S_{\ref{softgraphs}a}$ and $S_{\ref{softgraphs}b}$ are scaleless and
convert IR divergences in $S_{\ref{softgraphs}c,d}$ into UV
divergences (see for instance Ref.~\cite{Chay:2004zn} where this is worked out explicitly
in several cases). To integrate
$S_{\ref{softgraphs}c,d}$ we convert $d^{d-1}q = \pi^{1-\epsilon}/[2\Gamma(1\minus
\epsilon)]\, dq^+dq^- (q^+ q^-)^{-\epsilon} (q^+\plus
q^-)\theta(q^+)\theta(q^-)$. Evaluating the sum of diagrams we find
\begin{align} \label{softsum1}
  S_{\ref{softgraphs}a} \plus \ldots \plus  S_{\ref{softgraphs}g} 
  &=  \frac{C_F\alpha_s(\mu)}{\pi}\:
  \frac{\mu^{2\epsilon}\, (e^{\gamma_E})^\epsilon}{\epsilon\,\Gamma(1\minus\epsilon)} 
  \bigg[ \frac{\delta(\ell^-)\theta(\ell^+)}{(\ell^+)^{1\plus 2\epsilon}}
    + \frac{\delta(\ell^+)\theta(\ell^-)}{(\ell^-)^{1\plus 2\epsilon}} \bigg] .
\end{align}
This result is used in Eq.~(\ref{softsum}) of the text.

\section{Summary of Feynman diagrams in bHQET} \label{App:bHQETFeyn}

\begin{figure}[!t]
\begin{center}
\includegraphics[width=6.4in]{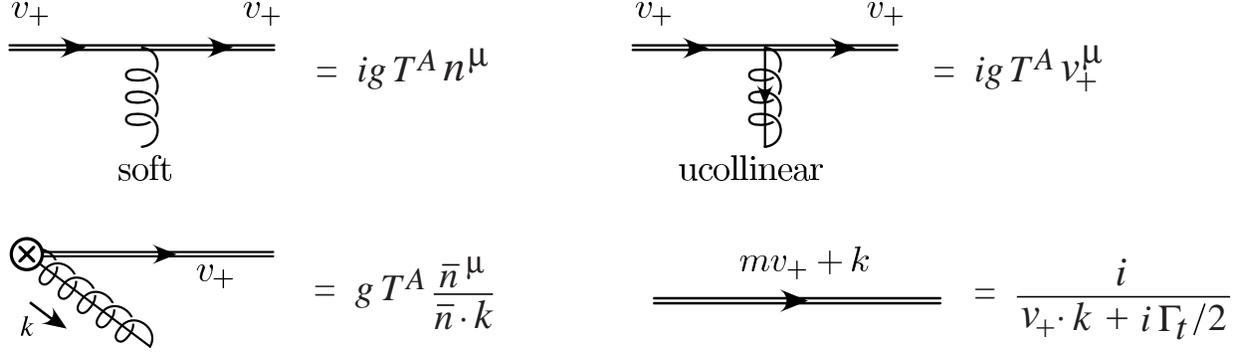}
{\caption[1]{bHQET Feynman rules for the top quarks annihilated by
    $h_{v_+}$: soft gluon coupling to the top quark, ucollinear gluon coupling
    to top quark, Wilson line ucollinear gluon coupling to the top quark, top
    propagator. Results for the antitop quarks annihilated by $\bar h_{v_-}$ are
    obtained by taking $v_+\to v_-$, $T^A\to \bar T^A=- (T^A)^T$, and $n\leftrightarrow \bn$.}
\label{bHQET_FR} }
\end{center}
\end{figure}

In this appendix we list results for the individual Feynman diagrams used in the
body of the paper in section~\ref{sect:bHQET}.  The velocity four vector $v^\mu _\pm$
and the momentum fluctuation four vector $k^\mu _\pm$ for the top and
antitop respectively are given by
\begin{eqnarray}
\label{BHQETres}
v^\mu_+ &=& \bigg( \frac{m}{Q} , \frac{Q}{m}, \mathbf{0}_\perp \bigg),
\qquad\quad
k^\mu_+ \sim  \Gamma\,\bigg(\frac{m}{Q}, \frac{Q}{m}, 1 \bigg),
\\
v^\mu_- &= & \bigg( \frac{Q}{m}, \frac{m}{Q} , \mathbf{0}_\perp \bigg), 
\qquad\quad
k^\mu_- \sim \Gamma\bigg(\frac{Q}{m}, \frac{m}{Q}, 1\bigg) 
\nonumber .
\end{eqnarray}
The Feynman rules for boosted HQET are summarized in Fig.~\ref{bHQET_FR}

{\it bHQET vertex graphs.}
First we give the results for the bHQET vertex graphs of Fig.~\ref{bHQETloops}.
These graphs are defined in terms of the integrals $V_{\ref{bHQETloops}a,b,c,d}$
as $\text{Fig}.\ref{bHQETloops}\kappa = V_{\ref{bHQETloops}\kappa} (
\>\bar{h}_{v_+}\Gamma ^\mu_i h_{v_-})$ for $\kappa=a,b,d$ and
$\text{Fig}.\ref{bHQETloops}\tau = V_{\ref{bHQETloops}\tau}$ for $\tau=c,e,f$.
The one loop integrals are given by
\begin{eqnarray}
\label{scetvertexint}
V_{\ref{bHQETloops}a}&=& -i C_F g^2 \tilde\mu^{2\epsilon}\int \frac{d^dk}{(2\pi)^d}
\Bigg \{ \frac{n\cdot v_- }{[n\cdot k][v_-\cdot (k + r _-)][k^2]} - \frac{n\cdot
  v_- }{[n \cdot k][\frac{1}{2} n\cdot v_- \bn\cdot k + v_{-}\cdot r _-][k^2]} 
  \Bigg \},\nn \\
V_{\ref{bHQETloops}b}&=&V_{\ref{bHQETloops}a} \>\text{with } 
  (\>n\leftrightarrow \bn , v_+ \leftrightarrow v_- , r _- \leftrightarrow r _+)
  \nn \\
V_{\ref{bHQETloops}c}&=& -C_F g^2 \tilde\mu^{2\epsilon}\int \frac{d^dk}{(2\pi)^d}
\frac{v_\pm \cdot v_\pm}{[k^2][v_\pm \cdot (k + l_\pm)]},
   \nn \\
V_{\ref{bHQETloops}d}&=&  i C_Fg^2 \tilde\mu^{2\epsilon}\int \frac{d^dk}{(2\pi)^d} 
   \frac{\bn\cdot n }{[k^2][\frac{1}{2}\bn \cdot v_+n\cdot k + v_+\cdot r _+]
   [\frac{1}{2}n\cdot v_- \bn \cdot k+ v_-\cdot r _-]} ,
 \nn\\
 V_{\ref{bHQETloops}e}&=& V_{\ref{bHQETloops}f} =0 \,,
\label{eq:B2}
\end{eqnarray}
where all the external legs are kept offshell by $(v_\pm \cdot r _\pm) \ne 0$ to
regulate infrared divergences.  As before, throughout this section all factors
in the denominator with square brackets are defined with a $+i0$ prescription.
Just as in SCET, the integrals $V_{\ref{bHQETloops}a,b}$ are defined with a
zero-bin subtraction~\cite{Manohar:2006nz} in order to avoid double counting the
region encoded in the soft loop in Fig.~\ref{bHQETloops}c.  The loop momentum
appearing in $V_{\ref{bHQETloops}a,b,d}$ has ultracollinear scaling as displayd
in Eq.~(\ref{BHQETres}). On the other hand, the loop momentum $k$ in
$V_{\ref{bHQETloops}c}$ has a homogenous soft scaling corresponding to the
exchange of a soft cross-talk gluon between the top and antitop sectors. Given
homogenous scaling of the soft loop momentum and the collinear scaling
of the velocity labels in Eq.~(\ref{BHQETres}), at leading order we have the
relations $v_+\cdot k = \frac{1}{2}(\bn \cdot v_+)(n\cdot k) $ and $v_-\cdot k =
\frac{1}{2} (n \cdot v_-)(\bn\cdot k) $ which have been used in
$V_{\ref{bHQETloops}c}$. Using $ v_\pm \cdot v_\pm =1,\>\bn \cdot n=2,
\>\bn\cdot v_+= n\cdot v_-= Q/m\>\text{and setting} \> (2v_\pm \cdot r
_\pm) =\Omega$ the integrals give:
\begin{eqnarray}\label{collgraphbhqet}
V_{\ref{bHQETloops}a} =  V_{\ref{bHQETloops}b}
  &=&  \frac{\alpha_s C_F}{4 \pi}   \bigg[ \frac{1}{\epsilon^2} +
 \frac{2}{\epsilon}\ln \Big(\frac{\mu}{-\Omega} \Big)
  +2\ln^2 \Big(\frac{\mu}{-\Omega} \Big)
  + \frac{5 \pi^2}{12}\bigg], \nn \\
V_{\ref{bHQETloops}c}&=& -i\, \Omega\, \frac{\alpha_s C_F}{4 \pi} \bigg[
\frac{1}{\epsilon}  + 2 \ln\Big( \frac{\mu}{-\Omega}\Big) + 2\bigg] \,,
  \nn \\ 
 V_{\ref{bHQETloops}d}  &=&  \frac{\alpha_s C_F}{4 \pi} \bigg[ -\frac{2}{\epsilon^2} -
 \frac{2}{\epsilon} \ln \Big( \frac{ \mu^2 Q^2}{-\Omega^2 m^2}  \Big)
 -  \ln^2  \Big( \frac{\mu^2 Q^2}{-\Omega^2 m^2}  \Big) - \frac{\pi^2}{2} \bigg]
.  
\end{eqnarray} 
From $V_{\ref{bHQETloops}c}$ for the wave function graph we get the $\overline{\text{MS}}$ wave function renormalization $Z_h$ and the corresponding residue $R_h$ as
\begin{align}
  Z_h &= 1 + \frac{\alpha_s C_F}{2 \pi \epsilon}, \>\>\>\> &R_h &= 1+
  \frac{\alpha _s C_F}{4\pi} \bigg[4 \ln\Big(\frac{\mu}{-\Omega}\Big) + 4 \bigg].
\end{align}
These results are used to obtain Eq.~(\ref{hqetvertex}) in the text.

{\it bHQET jet function.}
Next consider the graphs for the stable bHQET jet functions in
Fig.~\ref{forwardII}. Prior to taking the imaginary part they are
\begin{align}
\label{bHQETjetapp}
B_{\ref{forwardII}a}^{\Gamma =0}(\shat) =
B_{\ref{forwardII}b}^{\Gamma =0}(\shat) &= 
    -\frac{1}{\pi m\, \shat}\:  V_{\ref{bHQETloops}a}(\Omega \to \shat),
    \nn \\[3pt]
B_{\ref{forwardII}c}^{\Gamma =0}(\shat) &= 0 \,,
   \nn\\[3pt]
B_{\ref{forwardII}d}^{\Gamma =0}(\shat) &=\frac{-2i}{\pi m\, \shat^2}
   \: V_{\ref{bHQETloops}c}(\Omega \to \shat)  \,,
   \nn\\[3pt]
B_{\ref{forwardII}e}^{\Gamma =0}(\shat) &= -\frac{1}{\pi m} \: \frac{2\delta
  m}{\hat s^2} \,.
\end{align}
From Eqs.~(\ref{bHQETjetapp}) and
(\ref{collgraphbhqet}) we get \
\begin{align}
B_{\ref{forwardII}a}^{\Gamma =0}(\shat) =
B_{\ref{forwardII}b}^{\Gamma =0}(\shat)
  &=  - \frac{1}{\pi m\,\hat s} \,\frac{ \alpha_s C_F}{4\pi}
  \bigg[ \frac{1}{\epsilon ^2} 
   +\frac{2}{\epsilon}\> \text{ln}\Big ( \frac{\mu}{-\shat}\Big )
   + 2\> \text{ln}^2\Big(\frac{\mu}{-\shat}\Big ) + \frac{5\pi^2}{12} \bigg]
   \,, \nn \\
B_{\ref{forwardII}d}^{\Gamma =0} (\shat) &= -\frac{1}{\pi m\,\hat s}\,
  \frac{ \alpha_s C_F}{4\pi } \bigg[ \frac{2}{\epsilon } + 4\>
\text{ln}\Big(\frac{\mu}{-\shat}\Big ) + 4\bigg], 
\label{eq:B6}
\end{align}
where 
$\shat = \shat + i 0$. Adding
together $B_{\ref{forwardII}a}^{\Gamma =0}(\shat),\ldots,$
$B_{\ref{forwardII}e}^{\Gamma =0}(\shat)$ we arrive at Eq.~(\ref{bhqetpredisc})
in the main body of the paper.


{\it Zero-bin subtraction for the bHQET jet function.}  The $1/\epsilon$
singularities in Eq.~(\ref{eq:B6}) are UV divergences. This is ensured by the
zero-bin subtraction~\cite{Manohar:2006nz} for the graphs
$B_{\ref{forwardII}a,b}^{\Gamma =0}$ that are needed to avoid double-counting
with infrared regions already accounted for by the soft function (or the
contributions in the soft loop in $V_{\ref{bHQETloops}c}$). Using dimensional
regularization to regularize UV and IR divergences in Eqs.~(\ref{eq:B2}) this
may not be obvious since in this case the zero-bin subtraction is associated to
scaleless integrals. To illustrate the role of the zero-bin subtraction more
explicitly we reconsider the calculation of the anti-top jet function
$B_{\ref{forwardII}a}^{\Gamma =0}$ with a different IR regulator.  From the
definition of $B_{+}^{\Gamma=0}$ in the effective theory this computation
involves two terms, the naive loop integrand $\tilde
B_{\ref{forwardII}a}^{\Gamma=0}$ and a term induced by the zero-bin subtraction
on propagators $B_{\ref{forwardII}a(0)}^{\Gamma=0}$.  We use an explicit
$\rho$-term to regulate the soft and collinear IR divergences for $n\cdot
k=k^+\to 0$.  For the naive part of the result we have
\begin{eqnarray}
\label{regB9a}
\tilde B_{\ref{forwardII}a}^{\Gamma =0}
&=& \frac{i C_F g^2 \tilde\mu^{2\epsilon}}{\pi m\,\hat s} \,
\int\!\! \frac{d^dk}{(2\pi)^d}\,
\frac{n\cdot v_- }{[n\cdot k - \rho][v_-\cdot (k + r _-)][k^2]} \nn \\
&=& - \frac{1}{\pi m\,\hat s} \,
\frac{C_F g^2 \tilde\mu^{2\epsilon} }{4 \pi}\, (4 \pi )^\epsilon\, 
\,\Gamma(\epsilon)\, \Big(\frac{\bar{n}\cdot v_-}{n\cdot v_-}\Big)^{-\epsilon}\!\! \int
_{-\infty}^0\! \frac{d k^+}{2 \pi}\, \frac{ (k^+)^{-\epsilon}}{k^+
  -\rho}  \Big(k^+ + \frac{\hat s}{\bar{n} \cdot v_-} \Big)^{-\epsilon} 
\,.
\end{eqnarray}
For the graph $ B_{\ref{forwardII}a}^{\Gamma =0}$ the zero-bin subtraction is
obtained from the fact that the collinear propagators act as distributions and
induce a subtraction from the limit where $k^\mu\sim \Gamma\lambda^2$. Since
this subtraction is obtained from the Feynman rules of the same diagram as
Eq.~(\ref{regB9a}), it must have the same $\rho$-regulator for the $k^+\to 0$
limit. For this subtraction part we obtain 
\begin{eqnarray}
\label{regB9azerobin}
\tilde B_{\ref{forwardII}a(0)}^{\Gamma =0}
&=& \frac{i C_F g^2 \tilde\mu^{2\epsilon}}{\pi m\,\hat s} \,
\int \frac{d^dk}{(2\pi)^d} \frac{n\cdot
  v_- }{[n \cdot k -\rho]
[\frac{1}{2}\, n\cdot v_- \bn\cdot k + v_{-}\cdot r _-][k^2]} \nn \\
&=&  - \frac{1}{\pi m\,\hat s} \, 
\frac{C_F g^2 \tilde\mu^{2\epsilon}}{4 \pi} (4 \pi)^\epsilon \,\Gamma(\epsilon)
(\frac{\hat s}{n\cdot v_-})^{-\epsilon} \int _{-\infty}^0 \frac{d
  k^+}{2\pi} \frac{(k^+)^{-\epsilon}}{k^+ -\rho}  
\,.
\end{eqnarray}
The $\rho$ term in Eqs.~(\ref{regB9a}) and (\ref{regB9azerobin}) regulates
soft and collinear divergences, and these two results can be computed explicitly
in terms of this regulator. However it drops out in the difference of
Eqs.~(\ref{regB9a}) and (\ref{regB9azerobin}) which is IR finite:
\begin{align}
\label{regB9azerobinsum} 
&B_{\ref{forwardII}a}^{\Gamma =0}  = 
\tilde B_{\ref{forwardII}a}^{\Gamma =0} -
 B_{\ref{forwardII}a(0)}^{\Gamma =0} \\
&=  - \frac{1}{\pi m\,\hat s} \, 
\frac{C_F g^2 \tilde\mu^{2\epsilon}}{4 \pi} (4 \pi)^\epsilon 
\,\Gamma(\epsilon)\, 
\Big(\frac{\bar{n}\cdot v_-}{n\cdot v_-}\Big)^{-\epsilon}\!\!
  \int_{-\infty}^0\!\! \frac{d k^+}{2\pi} \: 
\frac{(k^+)^{-\epsilon}}{k^+ - \rho}\Big [
\Big(k^+ + \frac{\hat s}{\bar{n} \cdot v_-} \Big)^{-\epsilon} -
\Big(\frac{\hat s}{\bar{n} \cdot v_-} \Big)^{-\epsilon} \Big ]
\,. \nn 
\end{align}
For $k^+\to 0$ the term in the brackets is linear in $k^+$ and the
$\rho$-term can be dropped since the expression does not contain any
IR-divergences. Evaluating Eq.~(\ref{regB9azerobinsum}) we recover the result
shown in Eq.~(\ref{eq:B6}). Thus all $1/\epsilon$ singularities are indeed UV
divergences.

\section{Plus function and imaginary part identities}  \label{App:Im}

The plus function with an arbitrary exponent $1+\omega$ with $\omega<1$ is defined by
\begin{align} \label{dim-plus}
  \bigg[ \frac{\theta(x)}{(x)^{1+\omega}}\bigg]_+ &=  \lim_{\beta\to 0} \bigg[
  \frac{\theta(x\minus \beta)}{(x)^{1+\omega}} -
   \delta(x\minus \beta) \: \frac{\beta^{-\omega}}{\omega} \bigg].
\end{align}
A more general definition for the distribution $[\theta(x)/x^{1+\omega}]_+$ can
be defined, by first integrating $\theta(x)/x^{1+\omega}$ with a test function
for values of $\omega$ where the integrals converge, and then analytically
continuing the result to other values of $\omega$.  Expanding
Eq.~(\ref{dim-plus}) for small $\omega$ gives the definition for the
$\log^n(x)/x$ plus-functions for $n\ge 0$:
\begin{align} \label{plus}
  \bigg[ \frac{\theta(x)\ln^n x}{x}\bigg]_+ &\equiv  \lim_{\beta\to 0} \bigg[
  \frac{\theta(x\minus \beta)\ln^n x}{x} + 
   \delta(x\minus \beta) \: \frac{\ln^{n+1}\!\beta}{n+1} \bigg] \,.
\end{align}
The following rescaling identity for a dimensionless constant $\kappa$ is also
quite useful:
\begin{align} \label{rescale}
  \kappa \bigg[ \frac{\theta(x)\ln^n(\kappa x)}{\kappa x} \bigg]_+ &= 
   \frac{\ln^{n+1}(\kappa)}{n+1} \delta(x) + \sum_{k=0}^n \frac{n!}{(n-k)!\, k!}   
   \ln^{n-k}(\kappa) \bigg[ \frac{\theta(x)\ln^k(x)}{x}\bigg]_+ .
\end{align}
For example, for integrations over a finite range Eq.~(\ref{rescale}) allows
us to rescale the $+$-function to act only within the interval $[0,1]$, where
standard identites such as $\int_0^1 dx' g(x') [\theta(x')/x']_+ = \int_0^1 dx'
[g(x')-g(0)]/x'$ for a given test function $g$ can then be applied. This is
somewhat simpler than the corresponding general relation
$\int_0^\beta dx' g(x') [\theta(x')/x']_+ = \int_0^\beta dx' [g(x')-g(0)]/x'+
g(0)\ln(\beta)$.
Relation~(\ref{rescale}) has also been used to verify the multiplicative 
form of the consistency conditions in Eqs.~(\ref{CEver1}) and (\ref{CEver2}).

In computing the SCET and bHQET jet functions one has to take the imaginary
part of the forward scattering graphs. For this the following result for a
dimensionless variable $x$ is quite useful
\begin{align} \label{Imformula}
  {\rm Im}\bigg[ \frac{\ln^n(- x\minus i0)}{\pi(- x\minus i0)}\bigg]
   \! = \cos^2\!\Big(\frac{n\pi}{2}\Big) \frac{(-\pi^2)^{n/2}}{n+1} \delta(x) 
    +
    \sum_{j=0}^{\left[\frac{n-1}{2}\right]}
   \frac{(-1)^j\, n!\, \pi^{2j}}{(2j\plus 1)!
      (n\minus 2j\minus 1)!} \bigg[ \frac{ \theta(x) \ln^{n-2j-1}(x)}{x}\bigg]_+
   \! ,
\end{align}
where $[p]$ on the sum is the greatest integer not exceeding $p$, sometimes also
called the Gauss bracket of $p$. For the first few orders this gives
\begin{align}\label{discontinuities}
 &\frac{1}{\pi} \text{Im}\bigg[\, \frac{1}{x+i0}\,\bigg] = - \delta (x) 
  ,
  &\frac{1}{\pi} \text{Im}\bigg[\,\frac{\text{ln}(-x-i0)}{x+i0}\,\bigg] 
  &= - \Big[\frac{\theta(x)}{x}\Big]_+,
  \\[3pt]
 &\frac{1}{\pi} \text{Im}\bigg[\, \frac{1}{(x+i0)^2}\,\bigg] =  \delta'(x) 
 ,
 & \frac{1}{\pi} \text{Im}\bigg[\,\frac{\text{ln}^2(-x-i0)}{x+i0}\,\bigg] 
 & = \frac{\pi^2}{3}\delta (x) - 2\,\Big [\frac{\theta(x)\text{ln}(x)}{x} \Big ]_+ 
  .\nn
\end{align}
To compute the massive SCET jet function in Eq.~(\ref{Jnbare}) the following
identities were also used [with $s=s+i0 = x \kappa_1^2 +i 0$]:
\begin{align}
\frac{1}{\pi}\, \text{Im} \bigg\{ \frac{1}{s}
\text{Li}_2\Big(\frac{-s}{m^2}\Big) \bigg\}
  & 
   =- \frac{1}{s}\ln\Big(\frac{-s}{m^2}\Big) 
     \: \theta\big(\minus m^2\minus s\big)
   \,, \nn \\
\frac{1}{\pi}\, \text{Im} \bigg\{ \frac{1}{s}
  \ln\Big(\frac{m^2}{-s}\Big)\, \ln\Big(1+\frac{s}{m^2}\Big) \bigg\}
   &= \frac{1}{s} \,\ln\Big(\frac{m^2}{-s}\Big)\, \theta\big(\minus m^2\minus s\big)
  + \frac{1}{\kappa_1^2}  \Big[ \frac{ \theta(x)}{x}\Big ]_+
  \ln\Big(1+\frac{s}{m^2}\Big)
   \,, \nn \\
\frac{1}{\pi}\, \text{Im} \bigg\{
 \frac{m^2(m^2\plus 2s)}{s(s\plus m^2)^2} \, \ln\Big(\frac{m^2}{-s}\Big) 
  \bigg\}
   &=- \delta(s) \ln \Big ( \frac{m^2}{\kappa_1^2}\Big) 
      + \frac{1}{\kappa_1^2} \Big[ \frac{ \theta(x)}{x}\Big ]_+
  - \frac{\theta(s)\, s}{(s\plus m^2)^2}
   \nn \\
 &\ \ + m^2 \ln\Big(\frac{m^2}{-s}\Big) \delta'\big(s\plus m^2\big) \,.
\end{align}

\section{General RGE with plus and delta functions} \label{App:genRGE}

In this appendix we solve the general anomalous dimension equation
\begin{align} \label{Feqtn}
  \mu \frac{d}{d\mu} F(t,\mu) = \int_{-\infty}^{+\infty}\!\!\!\! dt'\: \gamma_F(t\minus t',\mu)\:
  F(t',\mu) \,,
\end{align}
where $t$ and $t'$ are variables of mass-dimension $j$, and $\gamma_F(t-t')$
involves a $+$-function and $\delta$-function. To motivate the general form for
$\gamma$ we consider a generic 1-loop amplitude which has the form
\begin{align}
  A^{\rm bare}(t) 
  & = \frac{1}{t\plus i0}+
  \frac{\alpha_s(\mu)}{4\pi}\frac{1}{t\plus i0} \Big( \frac{\mu^j}{-t\minus i0}\Big)^{2\epsilon/j} 
    \Big( \frac{\Gamma_0}{2\epsilon^2} + \frac{\gamma_0}{2\epsilon} + \ldots \Big)\,.
\end{align}
At one-loop the imaginary part of $A^{\rm bare}$ is renormalized by the
ultraviolet $Z$-factor
\begin{align}
  Z(t-t') = \delta(t\minus t') + \frac{\alpha_s(\mu)}{4\pi} \bigg\{
    \delta(t\minus t') \bigg[\frac{\Gamma_0}{2\epsilon^2}\plus 
   \frac{\Gamma_0}{2\epsilon} \ln\Big(\frac{\mu^2}{\kappa^2}\Big) \plus
   \frac{\gamma_0}{2\epsilon} \bigg] 
   -\frac{\Gamma_0}{j\kappa^j \epsilon} 
  \bigg[ \frac{\kappa^j \theta(t\minus t')}{t\minus t'}\bigg]_+
   \bigg\},  
\end{align}
where the numerical coefficients $\Gamma_0$ and $\gamma_0$ are the first terms
in the perturbative series for the anomalous dimensions
\begin{align} \label{GammaSeries2}
  & \Gamma[\alpha_s] = \frac{\alpha_s(\mu)}{4\pi} \Gamma_0 
     +  \Big[ \frac{\alpha_s(\mu)}{4\pi} \Big]^2\, \Gamma_1 + \ldots \,,
  & \gamma[\alpha_s] &=  \frac{\alpha_s(\mu)}{4\pi} \gamma_0 
     +  \Big[ \frac{\alpha_s(\mu)}{4\pi} \Big]^2\, \gamma_1 + \ldots \,.
\end{align}
At any order in $\alpha_s$ the anomalous dimension to be used in
Eq.~(\ref{Feqtn}) is
\begin{align} \label{gamma}
  \gamma_F(t\minus t',\mu) & = - \frac{2 \Gamma[\alpha_s]}{j\,\mu^j} \bigg[
  \frac{\mu^j \theta(t\minus t')}{t\minus t'}\bigg]_+ \!\! 
  + \gamma[\alpha_s] \: \delta(t-t')\nn\\
  & = -2 \Gamma[\alpha_s]\:
    \bigg\{ \frac{1}{j\,\kappa^j} \bigg[
  \frac{\kappa^j \theta(t\minus t')}{t\minus t'}\bigg]_+ \!\! - \delta(t\minus t')
   \ln\Big(\frac{\mu}{\kappa}\Big)  \bigg\}
  + \gamma[\alpha_s] \: \delta(t-t')   \,,
\end{align}
and depends on the dimension $j$, $t-t'$, and $\mu$.  For convenience we
introduced in Eq.~(\ref{gamma}) the mass scale $\kappa>0$ so that the plus
function has the dimensionless variables $t/\kappa^j$ and $t'/\kappa^j$. But
note that $\gamma_F(t-t',\mu)$ is independent of the choice of $\kappa$.
Here $\Gamma[\alpha_s]$ and $\gamma[\alpha_s]$ are
perturbative series in $\alpha_s(\mu)$ which start with a linear term as shown
in Eq.~(\ref{GammaSeries2}).
%

To solve Eq.~(\ref{gamma}) we use the Fourier transform method of
Ref.~\cite{Korchemsky:1993uz,Balzereit:1998yf}, including the improvements of
Ref.~\cite{Neubert:2004dd} which gives formulas that apply to all orders in
perturbation theory.  Our computation is a simple generalization of these
solutions to mass-dimension $j$ variables, which as we will see is key to
understanding how the renormalization group evolutions of the soft function and
the jet functions can combine to give local running. Taking a Fourier transform,
$\gamma(y) = \int dt\, \exp(-ity) \gamma(t)$ and $F(y)=\int dt\, \exp(-ity)
F(t)$ we have a simple multiplicative RGE
\begin{align} \label{gFy}
  \mu\frac{d}{d\mu} F(y,\mu) = \gamma_F(y,\mu) F(y,\mu) \,,\qquad\quad
   \gamma_F(y,\mu) = 
  \frac{2 \Gamma[\alpha_s]}{j} \ln(iy\,\mu^j\, e^{\gamma_E} ) 
   + \gamma[\alpha_s]
   \,,
\end{align}
where $y=y-i0$. Note that the form of the position space anomalous dimension in
Eq.~(\ref{gFy}) simply follows from locality applied to the bi-local vacuum matrix
element defining the jet function in position space. When translated to momentum
space this directly implies the convolution structure shown in
Eqs.~(\ref{rgeJS},\ref{rgeB}). Integrating Eq.~(\ref{gFy}) from $\mu_0$ to $\mu$
by changing variables to $\alpha_s$ with $d\ln\mu = d\alpha_s/\beta[\alpha_s]$
gives the solution
\begin{align} \label{lnF}
  \ln\Big[ \frac{F(y,\mu)}{F(y,\mu_0)} \Big]
    = \widetilde\omega(\mu,\mu_0) \: \ln\big( iy\, \mu_0^j\, e^{\gamma_E}\big) + {\widetilde
      K}(\mu,\mu_0)\,,
\end{align}
where
\begin{align} \label{wLfull}
  \widetilde\omega(\mu,\mu_0) &= \frac{2}{j}
  \int_{\alpha_s(\mu_0)}^{\alpha_s(\mu)} \frac{d\alpha}{\beta[\alpha]}\,
  \Gamma[\alpha] \,,
  \qquad\qquad
  \widetilde K_\gamma(\mu,\mu_0) =  \int_{\alpha_s(\mu_0)}^{\alpha_s(\mu)}
  \frac{d\alpha}{\beta[\alpha]}\: \gamma[\alpha] \,,
  \nn\\
  \widetilde K(\mu,\mu_0) &= \widetilde K_\gamma(\mu,\mu_0) 
  + 2 \int_{\alpha_s(\mu_0)}^{\alpha_s(\mu)}
  \frac{d\alpha}{\beta[\alpha]}\, \Gamma[\alpha]
  \int_{\alpha_s(\mu_0)}^\alpha \frac{d\alpha'}{\beta[\alpha']}  \,.
\end{align}
(Note that in the main body of this paper where we consider LL and NLL accuracy we 
write $\omega$, $K$, and $K_\gamma$ instead of $\widetilde \omega$, $\widetilde K$, and 
$\widetilde K_\gamma$, respectively.)  Thus the position space solution is $F(y,\mu)=
{\cal U}(y,\mu,\mu_0) F(y,\mu_0)$ with
\begin{align} \label{Uy}
  {\cal U}(y,\mu,\mu_0) &= e^{\widetilde K(\mu,\mu_0)} \:
 \big( iy\, \mu_0^j\, e^{\gamma_E}\big)^{\widetilde \omega(\mu,\mu_0)} \,.
\end{align}
The desired solution is the inverse transform, $F(t,\mu) = 1/(2\pi) \int\!\!
dy\, \exp(ity)\, F(y,\mu)$ so
\begin{align}\,
  F(t,\mu) = \frac{e^{\widetilde K}}{2\pi} \int\!\! dy \: e^{it y}\:
  F(y,\mu_0)\: 
  \big(iy \mu_0^j\, e^{\gamma_E}\big)^{\widetilde \omega} \,.
\end{align}
To simplify this result we use $F(y,\mu_0)= \int\! dt' \exp(-it'y)
F(t',\mu_0)$ and also the inverse transform 
\begin{align}
 (iy \mu^j)^\omega = \frac{1}{\Gamma(-\omega)}
\int\!  dt'\: \exp(-it'y\mu^j)\: \bigg[\frac{\theta(t')}{(t')^{(1+\omega)}}
  \bigg]_+ \,.
\end{align}
Doing the integrals over $y$ and $t'$ we obtain the final result
\begin{align}
  F(t,\mu) &= \int\!\! dt'\: U(t-t',\mu,\mu_0)\: F(t',\mu_0) \,, 
\end{align}
where the evolution kernel is
\begin{align} \label{U}
  U(t-t',\mu,\mu_0) &= \frac{e^{\widetilde K}\: \big(
  e^{\gamma_E}\big)^{\widetilde \omega}}{\mu_0^j\,\Gamma(-\widetilde \omega)} \:
    \bigg[ \frac{(\mu_0^j)^{1+\widetilde\omega}\theta(t\minus t')}{(t\minus
    t')^{1+\widetilde\omega}}\bigg]_+ \,.
\end{align}
This $+$-function is defined by Eq.~(\ref{dim-plus}) and $\widetilde K$ and
$\widetilde \omega$ are determined at what ever order one desires from
Eq.~(\ref{wLfull}). In section~\ref{sect:NLL} we carry out these integrals with
NLL accuracy. The above derivation also suffices to solve Eqs.~(\ref{gammaHQ})
and (\ref{gammaHm}) to obtain $U_{H_Q}$ and $U_{H_m}$ respectively. First we
note that $\gamma_{H_Q}(Q,\mu)$ has the same form as $\gamma_F(y,\mu)$ with
$j=2$ and $iye^{\gamma_E}\to 1/Q^2$, so the general solution is given by
Eq.~(\ref{Uy}) with the same substitutions, yielding the result in
Eq.~(\ref{UUU}). For $\gamma_{H_m}(Q/m,\mu)$ the cusp angle is fixed at
$m^2/Q^2$, so the solution is given by Eq.~(\ref{Uy}) with $\Gamma[\alpha_s]\to
0$ and $\gamma[\alpha_s]\to \Gamma_{H_m}[\alpha_s]
\ln(m^2/Q^2)+\gamma_{H_m}[\alpha_s]$. This yields $U_{H_m}$ in Eq.~(\ref{UUU}).


For the cases with convolutions a few additional identities are useful. The
evolution kernels obey 
\begin{align} \label{Uid1}
  \int\!\! dr' \ U(r-r',\mu,\mu_I)\: U(r'-r'',\mu_I,\mu_0) =  U(r-r'',\mu,\mu_0) \,,
\end{align}
which states that it is equivalent to evolve through an intermediate scale,
$\mu_0\to \mu_I\to \mu$, or directly from $\mu_0\to \mu$. To verify
Eq.~(\ref{Uid1}) one needs
\begin{align} \label{need1}
  \int\!\! dr'' \bigg[ \frac{(\mu_I^j)^{1+ \widetilde\omega_1}\theta(r\minus
    r'')}{(r\minus r'')^{1+ \widetilde\omega_1}} \bigg]_+ 
    \bigg[ \frac{(\mu_0^j)^{1+\widetilde\omega_2}\theta(r''\minus
    r')}{(r''\minus r')^{1+ \widetilde\omega_2}} \bigg]_+ \!\!
   &= \frac{\Gamma(-\widetilde\omega_1)\Gamma(-\widetilde\omega_2)}{(\mu_I^{-j})\Gamma(-\widetilde\omega')}
   \bigg(\frac{\mu_I^j}{\mu_0^j}\bigg)^{\!\widetilde\omega_1} 
  \bigg[ \frac{(\mu_0^j)^{1+ \widetilde\omega'}\theta(r\minus
    r')}{(r\minus r')^{1+ \widetilde\omega'}} \bigg]_+ \! ,\nn\\[5pt]
 \widetilde K(\mu,\mu_I) + \widetilde K(\mu_I,\mu_0)
  & = \widetilde\omega_1 \, \ln\bigg(\frac{\mu_0^j}{\mu_I^j}\bigg)\ + \widetilde K(\mu,\mu_0) \,,
\end{align}
where here $\widetilde\omega_1=\widetilde\omega(\mu,\mu_I)$, $\widetilde\omega_2=\widetilde\omega(\mu_I,\mu_0)$, and
$\widetilde\omega'=\widetilde\omega(\mu,\mu_0)=\widetilde\omega_1 + \widetilde\omega_2$. The first result in
Eq.~(\ref{need1}) is straightforward to derive using the Fourier transform.
Another useful identity simplifies the convolution of two $U$'s that have the
same renormalization scales, but variables with different mass-dimension, and
different anomalous dimension coefficients
\begin{align} \label{Uid2}
 & \int \!\! dr' \ 
   U\big(Q'(r\minus r'),\mu,\mu_0;j',\Gamma',\gamma',\widetilde\omega_1\big)\: 
   U\big(r'\minus r'',\mu,\mu_0;j,\Gamma,\gamma,\widetilde\omega_2\big) \nn\\
  &\qquad = \frac{1}{Q'}\, \bigg( \frac{(\mu_0)^{j'-j}}{Q'}\bigg)^{\widetilde\omega_1}\
    U\big(r\minus r'',\mu,\mu_0; 
          j,\Gamma'\plus \Gamma,\gamma'\plus \gamma,\widetilde\omega'\big)\,.
\end{align}
Here the variables after the semicolon denote parameter dependence, and $Q'$
simply denotes a variable with mass dimension $j'-j$. Also here
$\widetilde\omega_1=\widetilde\omega(\mu,\mu_0;\Gamma'/j')$ and
$\widetilde\omega_2=\widetilde\omega(\mu,\mu_0;\Gamma/j)$ are simply the
$\widetilde\omega$'s obtained from the other parameters, but this is not the
case for $\widetilde\omega'$ in the $U$ on the RHS where
$\widetilde\omega'\equiv \widetilde\omega_1+\widetilde\omega_2$.  The final
useful identity is
\begin{align} \label{Uid3}
  \lim_{\widetilde\omega'\to 0} U(r\minus r',\mu,\mu_0; j,\Gamma,\gamma,\widetilde\omega')
   &= e^{\widetilde K(\mu,\mu_0; \Gamma,\gamma)}\ \delta(r\minus r') \,,
\end{align}
which is easy to derive from Eqs.~(\ref{U}) and (\ref{dim-plus}). Using
Eqs.~(\ref{need1}) and (\ref{Uid2}) it is a straightforward exercise to verify
the consistency equations directly in the integral form given in
Eqs.~(\ref{cons2}) and (\ref{cons4}).

\section{Analytic results for $G_\pm$ in the peak \& tail cross-section} 
\label{App:Gfunc}

In this appendix we show how the functions $G_\pm$ defined in Eq.~(\ref{Gpm})
can be determined analytically.  As a first step we use
Eqs.~(\ref{bhqetjetstable}) and (\ref{Sdefsing}) with $\kappa_3=\mu_\Gamma$ and
$\kappa_2= \mu_\Gamma m_J/Q$ to compute the $\ell'$ integral to ${\cal
  O}(\alpha_s)$ [recall $m_J=m_J(\mu_\Gamma)$]
\begin{align}
\label{Epmdef}
& E_\pm^{\Gamma=0}\Big( \hat
s,\frac{Q}{m_J},\mu_\Gamma,\mu_\Lambda\Big) 
\, \equiv \,
\int_{-\infty}^{+\infty}\!\!\!
  d\ell^\prime\: B_\pm^{\Gamma=0}\Big(\hat s-\frac{Q}{m_J}\ell^\prime,
\mu_\Gamma\Big) 
\,\tilde S^{\rm part}(\ell^\prime,\mu_\Lambda,\delta_1)
 \nn\\[2mm]
 &= \ \
 \frac{1}{\mu_\Gamma\, m_J}\,\Bigg[\,
\delta(z) 
 + \frac{C_F\alpha_s(\mu_\Gamma)}{\pi}\,\bigg\{
  \Big(1-\frac{\pi^2}{8}\Big)\,\delta(z)
  + 2\Big[\frac{\theta(z)\ln z}{z}\Big]_+ 
  - \Big[\frac{\theta(z)}{z}\Big]_+
   \bigg\}
 \nn\\[2mm] 
 &\ \ \quad
+ \frac{C_F\alpha_s(\mu_\Lambda)}{\pi}\,\bigg\{
  \Big[\frac{\pi^2}{24}
   -\ln^2\Big(\frac{\mu_\Lambda\,Q}{\mu_\Gamma\,m_J}\Big)\Big]\,\delta(z)
  -2\Big[\frac{\theta(z)\ln z }{z}\Big]_+ 
  + 2\ln\Big(\frac{\mu_\Lambda\,Q}{\mu_\Gamma\,m_J}\Big)\,
     \Big[\frac{\theta(z)}{z}\Big]_+ \bigg\}
\nn\\[2mm] 
 &\ \ \quad
-\,\Big[\frac{ \delta_1(\mu_\Lambda)\, Q}{\mu_\Gamma\, m_J}
  +\frac{2\delta m_J(\mu_\Gamma)}{\mu_\Gamma} \Big]\,\delta^{\,\prime}(z)
\,\Bigg]
\,,
\end{align}
where $z =\hat s /\mu_\Gamma$ and $\delta_1$ is given in
Eq.~(\ref{delta1}).  The $\delta'(z)$ term in Eq.~(\ref{Epmdef}) contains the
residual mass correction $\delta m_J$ for the jet mass scheme in
Eq.~(\ref{dmjet}) and the subtraction $\delta_1$ for the
soft function from Eq.~(\ref{delta1}).  Using the relations
\begin{align}
\delta(z) \,\, & = -\frac{1}{\pi}\,\mbox{Im}\Big[\,\frac{1}{z\plus i0}\,\Big]\,,
& \Big(\frac{1}{z}\Big)_+ & 
  = -\frac{1}{\pi}\,\mbox{Im}\Big[\,\frac{\ln(-z\minus i 0)}{z\plus i 0}\,\Big]\,,
  \nn\\[2mm]
\Big[\frac{\ln z}{z}\Big]_+ &= -\frac{1}{\pi}\,\mbox{Im}\Big[\,
 \frac{\ln^2(-z\minus i 0)}{2(z\plus i 0)}
  +\frac{\pi^2}{6}\,\frac{1}{z\plus i 0}\,\Big]\,,
&\delta^\prime(z) \,\,& 
   = \frac{1}{\pi}\,\mbox{Im}\Big[\,\frac{1}{(z\plus i 0)^2}\,\Big]\,,
\end{align}
the result for $E_\pm$ can be rewritten as
\begin{align}
E_\pm^{\Gamma=0}\Big(\hat s,\frac{Q}{m_J(\mu_\Gamma)},\mu_\Gamma,\mu_\Lambda\Big) &=
\mbox{Im}\Big[{\cal E}_\pm^{\Gamma=0}\Big(\hat
s,\frac{Q}{m_J(\mu_\Gamma)},\mu_\Gamma,\mu_\Lambda\Big)
\,\Big]
\,,
\end{align}
with
\begin{align}
\label{Epmex}
 & {\cal E}_\pm^{\Gamma=0}\Big(\hat
s, \frac{Q}{m_J},\mu_\Gamma,\mu_\Lambda\Big)  
 = \frac{-1}{\pi m_J}\,\frac{1}{\hat s \plus i 0}\,
\Bigg\{\,
  1 + \frac{1}{\hat s\plus i 0}\,
\Big[2\,\delta m_J(\mu_\Gamma)+\frac{Q}{m_J}\delta_1(\mu_\Lambda)\Big]
 \\[2mm]
 &\quad 
 + \frac{C_F\alpha_s(\mu_\Gamma)}{\pi}\,\bigg[\,
 1 + \frac{5\pi^2}{24} 
 + \ln^2\!\Big(\frac{-\hat s\minus i 0}{\mu_\Gamma}\Big)
 - \ln\!\Big(\frac{-\hat s\minus i 0}{\mu_\Gamma}\Big)
\,\bigg]
\nn\\[2mm] 
 &\quad 
  + \frac{C_F\alpha_s(\mu_\Lambda)}{\pi}\,\bigg[
 -\!\frac{7\pi^2}{24} -\ln^2\Big(\frac{\mu_\Lambda\, Q}{\mu_\Gamma\, m_J}\Big)
 -\ln^2\!\Big(\frac{-\hat s\minus i 0}{\mu_\Gamma}\Big)
 +2 \ln\!\Big(\frac{\mu_\Lambda\, Q}{\mu_\Gamma\, m_J}\Big)
   \ln\!\Big(\frac{-\hat s\minus i 0}{\mu_\Gamma}\Big)
 \bigg] \Bigg\} .
\nn
\end{align}
Note that for $\mu_\Lambda\gtrsim\Lambda$ and $\mu_\Gamma \sim \hat s \sim
\Gamma \sim Q\Lambda/m$ there are no large logs in this expression, and
that the terms with $\delta m_J$ and $\delta_1$ are the same order in the power
counting.  Given Eq.~(\ref{Epmex}), doing the second integral in the variable
$\hat s^{\prime\prime}$ involving the Breit-Wigner function is simple since it
just results in a shift of the invariant mass variable into the positive complex
plane, as in Eq.~(\ref{factorizationGamma}):
\begin{align}
\int_{-\infty}^{+\infty}\!\! d\hat s^{\prime\prime}\
 & E_\pm^{\Gamma=0}\Big( \hat s-\hat s^{\prime\prime}, 
\frac{Q}{m_J},\mu_\Gamma,\mu_\Lambda\Big) 
\frac{\Gamma_t}{\pi(\hat s^{\prime\prime\, 2}+\Gamma_t^2)} 
 \nn\\[2mm]
 &\ \ 
  = \mbox{Im}\Big[ {\cal E}_\pm^{\Gamma=0}\Big(\hat s+i\Gamma_t, 
   \frac{Q}{m_J},\mu_\Gamma,\mu_\Lambda\Big)\Big]
 =  E_\pm\Big(\hat s, 
   \frac{Q}{m_J},\Gamma_t,\mu_\Gamma,\mu_\Lambda\Big)
\,.
\end{align}
For the final integration in the variable $\hat s^\prime$ in Eq.~(\ref{Gpm}) 
we have to convolute
the finite width version of the terms in Eq.~(\ref{Epmex}) with the evolution
kernel $U_{B}$. The relevant computations read
\begin{align}
\label{Gn}
\int_{-\infty}^{+\infty}
  d\hat s^\prime\: U_{B}(\hat s-\hat s^\prime,\mu_\Lambda,\mu_\Gamma)\,
\frac{\ln^n\Big(\frac{-\hat s^\prime-i\Gamma_t}{\mu_\Gamma}\Big)}
  {\hat s^\prime+i\Gamma_t}
& \equiv - G_n(\hat s,\Gamma_t,\mu_\Lambda,\mu_\Gamma) \,,
\end{align}
where 
\begin{align}
G_n(\hat s,\Gamma_t,\mu,\mu_0) & =
\frac{e^{K_3}\,(\mu_0 e^{\gamma_E})^{\omega_1}\: \Gamma(1\plus \omega_1) }
{(-\hat s-i\Gamma_t)^{1+\omega_1}}\ 
 I_n\Big(\frac{\hat s+i\Gamma}{\mu_0},\omega_1\Big) \,,
\end{align}
with $K_3=K_3(\mu,\mu_0)$ and $\omega_1=\omega_1(\mu,\mu_0)$ given in
Eqs.~(\ref{wL12}), and
\begin{align}
\label{Indef}
I_n(x,\omega)  
 & =
\left. \frac{d^n}{d\epsilon^n} \:
\frac{\Gamma(1-\epsilon+\omega)}{\Gamma(1-\epsilon)\: \Gamma(1+\omega)}
 \,(-x-i0)^{\epsilon}\   
\right|_{\epsilon=0}
\,.
\end{align}
For the terms we need
\begin{align}
I_0(x,\omega) & = 1\,, \nn\\[2mm]
I_1(x,\omega)\, & =  \ln(-x-i0)-H(\omega) \,,
\nn\\[2mm]
I_2(x,\omega) & =
\Big[H(\omega)-\ln(-x-i0)\Big]^2-\zeta_2+\Psi^\prime(1+\omega)
\,.
\end{align}
Here $H(\omega)$ is the harmonic number function and $\Psi^\prime(z)=d/dz
[\Gamma^\prime(z)/\Gamma(z)]$ is the derivative of the polygamma function.  Thus
the final result for the function $G_\pm$ is [$m_J=m_J(\mu_\Gamma)$]
\begin{align}
G_\pm\Big(\hat s,\frac{Q}{m_J},\Gamma_t\OMIT{,\mu_\Gamma},\mu_\Lambda\Big) & =
\mbox{Im}\bigg[\,
{\cal G}_\pm\Big(\hat s,\frac{Q}{m_J},
   \Gamma_t,\mu_\Gamma,\mu_\Lambda\Big) 
+\,
\delta {\cal G}_\pm\Big(\hat s,\frac{Q}{m_J},
  \Gamma_t,\mu_\Gamma,\mu_\Lambda\Big)
\,\bigg],
\end{align}
where taking $G_i=G_i(\hat s,\Gamma_t,\mu_\Lambda,\mu_\Gamma)$ we have
\begin{align}  \label{Fpmfin}
 {\cal G}_\pm\Big(\hat s,&
 \frac{Q}{m_J},\Gamma_t,\mu_\Gamma,\mu_\Lambda\Big) 
\, = \,
\frac{1}{\pi m_J}\,
\bigg\{\,
G_0 + \frac{C_F\alpha_s(\mu_\Gamma)}{\pi}\,\bigg[ G_2 -G_1 +
\Big(1+\frac{5\pi^2}{24}\Big)\,G_0 \bigg]
\nn\\[2mm] &
+ \frac{C_F\alpha_s(\mu_\Lambda)}{\pi}\,\bigg[
-G_2
+2\ln\Big(\frac{\mu_\Lambda Q}{\mu_\Gamma m_J}\Big)\,
G_1 -\Big(\frac{7\pi^2}{24}+
  \ln^2\Big(\frac{\mu_\Lambda Q}{\mu_\Gamma m_J}\Big)\Big)
  \,G_0
\,\bigg]
\,\bigg\}
\,,
\nn\\[3mm]
\delta {\cal G}_\pm\Big(\hat s,&
 \frac{Q}{m_J},\Gamma_t,\mu_\Gamma,\mu_\Lambda\Big) 
 = 
 -\frac{1}{\pi m_J}
 \Big[\frac{Q}{m_J} \delta_1(\mu_\Lambda) +2\delta m_J(\mu_\Gamma)\Big]\,
\frac{d}{d\hat s}\,
 G_0  \,.
\end{align}

\vskip 1cm

\section{Cross-section in the ultra-tail region} 
\label{App:SCETnum}

In the text we presented results for the cross section in the peak and tail
regions, where we assume $Q\gg m \gg \hat s$ and $m\gg \Gamma$. In this section
we analyze the cross section for $Q\gg m\sim \hat s$, so that $\hat s$ is far
above the peak region. This corresponds to $|M_{t,\bar t}-m_J|\sim m$, where the
top-antitop jet invariant mass double differential distribution can be described
by the SCET factorization formula of Eq.~(\ref{SFactThm}).  Writing this cross-section in
an analogous form to Eq.~(\ref{sigmaMM}) [$m_J=m_J(\mu_m)$] we have
\begin{align} \label{sigmaMMscet2}
  \frac{ d^2\sigma }{dM_t\, dM_{\bar t}} 
    & =\frac{4 M_t M_{\bar t}\,\sigma_0}{ (m_J\Gamma_t)^2}\
    F^{\rm SCET}\Big(M_t,M_{\bar t},Q,m_J\OMIT{,\mu_Q,\mu_m,\mu_\Lambda}\Big) \,.
\end{align}
where
\begin{align} \label{Gscetdef2}
  F&^{\rm SCET}\Big(M_t,M_{\bar t},Q, m_J\OMIT{,\mu_Q,\mu_m,\mu_\Lambda}\Big)  
  \,= \,(m_J\Gamma_t)^2\,
  H_Q(Q\OMIT{,\mu_Q},\mu_m) {\cal M}(m_J,\mu_m) U_{H_Q}^{(5)}(Q,\mu_m,\mu_\Lambda)
\nn\\[3mm] & \times
   \int_{-\infty}^\infty\!\!\!\! d\ell^+\, d\ell^- 
  G_n(s_t - Q\ell^+,Q,m_J,\Gamma\OMIT{,\mu_m},\mu_\Lambda)
  G_{\bar n}(s_{\bar t} -Q\ell^-,Q,m_J,\Gamma\OMIT{,\mu_m},\mu_\Lambda)
 \,S^{\rm mod}(\ell^+,\ell^-)\,.
\end{align}
Note that Eqs.~(\ref{sigmaMMscet2}) and (\ref{Gscetdef2}) are also appropriate
for describing the massless limit $m_J\to 0$ and the stable limit $\Gamma_t\to
0$.  To
obtain this result we manipulated the first form given in
Eq.~(\ref{sigmaMMscet}). Here, $S^{\rm mod}$ is the hadronic model function
given in Eq.~(\ref{SM1}), where we have suppressed its arguments $a$, $b$,
$\Lambda$.  The functions $G_{n,{\bar n}}$ can be written as
\begin{align} \label{Fnnb}
G_{n,{\bar n}}(s,& Q,m_J,\Gamma\OMIT{,\mu_m},\mu_\Lambda) \, \equiv\,
\int d s^\prime d s^{\prime\prime} d\ell^\prime\ 
U_{J}(s-s^\prime,\mu_\Lambda,\mu_m) 
\nn\\[3mm] &\, \times
J_{n,\bar n}(s^\prime -s^{\prime\prime} -Q\ell^\prime,m_J, \mu_m)
\, \tilde S^{{\rm part}}(\ell^\prime,\mu_\Lambda,\delta_1)\,
\frac{m_J \Gamma}{\pi(s^{\prime\prime 2}+m_J^2 \Gamma^2)}\,.
\end{align}
Here, $\tilde S^{\rm part}$ is the modified partonic soft function of
Eqs.~(\ref{Sdefsing}) and (\ref{Stilde}).  For consistency with our peak
cross-section results we continue to use the jet-mass scheme, by taking $\delta
m_J$ for the $\delta m$ term in Eq.~(\ref{Jrenm}).  Since it does not require
any technical effort, we include in $G_{n,{\bar n}}$ a constant width term for
the top quark through the convolution involving the variable $s^{\prime\prime}$.
We note, however, that away from the resonance region this width term leads to
power-suppressed effects, and, moreover, does not provide a consistent
description of the top quark decay. It is nevertheless convenient to introduce
the width term for practical purposes because it allows for an easy numerical
evaluation of the SCET factorization theorem for all values of $M_{t,\bar t}$
without running in to singularities for $M_{t,\bar t}$ close to the top quark mass.



We can carry out an analytic calculation of the functions $G_{n,\bn}$ defined in
Eq.~(\ref{Fnnb}). The calculation divides itself into two parts, the terms
singular for $s_{t,\bar t}\to 0$ which include the $\delta$-function and
$+$-functions in Eq.~(\ref{Jren}), and the non-singular $\theta$-function term
on the last line of Eq.~(\ref{Jren}).  The final result in the jet mass scheme
reads [$m_J=m_J(\mu_m)$]
\begin{align}
G_{n,{\bar n}}\Big(s,Q,m_J,\Gamma_t,\mu_\Lambda\Big) & =
\mbox{Im}\bigg[
{\cal G}_{n,{\bar n}}\Big(s,Q,m_J,\Gamma_t,\mu_m,\mu_\Lambda\Big) 
+
\delta {\cal G}_{n,{\bar n}}\Big(s,Q,m_J,\Gamma_t,\mu_m,\mu_\Lambda\Big)
\,\bigg] 
  \nn\\
 & \ \ + G^{\rm nonsing}_{n,{\bar n}}(s,Q,m_J,\mu_m) \,.
\end{align}
For the singular terms, ${\cal G}_{n,\bn}$ and $\delta {\cal G}_{n,\bn}$, a
computation can be carried out in close analogy to the computation of $G_\pm$
described in the App.~\ref{App:Gfunc}. Taking the singular terms in
Eq.~(\ref{Jren}) with $\kappa_1^2=\mu_m m_J$, and the results from
Eqs.~(\ref{Sdefsing}) and (\ref{Stilde}) with $\kappa_2=\mu_m\,m_J/Q$ we find
\begin{align}
 & {\cal G}_{n,{\bar n}} \Big(s,Q,m_J,\Gamma_t,\mu_m,\mu_\Lambda\Big)\nn\\
\, &\qquad = \,
\frac{1}{\pi}\,
\bigg\{\,
\tilde G_0
 + \frac{C_F\alpha_s(\mu_m)}{\pi}\,\bigg[\,
\tilde G_2 - \tilde G_1
+\Big( \ln^2\frac{\mu_m}{m_J}+\frac{1}{2}\ln\frac{\mu_m}{m_J}+
    \frac{\pi^2}{4}+2\Big) \tilde G_0
\,\bigg]
\nn\\[2mm] 
&\qquad\ 
 + \frac{C_F\alpha_s(\mu_\Lambda)}{\pi}\,\bigg[
  -\tilde G_2
  +2\ln\Big(\frac{\mu_\Lambda\, Q}{\mu_m\, m_J}\Big)\, \tilde G_1
  -\bigg(\frac{7\pi^2}{24} + \ln^2\Big(\frac{\mu_\Lambda\, Q}{\mu_m\, m_J}\Big) \bigg)
  \tilde G_0
  \,\bigg] \,\bigg\} 
\,, \nn
\\[3mm]
& \delta {\cal G}_{n,{\bar n}} \Big(s,Q,m_J,\Gamma_t,\mu_m,\mu_\Lambda\Big)
\nn\\ 
&\qquad\ 
= -\frac{1}{\pi} \Big[ Q \,\delta_1(\mu_\Lambda) +2m_J \delta m_J(\mu_m)\Big]\,
\frac{d}{d s}\, 
 \tilde G_0
\,,
\label{deltaFnn}
\end{align}
with
\begin{align}
\label{Stildedef}
\tilde G_n(s,Q,m_J,\Gamma_t,\mu_\Lambda,\mu_m) & \equiv \,
-\,\int_{-\infty}^{+\infty}\!\! d s^\prime\ U_{J}(s-s^\prime,\mu_\Lambda,\mu_m)\,
\frac{\ln^n(\frac{-s^\prime-i m_J \Gamma_t}{\mu_m m_J})}{s^\prime+i m_J\Gamma_t}
\nn\\[2mm]  & =
\frac{e^{K_1}\,(\mu_m^2 e^{\gamma_E})^{\omega_1}\,\Gamma(1\plus\omega_1)}
{(-s-i \,m_J \Gamma_t)^{1+\omega_1}}\,
  I_n\Big(\frac{s+i \,m_J\Gamma_t}{\mu_m m_J},\omega_1\Big) 
  \nn\\[3mm]
 &= \frac{1}{m_J}\, e^{K_1-K_3}\, \Big(\frac{\mu_m}{m_J}\Big)^{\omega_1}\,
   G_n\Big(\frac{s}{m},\Gamma_t,\mu_\Lambda,\mu_m\Big)
   \,.
\end{align}
Here, $\omega_1=\omega_1(\mu_\Lambda,\mu_m)$ and $K_1=K_1(\mu_\Lambda,\mu_m)$ are
given by Eq.~(\ref{wL1}) and the functions $I_n$ were defined in
Eq.~(\ref{Indef}). 
The term $\delta {\cal G}_{n,{\bar n}}$ arises from the jet mass definition (see
Eq.~(\ref{dmjet})) and from the subtraction that we carry out for the soft
function model (see Eq.~(\ref{delta1})). 

For the non-singular terms we note that the top-width effects represent ${\cal
  O}(\Gamma_t/m)$ power suppressed terms for any $s\ll Q^2$. For instance, when
$s\sim m^2$ the top width appears in the combination $s+im_J\Gamma_t$, and
when $s\sim m\Gamma$ the entire non-singular term is ${\cal O}(\Gamma_t/m)$.
Thus it is consistent to neglect the top quark width for the non-singular terms.
Setting $\Gamma_t=0$ in Eq.~(\ref{Fnnb}) and carrying out the $s'$ and $s''$
integrals gives 
\begin{align}
\label{deltaFns}
 {G}_{n,{\bar n}}^{\rm nonsing}& \Big(s,Q,m_J,\mu_m,\mu_\Lambda\Big)
\, \nn\\
&= \frac{\alpha_s(\mu_m)C_F}{\pi\, m_J^2}\:
  \frac{\theta(s)\, e^{K_1} }{\Gamma(-\omega_1)}\,
  \Big( \frac{\mu_m^2 e^{\gamma_E}}{s} \Big)^{\omega_1}\,
  \bigg[\, \frac{1}{\omega_1} \,
      {}_3F_2\Big(\{1,1,1\},\{2,1-\omega_1\}, \frac{-s}{m_J^2}\Big)
   \nn\\
 &\qquad   + \,\frac{s\,m_J^2}{4(1\minus \omega_1)\,(m_J^2\plus s)^2 }\,\bigg\{\,
  \frac{\omega_1^2\minus 1}{\omega_1}
  +\Big(\frac{s}{m_J^2}\minus \omega_1\Big)\,
   {}_2F_1\Big(1,1,2-\omega_1,\frac{-s}{m_J^2}\Big) \bigg\}\, \bigg] \,.
\end{align}

\section{Analytic results for the $P_{\rm T}$ function for thrust}
\label{app:PT}

In this appendix we derive an analytic result for ${\rm P_T}$, the
perturbative
corrections appearing in the thrust cross-section in
Eq.~(\ref{Tfactorization}).
Starting from Eq.~(\ref{PT}) for ${\rm P_T}$ with Eq.~(\ref{Pdef})
for ${\rm P}$
the key is to simplify the integral over the product of $G_+G_-$,
\begin{align}
G_{\rm T}(\hat s) &\equiv   \int_{-\infty}^{+\infty}\!\!\! d\hat s_d
\
     G_+\Big(\frac{\hat s+\hat s_d}{2},
\frac{Q}{m_J},\Gamma_t,\mu_\Lambda\Big)\:
     G_-\Big(\frac{\hat s+\hat s_d}{2},
\frac{Q}{m_J},\Gamma_t,\mu_\Lambda\Big)
   \\[4pt]
&\hspace{-1cm}
= \int\!\! d\hat s_s' \bigg[ \int\!\! \frac{d\hat s_d}{2} \:
     U_B\Big(\frac{\hat s\minus \hat s_s'-\hat
s_d'}{2},\mu_\Lambda,\mu_\Gamma\Big)
     U_B\Big(\frac{\hat s\minus \hat s_s'+\hat
        s_d'}{2},\mu_\Lambda,\mu_\Gamma\Big) \bigg]
   \int\!\! d\hat s_s'' \bigg[ \int\!\! d\hat s_d'' \:
\nn\\
&\hspace{-1cm}
\times
E_+^{\Gamma=0}\Big(\frac{\hat s_s'\minus \hat s_s''\minus \hat
s_d''}{2}\Big)
E_-^{\Gamma=0}\Big(\frac{\hat s_s'\minus \hat s_s''\plus \hat
s_d''}{2}\Big)
\bigg] \bigg[\! \int \!\! \frac{d\hat s_d}{2} \:
   m_J^2   B_+^{\rm tree}\Big(\frac{\hat s_s''\plus \hat
s_d}{2},\Gamma_t\Big)
   B_-^{\rm tree}\Big(\frac{\hat s_s''\minus \hat
s_d}{2},\Gamma_t\Big)\bigg].
\nn
\end{align}
where $B_\pm^{\rm tree}(\hat s,\Gamma_t)$ is simply a Breit-Wigner as shown in
Eq.~(\ref{eq:Btree}). This Breit-Wigner appears due to the factorization of
lifetime effects in section~\ref{sec:FactGamma}, and the perturbative
corrections to the jet-functions are part of $E_\pm^{\Gamma=0}$.

Here we rewrote the original integrations over symmetric variables ($\hat
s_s$'s) and antisymmetric variables ($\hat s_d$'s). Each of the integrations in
square brackets can be performed, with the help of Eq.~(\ref{Uid2}) and
(\ref{Epmdef}), to give the terms in the following result
\begin{align} \label{GT2}
G_{\rm T}(\hat s) &
=   \int\!\! d\hat s_s'\, d \hat s_s'' \Big[
   \tilde U_B(\hat s\minus \hat s_s',\mu_\Lambda,\mu_\Gamma)\:
   \Big]\Big[ \frac{2}{m_J}
   E_{\rm T}^{\Gamma=0}(\hat s_s'\minus \hat s_s'')\: \Big]
\Big[ m_J B_{+}^{\rm tree }(\hat s_s'',2\Gamma_t) \Big]
\end{align}
Here the Breit-Wigner has a width of $2\Gamma_t$, and the function $E_{\rm
  T}^{\Gamma=0}(\hat s)$ is identical to $E_+^{\Gamma=0}(\hat s)$ in
Eq.~(\ref{Epmdef}) but with the replacements $\{\alpha_s\to
2\alpha_s,\delta_1\to 2\delta_1,\delta m_J\to 2\delta m_J\}$. Finally the
evolution kernel is
\begin{align} \label{UBt}
\tilde U_B(\hat s,\mu , \mu_0) &=\frac{e^{2K_3}(
   e^{\gamma_E})^{2\omega_1}}{\mu_0\, \Gamma(- 2\omega_1)}
\bigg [ \frac{\mu_0^{1+2\omega_1} \theta (\hat s - \hat s')}{(\hat
s-\hat s')^{1+2\omega_1}} \bigg ]_+ \,,
\end{align}
which is equivalent to the kernel $U_B$ for the bHQET jet function but with the
anomalous dimensions doubled.  We can write the function $E_{\rm
  T}^{\Gamma=0}(\hat s)={\rm Im}\big[ {\cal   E}_{\rm T}^{\Gamma=0}(\hat
s)\big]$, where ${\cal E}_{\rm   T}^{\Gamma=0}(\hat s)$ has the same form as
${\cal E}_+^{\Gamma=0}(\hat s)$ in Eq.~(\ref{Epmex}) but with the replacements
$\{\alpha_s\to 2\alpha_s,\delta_1\to 2\delta_1,\delta m_J\to 2\delta m_J\}$.  The result in Eq.~(\ref{GT2}) has a structure such
that we can perform the last two integrations with the same techniques as in
Appendix~\ref{App:Gfunc}. We obtain $ G_{\rm T}(\hat s) = {\rm Im}\big[ {\cal
  G}_{\rm T}(\hat s) + \delta {\cal G}_{\rm   T}(\hat s)\big]$ where
\begin{align}
{\cal G}_{\rm T}(\hat s) &
\, = \,
\frac{2}{\pi m_J^2}\,
\bigg\{\,
G_0^T(\hat s)   + \frac{2C_F\alpha_s(\mu_\Gamma)}{\pi}\,\bigg[
G_2^T(\hat s)
-G_1^T(\hat s)   +
\Big(1+\frac{5\pi^2}{24}\Big)\,G_0^T(\hat s)   \bigg]
\nn\\[2mm] &
+ \frac{2C_F\alpha_s(\mu_\Lambda)}{\pi}\,\bigg[
\!-\! G_2^T(\hat s)
+2\ln\Big(\frac{\mu_\Lambda Q}{\mu_\Gamma m_J}\Big)\,
G_1^T(\hat s)   -\! \Big(\frac{7\pi^2}{24}+
   \ln^2\Big(\frac{\mu_\Lambda Q}{\mu_\Gamma m_J}\Big)\Big)
   \,G_0^T(\hat s)
\,\bigg]
\,\bigg\}
\,,
\nn\\[3mm]
\delta {\cal G}_{\rm T}(\hat s)
& =
-\frac{4}{\pi m_J^2}
\Big[\frac{Q}{m_J} \delta_1(\mu_\Lambda) +2\delta
m_J(\mu_\Gamma)\Big]\,
\frac{d}{d\hat s}\,
G_0^T  \,,
\end{align}
and $G_n^T(\hat s) = G_n^T(\hat s,\Gamma_t,\mu_\Lambda,\mu_\Gamma)$
is simply
$G_n(\hat s,\Gamma_t,\mu_\Lambda,\mu_\Gamma)$ from Eq.~(\ref{Gn}) but
with
$\omega_1\to 2\omega_1$, $K_3\to 2K_3$, and $\Gamma_t\to 2\Gamma_t$.

Including the prefactor from Eq.~(\ref{PT}) we have the final result
\begin{align}
  {\rm P_T}(\hat s,\mu_\Lambda) &= \frac{m_J Q^2}{2} \,
  H_{Q}(Q,\mu_Q)\, U_{H_Q}(Q,\mu_Q,\mu_m) \,
  H_m(m,\mu_m)\, U_{H_m}\Big(\frac{Q}{m_J},\mu_m,\mu_\Lambda\Big)
\nn\\
  &\quad \times
  G_{\rm T}\Big(\hat s, \frac{Q}{m_J}, \Gamma_t,\mu_\Lambda \Big)
  \,.
\end{align}

\vskip 0.5cm

\bibliography{topjet}

\end{document}